%% file: main.tex
\documentclass[11pt, a4paper, titlepage]{book} 
\usepackage[a-2b]{pdfx} % loads hyperref and xcolor
\usepackage[utf8]{inputenc} % UTF-8 encoding for input
\usepackage[T1]{fontenc} % Font encoding for international characters
\usepackage[paper=a4paper, inner=2cm, outer=3cm, top=3cm, bottom=3cm]{geometry} % Page geometry
\usepackage[nouppercase,swapnames]{frontespizio} % Frontespizio for the title page
\usepackage{graphicx} % For including logo
\usepackage{setspace} % Line spacing for the title page
\usepackage{amsmath, amssymb} % Advanced math symbols and environments
\usepackage{physics} % For differential operators, brackets, and Dirac notations
\usepackage{cancel} % For striking through math expressions
\usepackage{bbold} % For blackboard bold fonts (e.g., identity matrix)
\usepackage{graphicx} % For including figures
\usepackage{multirow} % For multi-row cells in tables
\usepackage{tabularx} % Flexible table environment
\usepackage{booktabs} % Enhanced table formatting (nicer horizontal lines)
\usepackage{float} % For the [H] specifier to force float placement
\usepackage{caption} % Custom captions for figures and tables
\usepackage{subcaption} % Subfigure and subtable environments
\usepackage{color} % Color support
\definecolor{darkgreen}{rgb}{0,0.5,0} % Custom color definitions
\definecolor{darkred}{rgb}{0.5,0,0} % Custom color definitions
\usepackage{parskip} % Controls paragraph spacing
\usepackage{comment} % For multi-line comments
\usepackage{verbatim} % For verbatim text environments
\usepackage{stackengine, scalerel} % Stacking and scaling text elements
\usepackage{array} % For advanced table features
\usepackage{float} % For managing float positioning
\numberwithin{equation}{section} % Equation numbering includes section numbers
\usepackage{cmds}
\usepackage[backend=bibtex, sorting=none, url=false, maxnames=100, minnames=100, style=numeric-comp, defernumbers=true]{biblatex} % Bibliography management
\usepackage{ragged2e} % For ragged right text in bibliography
\usepackage{breakurl} % To handle long URLs and DOIs
\AtEveryBibitem{
  \clearfield{archivePrefix}
  \clearfield{eprint}
  \clearfield{primaryClass}
  \clearfield{arxivId}
} % Custom bibliography settings to remove unnecessary fields

\DeclareBibliographyDriver{article}{%
  \printnames{author}
  \newunit\newblock
  \printfield{title}
  \newunit\newblock
  \printfield{journaltitle}
  \setunit{\addcomma\space}
  \printdate
  \newunit\newblock
  \printfield{doi}
  \newunit\newblock
  \usebibmacro{finentry}
}

\hypersetup{
    colorlinks=true,        % Enable colored links
    citecolor=blue,         % Citation link color
    linkcolor=darkred,          % Internal link color (e.g., for sections, tables, etc.)
    urlcolor=blue,          % External URL color
    linktoc=page,           % Link only page numbers in ToC
}
\usepackage[capitalise]{cleveref} % For automatic referencing with custom names
\Crefname{figure}{Fig.}{Figs.} % Custom reference names
\Crefname{section}{Sec.}{Secs.}
\Crefname{chapter}{Chap.}{Chaps.}
\Crefname{table}{Tab.}{Tabs.}

\usepackage{pdfpages} % For including entire PDF pages
\graphicspath{{figures/}} % Directory for figures

\title{Hamiltonian Lattice Gauge Theories:\\ emergent properties from tensor network methods}
\author{Giovanni Cataldi}

\begin{document}

\includepdf{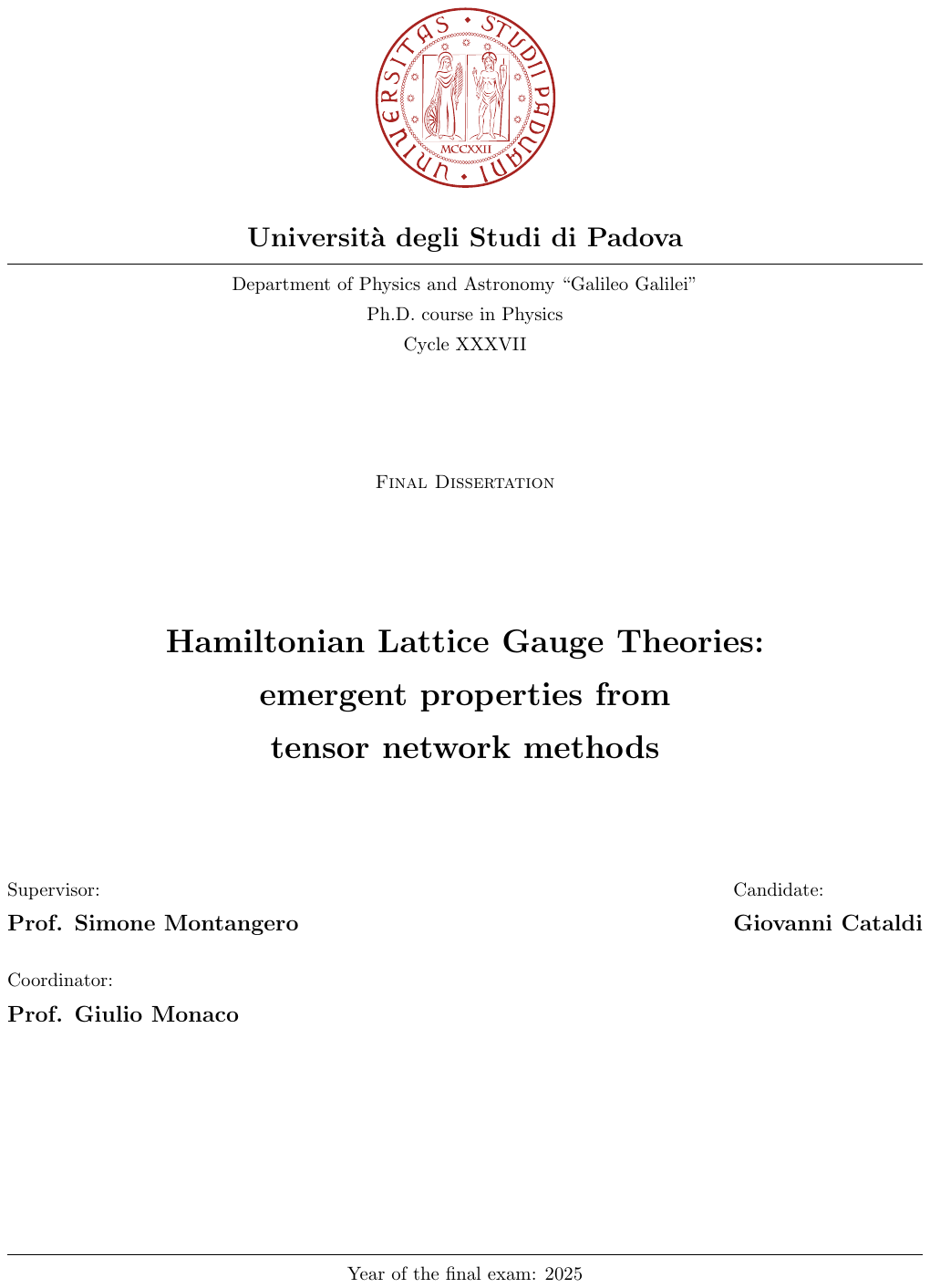}

\frontmatter
\begin{center}
    \section*{\huge Abstract}
\end{center}
This thesis develops advanced Tensor Network (TN) methods to address Hamiltonian Lattice Gauge Theories (LGTs), overcoming limitations in real-time dynamics and finite-density regimes. 
A novel dressed-site formalism is introduced, enabling efficient truncation of gauge fields while preserving gauge invariance for both Abelian and non-Abelian theories. 
This formalism is successfully applied to SU(2) Yang-Mills LGTs in two dimensions, providing the first TN simulations of this system and revealing critical aspects of its phase diagram and non-equilibrium behavior, such as a Quantum Many-Body (QMB) scarring dynamics.
A generalization of the dressed-site formalism is proposed through a new fermion-to-qubit mapping for general lattice fermion theories, revealing powerful for classical and quantum simulations. 
Numerical innovations, including the use of optimal space-filling curves such as the Hilbert curve to preserve locality in high-dimensional simulations, further enhance the efficiency of these methods. 
Together with high-performance computing techniques, these advances open current and future development pathways toward optimized, efficient, and faster simulations on scales comparable to Monte Carlo state-of-the-art.

This thesis\footnote{
    \textbf{Aknowledgements} Written with the financial support of MIUR through the PRIN2017 and PRIN2022 projects TANQU;
of the European Union via the QuantERA projects QuantHEP and T-NISQ, the Quantum Flagship project PASQuanS2, and the NextGenerationEU (PNRR) project CN00000013 - Italian Research Center on HPC, Big Data and Quantum Computing;
of the WCRI-Quantum Computing and Simulation Center of the University of Padova, Fondazione CARIPARO, and Progetti Dipartimenti di Eccellenza through the project Frontiere Quantistiche (FQ), and the INFN project QUANTUM.
We acknowledge the computational resources provided by Cloud Veneto, CINECA, the BwUniCluster, and the University of Padova Strategic Research Infrastructure Grant 2017: CAPRI (Calcolo ad Alte Prestazioni per la Ricerca e l’Innovazione).
} is based on the following preprints, publications, and codes to which the author significantly contributed\footnote{The authors with * equally contributed to the work.}:
\begin{refsegment}
    % Directly cite the references you want to include
    \nocite{Cataldi2021HilbertCurveVs,Cataldi2024Simulating2+1DSU2,Cataldi*2025QuantumManybodyScarring,Cataldi*2024DigitalQuantumSimulation, Magnifico2024TensorNetworksLattice,Cataldi2024Edlgt}
    % Print the bibliography for this segment only
    \printbibliography[segment=1,heading=none]
\end{refsegment}
\afterpage{\null\clearpage}

% Table of Contents
\tableofcontents
%\listoffigures % Prints the list of figures
%\listoftables % Prints the list of tables
\include{chapters/introduction}
\mainmatter
% Include chapters
\include{chapters/lgt}
\include{chapters/numerics}
\include{chapters/su2}

\include{chapters/scars}
\include{chapters/hubbard}
\include{chapters/conclusions}
% Bibliography
\printbibliography[heading=bibintoc]
\end{document}

%% file: chapters/introduction.tex
\chapter*{Introduction}
\phantomsection % Fix hyperlink location
\addcontentsline{toc}{chapter}{Introduction} % Add manually to ToC
Quantum field theory (QFT) is a foundational framework in particle physics, offering a comprehensive description of how particles interact. 
It integrates the principles of quantum mechanics with special relativity to provide insights into the behavior of particles under different forces. 
Gauge invariance is a fundamental principle in QFT, representing a special symmetry where the equations describing a physical system remain unchanged under local transformations of the fields. 
Then, a gauge theory is a framework that imposes these local symmetries as constraints, ensuring that certain transformations can occur independently at each point in space and time without altering the observable outcomes. 
The Standard Model (SM) of particle physics \cite{Gaillard1999StandardModelParticle} represents the current state-of-the-art in our understanding of gauge theories, as it successfully unifies the electromagnetic, weak, and strong interactions under a common theoretical framework, incorporating both Abelian and non-Abelian gauge symmetries. 
It has proven remarkably successful in explaining a wide range of experimental results and predicting new phenomena. 

However, non-Abelian gauge theories, particularly QCD, pose significant challenges, especially in the low-energy, non-perturbative regime. 
At high energies, where the gauge coupling constant is small, perturbation theory is an effective tool for making predictions, such as those concerning high-energy collisions in particle accelerators. 
However, at lower energies, typical of the hadronic scale (below 1 GeV), the gauge coupling constant becomes large, making perturbative methods inadequate.
To overcome these limitations, lattice gauge theories (LGTs) provide a non-perturbative framework for studying QCD and other gauge theories \cite{Kogut1979IntroductionLatticeGauge, Rothe2012LatticeGaugeTheories,Kogut1983LatticeGaugeTheory}, allowing numerical simulations to be performed. 
LGTs discretize spacetime into a lattice grid where matter and gauge fields occupy lattice sites and links respectively. 
One of the most successful applications of LGTs is lattice QCD \cite{Gupta1998IntroductionLatticeQCD,HernAndez2011LatticeFieldTheory, Nagata2022FinitedensityLatticeQCD}, which has been instrumental in computing the hadronic spectrum, such as the masses of protons and neutrons, and understanding phenomena like confinement \cite{Wilson1974ConfinementQuarks,Biswal2017ConfinementDeconfinementTransition,Bornyakov2018ConfinementdeconfinementTransitionDense,Engelhardt2000DeconfinementSUYangMills,Ambjorn1984StochasticConfinementDimensional,Ambjorn1984StochasticConfinementDimensional-1}, chiral symmetry breaking \cite{Mitter2015ChiralSymmetryBreaking,Gottlieb1987ChiralsymmetryBreakingLattice,Alford1999ColorflavorLockingChiral,Kogut1982ScalesChiralSymmetry}, and the role of topology in QCD at finite temperatures \cite{Creutz1980MonteCarloStudy,Creutz1982NumericalStudiesWilson,Kogut1985FurtherEvidenceFirstorder,Biswal2017ConfinementDeconfinementTransition}. 
In these terms, Monte Carlo (MC) simulations of LGTs play a critical role by providing statistically meaningful results for observables that are otherwise inaccessible through analytical methods \cite{Creutz1979MonteCarloStudy, Creutz1980MonteCarloStudy, Berg1981SULatticeGauge, Creutz1983MonteCarloComputations, Creutz1988LatticeGaugeTheory, Creutz1989LatticeGaugeTheories, Kieu1994MonteCarloSimulations, Ghobadpour2021MonteCarloSimulation, vanBemmel1994FixedNodeQuantumMonte, Xu2019MonteCarloStudy,Loan2003PathIntegralMonte, Lynn2019QuantumMonteCarlo}.
Despite these successes, MC simulations face challenges, particularly the sign problem, which hampers the simulation of a wide class of physical settings described by complex or negative actions (finite charge-density phases, fermions, real-time dynamics), whose numerical investigations remain -- to date -- an open problem \cite{Troyer2005ComputationalComplexityFundamental,Li2015SolvingFermionSign}. 

According to Feynman, all these challenges could have been solved by attacking quantum systems with a quantum computer, where traditional classical numerical methods could be circumvented, particularly in handling quantum correlations (\idest{} entanglement).
After his seminal proposal and the recent fast development of quantum hardware, there is growing interest in experimental platforms for simulating LGTs, such as ultra-cold atoms, optical lattices, and superconducting circuits. 
Recently, quantum-inspired strategies have been applied to LGTs, attempting to reproduce their quantum dynamics \cite{Martinez2016RealtimeDynamicsLattice,Schweizer2019FloquetApproachZ2,Yang2020ObservationGaugeInvariance,Zhou2022ThermalizationDynamicsGauge,Nguyen2022DigitalQuantumSimulation,Mildenberger2022ProbingConfinementMathbb}.

While awaiting noiseless and scalable quantum hardware, in the current \emph{Noisy Intermediate-Scale Quantum} (NISQ) era, MC problems have inspired the development of novel classical numerical algorithms based on Tensor Networks (TN) \cite{Verstraete2008MatrixProductStates,Orus2019TensorNetworksComplex,Montangero2018IntroductionTensorNetwork,Silvi2019TensorNetworksAnthology}. 
These techniques exploit the area-law entanglement bounds satisfied by a large class of quantum many-body (QMB) systems \cite{Eisert2010ColloquiumAreaLaws} and allow for an efficient representation of the low-energy sector by retaining in memory only the relevant states that contribute to equilibrium and non-equilibrium properties. 
This is achieved by compressing the exponentially large wavefunctions into a network of tensors interconnected through auxiliary indices with bond dimension $\chi$. 
The main TN anzätze for representing QMB states are Matrix Product States (MPS) for 1D systems \cite{Fannes1992FinitelyCorrelatedStates, Klumper1993MatrixProductGround,Chan2016MatrixProductOperators,Paeckel2019TimeevolutionMethodsMatrixproduct,Schollwock2011DensitymatrixRenormalizationGroup}, Projected Entangled Pair States (PEPS) \cite{Verstraete2006CriticalityAreaLaw, Verstraete2004RenormalizationAlgorithmsQuantumMany,Verstraete2008MatrixProductStates, Schuch2007ComputationalComplexityProjected,Cirac2019MathematicalOpenProblems,Cirac2021MatrixProductStates,Corboz2016ImprovedEnergyExtrapolation,Corboz2010SimulationStronglyCorrelated,Kraus2010FermionicProjectedEntangled,Lubasch2014AlgorithmsFiniteProjected,Lubasch2014UnifyingProjectedEntangled,Orus2014PracticalIntroductionTensor,Vanderstraeten2022VariationalMethodsContracting,Zohar2016ProjectedEntangledPair}, Tree Tensor Networks (TTN) \cite{Shi2006ClassicalSimulationQuantum,Gerster2014UnconstrainedTreeTensor,Silvi2010HomogeneousBinaryTrees,Silvi2019TensorNetworksAnthology,Qian2022TreeTensorNetwork} and Multiscale Entanglement Renormalization Ansatz (MERA) \cite{Qian2022TreeTensorNetwork,Vidal2007EntanglementRenormalization,Evenbly2009EntanglementRenormalizationTwo}, which can be defined in any dimension.

Recent advances in TN methods offered promising alternatives for studying LGTs without MC limitations like the sign problem. 
Indeed, TNs are particularly valuable for attacking non-perturbative LGT phenomena in the Hamiltonian formulation, accessing real-time dynamics and finite-density phases.
Exploiting TN algorithms, noteworthy results have been produced for Abelian LGTs in one \cite{Banuls2013MassSpectrumSchwinger,Rico2014TensorNetworksLattice,Kuhn2014QuantumSimulationSchwinger,Banuls2015ThermalEvolutionSchwinger,Buyens2016HamiltonianSimulationSchwinger,Buyens2017RealtimeSimulationSchwinger,Buyens2017FiniterepresentationApproximationLattice,Ercolessi2018PhaseTransitionsGauge,Magnifico2019MathbbGaugeTheories,Magnifico2019SymmetryprotectedTopologicalPhases,Funcke2020TopologicalVacuumStructure,Magnifico2020RealTimeDynamics,Rigobello2021EntanglementGenerationMathrm,Banuls2022QuantumInformationPerspective} and higher spatial dimensions \cite{Felser2020TwoDimensionalQuantumLinkLattice,Magnifico2021LatticeQuantumElectrodynamics,Emonts2023FindingGroundState}. 
As for non-Abelian LGTs, TN-based simulations were so far limited to one spatial dimension \cite{Silvi2017FinitedensityPhaseDiagram,Silvi2019TensorNetworkSimulation,Kadam2023LoopstringhadronFormulationSU}, but recently, they have been successfully applied to two dimensional systems \cite{Cataldi2024Simulating2+1DSU2}.

Despite all these achievements, one main challenge for numerical and quantum simulations of LGTs remains the finite-dimensional encoding of the continuous gauge fields, especially beyond one spatial dimension, where decoupling the gauge field's longitudinal component is required \cite{Bender2020GaugeRedundancyfreeFormulation}. 
Among known truncation recipes are Quantum Link Models (QLM) \cite{Horn1981FiniteMatrixModels,Orland1990LatticeGaugeMagnets,Chandrasekharan1997QuantumLinkModels,Brower1999QCDQuantumLink,Tagliacozzo2014TensorNetworksLattice}, an approach already considered for practical quantum simulation of LGTs \cite{Byrnes2006SimulatingLatticeGauge,Mathis2020ScalableSimulationsLattice,Davoudi2020AnalogQuantumSimulations,Mazzola2021GaugeinvariantQuantumCircuits,Kan2021InvestigatingMathrmTopological,Zohar2021QuantumSimulationLattice,Mariani2023HamiltoniansGaugeinvariantHilbert,Pomarico2023DynamicalQuantumPhase,Bauer2023QuantumSimulationFundamental,Bauer2023QuantumSimulationHighEnergy,Fontana2023QuantumSimulatorLink}, finite subgroups \cite{Ercolessi2018PhaseTransitionsGauge,Magnifico2020RealTimeDynamics,Haase2021ResourceEfficientApproach},
digitization of gauge fields \cite{Hackett2019DigitizingGaugeFields}, and fusion-algebra deformation \cite{Zache2023QuantumClassicalSpin}. 
Whatever the adopted solution, further effort is required to satisfy gauge symmetry at each local site, which involves the concurrent evaluation of all the gauge links and the attached matter site \cite{Zohar2018EliminatingFermionicMatter,Zohar2019RemovingStaggeredFermionic,Silvi2019TensorNetworkSimulation}. 
As it represents the highest energy-scale of the model, any violation of this constraint would prevent the correctness of simulations.

Another important issue to be faced in simulating LGTs is to account for the Fermi statistics of matter fields. 
In several TN methods as well on well-established conventional digital quantum simulation platforms (e.g. superconducting qubits, trapped-ions, Rydberg arrays, quantum dots) \cite{Arute2020ObservationSeparatedDynamics,Barends2015DigitalQuantumSimulation,Salathe2015DigitalQuantumSimulation,OMalley2016ScalableQuantumSimulation,Stanisic2022ObservingGroundstateProperties}, which are built on distinguishable, spatially localized qubits (or qudits), fermionic algebra (mutually-anticommuting operations) must be econded into a genuinely local algebra (mutually-commuting operations) of qudits. 
Standard fermion-to-qubit encodings, such as the Jordan-Wigner (JW) transformation \cite{Jordan1928UeberPaulischeAequivalenzverbot}, and other modern approaches \cite{Nielsen2006QuantumComputationGeometry,Bravyi2002FermionicQuantumComputation,Jiang2019MajoranaLoopStabilizer} make any fermionic Hamiltonian interaction inherently long-range and expensive from a computational perspective.
Together with the truncation of the gauge fields and the Gauss's law constraint, these challenges necessitate further advancements in numerical and quantum simulation strategies to accurately capture the ground-state and dynamical properties of LGTs.

In this thesis, we address these challenges: we discuss a theoretical and numerical formalism that provides a controllable gauge field truncation (in Casimir spectrum) and directly accesses the Hilbert subspace of gauge invariant states where gauge and matter fields are combined in a \emph{dressed-site} encoded with a bosonic statistics. 
Such a formalism is detailed and quantitatively discussed in \cite{Cataldi2024Simulating2+1DSU2, Magnifico2024TensorNetworksLattice} for both Abelian and non-Abelian LGT, yet successfully generalizes to lattice fermion theories without gauge invariance occurring in condensed matter \cite{Cataldi*2024DigitalQuantumSimulation}. 
We exploit this approach to perform numerical simulations of the SU(2) Yang-Mills LGT in and out of equilibrium, exploring compelling regimes in the phase diagram \cite{Cataldi2024Simulating2+1DSU2} and exotic behaviors like QMB scarring dynamics \cite{Cataldi*2025QuantumManybodyScarring}.
In parallel, we review the state-of-the-art TN methods for Hamiltonian LGTs, providing and testing optimal encodings for high-dimensional Hamiltonians \cite{Cataldi2021HilbertCurveVs} and figuring strategies for parallelized high-performance large-scale simulations of LGTs \cite{Magnifico2024TensorNetworksLattice}.

All the achievements conveyed in this work \cite{Cataldi2024Simulating2+1DSU2,Magnifico2024TensorNetworksLattice,Cataldi*2024DigitalQuantumSimulation,Cataldi*2025QuantumManybodyScarring,Cataldi2021HilbertCurveVs} not only advance the theoretical and numerical understanding of LGT by means of TN methods, but also provide valuable benchmarks for experimental quantum simulations \cite{Zohar2013ColdAtomQuantumSimulator,Banerjee2013AtomicQuantumSimulation,Wiese2013UltracoldQuantumGases,Tagliacozzo2013SimulationsNonAbelianGauge, Zohar2015QuantumSimulationsLattice, Mezzacapo2015NonAbelianSULattice,Banuls2020ReviewNovelMethods,Banuls2020SimulatingLatticeGauge,Atas2021SUHadronsQuantum,Klco2022StandardModelPhysics,Meurice2022TensorNetworksHigh,Atas2023SimulatingOnedimensionalQuantum,Davoudi2023GeneralQuantumAlgorithms,Zache2023FermionquditQuantumProcessors}, contributing to systematically identify the quantum advantage threshold~\cite{Zhou2020WhatLimitsSimulation,Ayral2023DensityMatrixRenormalizationGroup} towards a deeper understanding of the fundamental interactions governing particle and condensed matter physics. 

The thesis is structured as follows: in \cref{chap_LGT}, we provide a detailed overview of all the fundamental ingredients of an LGT. 
Starting from the basics of QFT, we discuss the main effects arising from the lattice discretization of the theory, with a special focus on fermion fields and energy cutoffs. 
The whole discussion is based on the Hamiltonian formalism, as it will be lately exploited for all the presented numerical results. 
The innovative part of this chapter starts from the dressed-site formalism discussed in \cref{sec_dressed_site_formalism}, a theoretical scheme that reveals extremely useful for simulating LGTs via TN and quantum hardware. 
Such an approach is detailed and quantitatively applied first to the challenging non-Abelian SU(2) Yang-Mills theory (\cref{sec_SU2_model}) and secondly to the Abelian U(1) case (\cref{sec_U1_model}). 
In both scenarios, we derive a pedagogical strategy to simulate the LGT in any spatial dimensions, with a controllable and scalable truncation of the gauge fields, where Gauss law is automatically satisfied. 
The resulting theories are made out of bosonic degrees of freedom, preventing simulations from the challenging sign problem that characterizes MC methods.

In \cref{chap_numerics}, we discuss the state-of-the-art numerical methods for attacking QMB Hamiltonians and, more in detail, the previously discussed LGTs.
We start reviewing Exact Diagonalization (ED) methods and the numerical strategies to exploit the global symmetries such as particle-number or momentum conservation via block diagonalization.
These features are currently implemented in the ED-LGT \textrm{Python} library \cite{Cataldi2024Edlgt}, developed throughout these three years for simulating and benchmarking QMB systems and dressed-site formulated LGTs such as the ones presented in \cref{chap_LGT}.
We then switch to reviewing TN methods, from the fundamental tensor operations to advanced algorithms for ground-state search and time evolution of QMB systems tailored for MPS and TTN ansätze.

A noteworthy section compares different ways of mapping high-dimensional Hamiltonians onto TN structures, demonstrating the Hilbert curve \cite{Cataldi2021HilbertCurveVs} as the optimal ordering for preserving distances and quantum correlations when simulating large scale high-dimensional QMB systems.
We conclude the chapter by discussing the status of TN methods dedicated to LGTs, providing current and future lines of development towards optimized, efficient, and faster simulations on scales comparable to MC state-of-the-art.
All these strategies are conveyed in a recent roadmap \cite{Magnifico2024TensorNetworksLattice} and stimulate benchmarking tests for quantum simulations of LGTs.

After discussing all theoretical and numerical tools that are necessary for simulating LGTs, we devote the second part of the thesis to the presentation of the most compelling results achieved and conveyed in the following recent scientific works: \cite{Cataldi2024Simulating2+1DSU2,Cataldi*2024DigitalQuantumSimulation,Cataldi*2025QuantumManybodyScarring}.
In \cref{chap_SU2_groundstate}, we present the first TN simulations of a SU(2) Yang-Mills LGT with flavorless fermionic dynamical matter in two spatial dimensions \cite{Cataldi2024Simulating2+1DSU2}.
The simulations rely on the \emph{hardcore-gluon} approximation, the smallest nontrivial electric-truncation of the SU(2) gauge fields discussed in \cref{sec_SU2_hardcoregluon}.
However, they provide a clear phase diagram of the model both at zero and finite baryon density, exploring challenging regimes of the couplings and addressing the continuum limit location.

In \cref{chap_scars}, we discuss the non-equilibrium properties of LGTs, with compelling regard to the non-ergodic behaviors observed in non-Abelian LGTs. 
After a pedagogical introduction to classical and quantum ergodicity, we expose the Eigenstate Thermalization Hypothesis (ETH) for isolated QMB systems concerning their spectral and dynamical properties. 
We review the first experimental and numerical observations of QMB systems with \emph{scarring} dynamics, an exotic behavior that weakly violates ETH. 
We highlight the strong connection between the main features of QMB scars and the fundamental mechanism appearing in Abelian LGTs.
Remarkably, we extend such a relation to non-Abelian LGTs by simulating a one-dimensional truncated SU(2) Yang-Mills LGT and observing all the spectral and dynamical properties of a scarring dynamics \cite{Cataldi*2025QuantumManybodyScarring}.
These phenomena are further explained through meson-baryon creation processes that are intrinsically due to the non-Abelian nature of the model.

Finally, \cref{chap_defermionization_lattice_fermion} focuses on the recent work \cite{Cataldi*2025QuantumManybodyScarring}, where we discuss an extension of the dressed-site formalism for generic lattice fermion theories without gauge invariance. 
In detail, we propose an efficient fermion-to-qubit mapping that is suitable for classical and (digital) quantum simulations.
We test its applicability on the spin-($1/2$) Fermi-Hubbard model in two spatial dimensions investigating both equilibrium and non-equilibrium properties via TN methods. 
We demonstrate the perfect equivalence with the original model and recover interesting features of the expected physics. 
The presented encoding is then compared in detail with other theoretical strategies to remove Fermi statistics, revealing efficient, scalable, and competitive.

%% file: chapters/lgt.tex
\chapter{Hamiltonian Lattice Gauge Theories}
\label{chap_LGT} 
As anticipated, gauge theories form the backbone of our understanding of fundamental interactions in particle physics. 
Central to gauge theories is the principle of gauge invariance, where physical laws remain unchanged under local transformations of certain fields. 
Different realizations of gauge theories correspond to different symmetry groups and interaction structures. 
In Quantum Electrodynamics (QED), the U(1) symmetry group governing electromagnetic interactions, with the photon as the associated gauge boson. 
Non-Abelian gauge theories, such as those based on SU(2) and SU(3) symmetry groups, describe the weak and strong interactions, respectively. 
These groups allow for more complex transformations, where the gauge bosons themselves interact with one another. 
For instance, in Quantum Chromodynamics (QCD), the SU(3) gauge theory of the strong force, gluons are the gauge bosons that mediate interactions between quarks and also interact among themselves due to the non-Abelian nature of the theory.

Lattice Gauge Theories (LGTs) \cite{Kogut1979IntroductionLatticeGauge, Rothe2012LatticeGaugeTheories,Kogut1983LatticeGaugeTheory} offer a non-perturbative framework for studying these models, particularly in regimes where perturbative methods fail, such as in low-energy QCD \cite{Gupta1998IntroductionLatticeQCD,HernAndez2011LatticeFieldTheory, Nagata2022FinitedensityLatticeQCD}. 
By discretizing spacetime into a lattice, LGTs enable numerical simulations of gauge theories, allowing us to explore compelling phenomena such as confinement \cite{Wilson1974ConfinementQuarks,Biswal2017ConfinementDeconfinementTransition,Bornyakov2018ConfinementdeconfinementTransitionDense,Engelhardt2000DeconfinementSUYangMills,Ambjorn1984StochasticConfinementDimensional,Ambjorn1984StochasticConfinementDimensional-1}, chiral symmetry breaking \cite{Mitter2015ChiralSymmetryBreaking,Gottlieb1987ChiralsymmetryBreakingLattice,Alford1999ColorflavorLockingChiral,Kogut1982ScalesChiralSymmetry}, the hadronic spectrum, and the dynamics of quarks and gluons \cite{Davoudi2022ReportSnowmass2021}. 

In traditional Monte Carlo (MC) approach to LGTs \cite{Creutz1979MonteCarloStudy, Creutz1980MonteCarloStudy, Berg1981SULatticeGauge, Creutz1983MonteCarloComputations, Creutz1988LatticeGaugeTheory, Creutz1989LatticeGaugeTheories, Kieu1994MonteCarloSimulations, Ghobadpour2021MonteCarloSimulation, vanBemmel1994FixedNodeQuantumMonte, Xu2019MonteCarloStudy,Loan2003PathIntegralMonte, Lynn2019QuantumMonteCarlo}, the action of a continuum gauge theory is regularized by working on a finite and discrete Euclidean spacetime, where both space and (imaginary) time are discretized \cite{Creutz1983MonteCarloComputations}.
Instead, Tensor Networks (TN) and quantum simulations typically rely on the Hamiltonian formalism, where time remains a real continuous variable, while space is discretized on a $D$-dimensional lattice. 
The Hamiltonian discretization of gauge theories explicitly breaks Lorentz invariance, which is however restored in the continuum limit. 
Simultaneously, Hamiltonian LGTs present features that distinguish them from other QMB systems commonly simulated via TN or quantum hardware \cite{Kogut1979IntroductionLatticeGauge}.
Next, we discuss these properties in more detail and outline the main steps that have to be taken to exploit TN algorithms.
Without any loss of generality, we focus on matter-coupled LGTs of the Yang-Mills type, routinely employed in high-energy physics to describe nature's fundamental interactions.

We start from a pedagogical review of lattice discretization in \cref{sec_lattice_discretization}, with special regard for fermion fields within the staggered fermion solution in \cref{sec_lattice_fermions} and gauge fields in \cref{sec_lattice_gaugefields}.
The original contribution of this chapter starts from \cref{sec_dressed_site_formalism}, where we present the \emph{dressed-site} formalism, a theoretical scheme that is suitable for attacking LGTs via TNs and quantum hardware directly within the gauge-invariant sector, where all the logical degrees of freedom undergo a bosonic statistics.
In \cref{sec_SU2_model,sec_U1_model}, we respectively detail two specific realizations of this formalism, SU(2) for the non-Abelian and U(1) for the Abelian scenario. 
Such an approach has been successfully applied to obtain all the numerical results detailed in \cref{chap_SU2_groundstate,chap_scars} (see also \cite{Cataldi2024Simulating2+1DSU2,Cataldi*2025QuantumManybodyScarring,Magnifico2024TensorNetworksLattice, Calajo2024DigitalQuantumSimulation}) and is generalized  to pure fermion theories (without gauge fields, \idest{} no LGTs) in \cite{Cataldi*2024DigitalQuantumSimulation} and \cref{chap_defermionization_lattice_fermion}.
% ============================================================================
\section{Lattice discretization}
\label{sec_lattice_discretization}
Let us consider continuous (D+1) spacetime domain ($\mathbb{R}^{D}\times \mathbb{R}$) identified by the (D+1)-dimensional point $r=(t,\vecpos)$, where $\vecpos=(\pos[1], \dots \pos[D])$. 
When performing spatial discretization, the time variable $t$ remains continuous, while the spatial domain $\mathbb{R}^{D}$ is replaced by a hyper-cubic D-dimensional lattice $\Lambda$, whose sites are equally spaced by lattice constant $\lspace$ (see \cref{fig_LGT_sketch}). 
Lattice sites are then labelled by coordinates $\vecsite=(\site[1],\dots,\site[D])$, where $\pos[k]=\lspace \site[k]$ and $\site[k]\in \qty{1\dots N_{k}}$, with $N_{k}$ denoting the number of lattice sites along the $k^{\text{th}}$ direction $\forall k=1,\dots,D$. 
Then, if $N=\prod_{k=1}^{D}N_{k}$ is the total number of lattice sites, with the corresponding lattice volume given by $V=\lspace^{D}N$.
Additionally, we define the unit lattice vectors $\pm\latvec[k]$ pointing along the $\pm k^{\text{th}}$ direction, assuming $\pm\latvec[k]=\latvec[\pm k]$.
We use the convention where summations over lattice vectors include only positive directions.
Then, the lattice link connecting two neighboring sites $\vecsite$ and $\siteplus_{k}$ along the direction $+\latvec[k]$, is denoted by the couple $(\genlink_{k})$ or equivalently with $(\vecsite+\latvec[k],-\latvec[k])$.
For future purposes, we refer to an \emph{even} (\emph{odd}) site, if $(-1)^{\vecsite}=(-1)^{\sum_{k}\site[k]}$ is even (odd).
% ============================================================================
\subsection{Preliminaries}
Lattice discretization of spatial dimensions affects all the mathematical objects: for a general scalar field $f(\vecpos)$ whose absolute value is \emph{square-integrable}
\begin{equation}
	\int_{\mathbb{R}^{D}} d\pos^{D} \abs{f(\vecpos)}^2 =L<\infty\,,
\end{equation} 
we have to consider the following replacements:
\begin{align}
    \int_{\mathbb{R}^{D}} d\pos^{D} &\longrightarrow \lspace^{D}\sum_{\vecsite}\,,&
    f(\vecpos)  &\longrightarrow \hat{f}_{\vecsite}\,,&
    \partial_{k}f(\vecpos) &\longrightarrow \frac{\hat{f}_{\vecsite+\latvec[k]}-\hat{f}_{\vecsite-\latvec[k]}}{2\lspace}\,.
    \label{eq_lattice_discretization}
\end{align}
Such a discretization prescription introduces an \emph{ultraviolet cut-off}, in energy and momentum, which depends on the inverse of the lattice spacing $\lspace$. 
Namely, for each spatial direction $k$, momentum is anisotropic and assumes $N_{k}$ discrete values $\mom[k](i)\in \qty[-\frac{\pi}{\lspace},\frac{\pi}{\lspace}]$, where $i \in \qty{1\dots N_{k}}$. 
The maximal momentum along each direction is then $\mom{\max}=\pi/\lspace$, and corresponds to the maximal energy $E_{\max}=\sqrt{\vecmom_{\max}^{2}+\mass[0]^{2}}$, where $\vecmom_{\max}=(\pi/\lspace,\pi/\lspace,\pi/\lspace)$. 
Correspondingly, assuming tha lattice $\Lambda$ with periodic boundary conditions, the continuous Fourier transform is adapted as follows:
\begin{align}
	f(\vecpos)&=\int_{\mathbb{R}^{D}}\frac{d\vecmom}{(2\pi)^{D}}\widetilde{f}(\vecmom)e^{i\vecpos\cdot \vecmom}
    &\longrightarrow&&
    f(\vecsite)&=\qty(\frac{\lspace}{2\pi})^{D}\sum_{\vecmom}^{BZ}\widetilde{f}_{\lspace}(\vecmom)e^{i\lspace \vecsite\cdot \vecmom}\,,
\end{align}
where BZ stands for the $1^{st}$ Brillouin zone $\qty[-\frac{\pi}{\lspace},\frac{\pi}{\lspace}]^{D}$ and $\widetilde{f}_{\lspace}(\vecmom_{\max})=\widetilde{f}_{\lspace}(-\vecmom_{\max})$. 
We can then express $\widetilde{f}_{\lspace}(\vecmom)$ in (finite) Fourier-series
\begin{equation}
	\widetilde{f}_{\lspace}(\vecmom)=\lspace^{D}\sum_{\vecsite}\hat{f}_{\vecsite}e^{-i\lspace \vecsite\cdot \vecmom}\,.
	\label{eq_Fourier_series}
\end{equation}
If we set $\hat{f}_{\vecsite}=(2\pi)^{-D}$, then \eqref{eq_Fourier_series} reduces to the Fourier series representation of the $\delta$ function in the $1^{st}$ BZ, namely
\begin{equation}
	\delta_{p}(\vecmom)=\qty(\frac{\lspace}{2\pi})^{D}\sum_{\vecmom}^{BZ}e^{-i\lspace \vecsite\cdot \vecmom}\,,
	\label{eq_delta_Fourier_series}
\end{equation}
where $p$ stands for \emph{periodic}. 
Note that, unlike the continuum limit, where $\lspace\to0$, here $\delta_{p}(\vecmom)$ has "non-vanishing" support in $\vecmom=0$ (modulo $2m\pi$ for $m\in\mathbb{N}$). 
Similarly, the Dirac $\delta$ function $\delta(\vecpos-\vecpos')$ becomes a Kronecker $\delta_{\vecsite\vecsite^{\prime}}$ (multiplied by a factor $\lspace^{-D}$)
\begin{equation}
	\delta_{\vecsite\vecsite^{\prime}}=\qty(\frac{\lspace}{2\pi})^{D}\sum_{\vecmom}^{BZ}e^{i\lspace \vecsite\cdot \vecmom}
	\label{eq_kronecker_Fourier_series}\,.
\end{equation}
% ============================================================================
\subsection{Continuum limit and Lorentz invariance}
The main physical consequence of the ultraviolet cutoff is the following: the discretized theory does not describe physics at scales (momenta) smaller (larger) than the lattice spacing $\lspace$ ($\pi/\lspace$), \idest{}, all the short-distance (high-energy) phenomena \cite{Kogut1979IntroductionLatticeGauge}.

Another effect of lattice discretization is the loss of the Lorentz invariance (the symmetry of special relativity), as the lattice spacing $\lspace$ introduces a preferred reference frame.
In the Lagrangian formalism, where both space and time are discretized, the spacetime lattice does not support continuous Lorentz transformations, but the theory retains a discrete version of spacetime symmetry, such as discrete rotations or translations.
On the contrary, the spatial discretization performed in Hamiltonian LGTs breaks the symmetry between space and time, which is a crucial part of Lorentz invariance.
In these terms, the Hamiltonian formalism is said to explicitly break Lorentz invariance.

Restoring the continuum theory and the Lorentz invariance requires to take the continuum limit, where $\lspace\to0$, $N\to \infty$, while keeping $\lspace\cdot N$ fixed \cite{Kogut1975HamiltonianFormulationWilson}. 
In this regime, maximum momentum $\vecmom_{\max}$ and energy $E_{\max}$ become infinite and the lattice approximates the continuous space. 
The continuum limit is typically associated with the process of \emph{renormalization}, where the theory is adjusted to correctly reproduce physical observables (which should be insensitive to the discretization) as the lattice spacing goes to zero. 
Unfortunately, the continuum limit location is non trivial and requires a lot of attention in understanding the relation between the lattice spacing and the Hamiltonian parameters. 
A comprehensive dissertation about the continuum limit would require a dedicated effort out of the purposes of the thesis.
Nonetheless, we will partially discuss it in \cref{sec_lattice_fermions,sec_lattice_gaugefields} by using simple dimensional analysis. 
For a deeper discussion, see \cite{Rothe2012LatticeGaugeTheories,Wilson1974ConfinementQuarks,Creutz1980MonteCarloStudy,Kogut1975HamiltonianFormulationWilson,Creutz1992QuantumFieldsComputer,Clemente2022StrategiesDeterminationRunning,Crippa2024DeterminingDimensionalQuantum}.
% ============================================================================
\subsection{Discretization of fermions fields}
\label{sec_lattice_fermions}
Discretization of more complex objects such as fermion fields, \idest{} spinors, requires more attention and encounters a set of problems whose solutions are not univocal nor perfect. 
Throughout this section, we will cover these challenges by focusing on one of the possible solutions: \emph{staggered fermions}.  
For a more general and complete discussion about fermion discretization, see \cite{Susskind1977LatticeFermions,Wilson1974ConfinementQuarks,Kogut1975HamiltonianFormulationWilson}.

Let us start from the continuum Dirac equation for a massive free fermion.
Assuming the spacetime coordinate $\pos=(t, \vecpos)$ and the metric $\eta^{\mu\nu}=\diag(+1,-1,-1,-1)$, we have:
\begin{equation}
    (i\hbar\gamma^{\mu}\partial_{\mu}-\mass[0]\lspeed)\psi(\pos)=
    \qty(i\frac{\hbar}{\lspeed}\gamma^{0}\partial_{t}-i\hbar\gamma^{k}\partial_{k}-\mass[0]\lspeed)\psi(\pos)=0\,,
    \label{eq_dirac_equation}
\end{equation}
where $\lspeed$ is the speed of light, $\hbar$ is the Planck constant, $\mass[0]$ is the mass parameter, and $\gamma^{\mu}$ are the $4\times4$ Dirac matrices satisfying Clifford's algebra
\begin{equation}
	\qty{\gamma^{\mu},\gamma^{\nu}}=2\eta^{\mu\nu}\,.
    \label{eq_clifford_algebra}
\end{equation}
The field $\psi(\pos)$ is a 4-dimensional spinor, whose Hermitian conjugate is $\overline{\psi}(\pos)=\psi^{\dagger}(\pos)\gamma^{0}$. 
The corresponding quantized version satisfies the canonical equal-time commutation relations
\begin{equation}
	\qty{\hpsi_{\alpha}(t,\vecpos),\hpsi^{\dagger}_{\beta}(t, \vecpos')}=i\hbar\delta_{\alpha\beta}\delta^{3}(\vecpos-\vecpos')\,,
\end{equation}
where the $\alpha$ and $\beta$ indices label spinor components.
In the Hamiltonian form, \cref{eq_dirac_equation} reads:
\begin{equation}
    -i\hbar \partial_{t}\hpsi(t,\vecpos)=\qty(i\lspeed\hbar \gamma^{0}\gamma^{k}\partial_{k}+\mass[0]\lspeed^{2}\gamma^{0})\hpsi(t,\vecpos)\,.
\end{equation}
Omitting the time dependence, the Dirac Hamiltonian in this form becomes 
\begin{equation}
    \ham_{\rm{Dirac}}
    =\int d\vecpos \ham(\vecpos)
    =\int d\vecpos \qty[\hpsi^{\dagger}(\vecpos)(i\lspeed\hbar\gamma^{0}\gamma^{k}\partial_{k}+\mass[0]\lspeed^{2}\gamma^{0})\hpsi(\vecpos)]\,,
    \label{eq_dirac_hamiltonian}
\end{equation}
When discretizing the space on a D-dimensional spatial lattice $\Lambda$, we replace $\hpsi(t,\vecpos)\to\hpsi_{\vecsite}(t)$, where $\vecpos=\lspace\vecsite$. 
Correspondingly, the canonical equal-time commutation relations become:
\begin{equation}
	\qty{\hpsi_{\vecsite, \alpha}(t),\hpsi^{\dagger}_{\vecsite^{\prime}, \beta}(t)}=i\hbar\delta_{\alpha\beta}\delta_{\vecsite,\vecsite^{\prime}}\,.
\end{equation}
Unless required, from now on, we will omit the temporal dependence, by simply focusing on $\hpsi_{\vecsite}$. 
Similarly, spatial derivatives $\partial_{k}\hpsi(\vecpos)$ are replaced by (symmetric) finite differences:
\begin{equation}
    \partial_{k}\hpsi(\vecpos) \longrightarrow \frac{1}{2\lspace}\qty[\hpsi_{\vecsite+\latvec[k]}-\hpsi_{\vecsite-\latvec[k]}]\,.
\end{equation}
By recalling \cref{eq_lattice_discretization}, the lattice Dirac Hamiltonian reads
\begin{equation}
    \ham_{\rm{Dirac}}^{\rm{latt}}=\lspace^{D}\sum_{\vecsite,k}\hpsi^{\dagger}_{\vecsite}\qty[i\lspeed\hbar\frac{\gamma^{0}\gamma^{k}}{2\lspace}(\hpsi_{\vecsite+\latvec[k]}-\hpsi_{\vecsite-\latvec[k]})] + \lspace^{D}\mass[0]\lspeed^{2}\sum_{\vecsite}\hpsi^{\dagger}_{\vecsite}\gamma^{0}\hpsi_{\vecsite}\,.
\end{equation}
We then redefine $\hpsi_{\vecsite}\to \lspace^{-D/2}\hpsi_{\vecsite}$ and obtain
\begin{equation}
    \ham_{\rm{Dirac}}^{\rm{latt}}=\sum_{\vecsite,k}\hpsi^{\dagger}_{\vecsite}\qty[i\lspeed\hbar\frac{\gamma^{0}\gamma^{k}}{2\lspace}(\hpsi_{\vecsite+\latvec[k]}-\hpsi_{\vecsite-\latvec[k]})] + \mass[0]\lspeed^{2}\sum_{\vecsite}\hpsi^{\dagger}_{\vecsite}\gamma^{0}\hpsi_{\vecsite}\,.
    \label{eq_dirac_equation_lattice}
\end{equation}
This choice makes the lattice fermion fields adimensional, even if their hidden dependence on the lattice spacing becomes important when considering the continuum limit location.
% ============================================================================
\subsubsection{Fermion Doubling Problem}
An artifact of the lattice discretization of fermions is the appearance of extra solutions of the Dirac equation,  known as \emph{doublers} \cite{Rothe2012LatticeGaugeTheories}.
In a $D$-dimensional space, these doublers lead to the appearance of $2^{D}-1$ extra fermion species with respect to the continuum theory: namely, each spatial direction $k$ contributes a factor of 2 to the number of solutions (one at $\mom[k]=0$ and one at $\mom[k]=\pi/\lspace$).

In order to formally visualizes these extra solutions, let us transform the lattice spinor field to the momentum space using the previously defined Fourier transform:
\begin{equation}
    \hpsi_{\vecsite}=\frac{1}{V}\sum_{\vecmom}e^{i\vecmom\cdot \vecsite \lspace}\hpsi(\vecmom)\,,
    \label{eq_spinor_latt_momentum}
\end{equation}
where $\vecmom$ is the lattice momentum, while $V$ is the lattice volume.
Substituing \cref{eq_spinor_latt_momentum} into the discretized derivative, we have:
\begin{equation}
    \frac{\hpsi_{\vecsite+\latvec[k]}-\hpsi_{\vecsite-\latvec[k]}}{2\lspace} =
    \frac{1}{V}\sum_{\vecmom}e^{i\vecmom\cdot \vecsite \lspace}\frac{e^{i\mom[k]\lspace}-e^{-i\mom[k]\lspace}}{2\lspace}\hpsi(\vecmom)=
    \frac{1}{V}\sum_{\vecmom}e^{i\vecmom\cdot \vecsite \lspace}\frac{i\sin(\mom[k]\lspace)}{a}\hpsi(\vecmom)\,.
\end{equation}
Then, the lattice Dirac Hamiltonian in the momentum space becomes
\begin{equation}
    \ham(\vecmom)=\hpsi(\vecmom)^{\dagger}\qty[-\sum_{k=1}^{3}\gamma^{0}\gamma^{k}\frac{\lspeed\hbar}{\lspace}\sin(\mom[k]\lspace)+\mass[0]\lspeed^{2}\gamma^{0}]\hpsi(\vecmom)\,.
\end{equation}
From the eigenvalue equation $H(\vecmom)\hpsi(\vecmom)=E(\vecmom)\hpsi(\vecmom)$, we have
\begin{equation}
    \qty[-\sum_{k=1}^{3}\gamma^{0}\gamma^{k}\frac{\lspeed\hbar}{\lspace}\sin(\mom[k]\lspace)+\mass[0]\lspeed^{2}\gamma^{0}]\hpsi(\vecmom)=E(\vecmom)\hpsi(\vecmom)
\end{equation}
Squaring both sides of the eigenvalue equation, we have:
\begin{equation}
    \begin{split}
        \gamma^{0}\qty[-\sum_{k=1}^{3}\frac{\lspeed\hbar \gamma^{k}}{\lspace}\sin(\mom[k]\lspace)+\mass[0]\lspeed^{2}]\gamma^{0}\qty[-\sum_{k'=1}^{3}\frac{\lspeed\hbar \gamma^{k'}}{\lspace}\sin(\mom[k']\lspace)+\mass[0]\lspeed^{2}]\hpsi(\vecmom)&=E^{2}(\vecmom)\hpsi(\vecmom)\\
        \qty[\sum_{k=1}^{3}\qty(\frac{\lspeed\hbar}{\lspace}\sin(\mom[k]\lspace))^{2}+\mass[0]^{2} c^{4}]\hpsi(\vecmom)&\underset{*}{=}
    \end{split}
\end{equation}
where, in the second passage, we canceled all the mixed terms by exploiting the following properties from the Cliffor's algebra in \cref{eq_clifford_algebra}:
\begin{align}
    \gamma^{0}\gamma^{0}&=\mathbb{1}&
    \gamma^{0}\gamma^{k}+\gamma^{k}\gamma^{0}&=0&
    \gamma^{j}\gamma^{k}+\gamma^{k}\gamma^{j}&=2\delta^{jk}\mathbb{1}\,.
\end{align} 
Clearly, when $\mom[k]\lspace\ll1$, then $\sin(\mom[k]\lspace) \sim \mom[k]\lspace$ and the energy reduces to the familiar dispersion relation of relativistic particles: $E(\vecmom) = \pm \sqrt{\lspeed^{2}\hbar^{2}\vecmom^{2}+\mass[0]^{2}c^{4}}$.
However, the sine function $\sin(\mom[k]\lspace)$ has an additional zero at $\mom[k]=\pi/\lspace$, which yields another low-energy solution at the border of the BZ.
Therefore, in $D$ spatial dimensions, we obtain $2^{D}$ fermion species. 
The extra degrees of freedom affect the extrapolation to the continuum limit such that the correct continuum results can not be recovered.
% ============================================================================
\subsubsection{Nielsen \& Ninomiya Theorem}
Handling and removing fermion doublers is neither easy nor without cost. 
Indeed, according to the Nielsen and Ninomiya no-go theorem \cite{Nielsen1981AbsenceNeutrinosLattice,Nielsen1981AbsenceNeutrinosLattice-1}, any \emph{local}, \emph{hermitian}, lattice fermion theory, with \emph{translational invariance} and \emph{chiral symmetry}, necessary displays fermion doublers. 
Hence, any attempt to remove doublers requires violating at least one of the four hypotheses of the theorem.
Choosing which of these characteristics to compromise allows for different strategies, with the Wilson fermion \cite{Wilson1974ConfinementQuarks} and staggered fermion \cite{Susskind1977LatticeFermions} methods being some of the most well-known approaches. 
While the former one being already studied for quantum simulations \cite{Bermudez2010WilsonFermionsAxion,Mazza2012OpticallatticebasedQuantumSimulator,Kuno2018GeneralizedLatticeWilson,Zache2018QuantumSimulationLattice}, throughout the whole thesis, we will rely on the second solution, staggered fermions, introduced by Kogut and Susskind in \cite{Kogut1975HamiltonianFormulationWilson}, which has been widely exploited in numerical and quantum simulations \cite{Silvi2019TensorNetworkSimulation,Felser2020TwoDimensionalQuantumLinkLattice,Magnifico2021LatticeQuantumElectrodynamics,Rigobello2021EntanglementGenerationMathrm,Rigobello2023HadronsHamiltonianHardcore, Cataldi2024Simulating2+1DSU2,Cataldi*2025QuantumManybodyScarring,Calajo2024DigitalQuantumSimulation}.
% ============================================================================
\subsubsection{Staggered Fermions}
Staggered fermions do not completely resolve the Fermion doubling problem, yet they represent a compromise that reduces the number of doublers by distributing the components of the Dirac spinor across multiple lattice sites.
This doubles the periodicity of the lattice, from $\lspace$ to $2\lspace$, and halves the Brillouin zone to $[-\frac{\pi}{2\lspace}, \frac{\pi}{2\lspace}]^{D}$.
Such a solution looses locality, but maintains a modified translational invariance with a checkerboard pattern and a remnant of chiral symmetry, which is important for maintaining some of the physical characteristics of massless lattice fermions \cite{Zache2018QuantumSimulationLattice}.
 
The practical realization of staggered fermions strongly depends on the spatial dimension of the lattice, as it affects the representation of spinors and gamma matrices \cite{Brauer1935SpinorsDimensions}. 
In detail, we need to adjust the matrices $\gamma^{0}$ and $\gamma^{0}\gamma^{k}$ for $k=1\dots D$ while maintaining all the Clifford's algebra properties.
% ============================================================================
\paragraph{(1+1)D Case} 
In one spatial dimension, $\vecsite=\site[1]$, and the four-component spinor reduces to a two-spinor with the $0^{th}$ (time) component and the $3^{rd}$ (space) component: $\qty(\begin{smallmatrix}\hpsi_{0}\\ \hpsi_{3}\end{smallmatrix})$.
Correspondingly, $\gamma^{0}$ and $\gamma^{0}\gamma^{1}$ can be any two Pauli matrices, for example $\sigma^{z}$ and $\sigma^{x}$. 
We have then:
\begin{equation}
    \ham^{\rm{1D}}_{\vecsite}=
    \begin{pmatrix}
        \hpsi_{0,\vecsite}^{\dagger}& \hpsi_{3,\vecsite}^{\dagger}
    \end{pmatrix}
    \begin{pmatrix}
        \mass[0]\lspeed^{2} & i\lspeed\hbar\partial_{1}\\
         i\lspeed\hbar\partial_{1}&-\mass[0]\lspeed^{2}\\
    \end{pmatrix}
    \begin{pmatrix}
        \hpsi_{0,\vecsite}\\
        \hpsi_{3,\vecsite}
    \end{pmatrix}\,.
\end{equation}
Then, in the staggered fermion solution, we decompose the spinor components by placing $\hpsi_{0}$ on \emph{even} sites (where $(-1)^{\site[1]}=+1$) and $\hpsi_{3}$ on \emph{odd} sites (where $(-1)^{\site[1]}=-1$).
Then, for \emph{even} sites, we have:
\begin{equation}
    \ham^{\rm{1D}}_{\rm{even}}=\mass[0]\lspeed^{2}\sum_{\vecsite}\hpsi_{\vecsite}^{\dagger}\hpsi_{\vecsite} 
    + \frac{i\lspeed\hbar}{2\lspace}\sum_{\vecsite}\hpsi_{\vecsite}^{\dagger}(\hpsi_{\vecsite+\latvec[1]}-\hpsi_{\vecsite-\latvec[1]})\,,
\end{equation}
whereas, for \emph{odd} sites, we have
\begin{equation}
    \ham^{\rm{1D}}_{\rm{odd}}=
    + \frac{i\lspeed\hbar}{2\lspace}\sum_{\vecsite}\hpsi_{\vecsite}^{\dagger}(\hpsi_{\vecsite+\latvec[1]}-\hpsi_{\vecsite-\latvec[1]})
    -\mass[0]\lspeed^{2}\sum_{\vecsite}\hpsi_{\vecsite}^{\dagger}\hpsi_{\vecsite} 
    \,.
\end{equation}
Notice that each of these two contributions contains the hermitian conjugate of the hopping term of the other one.
Therefore, combining even and odd sites, the (1+1)D Dirac Hamiltonian via staggered fermions reads:
\begin{equation}
    \begin{split}
        \ham_{\rm{Dirac}}^{\rm{\rm{1D}}}
        &=\sum_{\vecsite}\qty[\frac{i\lspeed\hbar}{2\lspace}\hpsi^{\dagger}_{\vecsite}\hpsi_{\vecsite+\latvec[1]}+\hc]+\mass[0]\lspeed^{2}\sum_{\vecsite}(-1)^{\vecsite}\hpsi^{\dagger}_{\vecsite}\hpsi_{\vecsite}\,,
    \end{split}
    \label{eq_dirac_ham_stag1D}
\end{equation}
% ============================================================================
\paragraph{(2+1)D Case} 
The case of two-spatial dimensions is less trivial, as there are multiple ways of defining the matrices $(\gamma^{0},\gamma^{0}\gamma^{1}, \gamma^{0}\gamma^{2})$. 
For instance, using two-spinors, there are two solutions: $(\gamma^{0},\gamma^{0}\gamma^{1}, \gamma^{0}\gamma^{2})\to (\sigma^{z},\sigma^{x}, \pm\sigma^{y})$.
Choosing the case with $+\sigma^{y}$, we obtain:
\begin{equation}
    \ham^{\rm{2D}}_{\vecsite}=
    \begin{pmatrix}
        \hpsi_{0,\vecsite}^{\dagger}& \hpsi_{3,\vecsite}^{\dagger}
    \end{pmatrix}
    \begin{pmatrix}
        \mass[0]\lspeed^{2} & i\lspeed\hbar (\partial_{1}-i\partial_{2})\\
         i\lspeed\hbar (\partial_{1}+i\partial_{2})&-\mass[0]\lspeed^{2}\\
    \end{pmatrix}
    \begin{pmatrix}
        \hpsi_{0,\vecsite}\\
        \hpsi_{3,\vecsite}
    \end{pmatrix}\,.
\end{equation}
The decomposition of the spinor components along the lattice is similar to the (1+1)D case: we place $\hpsi_{0}$ in \emph{even} sites (where $(-1)^{\site[1]+\site[2]}=+1$) and $\hpsi_{3}$ in \emph{odd} ones (where $(-1)^{\site[1]+\site[2]}=-1$). 
Then, for \emph{even} sites, we have:
\begin{equation}
    \ham^{\rm{2D}}_{\rm{even}}=\mass[0]\lspeed^{2}\sum_{\vecsite}\hpsi_{\vecsite}^{\dagger}\hpsi_{\vecsite} 
    + \frac{i\lspeed\hbar}{2\lspace}\sum_{\vecsite}\hpsi_{\vecsite}^{\dagger}(\hpsi_{\vecsite+\latvec[1]}-\hpsi_{\vecsite-\latvec[1]})
    + \frac{\lspeed\hbar}{2\lspace}\sum_{\vecsite}\hpsi_{\vecsite}^{\dagger}(\hpsi_{\vecsite+\latvec[2]}-\hpsi_{\vecsite-\latvec[2]})
\end{equation}
As for the \emph{odd} sites, we have
\begin{equation}
    \ham^{\rm{2D}}_{\rm{odd}}=-\mass[0]\lspeed^{2}\sum_{\vecsite}\hpsi_{\vecsite}^{\dagger}\hpsi_{\vecsite} 
    + \frac{i\lspeed\hbar}{2\lspace}\sum_{\vecsite}\hpsi_{\vecsite}^{\dagger}(\hpsi_{\vecsite+\latvec[1]}-\hpsi_{\vecsite-\latvec[1]})
    - \frac{\lspeed\hbar}{2\lspace}\sum_{\vecsite}\hpsi_{\vecsite}^{\dagger}(\hpsi_{\vecsite+\latvec[2]}-\hpsi_{\vecsite-\latvec[2]})
\end{equation}
Combining the two cases and noting the hermitian conjugate of the hopping terms, we find
\begin{equation}
    \begin{split}
    \ham^{\rm{2D}}=&\frac{\lspeed\hbar}{2\lspace}\sum_{\vecsite}
    \qty[\qty[i\hpsi_{\vecsite}^{\dagger}\hpsi_{\vecsite+\latvec[1]}
    +(-1)^{\site[1]+\site[2]}\hpsi_{\vecsite}^{\dagger}\hpsi_{\vecsite+\latvec[2]}]+\hc]
    +\mass[0]\lspeed^{2}\sum_{\vecsite}(-1)^{\vecsite}\hpsi_{\vecsite}^{\dagger}\hpsi_{\vecsite}\,.
    \end{split}
    \label{eq_dirac_ham_stag2D}
\end{equation}
% ============================================================================
\paragraph{(3+1)D Case}
In the three-dimensional case, we have to use the full 4-component spinor and the original definition of the gammma matrices:
\begin{align}
    \gamma^{0}&=\begin{pmatrix}
        \mathbb{1}_{2} & \\
        & -\mathbb{1}_{2}
    \end{pmatrix}&
    \text{and}&&
    \gamma^{k}&=\begin{pmatrix}
         & \sigma^{k} \\
        \sigma^{k}& 
    \end{pmatrix}\,.
\end{align}
Correspondingly, we have:
\begin{equation}
    \setlength{\arraycolsep}{1pt} % Adjust the spacing here
    \ham^{\rm{3D}}_{\vecsite}{=}
    \begin{pmatrix}
        \hpsi_{0,\vecsite}^{\dagger}& \hpsi_{1,\vecsite}^{\dagger} &
        \hpsi_{2,\vecsite}^{\dagger}& \hpsi_{3,\vecsite}^{\dagger}
    \end{pmatrix}
    \begin{pmatrix}
        \mass[0]\lspeed^{2} & & i\lspeed\hbar\partial_{3} & i\lspeed\hbar (\partial_{1}-i\partial_{2})\\
        & \mass[0]\lspeed^{2} & i\lspeed\hbar (\partial_{1}+i\partial_{2})& -i\lspeed\hbar\partial_{3}\\

        i\lspeed\hbar\partial_{3}& i\lspeed\hbar (\partial_{1}-i\partial_{2})&-\mass[0]\lspeed^{2} &\\
        i\lspeed\hbar (\partial_{1}+i\partial_{2})&-i\lspeed\hbar\partial_{3}&& -\mass[0]\lspeed^{2}\\
    \end{pmatrix}
    \begin{pmatrix}
        \hpsi_{0,\vecsite}\\
        \hpsi_{1,\vecsite}\\
        \hpsi_{2,\vecsite}\\
        \hpsi_{3,\vecsite}
    \end{pmatrix}.
\end{equation}
In this case, decomposing the spinor components along the lattice is less intuitive, but effordable. 
Namely, we place the component $\hpsi_{0}$ at site $\vecsite=(0,0,0)$: it couples with $\hpsi_{3}$ at sites where $\site[1]+\site[2]$ is odd and $\site[3]$ is even, with $\hpsi_{2}$ at sites where $\site[1]+\site[2]$ is even and $\site[3]$ is odd, with $\hpsi_{1}$ at sites where $\site[1]+\site[2]$ is odd and $\site[3]$ is odd, and finally with $\hpsi_{0}$ at
sites where $\site[1]+\site[2]$ is even and $\site[3]$ is even.
Correspondingly, $\ham^{\rm{3D}}$ decomposes into four separate sub-theories, with each theory touching a single fermion specie per site. 
In particular, $\hpsi_{0}$ and $\hpsi_{1}$ occupy even sites, while $\hpsi_{2}$ and $\hpsi_{3}$ occupy odd sites. 
As for the mass term, it acts positively on $\hpsi_{0}$ and $\hpsi_{1}$ (even sites), and negatively on $\hpsi_{2}$ and $\hpsi_{3}$ (odd sites).
As for the $x$-derivative, it acts positively everywhere, while the $y$- and $z$-derivatives act positively on $\hpsi_{0}$ and $\hpsi_{2}$ (where $\site[1]+\site[2]$ is even) and negatively on $\hpsi_{1}$ and $\hpsi_{3}$ (where $\site[1]+\site[2]$ is odd).
Summarizing, we obtain:
\begin{equation}
    \begin{split}
    \ham^{\rm{3D}}_{\rm{Dirac}}=&+\frac{\lspeed\hbar}{2\lspace}\sum_{\vecsite}
    \qty[\qty[
        i\hpsi_{\vecsite}^{\dagger}\hpsi_{\vecsite+\latvec[1]}
        +(-1)^{\site[1]+\site[2]}
        \hpsi_{\vecsite}^{\dagger}\hpsi_{\vecsite+\latvec[2]}
    +i(-1)^{\site[1]+\site[2]}
    \hpsi_{\vecsite}^{\dagger}\hpsi_{\vecsite+\latvec[3]}]+\hc]\\
    &+\mass[0]\lspeed^{2}\sum_{\vecsite}(-1)^{\vecsite}\hpsi_{\vecsite}^{\dagger}\hpsi_{\vecsite},
    \end{split}
    \label{eq_dirac_ham_stag3D}
\end{equation}
which perfectly matches \cite{Susskind1977LatticeFermions}.
We stress that the staggered phases in the hopping and the mass term are fundamental for recovering correct resultss. 
% =====================================================================
\subsection{Discretization of gauge fields}
\label{sec_lattice_gaugefields}
In QFT, the principle of gauge invariance underlies the interactions between matter fields (such as fermions) and gauge fields (such as bosons, \eg{} photons or gluons). 
Such an interaction is ruled by local (gauge) transformations from a corresponding gauge group. 
% =====================================================================
\subsubsection{Basics of group theory}
Let us consider a generic compact Lie group $\gaugegroup$ (\eg{}, SU(N) or U(1)), whose algebra has generators $T^{a}$, where $a\in\qty{1\dots \dim \gaugegroup}$.
In the fundamental representation of non-Abelian groups $\gaugegroup$ like SU(N), these generators could be the Gell-Mann matrices for SU(3), or Pauli matrices for SU(2). 
In general, the generators $T^{a}$ satisfy the following commutations rules
\begin{equation}
    [T^{a},T^{b}]=if^{abc}T^{c}\,,
    \label{eq_group_generators}
\end{equation}
where $f^{abc}$ are the \emph{structure constants} of the gauge group $\gaugegroup$. 
For a non-Abelian group, these constants are fully antisymmetric in the indices $a,b,c$, encoding the non-commuting relation between the generators (and correspondingly between the resulting transformations).
For instance, in the SU(2) case, $f^{abc}$ is the Levi-Civita symbol $\epsilon^{abc}$. 
Conversely, in the Abelian scenario, like U(1), the generator is unique and scalar $T=1$ and $f^{abc}=0$.

Then, any group element $\Omega(\pos)$ can vary from point to point in spacetime and is defined as
\begin{align}
    \Omega(\pos)&=\exp(i\theta^{a}(\pos)T^{a})&
    \text{with}&&
    \Omega(\pos) \Omega^\dagger(\pos) = \Omega^\dagger(\pos) \Omega(\pos) = \mathbb{1}\,,
    \label{eq_group_element}
\end{align}
where $\theta^{a}(\pos)$ are the local parameters of the gauge transformation.
For example, in the case of SU(2), $\Omega(\pos)$ could be a $2\times2$ unitary matrix with determinant 1. 
For U(1) (electromagnetism), $\Omega(\pos)$ would simply be a phase factor $\exp(i\theta(\pos))$.
% =====================================================================
\subsubsection{Matter fields}
Assuming a single matter flavor in the fundamental representation of the gauge group $\gaugegroup$, gauge transformations of spinors are generated by the charge operator $\colormatter^{a}(\pos)$, defined as:
\begin{equation}
   \colormatter^{a}(\pos) = \sum_{\alpha,\beta} \hpsi^\dagger_{\alpha}(\pos)T^{a}_{\alpha\beta} \hpsi_{\beta}(\pos).
\end{equation}
This operator represents how the matter fields interact with the gauge fields and underlies the dynamics of the gauge theory when multiple flavors are present.
Correspondingly, under a local gauge transformation $\Omega(\pos)$, matter fields transform as follows:
\begin{equation}
    \hpsi(\pos) \rightarrow \hpsi^{\prime}(\pos) = \Omega(\pos) \hpsi(\pos).
    \label{eq_gauge_transformations}
\end{equation}
% =====================================================================
\subsubsection{The covariant derivative}
The original Dirac equation of \cref{eq_dirac_equation} is clearly not invariant under local gauge transformations such as the ones in \cref{eq_gauge_transformations}; indeed when transforming the derivative, we find:
\begin{equation}
    \partial_{\mu} \hpsi(\pos) \rightarrow \partial_{\mu} \hpsi^{\prime}(\pos) 
    = \partial_{\mu} \qty(\Omega(\pos) \hpsi(\pos)) 
    = (\partial_{\mu} \Omega(\pos))\hpsi(\pos) + \Omega(\pos) \partial_{\mu} \hpsi(\pos)\,,
\end{equation}
so that the transformed Dirac equation reads
\begin{equation}
    \begin{split}
        0&=i\hbar\gamma^{\mu} \qty[(\partial_{\mu} \Omega(\pos)) \hpsi(\pos) + \Omega(\pos) \partial_{\mu} \hpsi(\pos)] - \mass[0]\lspeed \Omega(\pos) \hpsi(\pos)\\
        &=\Omega(\pos) \qty(i\hbar\gamma^{\mu} \partial_{\mu} \hpsi(\pos) - \mass[0]\lspeed \hpsi(\pos)) + i\hbar\gamma^{\mu} (\partial_{\mu} \Omega(\pos)) \hpsi(\pos).
    \end{split}
\end{equation}
To impose invariance under this transformation, the extra term $i\hbar\gamma^{\mu} (\partial_{\mu} \Omega(\pos)) \psi(\pos)$ must vanish. 
This would spontaneously happen just in case $\Omega(\pos)$ is constant (which corresponds to a \emph{global}, rather than local, gauge transformation). 
In general, to achieve gauge invariance, the ordinary derivative $\partial_{\mu}$ must be replaced by the covariant derivative $\hat{D}_{\mu}$, which transforms covariantly under gauge transformations thereby canceling the extra terms introduced by the local gauge transformation.
Namely, we have:
\begin{equation}
    \hat{D}_{\mu} \hpsi(\pos) = \qty(\partial_{\mu} - i \frac{\charge}{\hbar}\vecpot_{\mu}(\pos)) \hpsi(\pos)
    = \qty(\partial_{\mu} - i \coupling[0]\vecpot_{\mu}(\pos)) \hpsi(\pos)\,,
    \label{eq_covariant_derivative}
\end{equation}
where $\charge$ is the charge, $\coupling[0]=\charge/\hbar$ is the gauge coupling, while $\vecpot_{\mu}(\pos)=\vecpot_{\mu}^{a}(\pos)T^{a}$ is the full gauge field operator obtained by contracting the gauge field components $\vecpot_{\mu}^a(\pos)$ to the generators $T^{a}$ via the gauge group index $a$.
The gauge field operator transforms as follows:
\begin{equation}
    \vecpot_{\mu}(\pos) \rightarrow \vecpot_{\mu}^{\prime}(\pos) = \Omega(\pos) \vecpot_{\mu}(\pos) \Omega^\dagger(\pos) + \frac{i}{\coupling[0]} \Omega(\pos) \partial_{\mu} \Omega^\dagger(\pos)\,.
\end{equation}
Correspondingly, the continuum Dirac Hamiltonian in \cref{eq_dirac_hamiltonian} is made covariant as follows:
\begin{equation}
    \ham_{\rm{Dirac}}
    =\int d\vecpos \qty[\hpsi^{\dagger}(\vecpos)(i\lspeed\hbar\gamma^{0}\gamma^{k}(\partial_{k}-i\coupling[0]\vecpot_{k}) +\mass[0]\lspeed^{2}\gamma^{0})\hpsi(\vecpos)]\,.
    \label{eq_dirac_hamiltonian_cov}
\end{equation}
% =====================================================================
\subsubsection{The pure gauge Hamiltonian}
To make the Dirac equation gauge invariant, we needed to introduce an extra \dof{}, a gauge field, whose dynamics needs to be properly accounted and added to the Dirac Hamiltonian. 
In the continuum, the dynamics of gauge fields is captured by the field strength tensor $\hat{F}_{\mu\nu}(\pos)$, which for a generic non-Abelian gauge field $\vecpot_{\mu}(\pos)$ is given by:
\begin{equation}
    \begin{split}
        \hat{F}_{\mu\nu}(\pos) &= \partial_{\mu} \vecpot_{\nu}(\pos) - \partial_{\nu} \vecpot_{\mu}(\pos) - i\coupling[0]\qty[\vecpot_{\mu}(\pos), \vecpot_{\nu}(\pos)]\\
        \hat{F}_{\mu\nu}^{a}(\pos)T^{a} &=\qty[\partial_{\mu} \vecpot_{\nu}^{a}(\pos) - \partial_{\nu} \vecpot_{\mu}^{a}(\pos) + \coupling[0] f^{abc} \vecpot_{\mu}^{b}(\pos) \vecpot_{\nu}^{c}(\pos)]T^{a}.
    \end{split}
\end{equation}
To move towards the Hamiltonian formalism, we separate its electric and magnetic components:
\begin{subequations}
    \begin{align}
    \eleE_{k}^a(\pos)&= \hat{F}_{0k}^a(\pos) = \frac{1}{\lspeed}\partial_{t}\vecpot_{k}^a(\pos) - \partial_{k}\vecpot_0^a(\pos) - \coupling[0]f^{abc}\vecpot_{0}^b(\pos)\vecpot_{k}^c(\pos)\\
    \magn_{i}^a(\pos)&=-\frac{1}{2}\epsilon_{ijk}\hat{F}_{jk}^a(\pos) = \epsilon_{ijk} \qty(\partial_{j}\vecpot_{k}^a(\pos) - \frac{\coupling[0]}{2} f^{abc} \vecpot_{j}^b(\pos)\vecpot_{k}^c(\pos)),
    \end{align}
\end{subequations}
where $i,\spin,k$ are spatial indices.
In the Coulomb gauge, where $\vecpot^{0}(\pos) = 0$, the electric field reduces to $\eleE_{k}^a(\pos)= \hat{F}_{0k}^a(\pos) =\partial_{t}\vecpot_{k}^a(\pos)/\lspeed$. 
This choice does not change the physics, as it simply exploits an extra freedom in defining the gauge fields.  

The corresponding Hamiltonian of the pure gauge fields is
\begin{equation}
    \ham_{\rm{gauge}}=\int d\vecpos \qty[\frac{\permittivity}{2} \eleE_{k}^{a}(\pos)\eleE_{k}^{a}(\pos)+ \frac{1}{2\permeability}\magn_{k}^{a}(\pos)\magn_{k}^{a}(\pos)]\,.
    \label{eq_ham_pure_gauge_continuum}
\end{equation}
where $\permittivity$ and $\permeability = (\lspeed^{2}\permittivity)^{-1}$ correspond to the vacuum permittivity and permeability respectively. 
In dimensioned units (such as SI), the physical dimensions of the constants read
\begin{equation}
\begin{aligned}
    \qty[\permittivity] &=(\text{charge})^{2}(\text{length})^{2-D}(\text{energy})^{-1}\\
    \qty[\permeability]&=(\text{charge})^{-2}(\text{length})^{D-2}(\text{energy})\cdot (\text{length})^{-2}(\text{time})^{2}\\
    &= (\text{charge})^{-2}(\text{length})^{D-4}(\text{energy})(\text{time})^{2}\,.
    \end{aligned}
\end{equation}
Correspondingly, the electric and magnetic fields have the following dimensions:
\begin{equation}
    \begin{aligned}
        \qty[\eleE] &= (\text{charge})^{-1}(\text{length})^{-1}(\text{energy})\\
        \qty[\magn] &= (\text{charge})^{-1}(\text{length})^{-2}(\text{energy})(\text{time})\,.
    \end{aligned}
\end{equation}
As for the coupling constant $\coupling[0]$, its dimensional analysis reads
\begin{equation}
    \qty[\coupling[0]]=(\text{charge})(\text{energy})^{-1}(\text{time})^{-1}\,,
\end{equation}
while for the vector potential we have:
\begin{equation}
    \qty[\vecpot]=(\text{charge})^{-1}(\text{length})^{-1}(\text{energy})(\text{time})\,.
\end{equation}
% =====================================================================
\subsubsection{Discretized gauge fields}
When moving from the continuum to the lattice, each spatial component $k$ of the covariant derivative in \cref{eq_covariant_derivative} is replaced by the \emph{parallel transport}\footnote{Parallel transports have several names in literature. Sometimes they are also called \emph{connections} referring to a differential geometry framework. 
Sometimes, they are called \emph{comparators}.} $\Apara_{\vecsite, \latvec[k]}(t)$, a dimensionless link variable reflecting the gauge field’s influence along the link connecting neighboring lattice sites $\vecsite$ and $\vecsite +\latvec[k]$.
Omitting again the time dependence, we define the parallel transport as:
\begin{equation}
    \hat{D}_{k}=(\partial_{k}-i\coupling[0]\vecpot^{k}) \quad \longrightarrow \quad \Apara_{\vecsite, \latvec[k]} = \exp(i\coupling[0] \int_{\vecsite}^{\vecsite+\latvec[k]}\vecpot^{k}(\vecpos)d\ell)
    \sim \exp(i\coupling[0]\lspace \vecpot^{k})\,,
    \label{eq_parallel_transport}
\end{equation}
where in the second step we assumed $\vecpot^{k}(\pos)\sim \vecpot^{k}_{\vecsite}$ to be almost constant along the lattice link $(\genlink_{k})$ of length $\lspace$.
Assuming $\lspace$ to be sufficiently small, we can further expand the parallel transporter up to the first order and notice that:
\begin{equation}
    \begin{split}
        &\qty[\frac{i\hpsi^{\dagger}_{\vecsite}\Apara_{\vecsite, \latvec[k]}\hpsi_{\vecsite+\latvec[k]}}{2\lspace}+\hc]
        = i\qty[\frac{\hpsi^{\dagger}_{\vecsite}\Apara_{\vecsite, \latvec[k]}\hpsi_{\vecsite+\latvec[k]}-\hpsi^{\dagger}_{\vecsite+\latvec[k]}\Apara^{\dagger}_{\vecsite, \latvec[k]}\hpsi_{\vecsite}}{2\lspace}] \\
        &\qquad= i\qty[\frac{\hpsi^{\dagger}_{\vecsite}\Apara_{\vecsite, \latvec[k]}\hpsi_{\vecsite+\latvec[k]}-\hpsi^{\dagger}_{\vecsite}\Apara^{\dagger}_{\vecsite-\latvec[k], \latvec[k]}\hpsi_{\vecsite, -\latvec[k]}}{2\lspace}]\\
        &\qquad= i\hpsi^{\dagger}_{\vecsite}\frac{\qty[\Apara_{\vecsite, \latvec[k]}\hpsi_{\vecsite+\latvec[k]}-\Apara^{\dagger}_{\vecsite-\latvec[k], \latvec[k]}\hpsi_{\vecsite, -\latvec[k]}]}{2\lspace}\\
        &\qquad\sim \frac{i\hpsi^{\dagger}_{\vecsite}}{2\lspace}\qty[(1-i\coupling[0]\vecpot^{k}_{\vecsite}\lspace)\hpsi_{\vecsite+\latvec[k]}-(1+i\coupling[0]\vecpot^{k}_{\vecsite-\latvec[k]}\lspace)\hpsi_{\vecsite, -\latvec[k]}]\\
        &\qquad= i\hpsi^{\dagger}_{\vecsite}\qty[\frac{\hpsi_{\vecsite+\latvec[k]}-\hpsi_{\vecsite-\latvec[k]}}{2\lspace}
        -i\coupling[0]\vecpot^{k}_{\vecsite}
        \hpsi^{\dagger}_{\vecsite+\latvec[k]}]
        \xrightarrow{\lspace\to0}i\hpsi^{\dagger}(\vecpos)(\partial_{k}-i\coupling[0]\vecpot^{k}(\vecpos))\hpsi(\vecpos).
    \end{split}
\end{equation}
Then, the lattice version of the covariant Dirac Hamiltonian with staggered fermions is
\begin{equation}
    \begin{split}
        \ham_{\rm{Dirac}}=&\frac{\lspeed\hbar}{2\lspace} 
        \sum_{\genlink}
        \qty[s_{\genlink}\hpsi_{\vecsite} \Apara_{\genlink} \hpsi_{\vecsite+\latvec} + \hc]
        + \mass[0]\lspeed^2 \sum_{\vecsite}s_{\vecsite} \hpsi_{\vecsite}^{\dagger}\hpsi_{\vecsite, \latvec},
    \end{split}
\end{equation}
where $s_{\genlink}$ and $s_{\vecsite}$ are phases that arise when using staggered fermions.

Similarly to the vector potential, gauge transformations of the parallel transporter read:
\begin{equation}
    \Apara_{\vecsite, \latvec[k]} \rightarrow \Omega_{\vecsite} \Apara_{\vecsite, \latvec[k]}\Omega^\dagger_{\vecsite+\latvec[k]}\,,
\end{equation}
where $\Omega_{\vecsite}$ is the gauge transformation matrix at site $\vecsite$. 
In general, due to the non-commuting algebra in \cref{eq_T_generator}, the left $\Omega_{\vecsite}$ and the right $\Omega^{\dagger}_{\vecsite+\latvec[k]}$ gauge transformations are different and generated by the dimensionless conjugate fields $\eleL_{\genlink}^{a}$ and $\eleR_{\genlink}^{a}$ respectively, which satisfy the following relations:
\begin{subequations}
\label{eq:gauge_operator_algebra}
  \begin{align}
    \comm*{\eleL^{a}_{\genlink}}{\NApara_{\genlinkprime}} & = -\delta_{\vecsite\vecsite^{\prime}} \delta_{\latvec\latvec^{\prime}}\sum_{\gamma} T^{a}_{\alpha \gamma} \para[\gamma \beta] \,, \\
    \comm*{\eleR^{a}_{\genlink}}{\NApara_{\genlinkprime}} & = +\delta_{\vecsite\vecsite^{\prime}} \delta_{\latvec\latvec^{\prime}}\sum_{\gamma} \para[\alpha\gamma]T^{a}_{\gamma \beta}\,,
  \end{align}
\end{subequations}
where $\alpha,\beta,\gamma,\delta$ span the matrix indices of the group generators $T^{a}$.
In the Abelian U(1) case, where there is only one generator, there is only one electric field and the parallel transporter acts as a raising operator. 
Namely, we have:
\begin{align}
    \eleE_{\genlink}&=\eleL_{\genlink}=\eleR_{\genlink} &
    \comm*{\eleE_{\genlink}}{\Apara_{\vecsite, \latvec}} &= \Apara_{\vecsite, \latvec}\,.
    \label{eq_U1_electric_fields}
\end{align} 
% =====================================================================
\subsubsection{The pure gauge lattice Hamiltonian}
Once we have discretized all the gauge fields, we can build the corresponding lattice version of the pure gauge Hamiltonian in \cref{eq_ham_pure_gauge_continuum}. 
The electric energy density is given by the Casimir operator:
\begin{equation}
    \casimir_{\vecsite, \latvec} = \eleL_{\genlink}^{a}\eleL_{\genlink}^{a} = \eleR_{\genlink}^{a}\eleR_{\genlink}^{a}\,.
    \label{eq_casimir}
\end{equation}
To make the Casimir operator adimensional, the electric Hamiltonian is redefined as follows:
\begin{align}
    \ham_{\rm{elec}}&=\frac{\permittivity}{2} \int d^{D}\vecpos  \eleE_{k}(\pos)\eleE_{k}(\pos)&
    \to &&
    \ham_{\rm{elec}}^{\rm{latt}}&=\lspace^{D}\frac{\permittivity }{2}\sum_{\vecsite, k} \frac{\charge^{2}\lspace^{2-2D}}{\permittivity^{2}}\casimir_{\vecsite, \latvec[k]}=\sum_{\vecsite, k} \frac{\charge^{2}\lspace^{2-D}}{2\permittivity}\casimir_{\vecsite, \latvec[k]}\,.
\end{align}
We can recast the latter in dimensionless units, by redefining the gauge coupling constant as
\begin{equation}
\coupling  = \coupling[0]\sqrt{\frac{\hbar}{\lspeed \permittivity}}\lspace^{\frac{3-D}{2}}
            = \charge \frac{\lspace^{\frac{3-D}{2}}}{\sqrt{\lspeed\hbar \permittivity}}
            = \charge \lspace^{\frac{3-D}{2}}\sqrt{\frac{\lspeed \permeability}{\hbar}}\,.
    \label{eq_gauge_coupling}
\end{equation}
By doing so, we obtain
\begin{equation}
    \ham_{\rm{elec}}^{\rm{latt}}=\sum_{\vecsite, k} \frac{g^{2}\lspeed \hbar}{2\lspace}\casimir_{\vecsite, \latvec[k]}\,.
    \label{eq_H_electric}
\end{equation}
Notice that the new gauge coupling constant $\coupling$ is dimensionless yet depends on the lattice spacing. 
As discussed for the matter fields, such an hidden relation becomes important when discussing the continuum limit.

Correspondingly, the simplest way to describe the magnetic energy density on the lattice is by using Wilson loops \cite{Kogut1975HamiltonianFormulationWilson,Wilson1974ConfinementQuarks}, \idest{} gauge-invariant loops made out of parallel transporters $\Apara$.
On a hypercubic lattice, the smallest non-trivial loop is represented by a plaquette operator
\begin{equation}
    \plaq = \sum_{\alpha,\beta,\gamma,\delta}
    \NApara_{\vecsite, \latvec}
    \Apara_{\vecsite+\latvec,\latvec^{\prime}}^{\beta\gamma}
    \Apara_{\vecsite+\latvec^{\prime},\latvec}^{\gamma\delta\dagger}
    \Apara_{\vecsite,\latvec^{\prime}}^{\delta\gamma\dagger},
  \label{eq_plaquette_operator}
\end{equation}
where $\latvec$ and $\latvec^{\prime}$ span the plaquette's plane.
Since the parallel transporters in \cref{eq_parallel_transport} commutes when acting on different links (and so the vector potentials do), one can notice that:
\begin{equation}
    \begin{split}
        \plaq 
        = \exp(i\coupling[0]\oint_{\partial\square}\vecpot_{k}d\ell_{k}) 
        \underset{*}{=} \exp(i\coupling[0]\oint_{\square}\curl\vb{\vecpot}\cdot d\vb{s})
        = \exp(i\coupling[0]\oint_{\square}\vb{\magn}\cdot d\vb{s})
        \sim \exp(i\coupling[0]\lspace^{2}\magn)\,.
    \end{split}
\end{equation}
where in the $*$ we used the Stokes theorem, while in the last step we assumed the magnetic field to be almost constant in the plaquette of area $\lspace^{2}$.
Given these assumptions, we have:
\begin{equation}
    \begin{split}
        \Tr(\plaq+\plaq*)=2\cos(\coupling[0]\lspace^{2}\magn)
        \sim 2\qty[1-\frac{1}{2}\coupling[0]^{2}\lspace^{4}\magn^{2}],
    \end{split}
\end{equation}
from which we obtain that
\begin{equation}
    \magn^{2} \sim -\frac{1}{\coupling[0]^{2}\lspace^{4}}\Tr(\plaq+\plaq*).
\end{equation}
Then, the magnetic Hamiltonian can be discretized as follows:
\begin{align}
    \ham_{\rm{magn}}&=\frac{1}{2\permeability} \int d^{D}\vecpos  B_{k}(\pos)B_{k}(\pos)&
    \to &&
    \ham_{\rm{magn}}^{\rm{latt}}&=-\frac{1}{2\permeability\coupling[0]^{2}\lspace^{4-D}}\sum_{\square}\Tr(\plaq+\plaq*).
\end{align}
Finally, by using the dimensionless gauge coupling in \cref{eq_gauge_coupling}, we obtain:
\begin{align}
    \ham_{\rm{magn}}^{\rm{latt}}=-\frac{\lspeed\hbar}{2\coupling^{2}\lspace}\sum_{\square}\Tr(\plaq+\plaq*)\,.
    \label{eq_H_magnetic}
\end{align}
Notice that plaquette terms only exist in $D>1$, contributing to the increased complexity of quantum and TN simulations of LGTs in higher dimensions \cite{Zohar2021QuantumSimulationLattice}.
\subsubsection{The lattice gauge Hamiltonian with staggered fermions}
Summarizing and combining all the ingredients (matter and gauge fields), the general \emph{Kogut-Susskind Hamiltonian} via staggered fermions \cite{Kogut1979IntroductionLatticeGauge} reads:
\begin{equation}
\begin{split}
    \hamlgt =& \frac{\lspeed\hbar}{2\lspace} 
    \sum_{\genlink} \sum_{\alpha,\beta} 
    \qty[s_{\genlink}\hpsi_{\vecsite,\alpha} \NApara_{\genlink} \hpsi_{\vecsite+\latvec, \beta} + \hc]
    + \mass[0]\lspeed^2 \sum_{\vecsite,\alpha} s_{\vecsite} \hpsi_{\vecsite,\alpha}^{\dagger}\hpsi_{\vecsite, \latvec}\\
    & + \frac{\lspeed\hbar \coupling^{2}}{2\lspace} \sum_{\genlink} \casimir_{\vecsite, \latvec}
    - \frac{\lspeed\hbar}{2\coupling^{2}\lspace} \sum_{\square} \Tr(\plaq + \plaq*)\,,
\end{split}
\label{eq_Ham_LGT}
\end{equation}
where we expressed the dependence on the matrix indices $\alpha,\beta$ of the gauge algebra generators.
% =====================================================================
\section{Dressed-site model for Hamiltonian Lattice Gauge Theories}
\label{sec_dressed_site_formalism}
We have seen all the ingredients for discretizing an LGT with dynamical matter.
However, there remain fundamental challenges to be faced for attacking Hamiltonians like \cref{eq_Ham_LGT} with numerical methods. 

First of all, we need to achieve a finite-dimensional encoding of the continuous gauge fields such as U(1) or SU(N), especially beyond one spatial dimension, where decoupling the gauge field's longitudinal component is required \cite{Bender2020GaugeRedundancyfreeFormulation}. 
Among known truncation recipes are Quantum Link Models (QLM) \cite{Horn1981FiniteMatrixModels,Orland1990LatticeGaugeMagnets,Chandrasekharan1997QuantumLinkModels,Brower1999QCDQuantumLink,Tagliacozzo2014TensorNetworksLattice}, finite subgroups \cite{Ercolessi2018PhaseTransitionsGauge,Magnifico2020RealTimeDynamics,Haase2021ResourceEfficientApproach},
digitization of gauge fields \cite{Hackett2019DigitizingGaugeFields}, and fusion-algebra deformation \cite{Zache2023QuantumClassicalSpin}. 
Whatever the adopted solution, further effort is required to satisfy gauge symmetry at each local site, which involves the concurrent evaluation of all the gauge links and the attached matter site. 
In practice, an extensive number of local constraints (energy penalties), which corresponds to QMB operators concurrently acting on 2D+1 sites (2D gauge links and 1 matter site), where D is the number of spatial dimensions.
Moreover, since the Gauss law represents the fundamental conservation law underpinning the theory (the highest energy scale of the model), enforcing it is essential for the simulation to remain meaningful.
Any small violation of these local constraints would incur a significant energy cost, reflecting the critical role of gauge invariance in reproducing the correct physical behavior.

Another important issue to be faced in simulating LGTs is to account for the Fermi statistics of matter fields. 
In several TN methods as well on well-established conventional digital quantum simulation platforms (e.g. superconducting qubits, trapped-ions, Rydberg arrays, quantum dots) \cite{Arute2020ObservationSeparatedDynamics,Barends2015DigitalQuantumSimulation,Salathe2015DigitalQuantumSimulation,OMalley2016ScalableQuantumSimulation,Stanisic2022ObservingGroundstateProperties}, fermionic algebra (mutually-anticommuting operations) must be encoded into a genuinely local algebra (mutually-commuting operations) of qubits. 
Standard fermion-to-qubit encodings, such as the Jordan-Wigner (JW) transformation \cite{Jordan1928UeberPaulischeAequivalenzverbot}, and other modern approaches \cite{Nielsen2006QuantumComputationGeometry,Bravyi2002FermionicQuantumComputation,Jiang2019MajoranaLoopStabilizer,Verstraete2005MappingLocalHamiltonians, Corboz2010SimulationStronglyCorrelated, Kraus2010FermionicProjectedEntangled} make any fermionic Hamiltonian interaction inherently long-range and expensive from a computational perspective.

\begin{figure}
    \centering
    \includegraphics[width=1\textwidth]{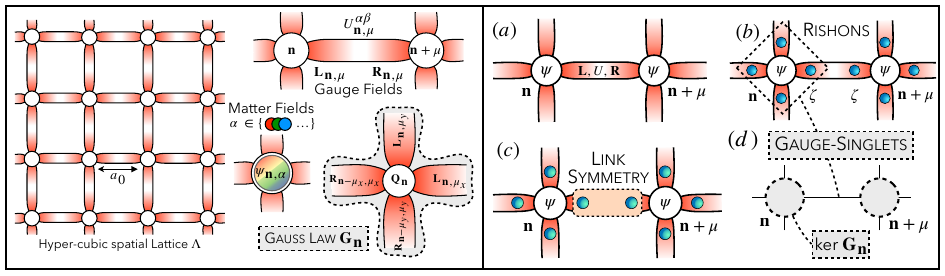}
    \caption{[Left] Graphical representation of the degrees of freedom of a 2D LGT:
    fermionic matter fields, defined on lattice sites,
    and gauge fields (the parallel transporter, and the chromo-electric fields, $\eleL_{\genlink}$ and $\eleR_{\genlink}$), living on lattice links. 
    A local gauge transformation at $\site$ acts on a matter site and all its attached links.
    [Right] Pictorial representation of the dressed site formalism adopted for TN simulations of LGTs:
    (a) we truncate the gauge link fields with an energy cutoff in the irreducible representation basis;
    (b) the truncated gauge link is split into two representations, one per half-link; each is equipped with a proper fermionic rishon mode $\rishon$.
    (c) All the half-links are absorbed into the attached matter site, forming a gauge-invariant dressed site (d) whose Hilbert space spans all the possible gauge singlets.
    }
    \label{fig_LGT_sketch}
\end{figure}

In numerical methods such as TNs, as well as in quantum computations, all these requirements significantly impact the computational resources needed for the simulations, as well as their efficiency. 
In this section, we present a theoretical scheme which is able to order achieve a finite and controllable gauge field truncation, avoid the Fermi statistics of matter fields and directly access the gauge invariant subspace, revealing suitable for TN methods as well as for quantum simulations of LGTs. 
Building on the work of \cite{Tagliacozzo2013SimulationsNonAbelianGauge, Silvi2014LatticeGaugeTensor}, we \emph{dress} every physical matter site with the information related to its adjacent gauge links. 
A pictorial scheme of this approach is outlined in \cref{fig_LGT_sketch}[Right]: (a) starting from the original description matter fields on sites and gauge fields on links in \cref{eq_Ham_LGT}, we truncate the gauge group imposing an energy cut-off on the Casimir operator. 
Then, (b) we express each truncated gauge link as a pair of fermionic rishon modes $\rishon$ and (c) constrain their link dynamics according to the original gauge group algebra. 
Ultimately, (d) we merge each of these modes to its adjacent matter site, ending up in a compact \emph{dressed-site} formalism \cite{Tagliacozzo2013SimulationsNonAbelianGauge, Silvi2014LatticeGaugeTensor, Zohar2018EliminatingFermionicMatter}.
The resulting effective Hamiltonian is made out of only bosonic operators and directly acts on the gauge invariant Hilbert sub-space.
Correspondingly, as done in Loop String Hadrons methods \cite{Raychowdhury2020LoopStringHadron}, the original (Abelian/non-Abelian) gauge invariance is exactly rewritten into an Abelian, nearest-neighbor, diagonal selection rule, and the explicit dependence on the fermionic matter is eliminated \cite{Felser2020TwoDimensionalQuantumLinkLattice, Zohar2018EliminatingFermionicMatter, Zohar2019RemovingStaggeredFermionic}. 
Such an approach is especially suitable when dealing with LGTs in high-dimensional lattices.

We will practically apply the dressed-site formalism on two paradigmatic examples that have been exploited for numerical simulations \cite{Cataldi2024Simulating2+1DSU2,Cataldi*2025QuantumManybodyScarring,Magnifico2024TensorNetworksLattice}: SU(2) in \cref{sec_SU2_model} and U(1) in \cref{sec_U1_model}.
% =====================================================================
\subsection{Gauge field truncation}
\label{sec_gauge_truncation}
The link Hilbert space is the space of square-integrable functions on the gauge group $\gaugegroup$, $L^2(\gaugegroup)$, which is infinite-dimensional for continuous groups such as U(1) or SU(N) \cite{Zohar2015FormulationLatticeGauge}.

In one space dimension, gauge degrees of freedom are unphysical (absence of transverse polarizations) and can thus be integrated out, albeit at the price of introducing non-local interactions \cite{Sala2018VariationalStudySU}.
Beyond one dimension, the removal is much more delicate, because it requires first decoupling the gauge field's longitudinal component \cite{Bender2020GaugeRedundancyfreeFormulation}.
When impossible or inconvenient to remove, gauge degrees of freedom might have to be truncated to perform TN or quantum simulation.
Among known truncation recipes are Quantum Link Models (QLM) \cite{Horn1981FiniteMatrixModels,Orland1990LatticeGaugeMagnets,Chandrasekharan1997QuantumLinkModels,Brower1999QCDQuantumLink,Tagliacozzo2014TensorNetworksLattice}, an approach already considered for practical quantum simulation of LGTs \cite{Byrnes2006SimulatingLatticeGauge,Mathis2020ScalableSimulationsLattice,Davoudi2020AnalogQuantumSimulations,Mazzola2021GaugeinvariantQuantumCircuits,Kan2021InvestigatingMathrmTopological,Zohar2021QuantumSimulationLattice,Mariani2023HamiltoniansGaugeinvariantHilbert,Pomarico2023DynamicalQuantumPhase,Bauer2023QuantumSimulationFundamental,Bauer2023QuantumSimulationHighEnergy,Fontana2023QuantumSimulatorLink}, finite subgroups \cite{Ercolessi2018PhaseTransitionsGauge,Magnifico2020RealTimeDynamics,Haase2021ResourceEfficientApproach},
digitization of gauge fields \cite{Hackett2019DigitizingGaugeFields}, and fusion-algebra deformation \cite{Zache2023QuantumClassicalSpin}.

Another approach is to truncate the spectrum of the electric energy density operator $\norm{\casimir}\leq\casimircutoff$ on each link \cite{Cataldi2024Simulating2+1DSU2,Rigobello2023HadronsHamiltonianHardcore}.
The cutoff is conveniently imposed in the irreducible representation (irrep) basis $\{\ket{\spin \mL \mR}\}$ \cite{Zohar2015FormulationLatticeGauge} of $L^2(\gaugegroup)$, where $\casimir$ is diagonal:
\begin{equation}
  \casimir \ket{\spin \mL \mR} = C_2(\spin) \ket{\spin \mL \mR}\,.
\end{equation}
Here, $\mL$ and $\mR$ are indices in the $\spin$-irrep of $\gaugegroup$ and $C_2(\spin)$ is the quadratic Casimir of $\spin$ \cite{Zohar2015FormulationLatticeGauge}.
In the strong coupling limit, where the electric energy term dominates $\hamlgt$, this truncation is equivalent to an energy cutoff \cite{Rigobello2023HadronsHamiltonianHardcore}.
% =====================================================================
\subsection{Gauss law and the dressed site}\label{sec:dressed_site}
The most distinctive feature of gauge theories is arguably the presence of local constraints, analogous to the Gauss law of classical electrodynamics, relating the configuration of the gauge field to the spatial distribution of charges \cite{Strocchi2013NonperturbativeFoundationsQuantum}.
At the quantum level, Gauss law is the statement that only gauge invariant states are physical, namely,
\begin{math}
  \hat{G}_{\vecsite}^{a} \ket*{\Psi_{\text{phys}}} = 0 \;\forall \vecsite, a
\end{math},
where $\hat{G}_{\vecsite}^{a}$ are the generators of local gauge transformations at $\vecsite$:
\begin{equation}
  \hat{G}_{\vecsite}^{a} = \colormatter^{a}_{\vecsite} + q^{a}_{\vecsite} + \sum_{\latvec}[\eleL^{a}_{\genlink}+\eleR^{a}_{\genlinkm}]\,,
  \label{eq_gauss_law_op}
\end{equation}
with $q^{a}_{\vecsite}$ representing eventual static background charges (typically vanishing).
Lattice Gauss law provides a set of \emph{vertex constraints}, each involving a lattice site and its 2D neighboring links.

Due to Gauss law, the physical Hilbert space of LGTs is much smaller than the tensor product of all local sites and link Hilbert spaces.
Properly exploiting gauge symmetries can thus significantly speed up numerical simulations \cite{Silvi2014LatticeGaugeTensor}.
Strategies that solve Gauss law by eliminating (partially or entirely) either the gauge fields or the matter fields have been developed.
Nonetheless, such approaches come with specific limitations: the range of interaction has to be extended, moreover, integrating-out gauge fields become problematic in $D>1$ \cite{Bender2020GaugeRedundancyfreeFormulation}, while the recipe for removing matter is a model (matter content) dependent \cite{Zohar2019RemovingStaggeredFermionic}.

Another possibility is to enforce Gauss law using a dressed site construction \cite{Tagliacozzo2013SimulationsNonAbelianGauge, Silvi2014LatticeGaugeTensor, Zohar2018EliminatingFermionicMatter, Zohar2019RemovingStaggeredFermionic}, sketched in \cref{fig_LGT_sketch} and outlined below.
Dressed sites have local dimensions typically larger than those resulting from the aforementioned approaches, but they are obtained from a model-independent prescription which has the advantage of preserving the locality of the interactions \cite{Silvi2014LatticeGaugeTensor,Rigobello2023HadronsHamiltonianHardcore}.

As a first step, we factorize each gauge link in a pair of modes, living at its edges.
Namely, every link is decomposed into a pair of left ($L$) and right ($R$) \emph{rishon} \dof{}, each associated with a Hilbert space spanned by the basis states $\ket{\spin m}$, identifying $\ket{j \mL \mR} \hookrightarrow \ket{j \mL} \otimes \ket{j \mR}$,
and writing parallel transporters as rishon bilinears \cite{Cataldi2024Simulating2+1DSU2}:
\begin{equation}
  \NApara_{\genlink} \to \sum_{i}
  \rishon^{L(i)\alpha}_{\vecsite,+\latvec}\rishon^{R(i)\beta\,\dagger}_{\siteplus,-\latvec} \,
  \label{eq_from_U_to_rishons}
  \,.
\end{equation}
Physical gauge link configurations, \idest{} those with the left and right rishons in the same irrep, are selected introducing a local link symmetry at the TN simulation level \cite{Cataldi2024Edlgt,Silvi2014LatticeGaugeTensor,Rigobello2023HadronsHamiltonianHardcore}.
Notice that such a constraint is always Abelian, regardless of the Abelian or non-Abelian nature of the gauge group.
% =====================================================================
\subsection{Projecting dressed site operators onto the gauge-invariant basis}
\label{sec_projecting_gaugeinvariant_ops}
Crucially, the gauge generators $\hat{G}_{\vecsite}^{a}$ now involve only the matter site at $\vecsite$ and its $2D$ neighboring rishons.
Fusing these degrees of freedom in a composite site, Gauss law becomes an internal constraint that singles out the dressed site Hilbert space as its gauge invariant subspace:
\begin{equation}\label{eq:dressed_site}
  \hildress = \ker G \subset \hilsite \otimes (\hilsemi)^{\otimes 2D}\,.
\end{equation}
Therefore, the effective operators of the resulting dressed-site Hamiltonian should be obtained by projecting the obtained ones on the subspace generated by a gauge-invariant basis $M$.
Namely, for any dressed-site operator $\obs$ among the previously defined, the corresponding effective operator $\obs^{\text{eff}}$ acting on gauge-invariant states reads:
\begin{align}
    \obs^{\text{eff}}&=M^{\dagger}\cdot \obs\cdot M\,,&
    \text{where}&&
    M^{\dagger}\cdot M&=\mathbb{1} &
    \text{and}&&
    \qty(M\cdot M^{\dagger})^{2}&=\qty(M\cdot M^{\dagger}) \,.
\end{align}
The practical computation of the gauge-invariant basis $M$ can be determined as the kernel of the Gauss Law operator of the corresponding LGT (see \cref{sec_SU2_model} for SU(2) and \cref{sec_U1_model} for U(1)).
The expansion of the gauge singlet basis states of $\hildress$ in terms of the matter and rishon bases is computed via Clebsch-Gordan decomposition \cite{Rigobello2023HadronsHamiltonianHardcore}.

The resulting operators $\obs^{\text{eff}}$, for any spatial dimension D and any value of the gauge truncation $\jmax$, are available for U(1) and SU(2) in the GitHub repository \href{https://github.com/gcataldi96/ed-lgt}{ed-lgt} \cite{Cataldi2024Edlgt}, which also allows for simulations of these LGTs via Exact Diagonalization. 
% =====================================================================
\subsection{Defermionization}
\label{sec_defermionization}
As aforementioned in the introduction, an important challenge to be faced in simulating LGTs is to account for the Fermi statistics of matter fields. 
In several TN methods and conventional digital quantum simulation platforms \cite{Arute2020ObservationSeparatedDynamics,Barends2015DigitalQuantumSimulation,Salathe2015DigitalQuantumSimulation,OMalley2016ScalableQuantumSimulation,Stanisic2022ObservingGroundstateProperties}, fermionic (anticommuting) algebra must be encoded into a genuinely local (commuting) algebra of qub(d)its. 
Standard fermion-to-qubit encodings, such as the Jordan-Wigner (JW) transformation \cite{Jordan1928UeberPaulischeAequivalenzverbot}, and other modern approaches \cite{Nielsen2006QuantumComputationGeometry,Bravyi2002FermionicQuantumComputation,Jiang2019MajoranaLoopStabilizer} make any fermionic Hamiltonian interaction inherently long-range and expensive from a computational perspective \cite{Verstraete2008MatrixProductStates}, especially for higher-dimensional LGTs.

Within the dressed-site formalism, it is additionally possible resolve this problem and effectively eliminate the Fermi statistics of matter fields.
This is possible for any gauge theory where the gauge field has a well-defined parity. 
Specifically, a local parity operator $\parity_{\genlink} = \parity_{\genlink}^{\dagger}$ such that $\parity_{\genlink}^2 = 1$ must satisfy $\{\Apara_{\genlink},\parity_{\genlink}\} = 0$, as it happens for $\mathbb{Z}_{2N}$, U(N), and SU(2N) \cite{Zohar2018EliminatingFermionicMatter, Zohar2019RemovingStaggeredFermionic, Cataldi2024Simulating2+1DSU2}. 
In these cases, it is possible to consider rishons with a Fermi statistics, and, as a result, all physical (gauge invariant) dressed site operators are genuinely local, \idest{} they mutually commute at a nonzero distance (as spins or bosons) \cite{Cataldi*2024DigitalQuantumSimulation}. 

As the resulting Hamiltonian in the dressed site formalims is made out of different fermions (matter fields and gauge rishons), it is convenient to rule the tensor product of general fermion operators.
For a fermionic QMB system with particles arbitrarily sorted along a certain path, any tensor product of fermionic operators should take into consideration the proper anti-commutation rules. 
Namely, a generic fermionic operator $\hat{F}_{\vecsite}$ acting on the $\vecsite^{th}$ position along the path reads:
\begin{equation}
    \hat{F}_{\vecsite}=\begin{pmatrix}
        \hat{f}_{11}&\dots&\hat{f}_{1N}\\
        \vdots&&\vdots\\
        \hat{f}_{N1}&\dots&\hat{f}_{NN}\\
    \end{pmatrix}_{F} 
	=\dots \otimes \parity_{\siteminus}\otimes \hat{F}_{\vecsite} \otimes \mathbb{1}_{\siteplus}\otimes \dots \,,
    \label{fermionic_qmb_op}
\end{equation}
where $\parity_{\vecsite}=\parity_{\vecsite}^{\dagger}=\parity_{\vecsite}^{-1}$ is a fermion parity operator that gets inverted after the action of a fermionic operator:
\begin{equation}
\qty{\parity_{\vecsite},\hat{F}_{\vecsite}}=0 \qquad 
\qty[\parity_{\vecsite},\hat{F}_{\vecsite^{\prime}\neq \vecsite}]=0 
\qquad \forall \vecsite,\vecsite^{\prime}\in \Lambda\,.
\label{fermion_parity_commutation}
\end{equation}
Therefore, matter fields admit a notion of parity satisfying \cref{fermion_parity_commutation}. For Dirac fermions
\begin{align}
  \hpsi_{\text{Dirac}} &= \qty( 
  \begin{array}{cc}
   0 & 1 \\
   0 & 0
   \end{array})_F&
   \parity_{\text{Dirac}} &= \qty( 
  \begin{array}{cc}
   +1 & 0 \\
   0 & -1
   \end{array})  \,,
   \label{eq_parity_dirac}
\end{align}
where the subscript $F$ is a reminder that the $\hpsi$ matrix is meant 'as a fermion', with the global action in \cref{fermionic_qmb_op}.
Similarly, as for Majorana fermions, we have:
\begin{align}
    \hat{\gamma}_{\text{Majorana}} &= \qty( 
  \begin{array}{cc}
   0 & 1 \\
   1 & 0
   \end{array})_F&
   \parity_{\text{Majorana}} &= \qty( 
  \begin{array}{cc}
   {+}1 & 0 \\
   0 &  -1
   \end{array})  \,.
   \label{eq_parity_majorana}
\end{align}
Similarly, as for the incoming (fermion) rishon-operators, we will provide an adequate notion of parity (see \cref{eq_SU2_rishonparity} for SU(2) and \cref{eq_U1_rishonparity} for U(1)).
% =====================================================================
\section{SU(2) Yang-Mills Lattice Gauge Theory}
\label{sec_SU2_model} 
In this section, we apply the dressed-site formalism to a non-Abelian scenario, focusing on the SU(2) Yang-Mills (YM) LGT, including dynamical matter. 
For simplicity we will focus on a (2+1)D system, where both the electric and magnetic gauge contributions play a non-trivial role.
Generalizations to the (3+1)D are trivial, provided to consider the correct phase factors in \cref{eq_dirac_ham_stag3D} due to staggered fermions.
Restrictions to the (1+1)D case are considered in \cref{sec_SU2_hardcoregluon} and applied in \cite{Silvi2019TensorNetworkSimulation,Calajo2024DigitalQuantumSimulation,Cataldi*2025QuantumManybodyScarring}.

\subsection{The model} 
Before introducing the SU(2) YM Hamiltonian, let us recall the main properties of the SU(2) gauge theory.
First, the SU(2) algebra generators are the Pauli matrices $\hat{T}^{a}=\hat{\sigma}^{a}/2$, while the structure constants are equal to the Levi-Civita tensor $f^{abc}=\varepsilon^{abc}$.
Correspondingly, left/right generators of SU(2) gauge transformations $\eleL_{\genlink}/\eleR_{\genlink}$ are hermitian and related each other as follows: for $a,b,c\in \qty{x,y,z}$
\begin{align}
    \qty[\eleL_{\genlink}^{a},\eleR_{\genlink}^{b}]&=0 &
    \qty[\eleL_{\genlink}^{a},\eleL^{b}_{\vecsite^{\prime},\vecsite^{\prime}+\latvec'}]&=
    i\delta_{\vecsite \vecsite^{\prime}} \delta_{\latvec \latvec'}
    \epsilon_{abc}\eleL_{\genlink}^{c}&
    \text{(same with $\eleR$)}&\,.
    \label{eq_SU2_gauge_algebra1}
\end{align}
Their trace yields the Casimir operator
\begin{equation}
    \eleE^{2}_{\genlink}=\sum_{a}\eleL_{\genlink}^{a}\eleL_{\genlink}^{a} = \sum_{a}\eleR_{\genlink}^{a}\eleR_{\genlink}^{a}\,,
    \label{eq_SU2_casimir}
\end{equation}
and the commutations rules with the parallel transporter $\NApara$ read
\begin{align}
    \qty[\eleL^{a}_{\genlink},\NApara_{\genlinkprime}]
    &=-\delta_{\vecsite\vecsite^{\prime}} \delta_{\latvec'} \sum_{\gamma} \frac{{\sigma}^{a}_{\alpha\gamma}}{2} \Apara_{\genlink}^{\gamma\beta},&
    [\eleR^{a}_{\genlink},\NApara_{\genlinkprime}]
    &=+\delta_{\vecsite \vecsite^{\prime}} \delta_{\latvec \latvec'} \sum_{\gamma}  \Apara_{\genlink}^{\alpha \gamma}\frac{{\sigma}^{a}_{\gamma\beta}}{2}\,.
    \label{eq_SU2_gauge_algebra2}
\end{align}
Then, when introducing dynamical matter, we focus on single-flavor matter fields with SU(2)-color 1/2, expressed in terms of SU(2)-color staggered (Dirac) fermions $\hpsi_{\vecsite, \alpha}$ \cite{Susskind1977LatticeFermions} and satisfying
\begin{align}
    \qty{\hpsi^{\dagger}_{\vecsite,\alpha}\hpsi_{\vecsite^{\prime},\beta}}
    &=i\hbar\delta_{\vecsite,\vecsite^{\prime}}\delta_{\alpha,\beta},&
    \text{where}&& 
    \alpha,\beta&\in\qty{\rla,\gla}&
    \text{are SU(2)-colors}&\,.
    \label{eq_SU2_matter_commutation_rules}
\end{align}
From \cref{eq_gauss_law_op} and assuming no background charges, Gauss Law operators are defined as 
\begin{align}
    \hat{G}^{a}_{\vecsite}&= \colormatter^{a}_{\vecsite} +\sum_{\latvec}\qty(\eleR^{a}_{\vecsite-\latvec,\vecsite} + \eleL^{a}_{\genlink})\,,&
    \text{where}&&
    \colormatter^{a}_{\vecsite} &=
    \sum_{\alpha\beta}
    \hpsi^{\dagger}_{\vecsite,\alpha}\frac{\sigma^{a}_{\alpha\beta}}{2}\hpsi_{\vecsite,\beta}
    \label{eq_SU2_gausslaw}
\end{align}
rotates the quark field at $\vecsite$, and, if squared, it yields the \emph{matter color density}, \idest{} the quadratic Casimir operator of the matter field gauge group transformations:
\begin{equation}
    \mattercasimir=\sum_{a}\colormatter^{a}_{\vecsite}\colormatter^{a}_{\vecsite}.
    \label{eq_SU2_matter_casimir}
\end{equation}
Correspondingly, $\eleL^{a}_{\vecsite-\latvec,\vecsite}$ and $\eleR^{a}_{\genlink}$ account for the transformation of the gauge links at its left and its right respectively \cite{Zohar2015QuantumSimulationsLattice}.

Then, the corresponding (2+1)D Hamiltonian with staggered fermions (see \cref{sec_lattice_fermions}) reads:
\begin{equation}
  \begin{aligned}
    \ham_{\rm{SU(2)}}^{\rm{YM}}=&+
    \frac{\lspeed\hbar}{2\lspace}\sum_{\alpha,\beta}\sum_{\vecsite} \Big[\text{-i} \hpsi^{\dagger}_{\vecsite, \alpha} \Apara_{\genlink_x}^{\alpha\beta} \hpsi_{\siteplus_x,\beta} 
    - (-1)^{\site[x]+\site[y]}\hpsi^{\dagger}_{\vecsite,\alpha} \Apara_{\genlink_y}^{\alpha\beta} \hpsi_{\siteplus_y,\beta} + \hc \Big]\\
    &+ \mass[0] \lspeed^{2} \sum_{\vecsite} (-1)^{\site[x]+\site[y]}\sum_{\alpha} \hpsi^{\dagger}_{\vecsite,\alpha} \hpsi_{\vecsite,\alpha} + \ham_{\text{pure}}\,.
  \end{aligned}
  \label{eq_H_SU2_full}
\end{equation}
As for the pure gauge Hamiltonian, we have
\begin{equation}
    \begin{aligned}
      \ham_{\text{pure}}=&\coupling^{2}\frac{\lspeed\hbar}{2\lspace} \sum_{\genlink}\eleE^{2}_{\genlink}
      -\frac{\lspeed\hbar}{2\lspace \coupling^{2}}\sum_{\square}\Tr(\plaq+\plaq*)\,,
    \end{aligned}
    \label{eq_H_SU2_pure}
  \end{equation}
where $\plaq$ is the plaquette operator defined in \cref{eq_plaquette_operator}, while the gauge coupling $\coupling(\charge,\lspace)\propto \charge \sqrt{\lspace}$ defined \cref{eq_gauge_coupling} is dimensionless, but scales nonetheless with the lattice spacing $\lspace$ to ensure that the color charge $\charge$ of a quark stays finite in the continuum limit.
% ===================================================================== 
\subsection{Truncating the SU(2) gauge group}
\label{sec_SU2_gaugetruncation}
\begin{figure}
    \centering
    \includegraphics[width=1\textwidth]{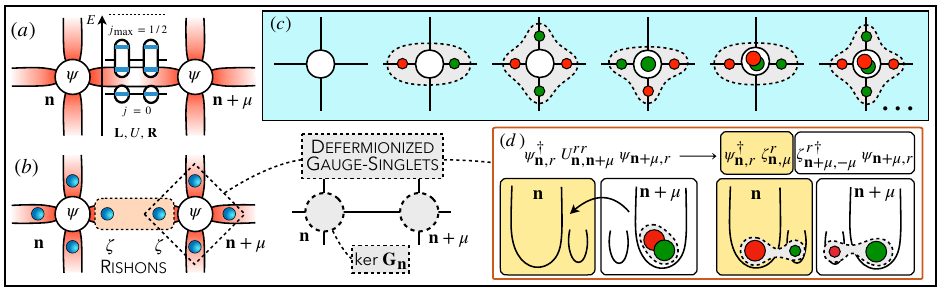}
    \caption{Sketched representation of the dressed-site model developed in \cref{sec_SU2_model} for SU(2): 
    (a) starting from the formulation with matter sites and truncated SU(2) gauge links, (b) we split the latter in pairs of rishon modes $\rishon$ defined in \cref{eq_SU2_general_rishon} and constrain their dynamics with the SU(2) link symmetry in \cref{SU2_linksymmetry}.
    We then (c) merge them with matter fields into dressed SU(2) gauge-singlets with a bosonic statistics. As a consequence, within the dressed-site formalism, the hopping always involves an even number of fermions, \idest{} the matter field plus a rishon mode (d).
    Figure inspired by \cite{Cataldi2024Simulating2+1DSU2, Magnifico2024TensorNetworksLattice}.}
    \label{fig_SU2_dressed_site_scheme}
    \end{figure}
As discussed in \cref{sec_gauge_truncation}, to make LGT Hamiltonians in \cref{eq_H_SU2_full,eq_H_SU2_pure} suitable for TN methods and quantum hardware, a finite-dimensional gauge-link Hilbert space is typically required. 
Here, we propose an effective truncation for the SU(2) gauge fields that can be applied to lattices of any spatial dimension and scalable to arbitrarily large truncations. 
In the limit of an infinitely large spin irreducible representation of the SU(2) gauge group, it recovers all the properties of the original SU(2) Yang-Mills LGT.  
Remarkably, the use of this approach in the smallest non-trivial truncation of SU(2), labeled as \emph{hardcore-gluon} approximation, has been used to achieve non-trivial results as the ones discussed in \cref{chap_SU2_groundstate} \cite{Cataldi2024Simulating2+1DSU2} and \cref{sec_scars_nonAbelianLGT} \cite{Calajo2024DigitalQuantumSimulation, Cataldi*2025QuantumManybodyScarring}. 
Of course, to accurately represent the full theory, for weak-$g$, larger gauge representations are required: this increases the computational challenges but it is still potentially accessible via TNs. 
Anyway, such a description of the gauge group is the first building block for building logical dressed-sites discussed in \cref{sec_dressed_site_formalism}, where Gauss Law is already satisfied (see \cref{fig_SU2_dressed_site_scheme}).

To truncate the continuous SU(2) gauge group, we express it in terms of the irreducible representation (irrep) basis \cite{Zohar2015FormulationLatticeGauge,Burgio2000BasisPhysicalHilbert}.
As SU(2) admits a \emph{quasi-real}-representation, where the fundamental and the anti-fundamental representations coincide, $\forall \genlink {\in} \Lambda$, the gauge Hilbert space of the $(\genlink)$ link can be written as:
\begin{align}
    \hillink&=\qty{\ket{\spin,\mL,\mR}},&
    \text{where}&&
    -\spin{\leq} m_{L(R)}{\leq}\spin
    \label{eq_gauge_hilbert_space}
\end{align}
is the corresponding third spin component associated with the left (right) side of the SU(2) link-irrep $\spin$.
Then, the parallel transporter $\NApara$ is given in terms of Clebsh-Gordan coefficients:
\begin{equation}
\bra{\spin^{\prime}\mL^{\prime}\mR^{\prime}}\NApara\ket{\spin\mL\mR}=
\sqrt{\frac{2\spin+1}{2\spin^{\prime}+1}}
\overline{C^{\spin \mL}_{\jonehalf,\alpha; \spin^{\prime}\mL^{\prime}} }
C^{\spin^{\prime} \mR^{\prime}}_{\spin \mR; \jonehalf,\beta}.
\label{eq_U_property1}
\end{equation}
Correspondingly, left and right generators are defined as:
\begin{subequations}
    \label{eq_LR_rishon_rotations}
    \begin{align}
        \label{eq_L_rishon_rotation}
        \langle \spin^{\prime} \mL^{\prime} \mR^{\prime}|\eleL^{a}| \spin \mL \mR \rangle & 
        =\delta_{\spin,\spin^{\prime}}S^{(\spin)a}_{\mL^{\prime},\mL}\delta_{\mR^{\prime},\mR},
        \\
        \label{eq_R_rishon_rotation}
        \langle \spin^{\prime} \mL^{\prime} \mR^{\prime} |\eleR^{a}| \spin \mL \mR \rangle & =
        \delta_{\spin,\spin^{\prime}}\delta_{\mL^{\prime},\mL} S^{(\spin)a}_{\mR^{\prime},\mR}\,.
     \end{align}
\end{subequations}
Then, the chromo-electric energy operator, \idest{} the quadratic Casimir in \cref{eq_SU2_casimir} reads \cite{Zohar2015QuantumSimulationsLattice}:
\begin{align}
    \eleE^{2}\ket{\spin\mL \mR } & =\spin(\spin+1)\ket{\spin \mL \mR }.
\end{align}
Then, the truncation of the gauge fields is applied by imposing a local energy cutoff $\Theta$ on the $\eleE^{2}$ spectrum, keeping the irreps $\spin\leq \jmax$  such that $C_{2}(\jmax)\leq\Theta$.
The truncation keeps the electric field operator $\eleE$ hermitian and protects the SU(2) gauge algebra rules of \cref{eq_SU2_gauge_algebra1,eq_SU2_gauge_algebra2}. 
Namely, truncated $\NApara$ fields satisfy the following left and right gauge transformations:
\begin{align}
  \qty[\eleL^{a}_{\jmax},\NApara]&=-\sum_{\gamma}\frac{\sigma^{a}_{\alpha\gamma}}{2}\Apara_{\gamma\beta} &
  \qty[\eleR^{a}_{\jmax},\Apara_{\alpha\beta}]&=+\sum_{\gamma}\Apara_{\alpha\gamma}\frac{\sigma^{a}_{\gamma \beta}}{2},
    \label{LR_generators}
\end{align}
where $\eleL^{a}_{\jmax}$ and $\eleR^{a}_{\jmax}$ are (truncated) generators of the left- and right-handed groups of SU(2) transformations. 
They can be expressed as the block-diagonal direct sum of spin matrices $S^{a}_{j}$ in consecutive $j$-representations from the smallest ($j=0$) to the largest one ($j=\jmax$):
\begin{equation}
\begin{aligned}
\eleL^{a}_{\jmax}&=\bigoplus_{j=0}^{\jmax}\qty(S^{a}_{j}{\otimes} \mathbb{1}_{j})=\diag(S^{a}_{0}{\otimes}\mathbb{1}_{0},\dots, S^{a}_{\jmax}{\otimes} \mathbb{1}_{\jmax})\\
\eleR^{a}_{\jmax}&=\bigoplus_{j=0}^{\jmax}\qty(\mathbb{1}_{j}{\otimes}S^{a}_{j})=\diag(\mathbb{1}_{0}{\otimes}S^{a}_{0},\dots,\mathbb{1}_{\jmax}{\otimes}S^{a}_{\jmax}).
\end{aligned}
\label{eq_LR_generator}
\end{equation}
Unfortunately, the truncation makes $\NApara$ unitary only as long as it acts on spin shells with $j<\jmax$ (it loses norm on the largest spin shell). 
Correspondingly, Wilson loops stay unitary as long as the outer spin shell $j=\jmax$ is nowhere populated. 
Namely, it implies that, $\forall j < \jmax$
\begin{equation}
\begin{aligned}
       {\sum_{\beta\gamma}}\NApara{\ket{0{;}0{,}0}}{\otimes} {\Apara_{\beta\gamma}}{\ket{j{;}\mL{,}\mR}}{\otimes} {\Apara_{\gamma\delta}}{\ket{0{;}0{,}0}}
\end{aligned}
\end{equation}
displays the same norm of $\Apara_{\alpha\delta}\ket{000}$ $\forall \alpha, \delta$, that is $1/\sqrt{2}$.
Ultimately, we require the parallel transporter to display \emph{spatial reflection symmetry}. 
Namely:
\begin{equation}
    \Apara^{\dagger}_{\alpha\beta}=-\hat{\mathcal{F}}\Apara_{\beta\alpha}\hat{\mathcal{F}},
\end{equation}
where $\hat{\mathcal{F}}$ is the \emph{swap operator} on a gauge link:
\begin{align}
\hat{\mathcal{F}}=\sum_{\spin,\mL,\mR}{\ket{\spin;\mL,\mR}}{\bra{\spin;\mR,\mL}}=\hat{\mathcal{F}}^{-1}=\hat{\mathcal{F}}^{\dagger}\,.
\end{align}
Whatever the gauge field truncation, in hopping terms, the action of $\Apara$ has to match the fundamental irrep of the matter field, whose Hilbert space in the Fock space, can be written as:
\begin{equation}
    \mathcal{H}_{\text{site}}=\qty{
    \ket{0}=\ket{\Omega},\;
    \ket{\rla}=\psi^{\dagger}_{\rla}\ket{\Omega},\;
    \ket{\gla}=\psi^{\dagger}_{\gla}\ket{\Omega},\;
    \ket{2}=\psi^{\dagger}_{\rla}
    \psi^{\dagger}_{\gla}\ket{\Omega}},
    \label{eq_matter_hilbert_space}
\end{equation} 
whose action in the irrep basis $\ket{\spin,m}$ corresponds to the following SU(2) charges:
\begin{equation}
\begin{split}
    \mathcal{H}_{\text{site}}=
    \qty{\ket{0,0},\ket{\mbox{$\frac{1}{2}$},\mbox{$\frac{1}{2}$}},\ket{\mbox{$\frac{1}{2}$};-\mbox{$\frac{1}{2}$}},\ket{0,0}}.
\end{split}
\end{equation}
% =====================================================================
\subsection{Rishon decomposition of SU(2) gauge fields}
\label{sec_SU2_rishondecomposition}
With all these premises, the resulting truncated SU(2) Yang-Mills Hamiltonian of \cref{eq_H_SU2_full}-\eqref{eq_H_SU2_pure} will act on the Hilbert space resulting from the tensor product of all the single matter Hilbert spaces and the corresponding ones of each truncated gauge field:
\begin{equation}
  \hil=\bigotimes_{\vecsite}\hilsite\bigotimes_{\genlink}\hil_{\rm{tr. link}}\,.
\end{equation}
Notice that matter and gauge fields display different statistics. 
In these terms, we expect $\Apara_{\genlink}^{\alpha\beta}$ to be \emph{mutually bosonic}, as it commutes with all the matter-fields operators:
\begin{equation}
  \qty[\Apara_{\genlink}^{\alpha\beta},\hpsi^{(\dagger)}_{\vecsite,\alpha}]=0
  \qquad \forall \vecsite,\forall \latvec, \forall \alpha,\beta
\end{equation}
and \emph{purely local}, as its link-algebra commutes with the one of any other link:
\begin{align}
  \qty[\NApara_{\genlink},\Apara_{\vecsite^{\prime},\vecsite^{\prime}+\latvec^{\prime}}^{\gamma\delta}]&=0 &
  \forall &\vecsite\neq \vecsite^{\prime},\latvec \neq \latvec^{\prime}, \forall \alpha,\beta,\gamma,\delta\,.
\end{align}
As anticipated in \cref{sec_dressed_site_formalism}, in the electric basis, the parallel transporter $\NApara$ admits a rishon-decomposition to arbitrary truncation of the maximum allowed spin shell $\jmax$.
Starting from a given spin shell $\spin$, we have to separately account for the action when both rishons are increased to shell $\spin+\frac{1}{2}$, and both are decreased to shell $\spin-\frac{1}{2}$. 
We can then decompose $\NApara$ as follows:
\begin{equation}
 \NApara_{\vecsite, \siteplus}=\rishon_{A,\genlink}^{\alpha} \rishon_{B,\siteplus,-\latvec}^{\beta\dagger}+\rishon_{B,\genlink}^{\alpha} \rishon_{A,\siteplus,-\latvec}^{\beta\dagger},
 \label{eq_SU2_U_definition}
\end{equation}
where the two $\rishon$-rishon species, A and B, act respectively as raising and lowering the spin shell of the SU(2) gauge irrep. Interestingly, they are related to each other as:
\begin{align}
    \rishon_{A}^{\alpha} &= i \sigma^{y}_{\alpha,\beta} \rishon_{B}^{\beta\dagger} &
    \rishon_{A}^ {\alpha\dagger} = i \sigma^{y}_{\alpha,\beta} \rishon_{B}^{\beta}\,.
\end{align}
We can then rewrite \cref{eq_SU2_U_definition} in terms of one species, \eg{} B. Renaming $\rishon_{B}^{\alpha}=\rishon^{\alpha}$, we obtain:
\begin{equation}
 \NApara_{\vecsite, \siteplus}= 
 i \sigma^{y}_{\alpha\gamma} \rishon_{\genlink}^{\gamma\dagger} \rishon_{\siteplus,-\latvec}^{\beta\dagger} 
 + i \sigma^{y}_{\beta\gamma} \rishon_{\genlink}^{\alpha} \rishon_{\siteplus,-\latvec}^{\gamma}\,,
\end{equation}
or equivalently
\begin{equation}
 \begin{aligned}
&\Apara^{\rla\rla}_{\vecsite, \siteplus}=\rishon_{\genlink}^{\gla\dagger} \rishon_{\siteplus,-\latvec}^{\rla\dagger} +
\rishon_{\genlink}^{\rla} \rishon_{\siteplus,-\latvec}^{\gla}\\
&\Apara^{\rla\gla}_{\vecsite, \siteplus} =\rishon_{\genlink}^{\gla\dagger} \rishon_{\siteplus,-\latvec}^{\gla\dagger} -
\rishon_{\genlink}^{\rla} \rishon_{\siteplus,-\latvec}^{\rla}\\
&\Apara^{\gla \rla}_{\vecsite, \siteplus}=-\rishon_{\genlink}^{\rla\dagger} \rishon_{\siteplus,-\latvec}^{\rla\dagger} + \rishon_{\genlink}^{\gla} \rishon_{\siteplus,-\latvec}^{\gla}\\
&\Apara^{\gla \gla}_{\vecsite, \siteplus}= -\rishon_{\genlink}^{\rla\dagger} \rishon_{\siteplus,-\latvec}^{\gla\dagger}- \rishon_{\genlink}^{\gla} \rishon_{\siteplus,-\latvec}^{\rla}\,.
 \end{aligned}
\end{equation}
For a chosen truncation $\jmax$ of the SU(2) irreducible representation, $\rishon$-rishons are defined as:
\begin{equation}
 \rishon^{\gla(\rla)}=\qty[\sum_{\spin=0}^{\jmax-\frac{1}{2}}\sum_{m=-\spin}^{\spin}\chi(\spin,m,\gla(\rla)) \ket{\spin,m}\bra{\spin+\mbox{$\frac{1}{2}$},m{\cmp}\mbox{$\frac{1}{2}$}}]_F\,,
 \label{eq_SU2_general_rishon}
\end{equation}
where the function $\chi(\spin,m,\alpha)$ reads
\begin{equation}
    \chi\qty(\spin,m,\gla(\rla))=\sqrt{\frac{\spin\cmp m+1}{\sqrt{(2j+1)(2j+2)}}}\,.
\end{equation}
It is possible to show that this construction is indeed compatible with the explicit form of the parallel transport reported in \cref{eq_U_property1}.
% =====================================================================
\subsubsection{SU(2) Rishon Parity}
\label{sec_SU2_rishon_parity}
Being fermions, $\rishon$-rishons anti-commute at different orbitals and with matter fields:
\begin{align}
  \qty{\rishon_{\genlink}^{\alpha},\rishon_{\siteplus,-\latvec}^{\beta}}&=0&
  \qty{\rishon_{\genlink}^{\alpha},\hpsi_{\vecsite,\beta}}&=0& \forall \alpha,\beta\,.
  \label{eq_SU2_zeta_commmutations}
\end{align}
To satisfy \cref{eq_SU2_zeta_commmutations}, we need to characterize them as fermion operators properly.
As discussed in \cref{sec_defermionization}, we can then introduce the notion of a parity for the SU(2) rishon modes. 
Similarly to \cref{eq_parity_dirac,eq_parity_majorana}, we define the SU(2) rishon parity operator $\parity_{\zeta}$ with an even ($+1$) parity sector on \emph{integer} irreps and odd ($-1$) sector on \emph{semi-integer} ones:
\begin{equation}
    \parity_{\zeta}=\diag(+1|-1,-1|+1,+1,+1|\dots)\,.
    \label{eq_SU2_rishonparity}
\end{equation} 
The action of the single rishon inverts the local parity: $\acomm*{\parity}{\rishon^\alpha} = 0$.
Correspondingly, the parallel transporter $\Apara_{\genlink}^{\alpha\beta}$ reads:
\begin{equation}
  \begin{split}
    \Apara_{\genlink}^{\alpha\beta}=& 
    i\sigma^{y}_{\alpha\gamma} \rishon_{\genlink}^{\gamma\dagger}\rishon_{\siteplus,-\latvec}^{\beta\dagger} 
    +i\sigma^{y}_{\beta\gamma} \rishon_{\genlink}^{\alpha} \rishon_{\siteplus,-\latvec}^{\gamma}\\
    =&+i\sigma^{y}_{\alpha\gamma}\qty(\rishon_{\genlink}^{\gamma\dagger}\cdot \parity_{\zeta, \genlink})\otimes \rishon_{\siteplus,-\latvec}^{\beta\dagger}
    +i\sigma^{y}_{\beta\gamma}\qty(\rishon_{\genlink}^{\alpha}\cdot \parity_{\zeta, \genlink})\otimes \rishon_{\siteplus,-\latvec}^{\gamma}\,.
  \end{split}
  \label{eq_U_rishons}
\end{equation}
Notice that the right-hand side of \cref{eq_U_rishons} preserves the SU(2) link parity, as desired.
% =====================================================================
\subsubsection{SU(2) Rishon Algebra}
\label{sec_SU2_rishon_algebra}
Instead of relying on two separate set of SU(2) generators, $\eleL^{a}$ and $\eleR^{a}$, $\rishon$-rishons have a unique gauge transformation algebra.  
The generators of SU(2) gauge rotations upon the $\rishon$-rishon space read:
\begin{equation}
\tgenerator^{a}_{\jmax}=\bigoplus_{\spin=0}^{\jmax}S^{a}_{\spin}=\diag(S^{a}_{0},S^{a}_{1},\dots S^{a}_{\jmax})\,.
\label{eq_T_generator}
\end{equation}
By construction, $\rishon$ operators are SU(2) covariant, as they transform as follows:
\begin{align}
  \qty[\rishon^{\alpha},\tgenerator^{a}]&=\frac{1}{2}\sum_{\beta}\hat{\sigma}^{a}_{\alpha\beta}\rishon^{\beta}&
  \qty[\tgenerator^{a},\rishon_{\alpha}^{\dagger}]&=\frac{1}{2}\sum_{\beta}\rishon^{\beta\dagger}\hat{\sigma}^{a}_{\beta\alpha}\,.
  \label{zeta_algebra}
\end{align}
Moreover, $\tgenerator^{a}$ is genuinely \emph{local}: for $\forall \vecsite\neq \vecsite^{\prime}$, $\forall \latvec\neq\latvec'$, and $\forall \alpha\in\qty{\rla,\gla}$:
\begin{equation}
\big[\tgenerator^{a}_{\vecsite, +\latvec},\rishon_{\vecsite^{\prime},+\latvec'}^{\alpha}\big]
=\big[\tgenerator^{a}_{\vecsite, +\latvec},\hpsi_{\vecsite^{\prime}, \alpha}\big]=0\,.
\end{equation}
We easily recover the left and right generators of the gauge field at link $(\genlink)$ as:
\begin{align}
  \eleL^{a}_{\vecsite,+\latvec}&=\tgenerator^{a}_{\vecsite,+\latvec}{\otimes} \mathbb{1}_{\siteplus,-\latvec}&
  \eleR^{a}_{\vecsite,+\latvec}&=\mathbb{1}_{\vecsite,+\latvec}{\otimes}\tgenerator^{a}_{\siteplus,-\latvec}\,.
  \label{eq_LR_rishon_generator}
\end{align}
Since in the SU(2) group the fundamental and the anti-fundamental representations coincide, the rishon formalism is meaningful as long as the quadratic Casimir operator of the two sides of the link coincide as in \cref{eq_casimir}: 
\begin{equation}
  \abs{\eleL^{a}_{\vecsite, +\latvec}}^{2}=\abs{\eleR^{a}_{\siteplus, -\latvec}}^{2}\,.
  \label{SU2_linksymmetry}
\end{equation}
Thanks to \cref{SU2_linksymmetry}, the two rishons of the link are in the same SU(2) irrep, and the parallel transport in \cref{eq_SU2_U_definition} coincides with the one in \cref{eq_U_property1}. 
Correspondingly, the Casimir operator of the $(\genlink)$ link in \cref{eq_casimir} can be expressed as:
\begin{equation}
  \begin{split}
  &\eleE^{2}_{\genlink}
  = \frac{1}{2}\qty[\abs{\eleL^{a}_{\vecsite,+\latvec}}^{2} +\abs{\eleR^{a}_{\siteplus,-\latvec}}^{2}]= \frac{1}{2}\qty[\abs{\tgenerator^{a}_{\vecsite,+\latvec}}^{2} +\abs{\tgenerator^{a}_{\siteplus,-\latvec}}^{2}]\,,
  \end{split}
  \label{eq_Electric_casimir}
\end{equation}
which looks explicitly symmetric under link reversal.
% =====================================================================
\subsection{Constructing dressed site operators}
\label{sec_SU2_dressedsite_operators}
We have then all the ingredients to build a \emph{dressed-site} compact representation. In the case of a 2D lattice, one possible pictorial description of \emph{dressed-site} states reads:
\begin{align}
    \ket{\begin{array}{ccc}
        &\rishon_{\vecsite,+\latvec[y]}&\\
        \rishon_{\vecsite,-\latvec[x]}&\qty(\hpsi_{\vecsite,\rla}\hpsi_{\vecsite,\gla})&\rishon_{\vecsite,+\latvec[x]}\\
        &\rishon_{\vecsite,-\latvec[y]}&\\
    \end{array}},&&\text{where}&&
    \ket{\begin{array}{ccc}
        &5&\\
        2&(0,1)&4\\
        &3&\\
    \end{array}}
    \label{eq_SU2_dressed_site}
\end{align}
is a possible internal ordering to be used as in \cref{fermionic_qmb_op} when constructing composite operators out of matter fields and rishons inside the dressed site. 

We can then rewrite the SU(2) Yang-Mills Hamiltonian in \cref{eq_H_SU2_full}-\eqref{eq_H_SU2_pure} via rishon modes.

\paragraph{Arrival operators}
Let us start with the hopping Hamiltonian term. Discarding all the pre-factors, we have\footnote{Just for practical implementation in numerical simulations, the definition of the arrival operators here recalls the parallel transporter definition in terms of the two rishon species. Of course, an equivalent definition in terms of only one of the two species is available.}:
\begin{equation*}
    \begin{split}
\ham^{\text{hopping}}_{\genlink}&=\sum_{\alpha,\beta}\hpsi^{\dagger}_{\vecsite ,\alpha} \Apara_{\genlink}^{\alpha\beta} \hpsi_{\siteplus,\beta}\\
    &=\sum_{\alpha,\beta}\hpsi^{\dagger}_{\vecsite,\alpha}
\qty[\rishon_{A,\genlink}^{\alpha} \rishon_{B,\siteplus,-\latvec}^{\beta\dagger} 
 {+}\rishon_{B,\genlink}^{\alpha} \rishon_{A,\siteplus,-\latvec}^{\beta\dagger}]
 \hpsi_{\siteplus,\beta}\\
    &=\qty[\hat{Q}^{\dagger}_{A,\genlink}\hat{Q}_{B,\siteplus,-\latvec}+\hat{Q}^{\dagger}_{B,\vecsite}\hat{Q}_{A,\siteplus,-\latvec}],
    \end{split}
\end{equation*}
where we defined two species of \emph{arrival} operators:
\begin{align}
\hat{Q}^{\dagger}_{A,\genlink}&=\sum_{\alpha}\hpsi^{\dagger}_{\vecsite,\alpha}\rishon_{A,\genlink}^{\alpha}&
\hat{Q}^{\dagger}_{B,\genlink}&=\sum_{\alpha}\hpsi^{\dagger}_{\vecsite,\alpha} \rishon_{B,\genlink}^{\alpha}.
\label{eq_arrival_operators}
\end{align}
These operators' practical construction must be consistent with the internal order in \cref{eq_SU2_dressed_site}.
\paragraph{Matter number density operators} 
Similarly, number density operators are expressed as
\begin{align}
        \hat{N}_{\vecsite,\rla}&
        =\psi^{\dagger}_{\vecsite}\psi_{\vecsite,\rla}
        \otimes\mathbb{1}_{\vecsite,\gla}\bigotimes_{\latvec} \mathbb{1}_{\genlink}&
        \hat{N}_{\vecsite,\gla}&
        =\mathbb{1}_{\vecsite,\rla}
    \otimes \psi^{\dagger}_{\vecsite}\psi_{\vecsite,\gla}\bigotimes_{\latvec} \mathbb{1}_{\genlink}
    \label{eq_number_operators1}
\end{align}
and give access to other observables like:
\begin{align}
    \hat{N}_{\vecsite,\text{tot}}&
    =\hat{N}_{\vecsite,\rla}+\hat{N}_{\vecsite,\gla}& 
    \hat{N}_{\vecsite,\text{pair}}&
    =\hat{N}_{\vecsite,\rla}\hat{N}_{\vecsite,\gla}&
    \hat{N}_{\vecsite,\text{single}}&=\hat{N}_{\vecsite,\text{tot}}-\hat{N}_{\vecsite,\text{pair}}\,,
\label{eq_number_operators2}
\end{align}
which respectively measure the total matter density and the corresponding occupancy of pairs or single particles.
\paragraph{Casimir operator}
As we assumed via \cref{eq_Electric_casimir} that each of the two $\rishon$-rishons equally contributes to the link-electric energy density, we can equivalently define a dressed-site operator summing the Casimir contributions from its attached rishons:
\begin{equation}
    \hat{\Gamma}_{\vecsite}=\frac{1}{2}\sum_{\latvec}(\tgenerator^{a}_{\genlink})^{2}\,.
    \label{eq_gamma_operators}
\end{equation}
\paragraph{Magnetic Operators}
Expressing $\Apara$ as in \cref{eq_SU2_U_definition}, the plaquette term gives rise to 16 contributions:
\begin{equation*}
    \begin{split}
        \Apara_{\square}&=\sum_{\alpha,\beta,\delta,\gamma}\qty[\Apara_{\genlink_x}^{\alpha\beta}  
        {\Apara_{\siteplus_x,\siteplus_x+\latvec[y]}^{\beta\gamma}}
        {\Apara_{\siteplus_x+\latvec[y],\siteplus_y}^{\gamma\delta}}
        {\Apara_{\siteplus_y,\vecsite}^{\delta\alpha}}]\\
        &=\sum_{\alpha,\beta,\delta,\gamma}
        \qty(\begin{array}{ccc}
        \ulcorner&\rishon^{\delta\dagger}_{B,+\latvec[x]} \rishon_{A,-\latvec[x]}^{\gamma}& \urcorner\\
        \rishon_{A,-\latvec[y]}^{\delta}&  & \rishon^{\gamma\dagger}_{B,-\latvec[y]}\\
        \rishon^{\alpha\dagger}_{B,+\latvec[y]}&  &\rishon_{A,+\latvec[y]}^{\beta}\\
        \llcorner& \rishon_{A,+\latvec[x]}^{\alpha} \rishon^{\beta\dagger}_{B,-\latvec[x]}&\lrcorner\\
        \end{array}) +\dots\\
        &\underset{*}= -[\qty(\begin{array}{ccc}
        C^{AA}_{-\latvec[y],+\latvec[x]}&\rule{0.4cm}{0.5 pt}& C^{AA}_{-\latvec[x],-\latvec[y]}\\
        |&  &|\\
        C^{AA}_{+\latvec[x],+\latvec[y]}&\rule{0.4cm}{0.5 pt}&C^{AA}_{+\latvec[y],-\latvec[x]}]\\
        \end{array}) +\dots,
    \end{split}
\end{equation*}
where, in $*$, we combined rishons in pairs to form \emph{corner operators} like the following:
\begin{equation}
\begin{aligned}
    \hat{C}^{AA}_{\vecsite,\latvec[1],\latvec[2]}&=\sum_{\alpha}
    \rishon_{A,\vecsite,\latvec[1]}^{\alpha}\rishon^{\alpha\dagger}_{A,\vecsite,\latvec[2]}
    =\sum_{\alpha, \kappa, \kappa'}i\sigma^{y}_{\alpha,\kappa}
    \rishon_{B,\vecsite,\latvec[1]}^{\kappa\dagger}i\sigma^{y}_{\alpha,\kappa'}\rishon^{\kappa'}_{B,\vecsite,\latvec[2]}\\
    \hat{C}^{BB}_{\vecsite,\latvec[1],\latvec[2]}
    &=\sum_{\alpha}\rishon_{B,\vecsite,\latvec[1]}^{\alpha}\rishon^{\alpha\dagger}_{B,\vecsite,\latvec[2]}=\hat{C}^{AA}_{\vecsite,\latvec[2],\latvec[1]}\\
    \hat{C}^{AB}_{\vecsite,\latvec[1],\latvec[2]}&=
    \sum_{\alpha}\rishon_{A,\vecsite,\latvec[1]}^{\alpha}\rishon^{\alpha\dagger}_{B,\vecsite,\latvec[2]}
    =\qty(\hat{C}^{BA}_{\vecsite,\latvec[2],\latvec[1]})^{\dagger}.
\end{aligned}
\label{eq_corner_operators}
\end{equation}
As for the previous dressed-site operators, the practical construction of corner operators has to be consistent with the internal ordering in \cref{eq_SU2_dressed_site}.

\paragraph{SU(2) Link Symmetry}
Within the dressed-site formalism, the condition in \cref{SU2_linksymmetry} requiring the two rishon of the links to display the same Casimir operator (\idest to be in the same spin shell $j$) is simply an Abelian $\mathbb{Z}_2$ Link Symmetry. 
It can be then easily encoded in ED and TN libraries employing symmetries \cite{Silvi2019TensorNetworksAnthology,Cataldi2024Edlgt}.

\subsubsection{Projecting the operators on the SU(2) gauge invariant subspace}
All the dressed-site operators previously derived must be projected in the SU(2) gauge-invariant basis $M$. 
The latter is made by SU(2) singlets and can be determined as the kernel of the Gauss Law operators $\hat{\vb{G}}_{\vecsite}^{a}$:
\begin{equation}
    \hat{\vb{G}}_{\vecsite}\cdot M=\qty[\mattercasimir+\sum_{\latvec}\sum_{a}\tgenerator_{\vecsite, \latvec}^{a}\tgenerator_{\vecsite, \latvec}^{a}]\cdot M=0,
\end{equation}
where $\tgenerator^{a}_{\latvec}$ are the $\rishon$-rishon generators along the $\latvec^{\rm{th}}$ direction and defined in \cref{eq_T_generator}, while $\mattercasimir$ is the matter color density introduced in \cref{eq_SU2_matter_casimir}.
In the minimal $\jmax=\frac{1}{2}$ truncation of SU(2), the gauge-invariant Hilbert space of every dressed site of the full Hamiltonian has 30 gauge invariant states, whereas, restricting to the pure theory, the local Hilbert space is 9-dimensional. 

\subsubsection{The operative defermionized Hamiltonian}
We are then ready to rewrite the SU(2) lattice Yang-Mills Hamiltonian in \cref{eq_H_SU2_full} making use of dressed-site operators in \cref{eq_arrival_operators,eq_corner_operators}. 
Namely, we have:
\begin{equation}
  \begin{split}
    {\ham}{=}&{-}\frac{\lspeed \hbar}{2\lspace}\sum_{\vecsite}
    \qty[\qty[i\qty[\hat{Q}^{\dagger}_{A,\vecsite,+\latvec[x]}\hat{Q}_{B,\siteplus_x,-\latvec[x]}]{+}( -1)^{\site[x]+\site[y]}\qty[\hat{Q}^{\dagger}_{A,\vecsite,+\latvec[y]}\hat{Q}_{B,\siteplus_y,-\latvec[y]}]{+}(A{\rightleftarrows}B)]+\hc]\\
    &+ \mass[0]\lspeed^{2}\sum_{\vecsite}(-1)^{\site[x]+\site[y]}\hat{N}_{\vecsite,\text{tot}} + \frac{\lspeed \hbar \coupling^{2}}{4} \sum_{\vecsite} \hat{\Gamma}_{\vecsite}
    \\
    &- \frac{\lspeed \hbar}{2\lspace\coupling^{2}}\sum_{\square}\Tr\qty[\qty[-\begin{matrix}
      \hat{C}^{AA}_{\ulcorner} &\hat{C}^{AA}_{\urcorner}\\
      \hat{C}^{AA}_{\llcorner} &\hat{C}^{AA}_{\lrcorner}
       \end{matrix}
       +\begin{matrix}
      \hat{C}^{BB}_{\ulcorner} &\hat{C}^{AB}_{\urcorner}\\
      \hat{C}^{AA}_{\llcorner} &\hat{C}^{AB}_{\lrcorner}
       \end{matrix}+\dots]+\hc].
   \end{split}
   \label{H_full_pt2}
\end{equation}
Not surprisingly, \cref{H_full_pt2} is completely \emph{bosonic}, as all the dressed-site operators are made out of pairs of fermions (matter field + rishon, pairs of matter fields, or rishon pairs). 
Then, fermionic degrees of freedom are completely hidden inside each dressed site and there is no more need to face anti-commutation rules (see \cref{fig_SU2_dressed_site_scheme}(d)).
% =====================================================================
\subsection{Hardcore-gluon model: minimally truncated gauge-fields}
\label{sec_SU2_hardcoregluon}
As an example, we consider the \emph{hardcore-gluon} approximation, corresponding to the smallest non-trivial representations of the gauge fields, obtained truncating the Casimir up to $\jmax=\frac{1}{2}$.
In analogy to cold quantum gases, the label \emph{hardcore-gluon} aims to stress that the only accessible local configurations are those states reachable from the bare vacuum with at most one application of $\NApara$.
Namely, we consider $(0{\otimes} 0){\oplus}(\frac{1}{2}{\otimes}\frac{1}{2})$ as the gauge field space, where $(j)$ is the irreducible representation of SU(2) \cite{Horn1981FiniteMatrixModels, Orland1990LatticeGaugeMagnets, Brower1999QCDQuantumLink, Yao2023SU2NonAbelianGauge, Muller2023SimpleHamiltonianQuantum}. 
Such a truncation is enough to guarantee the contribution of all the terms in the Hamiltonian \cref{eq_H_SU2_full,eq_H_SU2_pure}, and provides a good approximation of the untruncated low energy physics in the strong coupling limit $g\gg1$, where the gauge field energy term dominates the Hamiltonian.
Remarkably, it has been succesfully adopted in \cite{Cataldi2024Simulating2+1DSU2, Cataldi*2025QuantumManybodyScarring, Calajo2024DigitalQuantumSimulation} to extract interesting physics in and out of equilibrium (see \cref{chap_SU2_groundstate,sec_scars_nonAbelianLGT} respectively).

Within the \emph{hardcore-gluon} approximation, the gauge-link Hilbert space reduces to the states $\ket{\spin, m}$ with $\jmax=\frac{1}{2}$. 
In analogy to the fundamental representation of SU(2), we can parametrize the $m=\pm\frac{1}{2}$ values with the colors $\rla,\gla$ respectively:
\begin{equation}
  \hillink=
  \qty{\ket{0,0},\ket{\rla,\rla},\ket{\rla,\gla},
  \ket{\gla,\rla},\ket{\gla,\gla}}\,.
  \label{link_5D_Hilbert_space}
\end{equation}
We can then define the corresponding versions of the truncated gauge fields. 
As for the quadratic Casimir operator in \cref{eq_casimir}, we have:
\begin{equation}
  \casimir=\frac{3}{4}\diag(0|+1,+1,+1,+1)\,.
  \label{eq_SU2_casimir_matrix}
\end{equation}
Correspondingly, the truncated parallel transport reduces to \cite{Zohar2015FormulationLatticeGauge}:
\begin{equation}
    \NApara={\frac{1}{\sqrt{2}}}
    \qty(
      \begin{array}{c|cccc}
    0&{+}\delta_{\alpha\rla}\delta_{\beta\gla} 
     & -\delta_{\alpha\rla}\delta_{\beta\rla}  
     &{+}\delta_{\alpha\gla}\delta_{\beta\gla}
     & -\delta_{\alpha\gla}\delta_{\beta\rla}\\
    \hline
     -\delta_{\alpha\gla} \delta_{\beta\rla}&0&0&0&0\\
     -\delta_{\alpha\gla} \delta_{\beta\gla}&0&0&0&0\\
    {+}\delta_{\alpha\rla} \delta_{\beta\rla}&0&0&0&0\\
    {+}\delta_{\alpha\rla} \delta_{\beta\gla}&0&0&0&0\\
  \end{array})\,,
    \label{SU2_parallel_transport}
\end{equation}
where the $1/\sqrt{2}$ factor ensures that the hopping term preserves the state norm on its support. 
Similarly, also the rishon-decomposition in \cref{eq_SU2_U_definition} can be simplified as follows:
\begin{equation}
    \NApara_{\genlink}=
    \rishon_{\genlink}^{\alpha} \rishon_{\siteplus,-\latvec}^{\beta\dagger}\,,
\end{equation}
where $\rishon$-rishons in \cref{eq_SU2_general_rishon} reduce to:
\begin{equation}
    \begin{aligned}
        \rishon_{\rla} &=\frac{\ketbra{0}{\rla}{+}\ketbra{\gla}{0}}{\sqrt[4]{2}}
        =\frac{1}{\sqrt[4]{2}} \qty( 
        \begin{array}{c|cc}
            0 & 1 & 0 \\
            \hline
            0 & 0 & 0 \\
            1 & 0 & 0 \\  
        \end{array})_{F}&
        \rishon_{\gla} &=\frac{\ketbra{0}{\gla}{-}\ketbra{\rla}{0}}{\sqrt[4]{2}}
        =\frac{1}{\sqrt[4]{2}} \qty( 
        \begin{array}{c|cc}
            0 & 0 & 1 \\
            \hline
            {-}1 & 0 & 0 \\
            0 & 0 & 0 \\  
        \end{array})_{F}\,,
    \label{zeta_definition}
    \end{aligned}
\end{equation}
with the corresponding parity operator
\begin{equation}
    \parity_{\zeta} = +\op{0} - (\op{\rla} + \op{\gla}) = \diag(+1|-1,-1)
   \label{eq_SU2_rishon_parity}
\end{equation}
and SU(2) rishon-generators
\begin{align}
    \hat{T}_{1/2}^{x}&=\frac{1}{2}
    \qty(\begin{array}{c|cc}
        0& \\
        \hline
        & 0&1\\
        & 1&0\\
    \end{array})&
    \hat{T}_{1/2}^{y}&=\frac{1}{2}
    \qty(\begin{array}{c|cc}
        0& \\
        \hline
        & 0& -i\\
        & i&0\\
    \end{array})&
    \hat{T}_{1/2}^{z}&=\frac{1}{2}
    \qty(\begin{array}{c|cc}
        0& \\
        \hline
        & 1&0\\
        & 0& -1\\
    \end{array})\,.
    \label{eq_T1/2}
\end{align}
By definition, the spin-Hilbert space of every side of the link $(\genlink)$ is 3-dimensional:
\begin{equation}
    \mathcal{H}_{\vecsite,+\latvec}
    =\qty{\ket{0},\ket{\rla},\ket{\gla}}
    =\mathcal{H}_{\siteplus, -\latvec}.
\end{equation}
Correspondingly, the Hilbert space of the whole link $\mathcal{H}_{\text{link}}=\mathcal{H}_{\vecsite,+\latvec}\otimes \mathcal{H}_{\siteplus, -\latvec}$ has 9 states.
To recover the original 5-dimensional space in \cref{link_5D_Hilbert_space}, we must constrain the left and right rishons on each link to be in the same spin shell $\spin$ by selecting the even parity sector of a local $\mathbb{Z}_2$ symmetry defined in \cref{SU2_linksymmetry}.

\subsubsection{(1+1)D hardcore-gluon SU(2) LGT: a qudit model}
\label{sec_su2_qudit_model_1D}
In one spatial dimension, the Hamiltonian in \cref{eq_H_SU2_full} reduces to
\begin{equation}
    \begin{aligned}
    \ham_{\rm{SU(2)}}^{\rm{\rm{1D}}}=&
    \frac{\lspeed\hbar}{2\lspace}\sum_{\alpha,\beta}\sum_{\vecsite}\qty[-i\hpsi^{\dagger}_{\vecsite, \alpha} \NApara_{\genlink}\hpsi_{\siteplus,\beta}+\hc]
    + \mass[0] \lspeed^{2} \sum_{\vecsite,\alpha}(-1)^{\vecsite}\hpsi^{\dagger}_{\vecsite,\alpha} \hpsi_{\vecsite,\alpha}+\frac{\coupling^{2}\lspeed\hbar}{2\lspace}\sum_{\genlink}\eleE^{2}_{\genlink}.
    \end{aligned}
    \label{eq_SU2_1D_Hamiltonian}
\end{equation}
Correspondingly, the dressed-site formalism obtained in the \emph{hardcore-gluon} approximation results in a 6-dimensional local basis \cite{Calajo2024DigitalQuantumSimulation}:
\begin{equation}
    \qty{
    \ket{\alpha} =\ket{\mR^{\alpha}({\vecsite-\latvec,\vecsite}),\charge^{\alpha}(\vecsite),\mL^{\alpha}({\genlink})}
    }_{\alpha=1}^6
\end{equation}
which is made of the following states
\begin{equation}
    \begin{aligned}
        \ket{1} & =\ket{0,0,0}                                          , &
        \ket{2} & =\frac{\ket{\rla,0,\gla}-\ket{\gla,0,\rla}}{\sqrt{2}} ,   \\
        \ket{3} & =\frac{\ket{\gla,\rla,0}-\ket{\rla,\gla,0}}{\sqrt{2}} , &
        \ket{4} & =\frac{\ket{0,\rla,\gla}-\ket{0,\gla,\rla}}{\sqrt{2}} ,   \\
        \ket{5} & =\ket{0,2,0}                                          , &
        \ket{6} & =\frac{\ket{\rla,2,\gla}-\ket{\gla,2,\rla}}{\sqrt{2}} .
    \end{aligned}
    \label{eq_SU2_hardcoregluon_basis1D}
\end{equation}
Similarly to the procedure in \cref{sec_SU2_dressedsite_operators}, the construction of the dressed-site operators and their expression in the 6-dimensional qudit basis in \cref{eq_SU2_hardcoregluon_basis1D} follows: 
\begin{subequations}
    \begin{align}
    \hat{Q}_{\vecsite, L}=\hat{Q}_{\vecsite,+\latvec}=\sum_{\alpha}\rishon_{\vecsite,+\latvec}^{\dagger}\hpsi_{\vecsite}&=\sqrt{2}\ketbra{1}{4}+\ketbra{2}{3}+\ketbra{3}{6}+\sqrt{2}\ketbra{4}{5},\\
    \hat{Q}_{\vecsite,R}=\hat{Q}_{\vecsite,-\latvec}=-i\sum_{\alpha}\rishon_{\vecsite,-\latvec}^{\dagger}\hpsi_{\vecsite}&=\sqrt{2}\ketbra{1}{3}+\ketbra{2}{4}+\sqrt{2}\ketbra{3}{5}+\ketbra{4}{6},\\
    \hat{N}_{\vecsite}=\sum_{\alpha}\hpsi^{\dagger}_{\vecsite,\alpha} \hpsi_{\vecsite,\alpha}&=\ketbra{3}{3}+\ketbra{4}{4}+2\ketbra{5}{5}+2\ketbra{6}{6},\\
    \hat{\Gamma}_{\vecsite}=\frac{1}{2}\qty[\casimir_{\vecsite, +\latvec}+\casimir_{\vecsite, -\latvec}]&=2\ketbra{2}{2}+\ketbra{3}{3}+\ketbra{4}{4}+2\ketbra{6}{6}\,.
    \end{align}
\end{subequations}
For potential quantum simulation implementations, it is convenient to rewrite \cref{eq_SU2_1D_Hamiltonian} solely in terms of Hermitian operators, introducing \cite{Calajo2024DigitalQuantumSimulation}
\begin{equation}
    \begin{aligned}
        \vecpot^{(1)} & = \hat{Q}_{L} + \hat{Q}_{L}^{\dagger}, &
        \hat{B}^{(1)} & = \hat{Q}_{R} + \hat{Q}_{R}^{\dagger}, \\
        \vecpot^{(2)} & = i\qty[\hat{Q}_{L}-\hat{Q}_{L}^{\dagger}],&
        \hat{B}^{(2)} & = i \qty[\hat{Q}_{R} - \hat{Q}_{R}^{\dagger}].
    \end{aligned}
\end{equation}
Noting that $(\vecpot^{(1)}\otimes \vecpot^{(1)}+ \vecpot^{(2)}\otimes \vecpot^{(2)})=2(\hat{Q}_{L}\otimes\hat{Q}_{R}^{\dagger}+\hat{Q}_{L}^{\dagger}\otimes\hat{Q}_{R})$
Finally, we map \cref{eq_SU2_1D_Hamiltonian} to the following 6-dimensional qudit Hamiltonian \cite{Cataldi*2025QuantumManybodyScarring}:
\begin{equation}
    \ham_{\rm{SU(2)}}^{\rm{\rm{1D}}}=\frac{\lspeed\hbar}{4\lspace}\sum_{\vecsite}\sum_{k\in\{1,2\}}\qty[\hat{A}^{(k)}_{\vecsite}\hat{B}^{(k)}_{\vecsite+\latvec}]
    +\mass[0]\lspeed^{2}\sum_{\vecsite}(-1)^{\vecsite}\hat{M}_{\vecsite}+\frac{g^{2}\lspeed\hbar}{4\lspace}\sum_{\genlink}\hat{\Gamma}_{\vecsite}\,.
    \label{eq_SU2_Hamiltonian1D_dressed_site}
\end{equation}
Since the basis in \cref{eq_SU2_hardcoregluon_basis1D} is already gauge invariant, the Hamiltonian in \cref{eq_SU2_Hamiltonian1D_dressed_site} is tailored for TN simulations as well as to a six-level trapped-ion qudit quantum processor \cite{Ringbauer2022UniversalQuditQuantum}. 
In \cite{Calajo2024DigitalQuantumSimulation}, the authors discuss the experimental feasibility of using generalized Mølmer-Sørensen gates to efficiently simulate dynamics of \cref{eq_SU2_Hamiltonian1D_dressed_site}. 
A shallow circuit with these resources is demonstrated to be sufficient for implementing scalable digital quantum simulations of the model.
Numerical simulations of the corresponding dynamics manifest physically-relevant properties specific to non-Abelian field theories, such as baryon excitations \cite{Calajo2024DigitalQuantumSimulation}, and compelling non-ergodic behaviors compatible with quantum many-body scars \cite{Cataldi*2025QuantumManybodyScarring} and detailed in \cref{sec_scars_nonAbelianLGT}.

\subsubsection{(2+1)D hardcore-gluon SU(2) LGT}
In two spatial dimensions, where the logical site is structured as in \cref{eq_SU2_dressed_site}, the previous 6-dimensional dressed-site basis becomes a 30-dimensional Hilbert space.
The first 9 states are combinations of only gauge fields:
\begin{equation}
    \label{eq_2DSU2_dressed_basis_pure}
    \begin{aligned}
        \ket{0} & = \five{0}{0}{0}{0}{0}\\
		\ket{1} & = \frac{1}{\sqrt{2}}\five{0}{0}{0}{\rla}{\gla}-\frac{1}{\sqrt{2}}\five{0}{0}{0}{\gla}{\rla} \qquad \qquad \quad \quad
		\ket{2} = \frac{1}{\sqrt{2}}\five{0}{0}{\rla}{0}{\gla}-\frac{1}{\sqrt{2}}\five{0}{0}{\gla}{0}{\rla}\\
		\ket{3} & = \frac{1}{\sqrt{2}}\five{0}{\rla}{0}{0}{\gla}-\frac{1}{\sqrt{2}}\five{0}{\gla}{0}{0}{\rla} \qquad \qquad \quad \quad
		\ket{4} = \frac{1}{\sqrt{2}}\five{0}{0}{\rla}{\gla}{0}-\frac{1}{\sqrt{2}}\five{0}{0}{\gla}{\rla}{0}\\
		\ket{5} & = \frac{1}{\sqrt{2}}\five{0}{\rla}{0}{\gla}{0}-\frac{1}{\sqrt{2}}\five{0}{\gla}{0}{\rla}{0} \qquad \qquad \quad \quad
		\ket{6} = \frac{1}{\sqrt{2}}\five{0}{\rla}{\gla}{0}{0}-\frac{1}{\sqrt{2}}\five{0}{\gla}{\rla}{0}{0}\\
		\ket{7} & = \frac{1}{2}\qty[\five{0}{\rla}{\gla}{\rla}{\gla}-\five{0}{\rla}{\gla}{\gla}{\rla}
					+\five{0}{\gla}{\rla}{\rla}{\gla}+\five{0}{\gla}{\rla}{\gla}{\rla}]\\
		\ket{8} & = \frac{1}{2\sqrt{3}}
					\qty[2\five{0}{\rla}{\rla}{\gla}{\gla}
					-\five{0}{\rla}{\gla}{\rla}{\gla}
					-\five{0}{\rla}{\gla}{\gla}{\rla}
					-\five{0}{\gla}{\rla}{\rla}{\gla}
					-\five{0}{\gla}{\rla}{\gla}{\rla}
					+2\five{0}{\gla}{\gla}{\rla}{\rla}]\,,
    \end{aligned}
\end{equation}
Then, 12 states including single-occupied matter sites and gauge fields:
\begin{equation}
    \label{eq_2DSU2_dressed_basis_singlematter}
    \begin{aligned}
		\ket{9} & = \frac{1}{\sqrt{2}}\five{\rla}{0}{0}{0}{\gla}-\frac{1}{\sqrt{2}}\five{\gla}{0}{0}{0}{\rla} 
        \qquad \qquad \quad \quad 
		\ket{10} = \frac{1}{\sqrt{2}}\five{\rla}{0}{0}{\gla}{0}-\frac{1}{\sqrt{2}}\five{\gla}{0}{0}{\rla}{0}\\
		\ket{11} & = \frac{1}{\sqrt{2}}\five{\rla}{0}{\gla}{0}{0}-\frac{1}{\sqrt{2}}\five{\gla}{0}{\rla}{0}{0} 
        \qquad \qquad \quad \quad
        \ket{14} = \frac{1}{\sqrt{2}}\five{\rla}{\gla}{0}{0}{0}-\frac{1}{\sqrt{2}}\five{\gla}{\rla}{0}{0}{0}\\
		\ket{12} & = \frac{1}{2}\qty[
            \five{\rla}{0}{\gla}{\rla}{\gla}
            -\five{\rla}{0}{\gla}{\gla}{\rla}
			+\five{\gla}{0}{\rla}{\rla}{\gla}
            +\five{\gla}{0}{\rla}{\gla}{\rla}]\\
		\ket{13} & = \frac{1}{2\sqrt{3}}
            \qty[2\five{\rla}{0}{\rla}{\gla}{\gla}
            -\five{\rla}{0}{\gla}{\rla}{\gla}
            -\five{\rla}{0}{\gla}{\gla}{\rla}
            -\five{\gla}{0}{\rla}{\rla}{\gla}
            -\five{\gla}{0}{\rla}{\gla}{\rla}
            +2\five{\gla}{0}{\gla}{\rla}{\rla}]\\
        \ket{15} & = \frac{1}{2}\qty[
            \five{\rla}{\gla}{0}{\rla}{\gla}
            -\five{\rla}{\gla}{0}{\gla}{\rla}
            +\five{\gla}{\rla}{0}{\rla}{\gla}
            +\five{\gla}{\rla}{0}{\gla}{\rla}]\\
        \ket{16} & = \frac{1}{2\sqrt{3}}\qty[
            2\five{\rla}{\rla}{0}{\gla}{\gla}
            -\five{\rla}{\gla}{0}{\rla}{\gla}
            -\five{\rla}{\gla}{0}{\gla}{\rla}
            -\five{\gla}{\rla}{0}{\rla}{\gla}
            -\five{\gla}{\rla}{0}{\gla}{\rla}
            +2\five{\gla}{\gla}{0}{\rla}{\rla}]\\
        \ket{17} & = \frac{1}{2}\qty[
            \five{\rla}{\gla}{\rla}{0}{\gla}
            -\five{\rla}{\gla}{\gla}{0}{\rla}
            +\five{\gla}{\rla}{\rla}{0}{\gla}
            +\five{\gla}{\rla}{\gla}{0}{\rla}]\\
        \ket{18} & = \frac{1}{2\sqrt{3}}\qty[
            2\five{\rla}{\rla}{\gla}{0}{\gla}
            -\five{\rla}{\gla}{\rla}{0}{\gla}
            -\five{\rla}{\gla}{\gla}{0}{\rla}
            -\five{\gla}{\rla}{\rla}{0}{\gla}
            -\five{\gla}{\rla}{\gla}{0}{\rla}
            +2\five{\gla}{\gla}{\rla}{0}{\rla}]\\
        \ket{19} & = \frac{1}{2}\qty[
            \five{\rla}{\gla}{\rla}{\gla}{0}
            -\five{\rla}{\gla}{\gla}{\rla}{0}
            +\five{\gla}{\rla}{\rla}{\gla}{0}
            +\five{\gla}{\rla}{\gla}{\rla}{0}]\\
        \ket{20} & = \frac{1}{2\sqrt{3}}\qty[
            2\five{\rla}{\rla}{\gla}{\gla}{0}
            -\five{\rla}{\gla}{\rla}{\gla}{0}
            -\five{\rla}{\gla}{\gla}{\rla}{0}
            -\five{\gla}{\rla}{\rla}{\gla}{0}
            -\five{\gla}{\rla}{\gla}{\rla}{0}
            +2\five{\gla}{\gla}{\rla}{\rla}{0}]\,,
    \end{aligned}
\end{equation}
and finally 9 states with double-occupied matter sites coupled to gauge \dof{}:
\begin{equation}
    \label{eq_2DSU2_dressed_basis_doublematter}
    \begin{aligned}
        \ket{21} & = \five{2}{0}{0}{0}{0}\\
		\ket{22} & = \frac{1}{\sqrt{2}}\five{2}{0}{0}{\rla}{\gla}
                -\frac{1}{\sqrt{2}}\five{2}{0}{0}{\gla}{\rla} 
                \qquad \qquad \quad \quad 
		\ket{23} = \frac{1}{\sqrt{2}}\five{2}{0}{\rla}{0}{\gla}
                -\frac{1}{\sqrt{2}}\five{2}{0}{\gla}{0}{\rla}\\
		\ket{24} & = \frac{1}{\sqrt{2}}\five{2}{\rla}{0}{0}{\gla}
                -\frac{1}{\sqrt{2}}\five{2}{\gla}{0}{0}{\rla} 
                \qquad \qquad \quad \quad 
		\ket{25} = \frac{1}{\sqrt{2}}\five{2}{0}{\rla}{\gla}{0}
                -\frac{1}{\sqrt{2}}\five{2}{0}{\gla}{\rla}{0}\\
		\ket{26} & = \frac{1}{\sqrt{2}}\five{2}{\rla}{0}{\gla}{0}
                -\frac{1}{\sqrt{2}}\five{2}{\gla}{0}{\rla}{0} 
                \qquad \qquad \quad \quad 
		\ket{27} = \frac{1}{\sqrt{2}}\five{2}{\rla}{\gla}{0}{0}
                -\frac{1}{\sqrt{2}}\five{2}{\gla}{\rla}{0}{0}\\
		\ket{28} & = \frac{1}{2}
                \qty[\five{2}{\rla}{\gla}{\rla}{\gla}
                -\five{2}{\rla}{\gla}{\gla}{\rla}
				+\five{2}{\gla}{\rla}{\rla}{\gla}
                +\five{2}{\gla}{\rla}{\gla}{\rla}]\\
		\ket{29} & = \frac{1}{2\sqrt{3}}
					\qty[2\five{2}{\rla}{\rla}{\gla}{\gla}
					-\five{2}{\rla}{\gla}{\rla}{\gla}
					-\five{2}{\rla}{\gla}{\gla}{\rla}
					-\five{2}{\gla}{\rla}{\rla}{\gla}
					-\five{2}{\gla}{\rla}{\gla}{\rla}
					+2\five{2}{\gla}{\gla}{\rla}{\rla}]\,.
    \end{aligned}
\end{equation}
The resulting dressed-site basis of \cref{eq_2DSU2_dressed_basis_pure,eq_2DSU2_dressed_basis_singlematter,eq_2DSU2_dressed_basis_doublematter} is adopted in the simulations of \cite{Cataldi2024Simulating2+1DSU2} and detailed in \cref{chap_SU2_groundstate}.
% ========================================================================
\section{U(1) Lattice Gauge Theory}
\label{sec_U1_model}
In this section, we apply the dressed-site formalism developed in \cref{sec_dressed_site_formalism} to an Abelian scenario, focusing on the U(1) LGT including dynamical matter, \idest{} the lattice realization of quantum-electrodynamics (QED). 
Again, for simplicity we will consider a (2+1)D system, where all the gauge (electric and magnetic) contributions play a non-trivial role.
% ========================================================================
\subsection{The model}
Within the Wilson and Kogut-Susskind formulation of LGTs \cite{Wilson1974ConfinementQuarks,Kogut1975HamiltonianFormulationWilson}, the U(1) version of the general Hamiltonian in \cref{eq_Ham_LGT} reads:
\begin{equation}
    \begin{split}
        H_{\rm{U(1)}}=&
        \frac{\lspeed\hbar}{2\lspace}\sum_{\vecsite}\qty[-i\hpsi_{\vecsite}^{\dagger}\Apara_{\vecsite, \latvec[x]}\hpsi_{\vecsite+\latvec[x]}
        - (-1)^{\site[x]+\site[y]}\hpsi_{\vecsite}\Apara_{\vecsite,\latvec[y]}\hpsi_{\vecsite+\latvec[y]}+\hc]\\
        &+ \mass[0]\lspeed^{2} \sum_{\vecsite}(-1)^{\vecsite}\hpsi^{\dagger}_{\vecsite}\hpsi_{\vecsite}
        +\frac{\coupling^{2}\lspeed\hbar}{2\lspace}\sum_{\genlink}\casimir_{\genlink}
        - \frac{\lspeed\hbar}{2\coupling^{2}\lspace}\sum_{\square}\Tr(\plaq + \plaq*)\,,
      \end{split}
    \label{eq_U1_Hamiltonian}
\end{equation}
where, since there is only one generator, the electric field is unique $\eleL=\eleR=\eleE$ and the parallel transporter acts as a raising operator:
\begin{align}
    \qty[\Apara,\eleE]&=\Apara&
    \Apara^{\dagger}\Apara&=\mathbb{1}&
    \qty[\Apara, \Apara^{\dagger}]&=0\,.
    \label{eq_U1_algebra}
\end{align}
In the electric basis representation, where $\eleE$ is diagonal, \cref{eq_U1_algebra} can be obtained assuming the following behaviors: given the $\ket{\ell}$ electric state of the gauge link Hilbert space,
\begin{align}
    \begin{cases}
        \eleE\ket{\ell}=\ell\ket{\ell}\\
        \Apara\ket{\ell}=\ket{\ell-1}\\
        \Apara^{\dagger}\ket{\ell}=\ket{\ell+1}\\
    \end{cases}&&
    \text{so that}&&
    \begin{cases}
    \eleE\Apara\ket{\ell}=(\ell-1)\ket{\ell-1}\\
    \Apara\eleE\ket{\ell}=\ell\ket{\ell-1}\\
    \end{cases}\,.
    \label{eq_U1_algebradef1}
\end{align}
Correspondingly, U(1) Gauss law requires that $G_{\vecsite}\ket{\Psi_{\rm{phys}}}=0$ $\forall \vecsite$, where $G_{\vecsite}$ is the local generator of the U(1) gauge symmetry, satisfies $\comm*{G_{\vecsite}}{\ham}=0$, and is defined as follows:
\begin{equation}
    \hat{G}_{\vecsite}=\eleE_{\genlink}-\eleE_{\genlinkm}-\psi_{\vecsite}^{\dagger}\psi_{\vecsite}+\frac{1-(-1)^{\vecsite}}{2}\,.
    \label{eq_U1_gausslaw}
\end{equation}
% ========================================================================
\subsection{Truncating the U(1) gauge group}
\label{sec_U1_gaugetruncation}
To perform numerical simulations of \cref{eq_U1_Hamiltonian}, we need to perform a truncation of the infinite algebra of the U(1) gauge fields. 
As discussed in \cref{sec_gauge_truncation}, one possible solution is to express the gauge operators $\eleE$ and $\Apara$ directly in terms of spin operators in the $\spin$ representation of SU(2):
\begin{align}
    \eleE&=\hat{S}^{z}(\spin)&
    \Apara&=\frac{1}{\spin}\hat{S}^{-}(\spin)\,.
    \label{eq_U1_parallel_transport_spin}
\end{align}
Correspondingly, the gauge link Hilbert space $\mathcal{H}_{\rm{link}}$ is made out of $(2\spin+1)$ states $\ket{\ell}$, which can be labeled as $\ket{\spin,m}$, where $-\spin\leq m\geq \spin$ is the third component of the spin momentum $\spin$. 
According to the rules of angular momentum summation, we would have:
\begin{equation}
    \begin{cases}
        \eleE\ket{\spin,m}=S^{z}\ket{\spin,m}=m\ket{\spin,m}\\
        \Apara\ket{\spin,m}=\frac{S^{-}}{\spin}\ket{\spin,m}
        =\sqrt{(1-\frac{1}{\spin})-\frac{m}{\spin^{2}}(m-1)}\ket{\spin,m-1}\\
        \Apara^{\dagger}\ket{\spin,m}=\frac{S^{+}}{\spin}\ket{\spin,m}
        =\sqrt{(1+\frac{1}{\spin})-\frac{m}{\spin^{2}}(m+1)}\ket{\spin,m+1}\,,
    \end{cases}
    \label{EU_properties2}
\end{equation}
which yields to the following commutation relations:
\begin{equation}
    \begin{split}
        \qty[\Apara,\eleE]&
        =\frac{1}{\spin}\qty[\hat{S}^{-}(\spin),\hat{S}^{z}(\spin)]
        =\frac{\hat{S}^{-}(\spin)}{\spin}=\Apara\\
        \qty[\Apara,\Apara^{\dagger}]&
        =\frac{1}{\spin^{2}}\qty[\hat{S}^{-}(\spin),\hat{S}^{+}(\spin)]
        =-\frac{2\hat{S}^{z}(\spin)}{\spin^{2}}\underset{\spin\to\infty}{\longrightarrow}0.
    \end{split}
\end{equation}
Such a definition of parallel transport provides a uniform convergence to \cref{eq_U1_algebra} in the large-$j$ limit.
Similarly, we can define $\Apara=\diag_{-1}(+1, \dots, +1)$ as a ladder operator, which is perfectly unitary in the core, yet badly converges in the (upper and lower) truncated states. 
In the limit of $j\to\infty$, both solutions recover \cref{eq_U1_algebradef1}.
As we will discuss in \cref{sec_U1_basis_dimension}, the error induced by the truncation is particularly relevant in the small coupling limit, while remaining negligible in most scenarios \cite{Haase2021ResourceEfficientApproach, Verstraete2008MatrixProductStates}.
Choosing the ladder operator can reveal optimal for qubit base quantum hardware \cite{Paulson2021Simulating2DEffects}, while \cref{eq_U1_parallel_transport_spin} is preferable in case of quantum systems that allow for the implementation of interacting spin chains with large spins \cite{Senko2015RealizationQuantumIntegerSpin}.
For completeness, we mention that there are mappings of the parallel transporter preserving unitarity \cite{Kuno2015RealtimeDynamicsProposal, Kuno2016AtomicQuantumSimulation, Zhang2018QuantumSimulationUniversal,Notarnicola2020RealtimedynamicsQuantumSimulation}.
% ========================================================================
\subsection{Rishon decomposition of U(1) gauge fields}
\label{sec_U1_rishondecomposition}
The next step towards the dressed-site formalism is decomposing the $(2\spin+1)$ states of the link basis as a combination of two states, one per half-link: $\ket{\spin,m}_{\genlink}=\ket{\spin,m_{\vecsite,+\latvec},m_{\siteplus,-\latvec}}$. 
Omitting the $\spin$-label, the truncated link Hilbert space reads:
\begin{equation}
    \mathcal{H}_{\spin}=\qty{\ket{+\spin,-\spin},\ket{+\spin-1,-\spin+1},\dots, \ket{m,m},\dots \ket{-\spin+1,+\spin-1},\ket{-\spin,+\spin}}\,.
    \label{eq_U1_linkrishonspace}
\end{equation}
In the Euclidean basis, these gauge link states would read:
\begin{align}
    \ket{+\spin,-\spin}&=\begin{pmatrix}
        1\\
        0\\
        \vdots\\
        0\\
        0
    \end{pmatrix}&
    \ket{+\spin-1,-\spin+1}&=\begin{pmatrix}
        0\\
        1\\
        \vdots\\
        0\\
        0
    \end{pmatrix}&
    \dots&&
    \ket{-\spin+1,+\spin-1}&=\begin{pmatrix}
        0\\
        0\\
        \vdots\\
        1\\
        0
    \end{pmatrix}&
    \ket{-\spin,+\spin}&=\begin{pmatrix}
        0\\
        0\\
        \vdots\\
        0\\
        1
    \end{pmatrix}
    \label{eq_U1_linkbasis}
\end{align}
Then, we express the parallel transport $\Apara$ as the product of two rishon modes $\rishon_{L}$ and $\rishon_{R}$:
\begin{equation}
    \Apara_{\genlink}=\rishon_{A,\vecsite}\rishon_{B,\siteplus, -\latvec}^{\dagger}
    \label{eq_U1_parallel_transport_rishons}
\end{equation}
where, in full generality, we assumed the two rishon modes of the $(\genlink)$-link to belong to two different species, A and B \footnote{It is possible to prove that, in \emph{semi-integer} spin-representations, the two species coincide.}. 
A general definition for the rishon operators is
\begin{align}
    \rishon_{A}&=\begin{pmatrix}
        0&a_{1}&&&\\
        &0&a_{2}&&\\
        &&0&a_{3}&\\
        &&&0&\ddots\\
    \end{pmatrix}_{F}&
    \rishon_{B}&=\begin{pmatrix}
        0&b_{1}&&&\\
        &0&b_{2}&&\\
        &&0&b_{3}&\\
        &&&0&\ddots\\
    \end{pmatrix}_{F}\,,
\end{align}
where $\qty{a_{i}}_{i=1\dots 2s}$ and $\qty{b_{i}}_{i=1\dots 2s}$ are complex numbers to be determined in the specific spin-$s$ representation of U(1).
To obtain an operative Hamiltonian that is fully bosonic in every term, we require the rishon modes to satisfy a fermionic algebra. 
Namely, $\rishon$-rishons must anti-commute among themselves and with matter fields:
\begin{align}
  \qty{\rishon_{A,\genlink},\rishon_{B,\siteplus,-\latvec}}&=0
  &
  \qty{\rishon_{\genlink},\hpsi_{\vecsite}}&=0.
  \label{rishon_commmutations}
\end{align}
Being fermions, $\rishon$-rishon must also satisfy \eqref{fermion_parity_commutation}. 
We define the rishon parity operator as 
\begin{equation}
    P^{\rm{U(1)}}_{\rishon}=\diag(+1,-1,+1,-1,\dots)\,,
    \label{eq_U1_rishonparity}
\end{equation} where by convention we establish the first rishon state to be \emph{even}, while the remaining ones are determined by alternating the sign of parity. 
Then, the last state is even (odd) if the spin-$\spin$ representation is \emph{integer} (\emph{semi-integer}).

We can then properly express the parallel transport in \eqref{eq_U1_parallel_transport_rishons} as follows:
\begin{equation}
    \begin{split}
        \Apara_{\genlink}=&\rishon_{A,\vecsite, +\latvec}\rishon_{B,\siteplus, -\latvec}^{\dagger}
        =\qty[\rishon_{A,\vecsite,+\latvec} \otimes \mathbb{1}_{\siteplus, -\latvec}]
		\times \qty[\parity_{\zeta,\vecsite, +\latvec} \otimes \;\rishon^{\dagger}_{B,\siteplus, -\latvec}]\\
        =&\rishon_{A,\vecsite, +\latvec}\cdot \parity_{\zeta,\vecsite, +\latvec} \otimes \rishon^{\dagger}_{B,\siteplus, -\latvec}\\
        =&
        \underset{\vecsite}{
        \begin{pmatrix}
            0&-a_{1}&&&\\
            &0&+a_{2}&&\\
            &&0&-a_{3}&\\
            &&&&\ddots\\
        \end{pmatrix}}\otimes \underset{\siteplus}{
        \begin{pmatrix}
            0&&&&\\
            \overline{b}_{1}&0&&&\\
            &\overline{b}_{2}&0&&\\
            &&\overline{b}_{3}&0&\\
            &&&\ddots&\\
        \end{pmatrix}}.
    \end{split}
\end{equation}
The correct matrix elements $\qty{a_{i}}_{i=1\dots 2\jmax}$ and $\qty{b_{i}}_{i=1\dots 2\jmax}$ of the two $\rishon$-rishon modes in the specific spin-$\jmax$ representation for $U(1)$ are the ones matching the original choice of the parallel transporter (spin operator as in \cref{eq_U1_parallel_transport_spin} or ladder operator).
% =====================================================================
\subsubsection{Example: spin-1 representation of U(1)}
As an example, we derive the expression of the gauge fields (in terms of rishons) for the $\spin=1$-spin representation. 
Such a truncation is the smallest non-trivial \emph{integer} representation that gives access to non trivial features of the U(1) LGT \cite{Felser2021EfficientTensorNetwork, Magnifico2021LatticeQuantumElectrodynamics,Rigobello2021EntanglementGenerationMathrm}.

Within this truncation, the gauge fields read:
\begin{align}
    \eleE=\hat{S}^{z}&=\begin{pmatrix}
        +1&&\\
        &0&\\
        &&-1\\
    \end{pmatrix}&
    \Apara=\hat{S}^{-}&=\sqrt{2}\begin{pmatrix}
        0&&\\
        1&0&\\
        &1&0&\\
    \end{pmatrix}\,.
    \label{eq_U1_gaugefields_spin1}
\end{align}
The corresponding link Hilbert space $\hil^{\spin=1}$ defined in \eqref{eq_U1_linkbasis} is made out of the following states:
\begin{align}
    \ket{1,1}&=\begin{pmatrix}
        1\\
        0\\
        0
    \end{pmatrix}&
    \ket{0,0}&=\begin{pmatrix}
        0\\
        1\\
        0\\
    \end{pmatrix}&
    \ket{-1,-1}&=\begin{pmatrix}
        0\\
        0\\
        1\\
    \end{pmatrix}\,.
\end{align}
Then, from the definition of $\Apara$ in \eqref{eq_U1_gaugefields_spin1}, the only relevant matrix elements of $\rishon_{A}\rishon^{\dagger}_{B}$ are the ones that correspond to the action on sites belonging to \eqref{eq_U1_linkbasis}:
\begin{equation}
    \begin{aligned}
        \rishon_{A}\rishon^{\dagger}_{B}\ket{1,1}&= +a_{2}\overline{b}_{1}\ket{0,0}\\
        \rishon_{A}\rishon^{\dagger}_{B}\ket{0,0}&= -a_{1}\overline{b}_{2}\ket{-1,-1}\\
        \rishon_{A}\rishon^{\dagger}_{B}\ket{-1,-1}&=0
    \end{aligned} \qquad \text{whose solution could be} \qquad 
    \begin{aligned}
        a_{1}&=a_{2}=1\\
        b_{1}&=+\sqrt{2}=-b_{2}.
    \end{aligned}
\end{equation}
Summarizing, we have found
\begin{align}
    \rishon_{A}&=\begin{pmatrix}
        0&+1&\\
        &0&+ 1\\
        &&0\\
    \end{pmatrix}&
    \rishon_{B}&=\begin{pmatrix}
        0&+\sqrt{2}&\\
        &0&-\sqrt{2}\\
        &&0\\
    \end{pmatrix}\,.
    \label{rishons_spin1}
\end{align}
Notice that, by using the ladder definition for the parallel transporter, the two rishons read:
\begin{align}
    \rishon_{A}'&=\begin{pmatrix}
        0&+1&\\
        &0&+ 1\\
        &&0\\
    \end{pmatrix}_{F}&
    \rishon_{B}'&=\begin{pmatrix}
        0&+1&\\
        &0&-1\\
        &&0\\
    \end{pmatrix}{F}\,.
    \label{rishons_spin1_op2}
\end{align}
% =====================================================================
\subsection{Constructing the dressed-site operators}
We stress that the rishon decomposition is a general approach and can be applied to lattices of whatever dimension $D$. 
However, as discussed for SU(2), for a fixed $D$, the construction of the Hamiltonian operators requires attention in choosing an internal ordering of the dressed-site basis.
Working in a $D=2$-dimensional lattice, the single dressed-site basis can be sketched as:
\begin{equation}
    \ket{\begin{array}{ccc}
        &\rishon_{A,\vecsite,+\latvec[y]}&\\
        \rishon_{B,\vecsite,-\latvec[x]}&\psi_{\vecsite}&\rishon_{A,\vecsite,+\latvec[x]}\\
        &\rishon_{B,\vecsite,-\latvec[y]}&\\
    \end{array}}\quad \text{where the d.o.f. are ordered as:}\quad 
    \ket{\begin{array}{ccc}
        &4&\\
        1&0&3\\
        &2&\\
    \end{array}}\,.
    \label{eq_U1_dressed_site}
\end{equation}
Every dressed-site operator is then constructed as in \eqref{fermionic_qmb_op} according to the previous order.
We are then ready to express each Hamiltonian term in terms of dressed-site operators.
% =====================================================================
\paragraph{Hopping operators}
In the hopping Hamiltonian, the matter-gauge interaction (apart from the staggered factors) along the link $(\genlink)$ can be rewritten in terms of two bosonic \emph{arrival} operators. 
Namely
\begin{equation}
    \begin{split}
        \ham^{\text{hopping}}_{\genlink}
        &=\qty[\hpsi^{\dagger}_{\vecsite}\hat{U}_{\genlink}\hpsi_{\vecsite+\latvec}+\hc]\\
        &= \qty[\hpsi^{\dagger}_{\vecsite}\rishon_{A,\vecsite+\latvec}\rishon^{\dagger}_{B,\vecsite+\latvec,-\latvec}\hpsi_{\vecsite+\latvec}+\hc]\\
        &=\qty[\hat{Q}^{\dagger}_{\vecsite,+\latvec}\hat{Q}_{\vecsite+\latvec,-\latvec}+\hc]
    \end{split}
    \qquad \text{where} \qquad 
    \begin{aligned}
        \hat{Q}^{\dagger}_{\vecsite,+\latvec}=\hpsi^{\dagger}_{\vecsite}\rishon_{A,\vecsite,+\latvec}\\
        \hat{Q}^{\dagger}_{\vecsite,-\latvec}=\hpsi^{\dagger}_{\vecsite}\rishon_{B,\vecsite,-\latvec}.
    \end{aligned}
\end{equation}
Then, according to the internal ordering of dressed site defined in \eqref{eq_U1_dressed_site}, we have for instance:
\begin{equation}
    \begin{split}
        \hat{Q}^{\dagger}_{\vecsite,+\latvec[x]}=
        \hpsi^{\dagger}_{\vecsite}\rishon_{A,\vecsite,+\latvec[x]}=
        &\quad\hpsi^{\dagger}_{\vecsite}\otimes \mathbb{1}_{\vecsite,-\latvec[x]}\otimes \mathbb{1}_{\vecsite,-\latvec[y]}\otimes \mathbb{1}_{\vecsite,+\latvec[x]}\otimes \mathbb{1}_{\vecsite,+\latvec[y]}\\
        &\times \parity_{\psi,\vecsite}\otimes \parity_{\rishon,\vecsite, -\latvec[x]}\otimes \parity_{\rishon,\vecsite, -\latvec[y]}\otimes \rishon_{A,\vecsite,+\latvec[x]} \otimes  \mathbb{1}_{\vecsite,+\latvec[y]}\\
        =&\hpsi^{\dagger}_{\vecsite}\cdot \parity_{\psi,\vecsite} \otimes \parity_{\rishon,\vecsite, -\latvec[x]}\otimes \parity_{\rishon,\vecsite, -\latvec[y]}\otimes \rishon_{A,\vecsite,+\latvec[x]} \otimes \mathbb{1}_{\vecsite,+\latvec[y]}\,.
    \end{split}
    \label{eq_U1_arrival_operators}
\end{equation}
\paragraph{Number density operators} 
Similarly, number density operators can be expressed as:
\begin{equation}
    \begin{split}
        \hat{N}_{\vecsite}&=\hpsi^{\dagger}_{\vecsite}\psi_{\vecsite}
        =\hpsi^{\dagger}_{\vecsite}\cdot \hpsi_{\vecsite}\bigotimes_{\pm\latvec}\mathbb{1}_{\genlink}\,.
    \end{split}
    \label{eq_U1_number_op}
\end{equation}
% =====================================================================
\paragraph{Magnetic operators}
Let us rewrite the single plaquette interaction in terms of rishon modes. 
The idea is to perform a series of fermionic swaps to make the rishons of the same dressed site close in the internal ordering of the plaquette.
We have then:
\begin{equation}
    \begin{split}
        \plaq=&\qty[\Apara_{\genlink_{x}}\Apara_{\vecsite+\latvec[x],+\latvec[y]}
        \Apara^{\dagger}_{\vecsite+\latvec[y],+\latvec[x]}\Apara^{\dagger}_{\genlink_{y}}]\\
        =&\Big[\qty(\rishon_{A,\vecsite,+\latvec[x]}\rishon^{\dagger}_{B,\vecsite+\latvec[x],-\latvec[x]})\qty(\rishon_{A,\vecsite+\latvec[x],+\latvec[y]}\rishon^{\dagger}_{B,\vecsite+\latvec[x]+\latvec[y],-\latvec[y]})\times \\
        &\qquad \qty(\rishon_{A,\vecsite+\latvec[y],+\latvec[x]}\rishon^{\dagger}_{B,\vecsite+\latvec[x]+\latvec[y],-\latvec[x]})^{\dagger}\qty(\rishon_{A,\vecsite,+\latvec[y]}\rishon^{\dagger}_{B,\vecsite+\latvec[y],-\latvec[y]})^{\dagger}\Big]\\
        =&\Big[\qty(\rishon_{A,\vecsite,+\latvec[x]}\rishon^{\dagger}_{B,\vecsite+\latvec[x],-\latvec[x]})\qty(\rishon_{A,\vecsite+\latvec[x],+\latvec[y]}\rishon^{\dagger}_{B,\vecsite+\latvec[x]+\latvec[y],-\latvec[y]})\times \\
        &\qquad \qty(\rishon_{B,\vecsite+\latvec[x]+\latvec[y],-\latvec[x]}\rishon^{\dagger}_{A,\vecsite+\latvec[y],+\latvec[x]})\qty(\rishon_{B,\vecsite+\latvec[y],-\latvec[y]}\rishon^{\dagger}_{A,\vecsite,+\latvec[y]})\Big]\\
        =&\Big[\qty(\rishon_{A,\vecsite,+\latvec[x]}\rishon^{\dagger}_{A,\vecsite,+\latvec[y]})\qty(\rishon^{\dagger}_{B,\vecsite+\latvec[x],-\latvec[x]}\rishon_{A,\vecsite+\latvec[x],+\latvec[y]})\times \\
        &\qquad \qty(\rishon^{\dagger}_{B,\vecsite+\latvec[x]+\latvec[y],-\latvec[y]}\rishon_{B,\vecsite+\latvec[x]+\latvec[y],-\latvec[x]})\qty(\rishon^{\dagger}_{A,\vecsite+\latvec[y],+\latvec[x]}\rishon_{B,\vecsite+\latvec[y],-\latvec[y]})\Big]\\
        =&\qty[-\corner_{\vecsite, +\latvec[x],+\latvec[y]}\corner_{\vecsite+\latvec[x], +\latvec[y],-\latvec[x]}\corner_{\vecsite+\latvec[x]+\latvec[y], -\latvec[x],-\latvec[y]}\corner_{\vecsite+\latvec[y],-\latvec[y],+\latvec[x]}]\,,
    \end{split}
    \label{eq_U1_magnHamiltonian}
\end{equation}
where we merged rishon modes belonging to the same dressed site into \emph{corner} operators defined as $\corner_{\vecsite, \latvec[1],\latvec[2]}=\rishon_{\vecsite, \latvec[1]}\rishon^{\dagger}_{\vecsite,\latvec[2]}$. 
For instance, from the internal order of the dressed-site basis in \eqref{eq_U1_dressed_site}, we have for instance:
\begin{equation}
    \begin{split}
        \corner_{\vecsite,-\latvec[y],+\latvec[x]}=&\quad \rishon_{B,\vecsite,-\latvec[y]}\rishon^{\dagger}_{A,\vecsite,+\latvec[x]}\\
        =&\quad \parity_{\psi,\vecsite}\otimes \parity_{\rishon,\vecsite,-\latvec[x]}\otimes \rishon_{B,\vecsite,-\latvec[y]} \otimes \mathbb{1}_{\vecsite,+\latvec[x]}\otimes \mathbb{1}_{\vecsite,+\latvec[y]}\\
        =&\times \parity_{\psi,\vecsite}\otimes  \parity_{\rishon,\vecsite, -\latvec[x]} \otimes  \parity_{\rishon,\vecsite, -\latvec[y]} \otimes \rishon^{\dagger}_{A,\vecsite,+\latvec[x]}\otimes \mathbb{1}_{\vecsite,+\latvec[y]}\\
        =&\mathbb{1}_{\vecsite}\otimes \mathbb{1}_{\vecsite,-\latvec[x]}\otimes  \rishon_{B,\vecsite,-\latvec[y]}\cdot  \parity_{\rishon,\vecsite, -\latvec[y]}\otimes \rishon^{\dagger}_{A,\vecsite,+\latvec[x]}\otimes  \mathbb{1}_{\vecsite,+\latvec[y]}.
    \end{split}
    \label{eq_U1_corner_operators}
\end{equation}
\paragraph{Electric field operators}
In the dressed-site formalism, as with the parallel transporter, the electric field contribution is equally divided across the two half-links, with each half contributing half the total electric energy.
Namely, from \eqref{eq_U1_parallel_transport_spin} we have:
\begin{equation}
    \casimir_{\genlink}=\frac{1}{2}\qty[\hat{S}^{z}_{\vecsite,\latvec}(\spin)^{2}+\hat{S}_{\siteplus,-\latvec}^{z}(\spin)^{2}].
    \label{square_electric_field_integerrep}
\end{equation}
In the operative Hamiltonian, we combine the $2D$ half-links contribution to the electric energy onto the same single dressed-site operator:
\begin{equation}
    \hat{\Gamma}_{\vecsite}=\frac{1}{2}\sum_{\pm\latvec}\hat{S}^{z}_{\vecsite,\latvec}(\spin)^{2}.
\end{equation}
% =====================================================================
\subsubsection{Operative dressed-site U(1) Hamiltonian}
We can then rewrite the (2+1)D U(1) truncated Hamiltonian in \cref{eq_U1_Hamiltonian} as follows:
\begin{equation}
    \begin{split}
        H_{\rm{U(1)}}=&
        \frac{\lspeed\hbar}{2\lspace}\sum_{\vecsite}\qty[\text{-i}\hat{Q}_{\vecsite,+\latvec[x]}^{\dagger}\hat{Q}_{\vecsite+\latvec[x], -\latvec[x]}
        - (-1)^{\vecsite}\hat{Q}_{\vecsite,+\latvec[y]}^{\dagger}\hat{Q}_{\vecsite+\latvec[y], -\latvec[y]} + \hc]\\
        &+ \mass[0]\lspeed^{2} \sum_{\vecsite} (-1)^{\vecsite}\hat{N}_{\vecsite}
        +\frac{\coupling^{2}\lspeed\hbar}{4\lspace} \sum_{\vecsite}\hat{\Gamma}_{\vecsite}
        - \frac{\lspeed\hbar}{2 \coupling^{2}\lspace}\sum_{\square}\Tr(\plaq + \plaq*)\,.
    \end{split}
    \label{eq_U1_Hamiltonian2}
\end{equation}
% =====================================================================
\subsubsection{U(1) Link-Symmetry operators}
The equivalence between \cref{eq_U1_Hamiltonian} and \cref{eq_U1_Hamiltonian2} is correct as long as the link symmetries are satisfied, \idest{}, as long as the rishon states of each link have opposite electric fields. 
This constraint reduces to a $\mathbb{Z}_{2}$ symmetry on each link that can be easily resolved with dedicated libraries \cite{Cataldi2024Edlgt,Hauschild2018EfficientNumericalSimulations}, or imposed as two-body penalty terms (one per link) in the Hamiltonian, in such a way to behave as the largest energy scale of the problem.
% =====================================================================
\subsubsection{Open boundary conditions and half-integer representations}
In the case of open boundary conditions (OBC), we may want to freeze the half links on the bordering dressed sites of the lattice to the ground of the electric field. 
For \emph{integer} $\spin$-representations, this can be easily achieved by selecting the rishon configurations with $\ket{\spin, 0}$ on sites along the corresponding border $\latvec$. 
Equivalently, for each of these sites, we can add a penalty term like $\ham^{\text{border}}_{\genlink}=\alpha\qty(\eleE_{\genlink}-\spin)^{2}$.
As for \emph{semi-integer} representations, where the minimal electric field is degenerate, we can choose between one of the following penalties:
\begin{align}
    \ham^{\text{border}}_{\vecsite,-\latvec}&=\alpha\qty(\eleE_{\vecsite,-\latvec}-\lceil \spin\rceil)^{2}&
    \text{and}&&
    \ham^{\text{border}}_{\vecsite,+\latvec}&=\alpha\qty(\eleE_{\vecsite,+\latvec}-\lfloor \spin\rfloor)^{2}\,,
\end{align}
where $\lceil \spin\rceil$ ($\lfloor \spin\rfloor$) corresponds to the minimum upper (maximum lower) integer representation. 
% =====================================================================
\subsubsection{Projecting operators on the U(1) gauge-invariant dressed-site basis}
As discussed in \cref{sec_projecting_gaugeinvariant_ops}, all the previous operators need to be projected in the U(1) gauge-invariant subspace. 
Since we have chosen the staggered fermions formulation, we expect two gauge invariant bases, one for \emph{even} and one for \emph{odd} sites, respectively $M_{+}$ and $M_{-}$.
Each one coincides with the kernel of the corresponding U(1) Gauss Law operator $\hat{G}_{\vecsite,p}$:
\begin{align}
    \hat{G}_{\vecsite,p}M_{p}=\qty[\hat{N}_{\vecsite}+\sum_{\latvec}\hat{E}_{\latvec}-\frac{1-p}{2}\mathbb{1}_{\vecsite}]M_{p}&=0,
\end{align}
where $p=\pm1$ is the parity of the lattice site.
% =====================================================================
\subsection{Scaling of the local basis dimension}
\label{sec_U1_basis_dimension}
Of course, the larger the gauge field truncation (\idest{} the larger the spin-$\spin$ representation), the larger the effective dressed-site Hilbert space.
In \cref{tab_LGT_localdims}, we list the dressed-site dimension $d=\dim\hildress$ associated with the first few nontrivial gauge truncations of U(1) and SU(2), detailed in \cref{sec_SU2_model} in $D=2,3$ space dimensions.
Both the considered LGTs include dynamical matter, represented by one fermionic field multiplet in the fundamental representation of the gauge group.
The $\gtl$-th row of \cref{tab_LGT_localdims} is obtained keeping only the first $\gtl$ nonzero electric energy levels (\idest{}, using the $\gtl$-th smallest quadratic Casimir eigenvalue as cutoff $\casimircutoff$, see \cref{sec_gauge_truncation}).
As \cref{tab_LGT_localdims} shows, the local dimension increases rapidly with $\gtl$.
For (3+1)D non-Abelian LGTs, $d \sim O(10^4)$ is reached already within the first two truncations, making the study of the untruncated limit prohibitive.

As it will be discussed in \cref{sec_TN_roadmap}, differently from the models typically encountered in condensed matter physics, the local dimension of LGTs can thus be a limiting factor for Tensor Network simulations | especially when $d$ becomes comparable to commonly used TN bond dimensions ($100 \lesssim \chi \lesssim 500$ for TTN).
In these cases, crucially for high-dimensional LGTs, strategies aimed at further compressing the local computational basis are needed (see \cref{sec_TN_roadmap_opt_basistruncation}).
\begin{table}
    \centering
    \setlength{\tabcolsep}{4pt}
    \renewcommand{\arraystretch}{1.05}
    \begin{tabularx}{0.5\textwidth}{>{\centering\arraybackslash}X|rr|rr}
        \toprule
        $\gtl$ & \multicolumn{4}{c}{$d$}\\ \midrule
                & \multicolumn{2}{c|}{$(2+1)$-dimensions} & \multicolumn{2}{c}{$(3+1)$-dimensions}\\[1pt]
                & U(1) & SU(2) & U(1) & SU(2) \\
        \midrule
        1 & 35 & 30 & 267 & 178 \\
        2 & 165 & 168 & 3437 & 3670 \\
        3 & 455 & 600 & 18487 & 35280 \\
        4 & 969 & 1650 & 64953 & 214958 \\
        5 & 1771 & 3822 & 177155 & 967466 \\
        6 & 2925 & 7840 & 408421 & 3509062 \\
        7 & 4495 & 14688 & 835311 & 10828494 \\
        8 & 6545 & 25650 & 1561841 & 29473038 \\
        \bottomrule
    \end{tabularx}
    \caption{Dressed site Hilbert space dimension $d$ for increasing number $\gtl$ of allowed electric energy density levels in some (2+1)D and (3+1)D paradigmatic LGTs with dynamical matter and gauge groups U(1) and SU(2). 
    Table from \cite{Magnifico2024TensorNetworksLattice}.}
    \label{tab_LGT_localdims}
\end{table}

\begin{figure}
    \centering
    \includegraphics[width=0.5\textwidth]{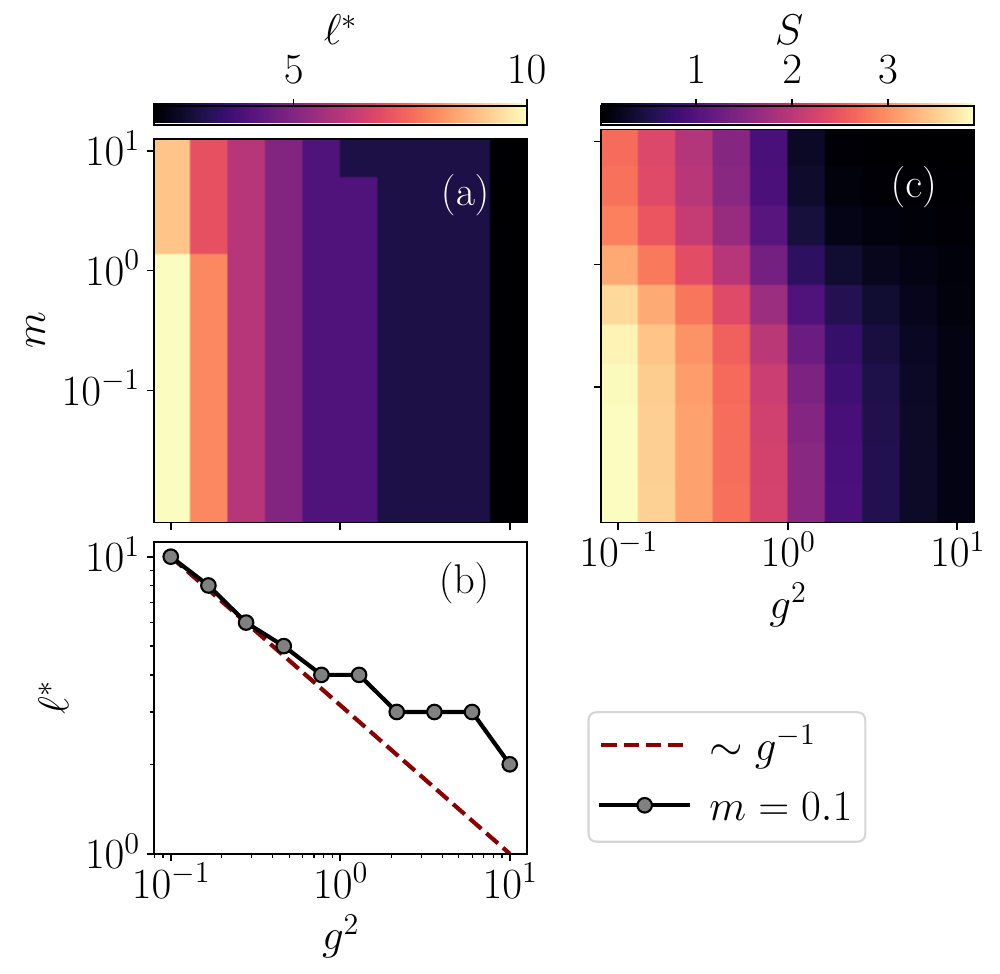}
    \caption{Exact diagonalization of a QED plaquette for a grid of masses $\mass\in [10^{-2},10^{1}]$ and couplings $\coupling^{2}\in [10^{-1},10^{1}]$.
    (a,b) Minimal gauge truncation $\gtlstar$ required to reach a precision $\gterr=10^{-5}$ in the magnetic energy $\ev*{\Re\plaq}$.
    (c) Corresponding entanglement entropy $S(\gtlstar)$ associated with a symmetric bipartition of the plaquette. 
    Figure from \cite{Magnifico2024TensorNetworksLattice}.}
    \label{fig_QED_convergence}
\end{figure}

Several numerical analyses suggest that, in some cases, a small-to-moderate truncation of the gauge group is enough for accurately approximating the continuum limits, at least for low-energy states \cite{Buyens2017FiniterepresentationApproximationLattice, Ercolessi2018PhaseTransitionsGauge, Rigobello2021EntanglementGenerationMathrm, Ciavarella2021TrailheadQuantumSimulation, Davoudi2021SearchEfficientFormulations, Tong2022ProvablyAccurateSimulation}.
However, the optimal gauge truncation depends on the Hamiltonian parameters $\mass[0]$ and $g$.

In order understand this behavior, we focus on a single QED plaquette in open boundary conditions, as it provides the minimal setting allowing for both electric and magnetic effects.
Then, to characterize the convergence in the gauge truncation $\gtl$, we consider a candidate observable $O$ and compute its ground state expectation value $\langle\gtobs\rangle_{\gtl}$ for increasing $\gtl$.
We iterate until the relative deviation between consecutive truncations drops below some threshold $\gterr$:
\begin{math}
  |\langle \gtobs\rangle_{\gtlstar} - \langle\gtobs\rangle_{\gtlstar-1}|
  <
  \gterr |\langle \gtobs \rangle_{\gtlstar}|
\end{math} for some $\gtlstar$.
\Cref{fig_QED_convergence}(a) shows the minimal truncation $\gtlstar$ at which the magnetic energy operator $\gtobs = \Re \plaq$ is converged to $\gterr=10^{-5}$.
We explore a grid of model parameters, whose extent has been chosen according to standard MC literature \cite{Fiore2005QED_3SpacetimeLattice, Raviv2014NonperturbativeBetaFunction, Svetitsky2015BetaFunctionThreedimensional, Xu2019MonteCarloStudy,Creutz1983MonteCarloComputations,Creutz1988LatticeGaugeTheory,Creutz1989LatticeGaugeTheories,Loan2003PathIntegralMonte,Rothe2012LatticeGaugeTheories,Funcke2023HamiltonianLimitLattice,Strouthos2008PhasesNonCompactQED,Bender2023QuantumClassicalMethods,Clemente2022StrategiesDeterminationRunning}. 
$\Re\plaq$ is used as a benchmark due to its relevance in the weak coupling regime, where the continuum limit of $D<3$ lattice QED is located \cite{Creutz1983MonteCarloComputations}.
As \cref{fig_QED_convergence}(b) shows, $\gtlstar$ depends heavily on the coupling, growing asymptotically like $\gtlstar\sim g^{-1}$ as $g$ is decreased, while $\mass[0]$ plays almost no role.
An analogous inverse dependence of the minimal gauge truncation on the coupling is expected for non-Abelian LGT in arbitrary dimensions.
Moreover, the continuum limit of non-Abelian LGTs in $D\leq3$ is also expected at $g\to0$ \cite{Creutz1983MonteCarloComputations}, further substantiating the need to compress the local dimension in TN simulations of LGT whenever extrapolation to the continuum is in order.
Developements along this directions represent an ongoing research line we are considering towards more accurate simulations.

Apart from the growth of the local dimension, extrapolation to the continuum is further complicated by the fact that the continuum limit of a lattice quantum field theory corresponds to a critical point of the underlying lattice model \cite{HernAndez2011LatticeFieldTheory}.
Close to criticality, quantum correlations are boosted and violations of the entanglement area law are expected \cite{Eisert2013EntanglementTensorNetwork}.
The higher entanglement entropy in the proximity of the continuum ($\coupling,\mass[0]\ll1$ regime) is already captured by the single-plaquette analysis of \cref{fig_QED_convergence}(c) (see \cite{Cataldi2024Simulating2+1DSU2} for the SU(2) case).
Continuum limits of LGTs are thus an area of potential advantage for quantum computation over classical methods, as the former is not limited by entanglement.
Nonetheless, quantum computation is also affected by the need to relax gauge truncations when $\coupling^{2} \to 0$, either by increasing the number of qubits used to encode a dressed site, which is at least $\lceil\log_2(d)\rceil$, or by using hybrid devices with both qubits and bosonic modes \cite{Kang2023LeveragingHamiltonianSimulation}.

\section{Summary}
In this chapter, we provided a detailed exploration of Hamiltonian Lattice Gauge Theories (LGTs) with an emphasis on the discretization process for both fermionic and gauge fields. 
We began by reviewing the lattice discretization of fermions, highlighting the staggered fermion approach to mitigate the fermion doubling problem. 
The discretization of gauge fields was then discussed, where the continuous gauge fields were replaced by parallel transporters, ensuring lattice gauge invariance. 
We detailed how the discretization of the electric and magnetic components of the gauge fields, leading to the construction of the lattice Hamiltonian. 
The dimensional analysis was carried out, emphasizing the role of the gauge coupling and its scaling with lattice spacing.

The second part of the chapter introduced the dressed-site formalism, which effectively integrates the gauge fields into matter sites, reducing the complexity of the gauge theory and making it more amenable to TN algorithms and quantum simulations.
This formalism has been thoroughly applied to two exemplary cases, such as the non-Abelian SU(2) Yang-Mills in \cref{sec_SU2_model} and the Abelian U(1) LGTs in \cref{sec_U1_model}.
The techniques and concepts introduced here for these two models will be foundational for understanding the numerical methods and physical results presented and investigated later in the subsequent chapters of this thesis.

%% file: chapters/numerics.tex
\chapter{Tensor Networks methods for LGTs}
\label{chap_numerics}
The rich phenomenology of complex quantum many-body (QMB) systems is often explored using advanced numerical techniques. 
For many quantum Hamiltonians, exact solutions become computationally infeasible as the system size increases, due to the exponential growth of the Hilbert space. 
For a system with $\Nsites$ particles and a local dimension $d$, the dimensionality of the Hilbert space grows as $\dim, \mathcal{H} = d^{\Nsites}$, leading to a $d^{\Nsites} \times d^{\Nsites}$ Hamiltonian matrix.
Exact diagonalization (ED), while valuable for small systems, quickly becomes impractical, requiring memory beyond the limits of most hardware as system size increases.

Given these challenges, alternative numerical approaches have emerged, with two key methods standing out: Quantum Monte Carlo (QMC) simulations \cite{Grotendorst2002QuantumSimulationsComplex} and Tensor Network (TN) methods \cite{Montangero2018IntroductionTensorNetwork}. 
QMC techniques use stochastic methods to solve the Schr\"dinger equation as a diffusion problem, while TN approaches leverage the entanglement properties of quantum systems to efficiently represent low-energy states, retaining only the most relevant contributions to equilibrium properties and real-time dynamics.

Both QMC and TN methods have significantly advanced our ability to simulate QMB systems, providing insights into ground state properties, phase transitions, and dynamic behaviors across various domains. 
Their applications extend beyond condensed matter physics, with relevance in fields such as nuclear physics, chemistry \cite{Lynn2019QuantumMonteCarlo}, material science, and even machine learning \cite{Montangero2018IntroductionTensorNetwork}. 
However, these methods also face significant challenges when dealing with strongly correlated fermionic systems. 
For instance, QMC suffers from the notorious \emph{sign problem} \cite{Loh1990SignProblemNumerical,Troyer2005ComputationalComplexityFundamental,Nagata2022FinitedensityLatticeQCD}, which stems from the antisymmetric nature of fermionic wave functions, leading to highly oscillatory integrals that are difficult to compute.

On the other hand, TNs provide a powerful framework for bypassing the sign problem, but they require careful handling of entanglement and dimensionality issues, particularly when applied to higher-dimensional systems. 
In these contexts, optimizations that preserve locality, such as the use of space-filling curves like the Hilbert curve, become crucial for maintaining efficiency in TN algorithms on high-dimensional lattices. 
Such curves help reducing the computational overhead by minimizing long-range interactions when mapping 2D lattices into effective 1D structures \cite{Cataldi2021HilbertCurveVs}.

The need for scalability becomes even more pressing when dealing with lattice gauge theories (LGTs), where large-scale simulations are necessary to access the continuum limit and explore complex phenomena such as non-perturbative dynamics and phase transitions. 
Here, TN methods, despite their successes, face challenges related to large bond dimensions and local basis truncations. 
Addressing these requires both theoretical innovations, such as optimized mappings and geometries, and computational improvements, like parallelization and hybrid techniques leveraging modern high-performance computing (HPC) infrastructures.

The structure of this chapter reflects the key elements necessary for an efficient exploration of QMB systems via ED and TN methods. 
In \cref{sec_ED}, we begin with an introduction to exact diagonalization, discussing the importance of symmetry-based techniques like block diagonalization to reduce the effective size of the Hilbert space and the associated computational load. 
Next, in \cref{sec_TN}, we turn to TN methods, outlining the fundamental concepts and the role of entanglement in TN algorithms \cite{Montangero2018IntroductionTensorNetwork}.
Loopless TN geometries, such as Matrix Product States (MPS) and Tree Tensor Networks (TTNs), are introduced in \cref{sec_TN_loopfree_geometries} as powerful tools for simulating one-dimensional and higher-dimensional finite-size systems. 
A key aspect of this section is the focus on the implementation of gauge symmetries within the TN structure, enabling a further reduction in computational costs. 
In \cref{sec_TN_gs_algorithm}, we detail the TTN variational ground-state search algorithm used for studying equilibrium properties, while \cref{sec_TN_time_evolution} outlines time evolution algorithms, such as Time-Evolved Block Decimation (TEBD) and the Time-Dependent Variational Principle (TDVP).

The original contributions of this chapter are in \cref{sec_TN_efficient_TN_geometry,sec_TN_roadmap}.
In the former, we analyze the effectiveness of different TN geometries in simulating high-dimensional systems, with a particular focus on 2D lattice models. 
Here, the importance of locality-preserving mappings like the Hilbert curve is highlighted, demonstrating their role in improving the accuracy and efficiency of TN simulations \cite{Cataldi2021HilbertCurveVs}. 
In the latter, we provide a roadmap for future developments, discussing both numerical and theoretical improvements that will enable TN methods to scale up to large LGT simulations, potentially reaching sizes and accuracies comparable to current Monte Carlo methods \cite{Magnifico2024TensorNetworksLattice}.
In conclusion, the chapter outlines a cohesive strategy for advancing the simulation of QMB systems, combining ED, efficient TN geometries, and a vision for overcoming the remaining computational barriers in large-scale LGT simulations.
% =================================================================
\section{Exact Representation of Quantum Many-Body Systems}
\label{sec_ED}
Consider a QMB system defined on a D-dimensional lattice with $\Nsites=\prod_{k}^{D}L_{k}$ sites.
A generic pure QMB state $\ket{\psiqmb}$ of the system lives in the tensor product of $\Nsites$ local Hilbert spaces $\mathcal{H}_{j}$, each one being of finite dimension $d_{j}$\footnote{For the moment, we are still considering the general case where system sites can have in principle different local dimension and different local Hilbert space. 
Later on, we will restrict to the case of $d_{j}=d$ $\forall j\in\qty{1,\dots, \Nsites}$.}: $\mathcal{H} = \mathcal{H}_1 \otimes \mathcal{H}_{2} \otimes \dots \otimes \mathcal{H}_N$.
The state $\ket{\psiqmb}$ can be exactly expanded in terms of a complete basis set of $\mathcal{H}$:
\begin{equation}
	\ket{\psiqmb}=\sum_{i_{1},\dots,i_{\Nsites}}\tensor_{i_{1},\dots,i_{\Nsites}}\ket{i_{1},\dots, i_{\Nsites}}\,,
	\label{eq_QMB_state}
\end{equation}
where $\ket{i_{1},\dots,i_{\Nsites}}=\ket{i_{1}}\otimes\dots\otimes\ket{i_{\Nsites}}$ represents the tensor product of the local basis vectors, while $\qty{i_{s}}$ is the canonical basis for the $s^{\text{th}}$ subsystem. 
As the state is pure, it is uniquely determined by the coefficient tensor $\tensor_{i_{1},\dots,i_{\Nsites}}$, whose entries are in general complex scalars and their number scales as $d^{\Nsites}$, \idest{}, exponentially with the system size $\Nsites$. 

In these terms, providing an exact description of the state entails knowing all the $d^{\Nsites}$ complex coefficients.
This is a fundamental limitation when solving a QMB problem on a classical computer, since an exponential scaling with the number of degrees of freedom implies that the exact representation described in \cref{eq_QMB_state} is computationally unfeasible for intermediate-large values of $\Nsites$, especially when the local basis becomes large as in the case of dressed-site formulated LGTs (see \cref{sec_dressed_site_formalism,sec_U1_basis_dimension}).
Nonetheless, ED simulations for small lattices are crucial for validating other methods such as Tensor Networks, and offer a concrete way to understand key aspects of quantum mechanics, especially the symmetry of many-body states \cite{Sandvik2010ComputationalStudiesQuantum}.
% =================================================================
\subsection{Block Diagonalization}
\label{sec_ED_block_diagonalization}
In the special case where the QMB system has a set of symmetries (as in LGTs), the computational cost of ED algorithms can be significantly reduced (while still remaining exponential in the system size) by using \emph{block diagonalization}, allowing for a more efficient numerical treatment of the problem. 
Indeed, when a symmetry group $G$ is present in the system, the total Hilbert space $\hil$ can be decomposed into a direct sum of smaller subspaces $\hil_{\alpha}$ labeled by an irreducible representation (or quantum number) $\alpha$.
Correspondingly, the QMB Hamiltonian, which is invariant under these symmetry transformations, can be written as follows:
\begin{equation}
	\ham=\bigoplus_{\alpha}\ham_{\alpha},
	\label{eq_symm_ham}
\end{equation}
where each $\ham_{\alpha}$ is a block corresponding to a specific symmetry sector $\alpha$.
Instead of diagonalizing the full Hamiltonian in the entire Hilbert space, one can diagonalize each block $\ham_{\alpha}$ independently. 
This drastically reduces the computational cost, as the size of each block is much smaller than the full Hilbert space.

In the following, we review two main cases of symmetries: U(1) invariance, related to particle-number conservation, and translational invariance, related to momentum convervation. 
Both symmetries are currently employed in the ED-LGT Python library \cite{Cataldi2024Edlgt} and exploitable for simulating any QMB system or LGT in the dressed-site formalism. 
\begin{figure}
	\centering
	\includegraphics[width=1\textwidth]{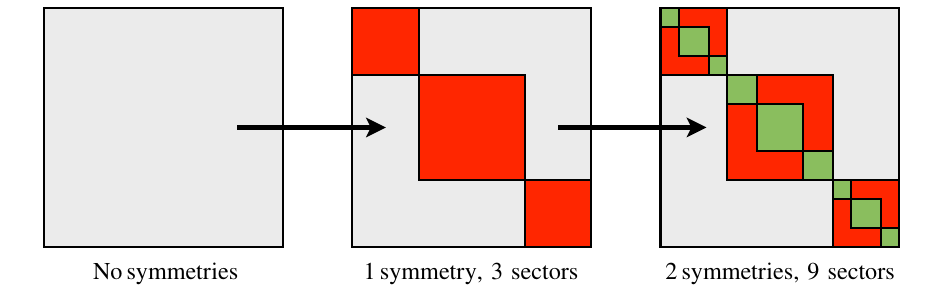}
	\label{fig_ED_blocks}
	\caption{Illustration of block diagonalization. In absence of symmetries, we have to diagonalize the whole Hamiltonian, which explores the whole Hilbet space $\sim d^{\Nsites}$.
	If the Hamiltonian has one symmetry, by constructing states labeled by the corresponding conserved quantum number, the Hamiltonian matrix breaks up into blocks that can be diagonalized independently of each other (middle). 
	Applying another symmetry, the blocks can be further broken up into smaller blocks (right) labeled by two different quantum numbers, etc \cite{Sandvik2010ComputationalStudiesQuantum}.}
\end{figure}
% =================================================================
\subsubsection{Particle-number conservation}
The presence of a global U(1) symmetry (not local as in LGTs) is associated with the conservation of a globally conserved quantity, such as the total number of particles, charge, or spin (along a certain direction) in a QMB system. 
Formally, this means that the Hamiltonian $\ham$ commutes with a total operator $\hat{A}$:
\begin{align}
\comm{\ham}{\hat{A}} &= 0&
\rm{where}&&
\hat{A} &= \sum_{\vecsite}\hat{a}_{\vecsite}
\end{align}
and $\hat{a}_{\vecsite}$ is the particle number operator on lattice site $\vecsite$. 
This implies that $\hat{A}$ is conserved under the time evolution generated by the Hamiltonian.
For instance, in the case of the Ising model without transverse field, the conserved quantity $\hat{A}$ is the magnetization along $z$ and $\hat{a}_{\vecsite}=\Sz_{\vecsite}$.
In bosonic or ferminic systems like in Hubbard Hamiltonians (see \cref{sec_hubbard_model}), the conserved quantity is the total particle number and $\hat{a}_{\vecsite}=\nop_{\vecsite}$ is the operator that counts the number of particles at site $\vecsite$. 
For fermions, $\nop_{\vecsite}$ can only take values 0 or 1 (due to the Pauli exclusion principle), while for bosons $\nop_{\vecsite}$ can take any non-negative integer value.
As for LGTs such as the ones described in \cref{chap_LGT} and simulated in \cref{chap_SU2_groundstate,chap_scars}, there is a global U(1) symmetry related to the invariants of the matter fields under rotations $\hpsi_{\vecsite}\to e^{i\theta}\hpsi_{\vecsite}$ and corresponds to the conservation of charge: baryon number for non-Abelian LGTs such as SU(2) Yang Mills in \cref{sec_SU2_model} and electric charge for Abelian ones such as QED in \cref{sec_U1_model}.
In the former case, $\hat{a}_{\vecsite}$ corresponds to the baryon number density $b$ defined in \cref{eq_SU2_b_density}; in the latter case, $\hat{a}_{\vecsite}=\hat{q}_{\vecsite}\equiv \hpsi^{\dagger}_{\vecsite}\hpsi_{\vecsite}-((1)^{\vecsite}-1)/2$ \cite{Magnifico2021LatticeQuantumElectrodynamics,Felser2020TwoDimensionalQuantumLinkLattice,Rigobello2021EntanglementGenerationMathrm}.
Due to the aforementioned sign-problem, accessing sectors with finite baryon number density $b$ or finite electric charge is challenging for MC based algorithms, while being straightforward for ED simulations (by exploiting block diagonalization) \cite{Cataldi2024Edlgt} or TN methods, where the U(1) sector is directly selected in the TN ansatz, ensuring that the symmetry is preserved throughout the simulation \cite{Singh2010TensorNetworkDecompositions,Singh2011TensorNetworkStates,Orus2019TensorNetworksComplex,Silvi2019TensorNetworksAnthology}.
% =================================================================
\subsubsection{Momentum conservation}
\label{sec_ED_momentum_conservation}
In a QMB system with translational symmetry, as in the case of periodic boundary conditions (PBC), the corresponding Hamiltonian $\ham$ is invariant under translations of particle positions or lattice sites. 
This means shifting all particles or lattice sites by a unit (or a set of units) leaves the physical properties of the system unchanged \cite{Sandvik2010ComputationalStudiesQuantum}.
For simplicity, let us focus on a 1D lattice, whose sites are labeled by $\site\in\qty{0,\dots,\Nsites}$.

We define a translation operator $\hat{T}$ which moves every particle from site $\site$ to site $\site+1$, \idest{}, $\hat{T}\ket{\site} = \ket{\site+1}$, with $\hat{T}^{\Nsites}=\mathbb{1}$.
Due to translational symmetry, $\hat{T}\ham=\ham\hat{T}$, meaning that $\ham$ and $\hat{T}$ admit a common set of eigenstates.
Thanks to Bloch's theorem, these eigenstates can be expressed as wave functions with a definite momentum, characterized by a wave vector $\mom=2\pi \site/\Nsites$:
\begin{equation}
	\ket{\varphi_{\mom}} = \frac{1}{\sqrt{\Nsites_{\varphi}}}\sum_{\site=0}^{\Nsites_{\varphi}-1} e^{-i\mom\site}\hat{T}^{\site}\ket{\varphi},
	\label{eq_momentum_vector}
\end{equation}
where $\ket{\varphi}$ is a reference state. 
To construct the momentum basis for given $\mom$, we need a set of reference states resulting in a complete set of normalizable orthogonal states. Clearly, for two states $\ket{\varphi_{\mom}}$ and $\ket{\varphi^{\prime}_{\mom}}$ to be orthogonal, the corresponding representatives must obey $\hat{T}^{r}\ket{\varphi}\neq \ket{\varphi^{\prime}}\, \forall r \in \qty{1,\dots, \Nsites}$.
Therefore, for each set of translated states $\{\ket{\varphi(r)} =\hat{T}^{r}\ket{\varphi}\}$, only one of them should be used as a reference state \cite{Sandvik2010ComputationalStudiesQuantum}.
The normalization factor $\Nsites_{\varphi}$ in \cref{eq_momentum_vector} takes into account the periodicity of the reference state $\ket{\varphi}$, which in general could be smaller to $\Nsites$: $\hat{T}^{\Nsites_{\varphi}}\ket{\varphi}=\ket{\varphi}$ for some $\Nsites_{\varphi}<\Nsites$.

Once we have selected an orthonormal set of reference states, we can build the corresponding momentum basis via \cref{eq_momentum_vector} and express the Hamiltonian as follows:
\begin{equation}
	\ham_{\mom\mom^{\prime}}=\mel{\psi_{\mom}}{\ham}{\psi_{\mom^{\prime}}},
\end{equation}
where, due to the translational invariance, $\ham_{\mom\mom^{\prime}}=0$ for $\mom\neq\mom^{\prime}$.
Eigenvalues and eigenstates of the Hamiltonian in a particular momentum sector $\mom$ are obtained diagonalizing the block $\ham_{\mom\mom}$. 

When dealing with QMB systems where translational invariance involves a logical basis of $N^{b}>1$ sites, the translational symmetry unit is expanded from a single site to $N^{b}$ sites.
Correspondingly, we have to update the momentum vetor definition in \cref{eq_momentum_vector} with a normalization factor that matches the total number of unit cells $L=\frac{N}{N^{b}}$ or a smaller periodicity $L_{\varphi}$:
\begin{equation}
	\ket{\varphi_{\mom}} = \frac{1}{\sqrt{L_{\varphi}}}\sum_{\site=0}^{L_{\varphi}-1} e^{-i\mom\site}\hat{T}^{N^{b}\site}\ket{\varphi}.
\end{equation}
For instance, in LGTs employing the staggered fermion solution, the corresponding Hamiltonian is invariant under translations of two sites along each spatial direction.
For instance, ED simulations of the (1+1)D SU(2) Yang-Mills LGT in PBC (see \cref{sec_scars_nonAbelianLGT} and \cite{Cataldi*2025QuantumManybodyScarring}) have been obtained in the zero momentum, since the scarred initial states of the dyanmics belong to that sector.
% =================================================================
\subsection{Basic Tensor Operations}
\label{sec_TN_operations}
\begin{figure}
\centering
\includegraphics[width=\textwidth]{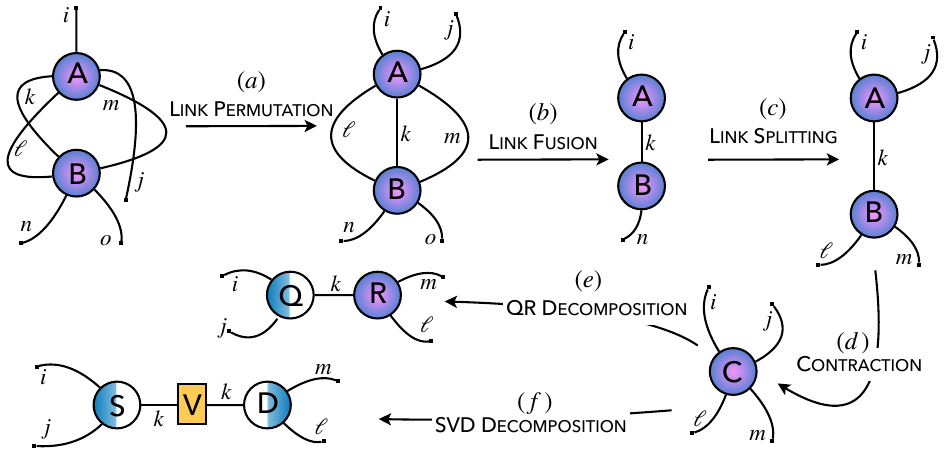}
\caption{Pictorial representation of the main manipulations (a)-(c), contractions (d), and decompositions (e)-(f) which can be performed on the tensor coefficient in \cref{eq_QMB_state}.}
\label{fig_TN_operations}
\end{figure}

Aside by block diagonalization and the use of symmetries, by accurately manipulating the coefficient tensor $\tensor$ in \cref{eq_QMB_state}, we can further extract a lot of information, such as quantum correlations, entanglement, and few-body observables, sometimes accessing only a portion of all its entries. 
The main operations which can be performed over $\tensor_{i_{1},\dots,i_{\Nsites}}$ involve the indices labelling its links. 
In particular, it is possible to distinguish among at least three different kinds of operations affecting links: manipulations, contractions, and decompositions. Below, we briefly discuss each of them. 
For an equivalent pictorial representation, see \cref{fig_TN_operations}.
% =================================================================
\subsubsection{Manipulations}
Link manipulations involve operations like permutation, which reorders tensor indices, fusion, which combines adjacent links into a single index, and splitting, which breaks down a link into multiple smaller indices. 
% =================================================================
\paragraph{Link permutation}
It reorders the links $i_{1}\dots i_{\Nsites}$ of the tensor $\tensor$ by moving the link at position $\ell$ into $k=\sigma(\ell)$ of the permuted tensor $\tensor^{[p]}$:
\begin{equation}
	\tensor^{[p]}_{i_{\sigma(1)}\dots i_{\sigma(\Nsites)}}=\tensor_{i_{1}\dots i_{\Nsites}}.
\end{equation}
% =================================================================
\paragraph{Link fusion} 
It combines $m-k+1$ adjacent links ($i_{k}\dots i_{m}$) of dimensions $(d_{k}\dots d_{m})$ into a fused link $j$ defined as follows:
\begin{equation}
	\begin{split}
		j=\text{fuse}(i_{k}\dots i_{m})&=1+\sum_{r=k}^{m}(i_{r}-1)\prod_{j=1}^{r-k}d_{j}\quad \text{where} \quad j\in\mathbb{J}=\mathbb{1}_{k}\times \dots \times \mathbb{1}_{m}\\
		&=1+(i_{k}-1)d_{k}^{0}+(i_{k+1}-1)d_{k+1}+\dots + (i_{m}-1)d^{m-k}_{m}.
	\end{split}
\end{equation}
% =================================================================
\paragraph{Link splitting} It replaces a given link $k$ with index $j_{k}$ and dimension $d_{k}$ with $m-k+1$ links $i_{k}\dots i_{m}$ of dimensions $d_{k^{\prime}}\dots d_{m^{\prime}}$ such that $d_{k}=d_{k^{\prime}}\times \dots \times d_{m^{\prime}}$:
\begin{equation}
	\tensor^{\prime}_{i_{1}\dots i_{k-1} (i_{k+1}\dots i_{m})i_{m+1}\dots i_{\Nsites}}=\tensor_{i_{1}\dots i_{k-1}, \;j_{k}, \;i_{m+1}\dots i_{\Nsites}}.
\end{equation}
Link fusion and splitting are very useful for decomposing large tensors into manageable parts.
% =================================================================
\subsubsection{Contractions}
Whenever two tensors A and B, with links $\qty{a_{1},\dots a_{n}}$ and $\qty{b_{1}\dots b_{m}}$ of dimensions respectively $\qty{d^{A}_{1\dots n}}$ and $\qty{d^{B}_{1\dots m}}$, have a set of $k$ common links $s_{j}$ (where $\{d^{A}_{n-k+j}\}=\{d^{B}_{j}\}=\{d^{s}_{j}\}$ for $j=1\dots k$), we can contract them over $\qty{s_{1}\dots s_{k}}$, obtaining a single tensor C:
\begin{equation}
	C_{ab}\underset{split}{=}C_{a_{1},\dots a_{n-k}b_{k+1}\dots b_{m}}=\sum_{s_{1}\dots s_{k}}A_{a_{1},\dots a_{n-k}s_{1}\dots s_{k}}B_{s_{1}\dots s_{k}b_{k+1}\dots b_{m}}\underset{fuse}{=}\sum_{s}A_{as}B_{sb}\,,
\end{equation}
where in the $1^{st}$ and in the $3^{rd}$ equalities we exploited the link splitting and fusing operations respectively, obtaining the single links $s$, $a$, and $b$, with dimensions respectively:
\begin{align}
d_{s}&=d_{1}^{s}\times \dots \times d_{k}^{s}&d_{a}&=d_{1}^{a}\times \dots \times d_{n-k}^{a}& d_{b}&=d_{k+1}^{b}\times\dots \times d_{m}^{b}.
\end{align}
% =================================================================
\subsubsection{Decompositions}
\label{sec_TN_decompositions}
A tensor decomposition replaces a single input tensor with two or more output tensors, connected through a certain number of auxiliary links; correspondingly, it preserves with a given distribution, all the physical links of the input tensor. 
To perform a decomposition we need to:
\begin{enumerate}
	\item choose a \emph{bipartition} of the physical links of the input tensor $\tensor$: $(i_{1}\dots i_{r})\;(i_{r+1}\dots i_{\Nsites})$.
	\item fuse these latter into two single row and column links: $\tensor_{(i_{1}\dots i_{r})(i_{r+1}\dots i_{\Nsites})}=\tensor_{ab}$
	\item decompose the obtained tensor into two factor matrices and (eventually) a diagonal one:
	\begin{equation}
		\tensor_{ab}=\sum_{k=1}^{d_{k}}A_{ak}\lambda_{k}B_{kb};
	\end{equation}
	\item split the links initially fused to restore the initial network geometry.:
	\begin{equation}
		\tensor_{(i_{1}\dots i_{\Nsites})}=\sum_{k}A_{(i_{1}\dots i_{r}), k}\lambda_{k} B_{k,(i_{r+1}\dots i_{\Nsites})}.
	\end{equation}
\end{enumerate}
Noticeable examples are the QR decomposition and the SVD decomposition.
% =================================================================
\paragraph{QR decomposition}
We can split $\tensor$ into two tensors, namely $Q$ and $R$:
\begin{equation}
	\tensor_{(i_{1}\dots i_{r})(i_{r+1}\dots i_{\Nsites})}=\tensor_{ab}\underset{\rm{QR}}{=}\sum_{k=1}^{d_{k}}Q_{ak}R_{kb}=\sum_{k=1}^{d_{k}}Q_{(i_{1}\dots i_{r})k}R_{k(i_{r+1}\dots i_{\Nsites})}.
\end{equation} The matrix $R_{kb}$ is an \emph{upper trapezoidal matrix}, whereas $Q_{ak}$ is a \emph{semi-unitary matrix} w.r.t. $k$, that is $\sum_{a}Q_{ak}Q^{*}_{ak^{\prime}}=\delta_{kk^{\prime}}$. The minimal dimension $d_{k}$ is such that 
\begin{equation}
	d_{k}=\min\qty{a,b}=\min\qty{(i_{1}\cdot\dots \cdot i_{r}),(i_{r+1}\cdot\dots \cdot i_{\Nsites})}.
	\label{eq_TN_QR_decomposition}
\end{equation} 
% =================================================================
\paragraph{Singular value decomposition (SVD)} 
In this case, we split $\tensor$ into three tensors:
\begin{equation}
	\tensor_{(i_{1}\dots i_{r})(i_{r+1}\dots i_{\Nsites})}=\tensor_{ab}\underset{\rm{SVD}}{=}\sum_{k=1}^{d_{k}}S_{ak}V_{kk}D_{kb}=\sum_{k=1}^{d_{k}}S_{ak}\lambda_{k}D_{kb}=\sum_{k=1}^{d_{k}}S_{(i_{1}\dots i_{r})k}\lambda_{k}D_{k(i_{r+1}\dots i_{\Nsites})},
	\label{eq_TN_SVD_decomposition}
\end{equation}
where $V$ is a diagonal and positive real matrix, whereas $S$ and $D$ are \emph{semi unitary matrices} ($S^{\dagger}S=\mathbb{1}=DD^{\dagger}$).
This decomposition is quite useful to mention the possibility of \emph{truncation} for an index. Indeed, since the diagonal elements of V are real, positive, and hence lower bounded from below by 0, they can be ordered (for instance) in \emph{descending order}: 
\begin{equation}
	\qty[V_{11}=\lambda_{1}\geq V_{22}=\lambda_{2}\dots \geq0].
\end{equation} 
Moreover, if $\lambda_{k}<\epsilon$ $ \forall k > \chi$ for $\chi \in\qty{1,\dots, d_{k}}$, it is possible to disregard some of the singular values $\lambda_{k}$ with a precision $\epsilon$. 
In other words, whenever the singular values decay fast enough after some $\lambda_{\chi}$, we can neglect them. 
Correspondingly, the matrices $S$ and $D$ can be truncated, and $d_{k}$ reduced to $\chi$. 
Such a reduction introduces an error in the original representation of the tensor $\tensor_{ij}$, which can be estimated as follows:
\begin{equation}
	\norm{\tensor_{ij}-\sum_{k=1}^{\chi}S_{ik}\lambda_{k}D_{kj}}=\norm{\sum_{k=m+1}^{d_{k}}S_{ik}\lambda_{k}D_{kj}}<\sum_{k=\chi+1}^{d_{k}}\norm{S_{ik}D_{kj}}<\epsilon C
	\label{eq_TN_bond_truncation}
\end{equation}
for a finite constant C. 
As we will see in \cref{sec_TN}, in the context of Tensor Networks, the parameter $\chi$ will correspond to the \emph{bond dimension}.
% =================================================================
\paragraph{Schmidt Decomposition}
Let us consider a bipartition according to which the first $\Nsites_{A}$ particles belong to a subsystem $A$ with Hilbert space $\hil_{A}=\bigotimes_{k=1}^{\Nsites_{A}}\hil_{k}$, whereas, the remaining $\Nsites_{B}=\Nsites-\Nsites_{A}$ particles belong to the subsystem $B$ with Hilbert space $\hil_{B}=\bigotimes_{k=1}^{\Nsites_{B}}\hil_{k}$; globally, we have $\hil=\hil_{A}\otimes\hil_{B}$. 
Assuming all system sites with the same $d$-dimensional Hilbert space, \idest{} $\dim \hil_{k} = d\, \forall k$, then the QMB state and its coefficient tensor can be arranged as a \textit{linear superposition} of products of states of the two sub-partitions:
\begin{equation}
	\ket{\psiqmb}=\sum_{n=1}^{\Nsites_{A}}\sum_{m=1}^{\Nsites_{B}}C_{nm}\ket{a_{n}}\ket{b_{m}}\,,
	\label{eq_TN_psi_bipartite1}
\end{equation} 
where $\qty{\ket{a_{n}}}_{n=1}^{\Nsites_{A}}\in\hil_{A}$ and $\qty{\ket{b_{m}}}_{m=1}^{\Nsites_{B}}\in\hil_{B}$ are orthonormal basis of A and B. 
The corresponding density matrix $\rho$ associated to \eqref{eq_TN_psi_bipartite1} reads
\begin{equation}
	\begin{split}
		\rho=\ket{\psiqmb}\bra{\psiqmb}&=\sum_{n,n^{\prime}}\sum_{m,m^{\prime}}C_{nm}C^{*}_{n^{\prime}m^{\prime}}\ket{a_{n}}\ket{b_{m}}\bra{a_{n^{\prime}}}\bra{b_{m^{\prime}}}\\
		&=\sum_{n,n^{\prime}}\sum_{m,m^{\prime}}\rho_{nm}^{n^{\prime}m^{\prime}}\ket{a_{n}}\bra{a_{n^{\prime}}}\otimes\ket{b_{m}}\bra{b_{m^{\prime}}}\,,
	\end{split}
	\label{eq_TN_rho_bipartite1}
\end{equation} 
where we grouped the coefficient matrix $C_{nm}$ and its complex conjugate $C^{*}_{n^{\prime}m^{\prime}}$ into a single $4-$order coefficient tensor $\rho_{nm}^{n^{\prime}m^{\prime}}$.
The \emph{reduced density matrix} of subsystem A is then given by tracing $\rho$ over all the states of B:
\begin{equation}
\begin{split}
\rho_{A}=\sum_{\ell=1}^{\Nsites_{B}}\mel{b_{\ell}}{\rho}{b_{\ell}}
	&=\sum_{n,n^{\prime}}\sum_{\ell}\sum_{m,m^{\prime}}\rho_{nm}^{n^{\prime}m^{\prime}}\ket{a_{n}}\bra{a_{n^{\prime}}}\otimes\braket{b_{\ell}}{b_{m}}\braket{b_{m^{\prime}}}{b_{\ell}}\\
	&=\sum_{n,n^{\prime}=1}^{\Nsites_{A}}\sum_{\ell=1}^{\Nsites_{B}}\rho_{n\ell}^{n^{\prime}\ell}\ket{a_{n}}\bra{a_{n^{\prime}}}\,,
\end{split}
\end{equation} 
where we recall the partial trace not to depend on the choice of the basis of the partition we trace out. 
Then, if $\lambda_{1}, \lambda_{2},\dots \lambda_{\Nsites_{A}}$ are the eigenvalues of $\rho_{A}$ and $\ket{\alpha_{1}} \dots \ket{\alpha_{\Nsites_{A}}}$ are the corresponding eigenvectors, we can rewrite $\ket{\psiqmb}$ in terms of $\qty{\ket{\alpha_{n}}}_{n=1}^{\Nsites_{A}}$ as follows:
\begin{equation}
	\ket{\psiqmb}=\sum_{n,m}\mathcal{C}_{nm}\ket{\alpha_{n}}\ket{b_{m}}\,,
	\label{eq_TN_schmidt0}
\end{equation}
where $\mathcal{C}_{nm}$ is a new coefficient tensor.
We can then build up a new basis also for the subsystem B, in such a way that each vector is a clever superposition of vectors of the old basis:
\begin{equation}
	\ket{\beta_{k}}=\sum_{m=1}^{\Nsites_{B}}\frac{\mathcal{C}_{km}}{\lambda_{k}}\ket{b_{m}}.
	\label{eq_TN_change_basis}
\end{equation}
Since the matrix $\mathcal{C}_{km}/\lambda_{k}$ is unitary for each $k$, then $\qty{\ket{\beta_{k}}}$ is an orthonormal basis, and we can rewrite \eqref{eq_TN_psi_bipartite1} and \eqref{eq_TN_rho_bipartite1} in terms of the \emph{Schmidt decomposition}:
\begin{equation}
	\ket{\psiqmb}=\sum_{n,m}\mathcal{C}_{nm}\ket{\alpha_{n}}\ket{b_{m}}=\sum_{n}\lambda_{n}\ket{\alpha_{n}}\ket{\beta_{n}}\qquad \rho=\sum_{n,m}\lambda_{n}\lambda^{*}_{m}\ket{\alpha_{n}}\bra{\alpha_{m}}\otimes \ket{\beta_{n}}\bra{\beta_{m}}.
	\label{eq_TN_schmidt_decomposition}
\end{equation} 
It is then clear that both the reduced density matrices $\rho_{A}$ and $\rho_{B}$ are diagonal in the Schmidt basis, and display the same positive spectrum:
\begin{subequations}
\begin{align}
	\rho_{A}=\sum_{q}\mel{\beta_{q}}{\rho}{\beta_{q}}=\sum_{n,m}\sum_{q}\lambda_{n}\lambda_{m}^{*}\bra{\beta_{q}}\ket{\beta_{n}}\ket{\alpha_{n}}\bra{\alpha_{m}}\bra{\beta_{m}}\ket{\beta_{q}}&=\sum_{q}\abs{\lambda_{q}}^{2}\ket{\alpha_{q}}\bra{\alpha_{q}}\\
	\rho_{B}=\sum_{k}\mel{\alpha_{k}}{\rho}{\alpha_{k}}&=\sum_{k}\abs{\lambda_{k}}^{2}\ket{\beta_{k}}\bra{\beta_{k}}.
\end{align}
\label{eq_TN_rho_schmidt_decomposition}
\end{subequations}
Such a Schmidt decomposition is very exploitable when computing quantum correlations between bipartitions of a system in a pure state.
Conversely, it reveal useless for multipartite descriptions or when applied to mixed states, where no expression in terms of single partitions is available. 
% =================================================================
\subsubsection{Physical perspective}
\label{sec_TN_physical_meaning}
What may appear as just a series of algebraic or technical manipulations on purely mathematical objects, is firmly connected to the control of quantum correlations and \emph{entanglement} in a QMB system \cite{Plenio2006IntroductionEntanglementMeasures,Amico2008EntanglementManybodySystems}. 
From a purely physical perspective, if the coefficient tensor $\tensor_{(i_{1}\dots i_{\Nsites})}$ describes the state of a $\Nsites$-body system, then, every decomposition or contraction allows to move from a many-body description (with a single large tensor), to a single site or partite description (made of several small physical tensors) and the other way around, with eventual intermediate steps.
The auxiliary link $k$ along which decompositions/contractions are performed deserves special regard: in absence of loops in the TN structure, its dimension $d_{k}$ quantifies the number of shared degrees of freedom between the arisen subsystems (e.g. QR in \cref{eq_TN_QR_decomposition}, SVD in \cref{eq_TN_SVD_decomposition}, or Schmidt in \cref{eq_TN_schmidt_decomposition} decompositions). 
In turn, these shared d.o.f. provide the amount of quantum correlations through which the partitions are mutually entangled. 
Therefore, the larger the truncation performed on $d_{k}\to \chi$ (see \cref{eq_TN_bond_truncation}), the smaller the entanglement between the partition we keep in track. 
For instance, exploiting the entanglement entropy \cite{Eisert2010ColloquiumAreaLaws}, we have:
\begin{equation}
	\entropy_{A}=-\sum_{k=1}^{\chi}\lambda_{k}\log \lambda_{k}.
	\label{eq_entanglement_entropy}
\end{equation}
% =================================================================
\subsection{Entanglement area law}
\label{sec_TN_area_law}
As anticipated in \cref{sec_ED}, ED simulations of a QMB system, even resolving all its symmetries, are not feasible for a large number system size $\Nsites$, as the total Hilbert space grows exponentially with $\Nsites$. 
Nonetheless, a great variety of natural QMB systems happens to be described by ground and thermal equilibrium states that own little-to-moderate entanglement content.
The physical states of these systems live in a small corner of the total Hilbert space, which can be efficiently targeted and parameterized without the exponential cost required by an exact representation.

This property is formally described as \emph{entanglement area law}, fulfilled by low-energy states of local QMB Hamiltonians \cite{Eisert2010ColloquiumAreaLaws}: the entanglement between a partition of the system and the rest is proportional to the area of the boundary between them, instead of its volume, as happens for the majority of states in the Hilbert space.
Thus, a state obeying area law contains much fewer quantum correlations than expected for a generic (or random) QMB state\footnote{Small corrections to the area law exist for instance close to a transition point for one-dimensional quantum systems (logarithmic corrections).
However, the entanglement remains overall moderate \cite{Calabrese2004EntanglementEntropyQuantum}.}.
From a theoretical point of view, the entanglement area law has been rigorously proven for (i) one-dimensional gapped local Hamiltonians, where the locality means that a lattice site interacts only with neighboring sites, without two-body all-to-all interactions \cite{Hastings2007AreaLawOnedimensional,Kuwahara2020AreaLawNoncritical, Cho2018RealisticAreaLawBound}; (ii) for quantum states at thermal equilibrium, independently from the dimensionality of the system \cite{Wolf2008AreaLawsQuantum}.
Even though rigorous proof for QMB systems in higher dimensions is lacking, several numerical and phenomenological shreds of evidence suggest that area law still holds in the presence of local interactions \cite{Masanes2009AreaLawEntropy, Hamza2009ApproximatingGroundState, Eisert2010ColloquiumAreaLaws, Kastoryano2019LocalityBoundaryImplies, Cirac2019MathematicalOpenProblems}.
% =================================================================
\section{Tensor Networks Methods}
\label{sec_TN}
We are then ready to present a general overview of Tensor Networks (TN), which are based on controlled wave-function variational ansätze exploiting the area-law entanglement bounds satisfied by locally interacting QMB systems. 
Thus, TN admit an efficient representation of the low-energy sectors contributing to the equilibrium properties and (low-entangled) time evolution \cite{Eisert2010ColloquiumAreaLaws}, without suffering from the aforementioned sign problem \cite{Loh1990SignProblemNumerical}.
Before looking at specific TN realizations, it is convenient to provide the general idea behind a TN geometry.
% =================================================================
\subsection{Tensor Network state}
The area law described in \cref{sec_TN_area_law} has important implications on the numerical simulation of QMB models: indeed, for ground-states and first excited states of local Hamiltonians where the entanglement content is low-to-moderate, it is possible to obtain an approximate but efficient representation capable of describing the main properties of these states \cite{Eisert2010ColloquiumAreaLaws}.
TNs give a natural language for this representation, by replacing the complete rank-$\Nsites$ tensor $\tensor_{i_{1},\dots,i_{\Nsites}}$ of \cref{eq_QMB_state} with a chain of $Q$ smaller tensors $\tensor^{[q]}$ with $q\in\qty{1,\dots,Q}$ connected via \emph{auxiliary links} $\qty{\alpha}_{q}$:
\begin{equation}
	\tensor_{i_{1}\dots i_{\Nsites}}=\sum_{\alpha_{1},\dots\alpha_{L}}\tensor^{[1]}_{\qty{i}_{1},\qty{\alpha}_{1}}\tensor^{[2]}_{\qty{i}_{2},\qty{\alpha}_{2}}\dots \tensor^{[Q]}_{\qty{i}_{Q},\qty{\alpha}_{Q}}.
\label{eq_TN_tensor_decomposition}
\end{equation}
Each sigle tensor $\tensor^{[q]}$ has a set of physical indices $\{i\}_{q}$ of dimension $d$ (one for each lattice site), whereas the dimension of the auxiliary indices $\qty{\alpha}_{q}$ is upper bounded by a control parameter $\chi$ that is called \emph{maximal bond dimension}.

The advantage of passing from the exact representation of \cref{eq_QMB_state} to a TN representation like \cref{eq_TN_tensor_decomposition} is that the number of TN parameters is of the order $O(\mathrm{poly}(d)\mathrm{poly}(N)\mathrm{poly}(\chi))$, \eg{}, $\mathcal{O}(\Nsites \chi \max(\chi, d)^2)$ for the TTN \cite{Magnifico2024TensorNetworksLattice}. 
The scaling with the system size and consequently the computational complexity is now \emph{polynomial} and not exponential, as in the exact representation of the QMB state.
It is worth noting that the bond dimension $\chi$ determines the degree of entanglement and quantum correlations encoded in the TN.
Namely, for $\chi=1$, the TN describes a \emph{separable} state (product state, no entanglement), whereas one recovers the exact but inefficient representation in the limit $\chi {\lesssim} d^{N}$ assuming the system to be strongly entangled.
Tuning $\chi$ properly allows interpolating between these two extreme regimes, efficiently reproducing the entanglement of the quantum state by exploiting decompositions like Schmidt/QR/SVD discussed in \cref{sec_TN_operations} \cite{Cirac2021MatrixProductStates, Silvi2019TensorNetworksAnthology}.
For example, dynamical problems characterized by a slow entanglement entropy growth, as in the many-body scars dynamics observed in \cref{sec_scars_nonAbelianLGT}, can be efficiently and accurately described by MPS ansatz with small or intermediate bond dimensions \cite{Verstraete2008MatrixProductStates, Schollwock2011DensitymatrixRenormalizationGroup}.
% =================================================================
\subsubsection{Gauge invariance}
More in general, the employment of auxiliary links clearly adds in the TN state some \emph{information redundancy} with respect to the quantum state it describes. 
Such a redundancy corresponds to a set of linear transformations of the TN, which leave the quantum state unchanged. 
We usually refer to these transformations as \emph{gauge transformations}, as they corresponds to \emph{local and invertible manipulations} on the auxiliary TN links $\qty{\alpha}_{q}$ without affecting the physical degrees of freedom. 

To give an example, let us focus on the following portion of \cref{eq_TN_tensor_decomposition}
\begin{equation}
\tensor^{\qty[q+q^{\prime}]}_{\qty{i},\qty{i^{\prime}}}=\sum_{\alpha_{\eta}}\tensor^{[q]}_{\qty{i},\qty{\alpha}_{\vert_{\eta}},\alpha_{\eta}}\tensor^{[q^{\prime}]}_{\qty{i^{\prime}},\qty{\alpha^{\prime}}_{\vert_{\eta}},\alpha_{\eta}},
\label{eq_TN_gauge_example1}
\end{equation} 
where $\qty{\alpha}_{\vert_{\eta}}$ collects the auxiliary links of the tensor except for $\eta$ connecting the nodes $q$ and $q^{\prime}$ with dimension $\chi_{\eta}$. 
Then, given a $\chi^{\prime}_{\eta}\times \chi_{\eta}$ left-invertible matrix $Y_{\alpha_{\eta}\alpha^{\prime}_{\eta}}$, such that
\begin{equation}
	\sum_{\alpha^{\prime}}\big(Y^{-1}\big)_{\alpha \alpha^{\prime}}Y_{\alpha^{\prime} \alpha^{\prime\prime}}
	=\delta_{\alpha \alpha^{\prime\prime}}
	\neq \sum_{\alpha^{\prime}}Y_{\alpha \alpha^{\prime}}\big(Y^{-1}\big)_{\alpha^{\prime} \alpha^{\prime\prime}},
\end{equation} 
the TN state in \cref{eq_TN_gauge_example1} is left unchanged if the tensors $\tensor^{[q]}$ and $\tensor^{[q^{\prime}]}$ transform as follows:
\begin{equation}
	\begin{aligned}
		\tensor^{[q]}_{\qty{i},\qty{\alpha}_{\vert_{\eta}},\alpha_{\eta}}&\longrightarrow \tilde{\tensor}^{[q]}_{\qty{i},\qty{\alpha}_{\vert_{\eta}},\alpha_{\eta}}=\sum_{\beta}Y_{\alpha_{\eta}b_{\eta}}\tensor^{[q]}_{\qty{i},\qty{\alpha}_{\vert_{\eta}},b_{\eta}}\\
		\tensor^{[q^{\prime}]}_{\qty{i^{\prime}},\qty{\alpha^{\prime}}_{\vert_{\eta}},\alpha_{\eta}}&\longrightarrow \tilde{\tensor}^{[q^{\prime}]}_{\qty{i^{\prime}},\qty{\alpha^{\prime}}_{\vert_{\eta}},\alpha_{\eta}}=\sum_{\gamma}\big(Y^{-1}\big)_{\alpha_{\eta}\gamma_{\eta}}\tensor^{[q^{\prime}]}_{\qty{i^{\prime}},\qty{\alpha^{\prime}}_{\vert_{\eta}},\gamma_{\eta}}.
	\end{aligned} 
\label{eq_TN_gauge_transformations1}
\end{equation}
Overall, these contractions on the two edges of the link $\eta$ leaves the composite tensor $\tensor^{\qty[q+q^{\prime}]}$ in \cref{eq_TN_gauge_example1} invariated. 
Therefore, since the state $\ket{\psiqmb}$ in \cref{eq_TN_tensor_decomposition} depends on $\tensor^{\qty[q]}$ and $\tensor^{\qty[q^{\prime}]}$ via $\tensor^{\qty[q+q^{\prime}]}$, it is insensitive to such a transformation.
The intrinsic presence of gauge invariances arising from the TN prescription is going to be numerically exploited when contracting $\ket{\psiqmb}$ with its corresponding $\bra{\psiqmb}$ through the Hamiltonian or other many-body operators.
% =================================================================
\subsection{Loopless Tensor Networks ansätze}
\label{sec_TN_loopfree_geometries}
\begin{figure}
	\centering
	\includegraphics[width=\textwidth]{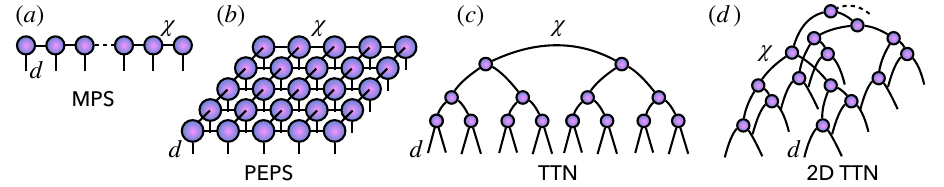}
	\caption{Examples of tensor network structures \cite{Magnifico2024TensorNetworksLattice}: (a)~Matrix Product States (MPS); (b)~Projected Entangled Pair States (PEPS); (c)~Tree Tensor Networks (TTN) for an underlying one-dimensional system; (d)~TTN for an underlying two-dimensional square lattice. 
	The physical links with local dimension $d$ and the auxiliary links with bond dimension $\chi$ are highlighted in all the figures.}
	\label{fig_TN_architectures}
\end{figure}
The most widely used TN architectures are represented in \cref{fig_TN_architectures}.
For 1D systems, Matrix Product States (MPS) \cite{Brockt2015MatrixproductstateMethodDynamical,Chan2016MatrixProductOperators,Cirac2021MatrixProductStates,Kliesch2014MatrixProductOperatorsStates,Klumper1993MatrixProductGround} represent the landmark of TN simulations, and are largely employed for Density Matrix Renormalization Group's (DMRG) techniques \cite{White1992DensityMatrixFormulation,White1993DensitymatrixAlgorithmsQuantum}. 
Beyond one dimension, there are Projected Entangled Pair States (PEPS) \cite{Orus2014PracticalIntroductionTensor,Verstraete2008MatrixProductStates,Vanderstraeten2016GradientMethodsVariational,Vanderstraeten2022VariationalMethodsContracting,Lubasch2014AlgorithmsFiniteProjected,Lubasch2014UnifyingProjectedEntangled,Tepaske2021ThreedimensionalIsometricTensor}, Tree Tensor Networks (TTN) \cite{Silvi2010HomogeneousBinaryTrees,Silvi2019TensorNetworksAnthology,Shi2006ClassicalSimulationQuantum,Gerster2014UnconstrainedTreeTensor} and Multi-scale Entanglement Renormalization Ansatz (MERA) \cite{Vidal2007EntanglementRenormalization,Evenbly2009EntanglementRenormalizationTwo,Cincio2008MultiscaleEntanglementRenormalization}.

In this thesis, we focus on TTNs, which together with MPS, are \emph{loopless Tensor Networks} \cite{Silvi2019TensorNetworksAnthology}, a special class of TN geometries which have found increasing applications for QMB and LGT simulations on finite lattices \cite{Verstraete2008MatrixProductStates,Orus2019TensorNetworksComplex,Cataldi2021HilbertCurveVs,Ferrari2022AdaptiveweightedTreeTensor,Rico2014TensorNetworksLattice,Kuhn2014QuantumSimulationSchwinger,Zohar2021QuantumSimulationLattice,Meurice2022TensorNetworksHigh,Tagliacozzo2014TensorNetworksLattice, Felser2020TwoDimensionalQuantumLinkLattice, Magnifico2021LatticeQuantumElectrodynamics,Cataldi2024Simulating2+1DSU2}. 
Of course, the main concepts and ideas of these two ansätze can be easily generalized and applied to other TN structures, such as PEPS or MERA, which we briefly outline in \cref{sec_TN_loop_geometries}.
% =================================================================
\subsubsection{Loopless Tensor Network attributes}
Before exploring MPS and TTN geometries, it is convenient to review all the TN attributes linked to an intrinsic loopless geometry, which hold for the all the loopless TN realizations.
As the name says, loopless TNs can be thought as network graphs which contain no circles among the single tensors. 
The main consequences of the absence of loops for TNs are the followings: first, they can exploit the full power of gauge transformations in \cref{eq_TN_gauge_transformations1}, reducing the number of needed contractions and enhancing the computational speed.
Secondly, the maximal entanglement between two given subsystems of the TN depends on the chosen bipartition of the system, and hence on the bond dimension $\chi$ of the auxiliary link over which the bipartition is performed. 
Such a feature is known as \emph{entanglement clustering} \cite{Eisert2013EntanglementTensorNetwork}.
More in detail, any loopless TN geometry admits the following attributes.
% =================================================================
\paragraph{A distance} $dist(a,b)$ between two tensor nodes $a, b$ of the network: it corresponds to the number of links encountered along the \emph{shortest path} between them. For a loopless TN, such a path is unique: all the nodes which share the same distance $d$ w.r.t. a central node $c$ form a \emph{level} $L(d;c)$. 
This gives the loopless TN geometry a well defined layered structure w.r.t. each tensor.
% =================================================================
\paragraph{A bipartition} of the QMB system: every (auxiliary) link $\eta$ between two nodes $q$ and $q^{\prime}$ induces a bipartition of state of the whole model. 
The Hilbert states of the two partitions are spanned by the canonical basis states 
\begin{equation}
	\qty{\ket{i_{s}}: dist(s,q)<dist(s,q^{\prime})}\qquad\text{and}\qquad \qty{\ket{i_{s}}: dist(s,q)>dist(s,q^{\prime})}
\end{equation}
respectively, that is by fusing all links $s$ ultimately connected either to $q$ or $q^{\prime}$ once removed the bipartition link from the network. 
Then, if A and B are the resulting TN subportions, we can rewrite any pure state $\ket{\psiqmb}$ using the Schmidt decomposition in \cref{eq_TN_schmidt_decomposition}:
\begin{equation}
	\ket{\psiqmb}=\sum_{k=1}^{\chi_{\eta}}\ket{A_{k}}\ket{B_{k}}\lambda_{k}
	\label{schmidt}
\end{equation} 
for $\lambda_{k}$ sorted in a descending order. 
In these terms, a loopless TN can be seen as a simultaneous Schmidt decomposition over all the auxiliary links.
For each network bipartition over a link $\eta$, the bond dimension $\chi_{\eta}$ provides an \emph{upper bound} to the Schmidt rank, which in turn contributes to define the \emph{bipartite entanglement} between the two subsystems (see \cref{eq_entanglement_entropy}). 
Obviously, the larger the bond dimension, the larger is the amount of quantum correlations encoded by the TN structure for the given bipartition. 
In these terms, by exploiting the SVD in \cref{eq_TN_SVD_decomposition}, it is possible to reduce the dimension of the auxiliary link $\eta$ simply by discarding the smallest values $\lambda_{k}$ and truncating the sum up to a given threshold $\chi_{\eta}^{\prime}$, obtaining a truncated state:
\begin{align}
	\ket{\psiqmb^{\rm{trunc}}}&=\frac{1}{\mathcal{N}_{kept}}\sum_{k=1}^{\chi_{\eta}^{\prime}}\ket{A_{k}}\ket{B_{k}}\lambda_{k},&
	\rm{where}&& 
	\mathcal{N}_{kept}&\equiv \sqrt{\sum_{k=1}^{\chi_{\eta}^{\prime}}\lambda_{k}}.
\end{align}
The error provided by the truncation, is estimated as in \cref{eq_TN_bond_truncation}.

The the computational complexity of a given TN ansatz is then expressed as a polynomial function of the \emph{maximal bond dimension} $\chi=\max_{\eta}\chi_{\eta}$ among all the TN auxiliary links, which represents the maximal level of entanglement between any two partitions formed over an auxiliary link. 
In these terms loopless TN ansätze display a smother scalings ($\mathcal{O}(\chi^{3})$ for MPS and $\mathcal{O}(\chi^{4})$ for TTN) w.r.t. PEPS \cite{Cirac2021MatrixProductStates} or MERA \cite{Qian2022TreeTensorNetwork}, where the presence of loops strongly increase computational costs, at least with $\mathcal{O}(\chi^{8})$.
% =================================================================
\subsubsection{Gauges in loopless Tensor Networks}
The absence of loops in the TN geometry allows for a strong speed up of tensor contractions between the network and its conjugate (like the ones in expectation values $\bra{\psiqmb}{\obs}\ket{\psiqmb}$ for any observable $\obs$) by installing and exploiting gauge transformations like \cref{eq_TN_gauge_transformations1}. 

\begin{figure}
	\centering
	\includegraphics[width=1\textwidth]{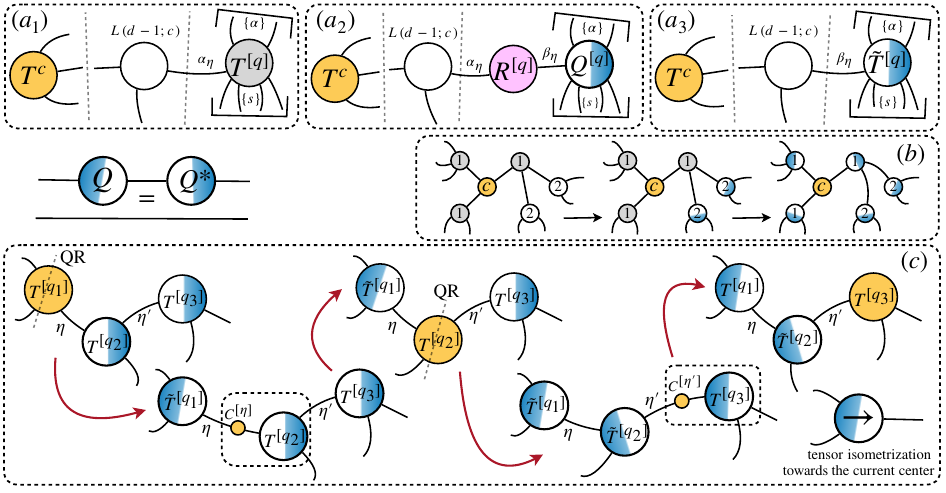}
	\caption{Pictorial representation of the employment of unitary gauges on a TN structure making use of QR decompositions. 
	(a) QR decomposition of a node $q$ of the level $L(d;c)$. 
	(b) Iterative gauge transformations of TN nodes with respect to the chosen (red) center one from the highest to the lowest level L of the TN structure. 
	(c) Iterative gauge transformation of the TN tensors in between of two successive center nodes.}
	\label{fig_TN_gauge_operations}
\end{figure}

Among all the possible choices of local symmetries, we select the \emph{unitary gauge}, which will be exploited in the ground state search algorithm of \cref{sec_TN_gs_algorithm}. 
The unitary gauge is defined w.r.t. a single node $\tensor^{[c]}$ of the TN, named \emph{current gauge center}. 
Starting from nodes of \emph{maximal distance} $d=d_{max}$ from $T^{[c]}$, we can install unitary gauges on each level $\qty{L(d;c):\; d=1\cdots d_{max}}$ of the network as follows:
\begin{enumerate}
	\item for each node $q \in L(d;c)$, perform a QR decomposition with respect to the link $\eta$ which leads towards the center node $c$ (see \cref{fig_TN_gauge_operations}a):
	\begin{equation}
		\tensor^{[q]}_{\qty{s}\qty{\alpha}_{\vert_{\eta}},\alpha_{\eta}}=\sum_{b_{\eta}=1}^{\chi_{\eta}}Q^{[q]}_{\qty{s}\qty{\alpha}_{\vert_{\eta}},b_{\eta}}R^{[q]}_{b_{\eta}\alpha_{\eta}}.
		\label{eq_TN_unitary_gauge}
	\end{equation} 
	\item update $\tensor^{[q]}\longrightarrow \tilde{\tensor}^{[q]}\equiv Q^{[q]}$ with the \emph{semi unitary isometry} of the QR decomposition, and contract the upper trapezoidal matrix $C^{[\eta]}\equiv R^{[q]}$ over the corresponding auxiliary link $\eta$ into the adjacent tensor $\tensor^{[q^{\prime}]}$ from the inner layer $L(d-1;c)$ (see \cref{fig_TN_gauge_operations}a). 
	Do it for all the nodes of the level $L(d;c)$.
	\item advance inwards $d\longrightarrow d-1$ and repeat the previous steps for all the tensors of $L(d-1;c)$, up to reach $d=1$ (see \cref{fig_TN_gauge_operations}b). 
	Notice that this iterative procedure implements a \emph{global gauge transformation} which does not change the physical state of the network. 
	The computational complexity of such a proceeding (QR decompositions plus R contractions over auxiliary links) is dominated by the QR decomposition $\sim \mathcal{O}(\chi^{Z+1})$, where Z is the maximal number of links of a single tensor in the network (maximal degree of a tensor).
	\item move the center from a node $c$ to a node $c^{\prime}$. 
	This corresponds to the implementation of steps 1 and 2 for each node along the path connecting $c$ to $c^{\prime}$. 
	Obviously, the non unitary matrix $C^{[\eta]}$ out coming from the QR decomposition of a node close to $c$ has to be contracted with the next node on the path towards $c^{\prime}$. 
	For example, if $c=q_{1}$ and $c^{\prime}=q_{3}$ as in \cref{fig_TN_gauge_operations}c, we should have the following steps
	\begin{equation}
		\begin{aligned}
			&\tensor^{[q_{1}]}_{\qty{s_{1}}\qty{\alpha_{1}}_{\vert_{\eta}}\alpha_{\eta}}\underset{\rm{QR}}{\longrightarrow}\sum_{\alpha_{\eta}}\tilde{\tensor}^{[q_{1}]}_{\qty{s_{1}}\qty{\alpha_{1}}_{\vert_{\eta}}\alpha_{\eta}}C^{[\eta]}_{\alpha_{\eta}b_{\eta}} \\
			&\sum_{b_{\eta}}C^{[\eta]}_{\alpha_{\eta}b_{\eta}}\tensor^{[q_{2}]}_{\qty{s_{2}}\qty{\alpha_{2}}_{\vert_{\eta,\eta^{*}}}b_{\eta}\alpha_{\eta^{*}}}\longrightarrow \tensor^{[q_{2}]}_{\qty{s_{2}}\qty{\alpha_{2}}_{\vert_{\eta,\eta^{*}}}\alpha_{\eta}\alpha_{\eta^{*}}} \\
			&\tensor^{[q_{2}]}_{\qty{s_{2}}\qty{\alpha_{2}}_{\vert_{\eta^{*}}}\alpha_{\eta^{*}}}\underset{\rm{QR}}{\longrightarrow}\sum_{\alpha_{\eta^{*}}}\tilde{\tensor}^{[q_{2}]}_{\qty{s_{2}}\qty{\alpha_{2}}_{\vert_{\eta^{*}}}\alpha_{\eta}^{*}}C^{[\eta^{*}]}_{\alpha_{\eta^{*}}b_{\eta^{*}}}\\
			&\sum_{b_{\eta^{*}}}C^{[\eta^{*}]}_{\alpha_{\eta^{*}}b_{\eta^{*}}}\tensor^{[q_{3}]}_{\qty{s_{3}}\qty{\alpha_{3}}_{\vert_{\eta^{*}}}b_{\eta^{*}}}\longrightarrow \tensor^{[q_{3}]}_{\qty{s_{3}}\qty{\alpha_{3}}_{\vert_{\eta^{*}}}\alpha_{\eta^{*}}}
		\end{aligned}
	\end{equation} where $C^{[\eta]}$ and $C^{[\eta^{*}]}$ are the non unitary upper trapezoidal matrices respectively defined on any two links $\eta$ and $\eta^{*}$ connecting $q_{1}\to q_{2}$ and $q_{2}\to q_{3}$.
\end{enumerate}
Aside of the gauge isometrization w.r.t a single tensor node, any loopless TN ansatz admits a \emph{canonical form} where all the tensors are concurrently isometrized, allowing for multipartite representation of QMB states (see \cite{Vidal2003EfficientClassicalSimulation,Vidal2004EfficientSimulationOneDimensional, Perez-Garcia2007MatrixProductState}).
However, such a canonical form is not suitable for the practical TN contractions that are performed during the variational algorithms, where the whole TN structure is sequentially optimized w.r.t each single tensor (see \cref{sec_TN_gs_algorithm}), which can be more conveniently chosen as the unique gauge center.

Aware of all these features of loopless TN geometries, we can look at two specific realizations: Matrix product states in \cref{sec_TN_MPS} and (binary) Tree Tensor Networks in \cref{sec_TN_TTN}.
% =================================================================
\subsection{Matrix product states}
\label{sec_TN_MPS}
The simplest yet most diffused and powerful loopless TN ansatz in one dimension, is represented by Matrix product states (MPS), which reproduces the structure of the lattice with a network of tensors, one for each lattice site \cite{Perez-Garcia2007MatrixProductState}.
As shown in \cref{fig_TN_architectures}(a), each tensor in the bulk of the network has three indices: one physical leg $i$ of dimension $d$, representing the local degrees of freedom, and two auxiliary legs $\alpha$s of dimension $\chi$ connected to the neighboring sites. 

More formally, the MPS representation consists in reshaping the generic QMB pure state $\ket{\psiqmb}$ in \cref{eq_QMB_state} as follows \cite{Fannes1992FinitelyCorrelatedStates, Klumper1991EquivalenceSolutionAnisotropic,Klumper1993MatrixProductGround}:
\begin{equation}
  \ket{\psiqmb}=
	\sum_{i_{1},\dots, i_{\Nsites}=1}^{d}
	\sum_{\alpha_{1},\dots, \alpha_{\Nsites}=1}^{\chi}
	A^{[1],i_{1}}_{\alpha_{1},\alpha_{2}} A^{[2],i_{2}}_{\alpha_{2},\alpha_{3}},\dots, A^{[\Nsites-1],i_{\Nsites-1}}_{\alpha_{N-2},\alpha_{\Nsites}}A^{[\Nsites],i_{\Nsites}}_{\alpha_{N-1},\alpha_{1}} \ket{i_{1},i_2,\dots,i_{\Nsites}},
	\label{eq_TN_MPS_state}
\end{equation}
where each 3-rank tensor $A^{[k],i_{k}}_{\alpha_{k},\alpha_{k+1}}$ describes the local $k^{\rm{th}}$ lattice site with the index $i_{k}$ and the amount of quantum correlations shared its neighbors through the indices $\alpha_{k}, \alpha_{k+1}$.
In open boundary conditions (OBC), the tensors at the boundaries of an MPS have one trivial auxiliary link $\alpha_{1}=\alpha_{\Nsites}=1$.
Fixing the physical index $i_{k}$, the tensor $A^{[k],i_{k}}_{\alpha_{k},\alpha_{k+1}}$ is a $\chi\times\chi$ complex matrix.
In this perspective, \cref{eq_TN_MPS_state} is a sum of basis elements weighted by matrix products.

In principle, every state of a one-dimensional QMB system can be written exactly in the MPS form, with bond dimension $\chi=d^{\Nsites/2}$ growing exponentially with the lattice sites $\Nsites$ \cite{Vidal2003EfficientClassicalSimulation}. 
However, since MPS intrinsically satisfies area law, when dealing with low-energy states of local Hamiltonians, and their real-time dynamics for small-to-intermediate time, an approximate but efficient representation at fixed maximal bond dimension is usually very accurate in capturing the essential features of the systems \cite{Orus2014PracticalIntroductionTensor,Eisert2010ColloquiumAreaLaws}. 

In this case, the number of parameters in the MPS representation grows only polynomially with the system size, drastically reducing the overall computational complexity.
More in detail, the MPS ground state searching algorithm (based on DMRG) scales like $\mathcal{O}(\Nsites d \chi^3 + \Nsites d^{2} \chi^{2})$ or $\mathcal{O}(\Nsites d^3\chi^3)$, depending if the algorithm optimizes a single MPS tensor at a time, or two-tensors simultaneously \cite{Schollwock2011DensitymatrixRenormalizationGroup, Chan2016MatrixProductOperators}.
Two-site optimization is usually important to avoid local minima or meta-stable configurations during the energy variational minimization \cite{Gleis2023ControlledBondExpansion}. The subspace expansion is an intermediate approach with the benefits of the two-tensor update and a tunable computational cost between the two approaches \cite{Silvi2019TensorNetworksAnthology}.
% =================================================================
\subsection{Tree Tensor Networks}
\label{sec_TN_TTN}
The other important family of loopless TNs is represented by Tree Tensor Networks (TTN), in which the wave function is decomposed into a hierarchical network of tensors that do not contain internal loops \cite{Shi2006ClassicalSimulationQuantum, Silvi2019TensorNetworksAnthology}.
This way, the network can be efficiently contracted and manipulated in polynomial time by exploiting the unitary gauge transformations detailed in \cref{sec_TN_loopfree_geometries}.

A particular class of TTN is represented by binary TTN (bTTN), reported in \cref{fig_TN_architectures}c and \cref{fig_TN_architectures}d for one- and two-dimensional lattices.
In these structures, tensors in the lowest layer have two physical legs of dimension $d$ (representing two lattice sites) and an auxiliary leg of dimension $\chi$, whereas, in the upper layers, they have three auxiliary legs of dimensions up to $\chi$.
The network intrinsically encodes a renormalization procedure, in which, at each layer, two sites are mapped into a single effective one.

More in detail, bTTN are quite useful for a series of reasons:
\begin{enumerate}
	\item thanks to their loopless TN geometry, in finite-range models, ground searching algorithms for bTTNs display a numerical complexity of the order $\mathcal{O}(\Nsites d^2 \chi^2 + N \chi^4)$ \cite{Silvi2019TensorNetworksAnthology} (see also \cref{sec_TN_gs_algorithm}).
	This is a much more favorable scaling concerning equivalent algorithms for other TN structures, and allows reaching quite large values of bond dimensions ($\chi \approx 500$) \cite{Qian2022TreeTensorNetwork} to keep the variational TTN optimization under control.
	\item similarly to MPS, they provide an efficient basis for DMRG-based algorithms, which represent the state-of-the-art technique for the numerical simulation of 1D QMB systems with both short and long-range interactions \cite{White1992DensityMatrixFormulation,White1993DensitymatrixAlgorithmsQuantum}.
	\item they can handle both open (OBC) and periodic (PBC) boundary conditions, without the typical limitations of DMRG for periodic systems.
	\item any D-dimensional realization of a bTTN admits an equivalent realization in 1D \cite{Felser2020TwoDimensionalQuantumLinkLattice,Cataldi2021HilbertCurveVs,Cataldi2024Simulating2+1DSU2,Magnifico2021LatticeQuantumElectrodynamics}.
	For instance, to move from a 2D bTTN to the corresponding 1D version (see \cref{fig_TN_architectures}), we just need to rotate each layer of the tree with alternate clockwise and anti-clockwise $\pi/2$ rotations in the $xy$ plane. 
	This procedure requires the application of a 2D-1D site-ordering that generates a mapping between the 2D physical system and a 1D chain \cite{Cataldi2021HilbertCurveVs} (see \cref{sec_TN_efficient_TN_geometry}).
	Once the 2D physical model is modeled on a 1D chain, one can apply the TTN variational algorithm discussed in \cref{sec_TN_gs_algorithm} by using 1D bTTNs.
	\item due to their hierarchical structure, bTTNs offer a strong connectivity: rather than MPS, the distance between two physical lattice sites is \emph{logarithmically} rescaled in the network. 
	As a results, they reveal particularly advantageous to reproduce properly long-range interactions (see also \cref{sec_TN_efficient_TN_geometry}). 
\end{enumerate}

\subsubsection{Augmented Tree Tensor Networks}
The drawback of loopless structures like binary TTN, is that the area law may not be explicitly reproduced in dimensions higher than one \cite{Ferris2013AreaLawRealspace}, which becomes a limiting factor when large systems are addressed.
Nonetheless, it is possible to explicitly encode the area law of high dimensional systems in the TTN ansatz by introducing an additional layer of independent disentanglers, acting on different couples of lattice sites and connected to the corresponding physical legs.
This process augments the expressive power of TTN, and the resulting ansatz is known as augmented Tree Tensor Network (aTTN) \cite{Felser2021EfficientTensorNetwork}.
The computational complexity of variationally optimizing an aTTN structure, which means optimizing both the tensors and the disentanglers, is of the order
$\mathcal{O}(\Nsites \chi^{4} d^{4} + \Nsites \chi d^{7})$.
We point out that the scaling of the computational costs with the local dimension $d$ is particularly severe in the case of aTTN due to the presence of the disentanglers layer.
% =================================================================
\subsection{Tensor Network ansätze with loops}
\label{sec_TN_loop_geometries}
Beyond loopless TN structures like MPS and TTN, there exist other ansätze, which despite their higher computational costs in contractions, offer unique advantages for describing complex QMB systems beyond one dimensions. 
Below we briefly outline two realizations of these TN ansätze.
% =================================================================
\paragraph{Projected Entanglement Pair States (PEPS)}
displayed in \cref{fig_TN_architectures}b, represent a generalization of the MPS ansatz to two- or higher-dimensional lattices which directly encode in their structure the area law of entanglement \cite{Cirac2021MatrixProductStates}.
In this case, each tensor in the bulk has a physical leg of dimension $d$, and a number of $\chi$-dimensional auxiliary legs depending on the coordination number of the considered lattice.
For example, this coordination number is four in the case of a two-dimensional square lattice, as shown in \cref{fig_TN_architectures}(b). 
Despite the area law implementation, due to the presence of loops in the TN geometry, an exact contraction of PEPS is an exponentially hard problem, meaning that PEPS can not be efficiently contracted for numerical computing, \eg{} scalar products of states or physical observables \cite{Schuch2007ComputationalComplexityProjected}.
To circumvent this problem, approximate contraction methods have been developed during the last years, and are still at the center of current research efforts \cite{Cirac2021MatrixProductStates}.
But even with exploiting these approximate techniques, the computational complexity for ground state optimization remains quite high, \eg{} of the order of $\mathcal{O}(\Nsites d^2\chi^8)$ \cite{Eisert2013EntanglementTensorNetwork, Vanderstraeten2022VariationalMethodsContracting}, limiting the maximum reachable bond dimensions (typical values are of the order $\chi \approx 10$).
% =================================================================
\paragraph{Multi-scale Entanglement Renormalization Ansatz (MERA)}
displayed in \cref{fig_TN_architectures}, extends the capabilities of MPS and TTN by efficiently representing QMB systems with long-range correlations, particularly at criticality. 
While TTN can capture certain entanglement structures beyond what MPS can handle, they are limited by their tree-like structure, which lacks the loops necessary to efficiently encode the entanglement scaling of critical systems. 
MERA addresses this limitation by introducing a layered, hierarchical network with loops, allowing for a logarithmic scaling of entanglement entropy and revealing suitable for critical ground states \cite{Vidal2007EntanglementRenormalization,Evenbly2009EntanglementRenormalizationTwo}.

Unlike TTN, where contractions can be performed in a straightforward and computationally inexpensive manner as detailed in \cref{sec_TN_loopfree_geometries,sec_TN_gs_algorithm}, MERA introduces additional complexity through its disentanglers and isometric tensors, which are necessary for accurately capturing the multi-scale entanglement in critical systems.
The presence of loops in MERA increases the complexity of the network and consequently the cost of tensor manipulations. 
For 1D systems, the scaling of MERA constractions and optimizations is accessible $\mathcal{O}(\Nsites \chi^4)$ but more expensive than TTNs. 
Beyond one dimension, \eg{} for 2D systems, optimizing MERA is much more challenging and scales with $\mathcal{O}(\Nsites \chi^8)$.
Current research is focusing on developing more efficient contraction and optimization algorithms to enhance its applicability to complex QMB systems \cite{Cincio2008MultiscaleEntanglementRenormalization}. 
Nonetheless, despite its higher computational, MERA’s ability to capture the intricate entanglement patterns of critical systems makes it a powerful tool in the study of QMB physics.

\section{Efficient Tensor Network geometries beyond one dimension}
\label{sec_TN_efficient_TN_geometry}
If MPS are the established TN ansatz for one-dimensional systems, the development of efficient TN algorithms for higher-dimensional systems is still ongoing \cite{Phien2015InfiniteProjectedEntangled,Vanderstraeten2016GradientMethodsVariational,Fishman2018FasterMethodsContracting,Xie2017OptimizedContractionScheme,Corboz2018FiniteCorrelationLength,Kshetrimayum2019TensorNetworkAnnealing}. 
On one side, a possible and frequently used approach lies in the application of consolidated numerical methods based on DMRG (MPS) by suitably mapping the high-dimensional problem onto an equivalent one-dimensional one. 
Indeed, the DMRG algorithm is often used for analyzing quasi-one-dimensional systems, such as ladder structures \cite{Schollwock2011DensitymatrixRenormalizationGroup}, and two-dimensional systems, obtaining reliable numerical results in a wide variety of condensed-matter problems \cite{Ali2021OrderingSitesDensity,Ghelli2020TopologicalPhasesTwolegged,Kantian2019UnderstandingRepulsivelyMediated,Xiang2001TwodimensionalAlgorithmDensitymatrix}.

However, using MPS/DMRG on high-dimensional lattices generally involves a significant increase in computational complexity. 
When adapting an MPS to two dimensions, it requires a careful arrangement of the 2D lattice into a 1D structure.
Two main strategies can be implemented: the first one consists in considering the $n \times n'$ 2D lattice as an $n'$-legs ladder, then grouping all the $n'$ sites of each column in a single numerical site, and finally applying the one-dimensional algorithms to the resulting chain \cite{Tschirsich2019PhaseDiagramConformal}. 
The cost of this approach is the exponential growth with $n'$ of the local basis dimension, which usually limits the application of this strategy to quasi-two-dimensional systems, in which the number of ladder legs is small with respect to the longitudinal extension.

The second strategy, which we discuss here, lies in covering the 
two-dimensional lattice by a one-dimensional curve and then applying one-dimensional algorithms to the resulting effective chain. 
The drawback at this approach is that even if the original model contains only nearest-neighbor interactions, the resulting model on the effective one-dimensional chain shows long-range interactions, that have an influence on the numerical approach efficiency \cite{Yamada2011ParallelizationDesignMulticore}.
In detail, the latter strategy induces a specific one-dimensional site-ordering in the 2D lattice, generating a mapping between the 2D physical system and a 1D chain. 
This procedure can be defined for any mapping
\begin{equation}
	\label{eq_2D1D_mapping}
	\mathcal{M}:\,i\in [1,n]\times [1,n'] \rightarrow \mu \in [1,nn']
\end{equation}
from a $n\times n'$ lattice onto a chain with $n\,n'$ sites, through which it is possible to translate a two-dimensional model into a one-dimensional one. 
Finally, one can apply 1D TN algorithms, such as DMRG or the TTN variational one discussed in \cref{sec_TN_gs_algorithm}, while accounting for the long-range interactions introduced by the mapping. 
Long-range interactions eventually require a larger bond dimension to properly describe the system. 
It follows that the performance of the simulation depends on the capability of the mapping to preserve the locality of the interactions from the original 2D Hamiltonian to the effective new 1D one.

In this section, we quantify the importance of the choice of the site-ordering curve based on the numerical results obtained in \cite{Cataldi2021HilbertCurveVs}.
In particular, we consider a square $n\times n$ lattice and compare the numerical precision achieved using the standard snake curve (see \cref{fig_hilbertcurve}d) and the Hilbert curve (see \cref{fig_hilbertcurve}c) detailed in \cref{sec_hilbert_curve}.
We demonstrate that, compared to other space-filling curves, the locality-preserving properties of the Hilbert curve leads to a clear improvement of numerical precision, especially for large system sizes, and turns out to provide the best performances for the simulation of 2D lattice systems via 1D TN structures.

\subsection{Preserving the locality of interactions}
Before effectively comparing different $2D\to1D$ mappings, let us deepen the concept of preserving the interaction range (\idest{} locality). 
In the context of numerical simulations with TN algorithms, it would be extremely useful to map a 2D lattice into a 1D chain in a way that preserves locality at most, avoiding long-range interactions of sites that are very far apart from each other. 
However, it is straightforward to prove that it is impossible to map a $D$-dimensional lattice into a $D^{\prime}$-dimensional one, with $D^{\prime} < D$, so that two neighboring sites in the $D$-lattice are always close together in the $D^{\prime}$-lattice. 
Let us consider, for instance, nearest neighbour sites: each point of the $D$-dimensional square 
lattice has $2D$ nearest neighbours, while only $2D^{\prime}$ nearest neighbours in the $D^{\prime}$-dimensional one. Thus, after the mapping, at least $2(D-D^{\prime})$ sites will be placed at a distance larger than one unit. 
For the specific case of a 2D square lattice and a 1D chain, this implies that the optimal solution would be a mapping such that two out of four nearest neighbours in the 2D lattice remain nearest neighbours in the 1D chain. 
If we strictly follow this argument, the best choice would be the frequently used snake curve, shown in \cref{fig_hilbertcurve}d. 
However, in this way, the other two nearest neighbours of a generic site of the 2D lattice would be rather far away from each other in the one-dimensional setup, with their distance increasing up to $2n$. 
If our goal is to preserve the locality also for these two other nearest neighbours, the curve that tends to globally satisfy these constraints, outperforming other types of mapping, is exactly the Hilbert curve. 

\subsection{Hilbert curve}
\label{sec_hilbert_curve}
Without going into much detail (see \cite{Jagadish1990LinearClusteringObjects,Abel1990ComparativeAnalysisTwodimensional,Moon2001AnalysisClusteringProperties} for an exhaustive description), the Hilbert curve is a 1D fractal-like self-similar \emph{space-filling} curve with Hausd\"orff dimension $\delta=2$\footnote{The fractal or Hausd\"orff dimension $\delta$ is a measure of roughness (or smoothness) for a smooth, differentiable $D$-dimensional surface $S\in \mathbb{R}^{D+1}$ with topological dimension $D$.
If the surface is nondifferentiable and rough, the fractal dimension $\delta$ of $S$ takes values between the topological dimension $D$ and $D+1$ \cite{Gneiting2012EstimatorsFractalDimension}. 
The closer $\delta$ to $D+1$, the better the ability of $S$ in covering the $\mathbb{R}^{D+1}$ manifold space where the surface is embedded.} introduced for the first time in 1891 by the mathematician David Hilbert \cite{Hilbert1891UeberStetigeAbbildung}. 
Notably, among other interesting properties, it allows to create a mapping between a 1D and a 2D space by keeping and preserving the locality: this implies that two points that are close to each other in the 1D space are close to each other also after folding on the 2D space. The converse is not always strictly true, as unavoidable when passing 
from two-dimensions to one-dimension. However, even in this case, the curve shows a tendency to preserve the locality as much as possible \cite{Abel1990ComparativeAnalysisTwodimensional,Jagadish1990LinearClusteringObjects,Moon2001AnalysisClusteringProperties}.
For this reason, it finds several applications in computer science \cite{Lemire2011ReorderingColumnsSmaller} and bioinformatics \cite{Anders2009VisualizationGenomicData}.

\begin{figure}
  \centering
  \includegraphics[width=\textwidth]{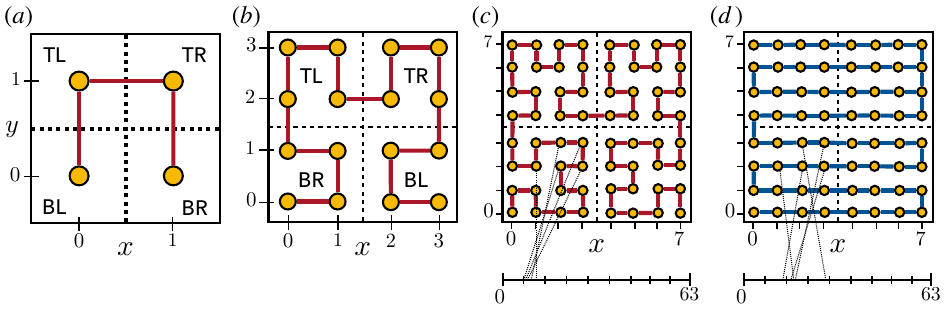}
  \caption{Iterative construction of the space-filling Hilbert curve on an $n=4,8,16$ (a)-(c) square lattice. 
	Examples of nearest-neighboring 2D lattice points mapped into 1D points making use of the Hilbert curve (c) and the Snake curve (d) respectively. 
	Figure from \cite{Cataldi2021HilbertCurveVs}.}
	\label{fig_hilbertcurve}
\end{figure}

\paragraph{Effective construction}
The basic element of the Hilbert curve is obtained by connecting the elements of a $2\times2$ lattice starting from the bottom-left (BL) to the bottom-right (BR) corner as shown in \cref{fig_hilbertcurve}a. 
From this we can now construct the $n = 4$ Hilbert curve via the following procedure: draw the $n = 2$ Hilbert curve into the top-left (TL) an top-right (TR) quadrants, while rotating it clockwise and counterclockwise by 90 degrees in the BL and BR quadrants respectively. 
Then, by joining these different curve replicas, we end up with the $n = 4$ curve shown in \cref{fig_hilbertcurve}b.
The generalization to the level $n$ for covering the $n \times n$ lattice 
is then straightforward: the $2n$ curve is obtained by drawing the $n$ curve in the four main quadrants of the $2n \times 2n$ lattice and by applying the just mentioned rules for the rotations in the BL and BR quadrants. 
In \cref{fig_hilbertcurve}c, for example, we report the Hilbert curve for the $8 \times 8 $ lattice. 

\subsubsection{Comparing space filling curves}
To get an intuition of how different curves preserve locality, we report in \cref{fig_hilbertcurve}c the Hilbert mapping of a generic site of the 2D lattice (within a given quadrant) and its four nearest neighbours: it is evident that all these points that are close on the two-dimensional plane remain fairly close on the one-dimensional chain. 
On the contrary, this property does not hold for the standard snake mapping, where, as expected, two out of four nearest neighbours are almost always mapped at large distance, as shown in \cref{fig_hilbertcurve}d.
The only regions in which the Hilbert mapping does not preserve the locality are the inner borders of the main quadrants. 
In this case, nearest neighbor sites belonging to two adjacent quadrants turn out to be mapped quite far away from each other.
Nonetheless, in the limit of large lattices, the number of these points becomes negligible $\sim \mathcal{O}(n)$ w.r.t the total number of sites. 

\subsection{Numerical evidences}
In \cite{Cataldi2021HilbertCurveVs}, we compare the performance obtained from the Hilbert and the snake curve on MPS and TTN simulations.
We focus on the ground state properties of the 2D Ising model at zero temperature in presence of a uniform transverse magnetic field along the z-direction \cite{Suzuki2013QuantumIsingPhases} with the following Hamiltonian:
\begin{equation}
	\ham= J \sum_{\avg{i,j}}\Sx_{i}\Sx_{j} + \lambda \sum_{i} \Sz_{i}, 
	\label{eq_ising_ham}
\end{equation}
where $\avg{i,j}$ refers to the nearest neighbour sites and $\hat{\sigma}_{i}^{\alpha}$ is the Pauli matrix in the $\alpha$-direction defined on the site $i$. 
The coefficient $J$ represents the strength of the interaction between nearest neighbour spins along the $x$-axis, whereas $\lambda$ determines the strength of the external magnetic field along the $z$-axis. 
In the following, we consider the antiferromagnetic (AF) scenario 
for $J>0$, and, without loss of generality, we fix the energy scale by setting~$J=1$.

\subsubsection{Ground-state energy density}
\begin{figure}
	\centering
	\includegraphics[width=\textwidth]{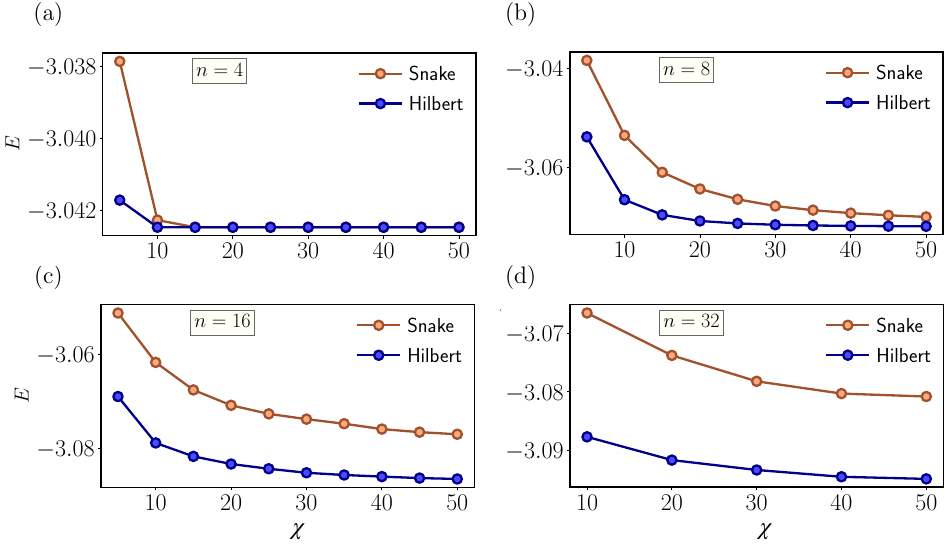}
	\caption{Numerical simulations of the 2D quantum Ising Hamiltonian at $\lambda=2.9$, i.e. close to the quantum critical point.
	Ground-state energy density measurements obtained via TTN at different system sizes $n$ employing the Snake and the Hilbert ordering for incremental values of the bond dimension $\chi$. 
	Similar results are obtained via MPS and discussed in \cite{Cataldi2021HilbertCurveVs}.}
	\label{fig_TN_ising_simulations}
\end{figure}

We simulate the Hamiltonian in \cref{eq_ising_ham} on a square lattice $n\times n$ with different sizes, up to $n=64$, and both boundary conditions. 
To compare the Hilbert and the snake mappings, we express \cref{eq_ising_ham} as a long-range, one-dimensional Hamiltonian $\ham_{S(H)}$ where sites are sorted according to the snake $(S)$ or the Hilbert ($H$) curve.
As shown in \cite{Cataldi2021HilbertCurveVs}, for both MPS and TTN simulations, the Hilbert curve mapping provides more accurate ground state energies, especially close to the second order critical point \cite{Friedman1978IsingModelTransverse,Li2018QuantumPhaseTransition}, stressing the relevance of our analysis for the study of quantum phase transitions.
More in detail, as shown in \cref{fig_TN_ising_simulations}, the Hilbert curve shows a faster convergence of the energy as the bond dimension $\chi$ is increased. 

\subsubsection{Local Magnetization}
\begin{figure}
	\centering
	\includegraphics[width=\textwidth]{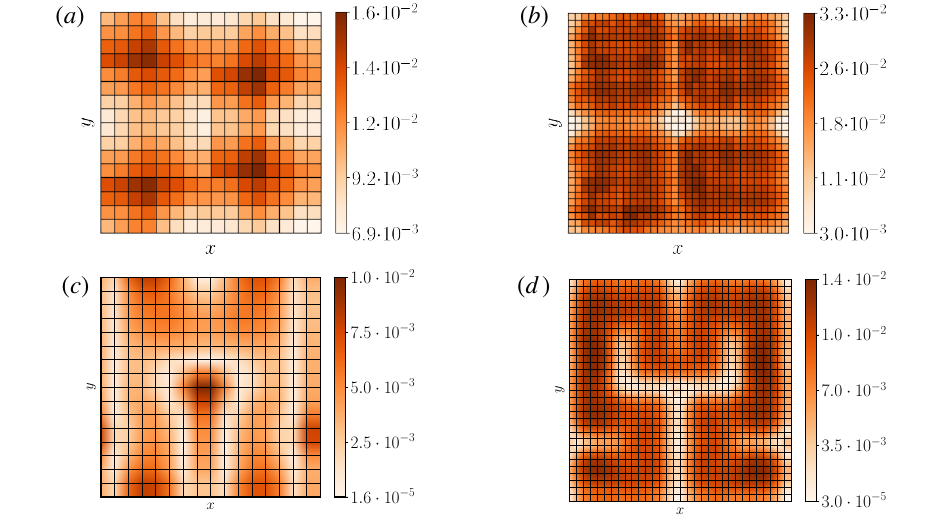}
	\caption{Local magnetization difference between the Hilbert and the snake ground states for $n = 16$ (left panel) and $n = 32$ (right panel) computed by using TTNs (upper row) and MPS (lower row) for $\lambda = 2.9$, $\chi = 50$, and PBC. 
	The regions with similar expectation values (\idest{} the quadrants borders) are those where neither the Hilbert nor the snake curve are able to properly preserve the locality of the interactions.}
	\label{fig_TN_hilbert_magnetization}
\end{figure}

Additionally, it is possible to understand which regions of the lattice contribute to the energy improvement by studying the expectation values of the local magnetization. 
We measure the local expectation values of the magnetization $\qty{\avg{\Sz_{k}}}_{1\leq k \leq n^2}$ on both the ground states of the two Hamiltonians $\ham_{\mathcal{H}}$ and $\ham_{\mathcal{S}}$ and compute the difference $\Delta \Sz_{k}=|\avg{\Sz_{k}}_\mathcal{H}-\avg{\Sz_{k}}_\mathcal{S}|$. 
As shown in \cref{fig_TN_hilbert_magnetization}, $\Delta \Sz_{k}$ seems to be significant all over the lattice quadrants, while vanishing on their edges (the black lines in \cref{fig_hilbertcurve}c), where the magnetization values obtained from the two methods are similar. 
This result aligns with the fact that the Hilbert curve excels at preserving the locality of interactions within the core of the quadrants, while its performance near the edges becomes comparable to that of other orderings, such as the snake.
In these terms, we expect to observe similar results also when considering other observables, such as two-point correlation functions.

\subsection{Distance distributions}
\label{sec_TN_distances}
For a more rigorous explanation of the previous numerical results, let us introduce two distances $d_{\mathrm{MPS}}$ and $d_{\mathrm{TTN}}$, defined as the number of links connecting two different lattice sites within the MPS and TTN network geometries respectively. 
Namely, we consider the interaction terms of the 1D Hamiltonians $\ham_{\mathcal{S}}$ and $\ham_{\mathcal{H}}$ of \cref{eq_ising_ham} and compute the distances $d_\mathrm{MPS}$ and $d_\mathrm{TTN}$ between all the relative couples of sites.
We obtain then four sets of distances, each of them labeled by the relative mapping (snake and Hilbert) and the TN geometry (MPS and TTN).
\begin{figure}
  \centering
	\includegraphics[width=\textwidth]{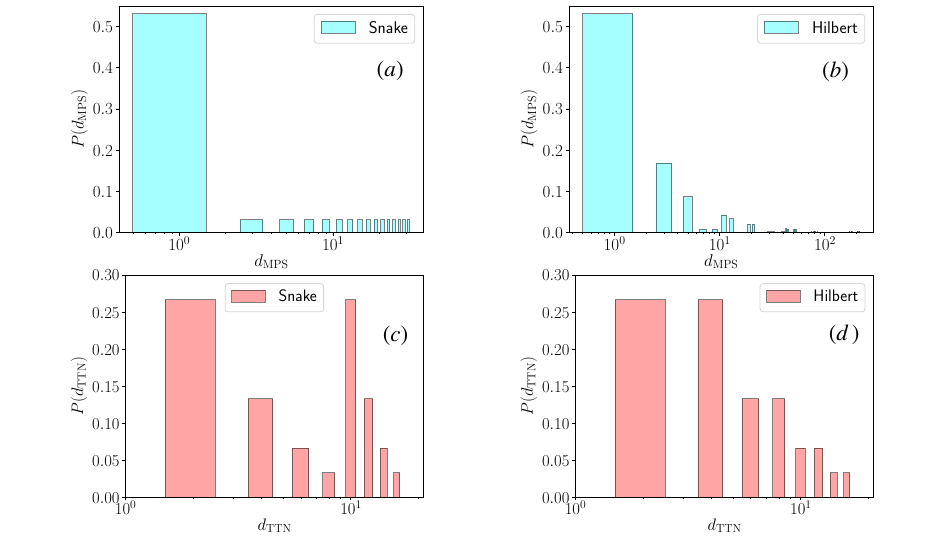}
	\caption{Distributions of the number of TN links separating physically adjacent lattice sites in the 2D lattice $d_\mathrm{MPS}$ (first row) and $d_\mathrm{TTN}$ (second row) relative to the pairs of sites connected by an interaction term in the Hamiltonians $\ham_\mathcal{S}$ (fist column) and $\ham_\mathcal{H}$ (second column), for $n = 16$.}
	\label{fig_TN_hilbert_distances_curves}
\end{figure}
\begin{figure}
	\centering
	\includegraphics[width=\textwidth]{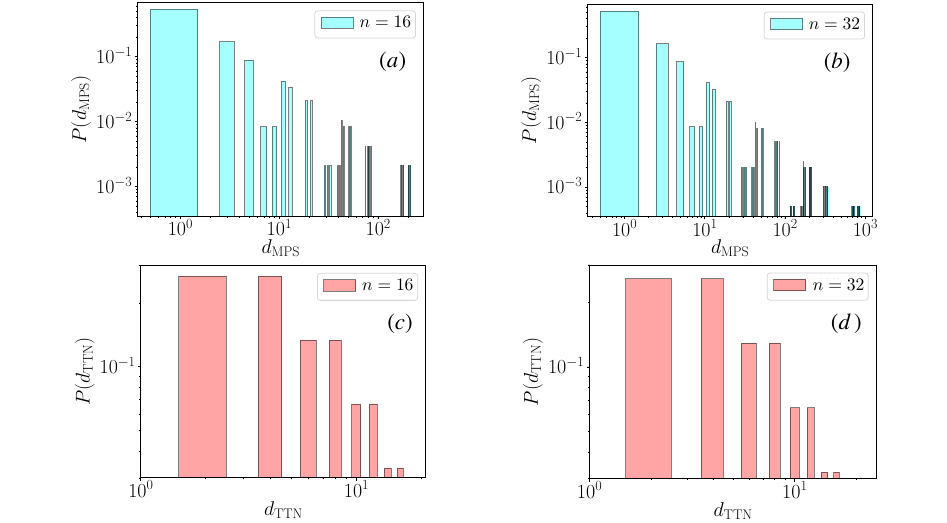}
	\caption{Distributions of the distances $d_\mathrm{MPS}$ (first row) and $d_\mathrm{TTN}$ (second row) relative to the pairs of sites connected by an interaction term in the Hamiltonian $\ham_\mathcal{H}$ computed for $n = 16$ (first column) and $n = 32$ (second column).}
	\label{fig_TN_hilbert_distances_methods}
\end{figure}

\paragraph{Different curves}
In \cref{fig_TN_hilbert_distances_curves}, we fix $n=16$ and plot the distributions of $d_\mathrm{MPS}$ and $d_\mathrm{TTN}$ for the snake (left column) and the Hilbert (right column) mappings: one can qualitatively see that the probabilities at large distances are larger for the snake mapping than for the Hilbert one. 
This feature is particularly evident in the TTN case, due to the logarithmic scaling of distances within the network. 

\paragraph{Different Tensor Network Geometries} 
Remarkably, when focusing on the same space filling curve (\eg{} the Hilbert curve) we can compare the MPS and TTN method the distributions of $d_\mathrm{MPS}$ and $d_\mathrm{TTN}$ on different values of $n$.
As shown in \cref{fig_TN_hilbert_distances_methods}, the fraction of interaction terms with largest distances changes when we move from $n=16$ to $n=32$ and decreases faster in the MPS case than in the TTN one.
This explains why the improvement achievable with the MPS and the Hilbert curve is larger than the one achievable with TTNs, even though the TTN simulations are overall more accurate than the MPS ones.

\subsection{Conclusions}
These results are particularly interesting, as they provide useful hints for simulating QMB systesm and LGTs beyond one spatial dimension. 
In particular, the choice of an optimal space filling curve, such as the Hilbert one, to map a $d-$dimensional lattice onto a one-dimensional one, reveals fundamental to minimize long-range interactions on both MPS and TTN ansätze.

While the locality-preserving properties of the Hilbert ordering are directly mapped onto the MPS chain, in the case of TTNs they are enhanced by the logarithmic scaling of distances within the TTN structure.
By recalling that the standard TTN methods for two-dimensional systems rely on the construction of a 2D bTTN \cite{Felser2021EfficientTensorNetwork}, it is worth noting that the 2D bTTN and the Hilbert curve mapping generate the same long-range interactions within the corresponding networks. 
Therefore, the former benefits from the mathematical locality-preserving properties of the Hilbert curve. 

We have then obtained a systematic strategy to evaluate, given a QMB system defined on a two-dimensional lattice, the more efficient site-ordering to perform simulations by using TNs, based on the analysis of the effective, one-dimensional system generated by the mapping. 
Our results suggest that 1D variational ground state searching algorithms, such as DMRG, strongly benefits from the choice of the Hilbert curve for treating 2D systems. 
This can be easily extended to rectangular lattices, such as long narrow cylinders, or on 3D cubic lattice, since it is possible to generalize the Hilbert curve definition also in these cases. 
Other possible outlooks include the application of this strategy to the study of systems characterized by non-trivial lattice structures and interactions, as for example in spin glasses \cite{Mydosh2015SpinGlassesRedux} and quantum chemistry \cite{Cao2019QuantumChemistryAge}.

% =================================================================
\section{Variational ground state search algorithm}
\label{sec_TN_gs_algorithm}
We now present the ground-state search algorithm for TTN simulations which has been used for simulating LGTs \cite{Felser2020TwoDimensionalQuantumLinkLattice,Magnifico2021LatticeQuantumElectrodynamics,Cataldi2024Simulating2+1DSU2} such as the (2+1)D SU(2) theory in \cref{chap_SU2_groundstate}. 
A pictorial explanation of the algorithm is sketched in \cref{fig_TN_variational}a.
Let us suppose to search for the ground state properties of a QMB Hamiltonian $\ham_{\rm{QMB}}$. 
Formally, if $\ket{\psiqmb}$ is the target state, any observable requires the contraction of the Hamiltonian tensor operator with the TN state and its hermitian conjugate: $\expval{\ham}{\psiqmb}$. 
To reduce the complexity of the minimization problem, we use a \emph{variational} approach that optimizes a single tensor $\tensor_{\rm{var}}$ of $\ket{\psiqmb}$ per step\footnote{In practice, we exploit the Krylov sub-space expansion technique to numerically solve the local eigenvalue problem for each tensor \cite{Silvi2019TensorNetworksAnthology}.}. 
We fix a $\tensor_{\rm{var}}$ as the \emph{optimization center} and treat all its parameters as variational. 
Correspondingly, all the other tensors of the TN are kept fixed to form the so-called \emph{TN environment}. 
We then contract the environment with the Hamiltonian operator, obtaining an \emph{effective Hamiltonian} $\ham_{\rm{eff}}$ which acts only on $\tensor_{\rm{var}}$. 
Properly, this is just a \emph{local optimization problem}, which involves only two single tensors, respectively $\tensor_{\rm{var}}^{*}$ from $\bra{\psiqmb}$ and $\tensor_{\rm{var}}$ from $\ket{\psiqmb}$
\footnote{This step is efficiently carried out by applying the Arnoldi method of the ARPACK library \cite{Lehoucq1998ARPACKUsersGuide}}. 
Then, if we sequentially solve the reduced optimization problem for each TN node, (\idest{} by moving the optimization center following the numbered tensors in \cref{fig_TN_variational}a), the energy expectation value is gradually reduced up to a certain fixed minimal value.
The whole series of local optimizations in $\ket{\psiqmb}$ is also known as \emph{sweep sequence}. 
% =================================================================
\subsection{High level operations} 
\begin{figure}
	\centering
	\includegraphics[width=1\textwidth]{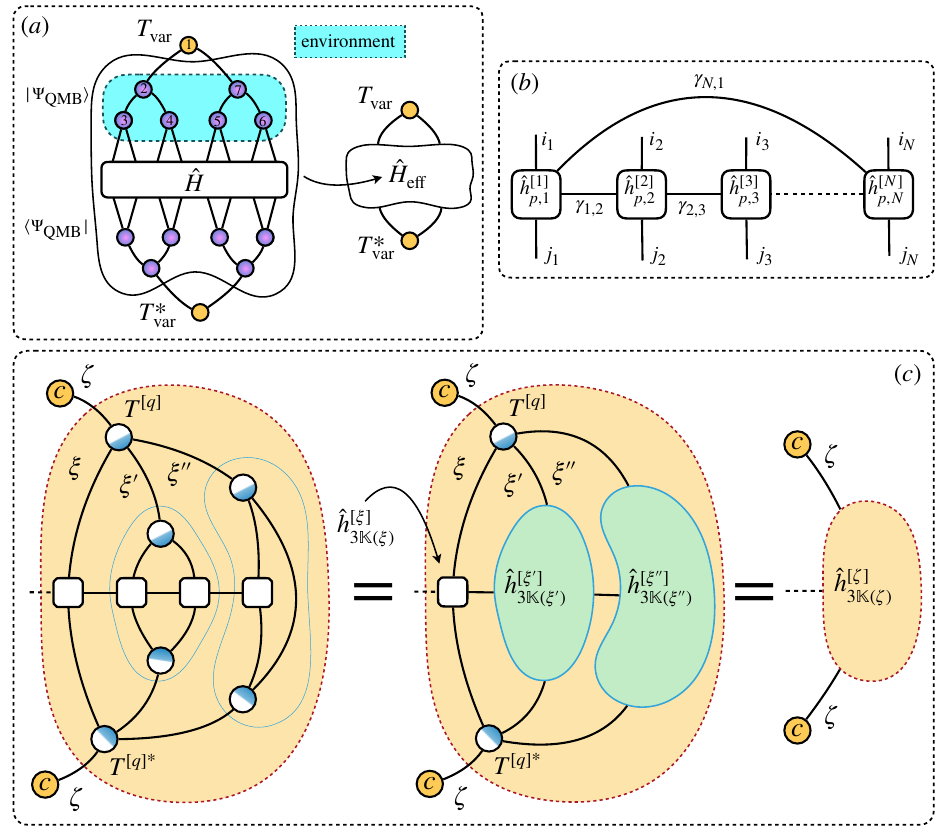}
	\caption{(a) Pictorial representation of the variational ground-state search algorithm w.r.t a given node of the TTN geometry. 
	By contracting the environments of $\ket{\psiqmb}$ and $\bra{\psiqmb}$ with the Hamiltonian TPO, one obtains an \emph{effective Hamiltonian} and a \emph{local optimization problem} with respect to $\tensor^{\rm{var}}$. 
	Iterating this procedure along the black arrow, one globally minimizes the TPO Hamiltonian with respect to the whole TN structure.
	(b) Representation of a 1D TPO Hamiltonian in PCB. 
	Each single-site operator $\hop_{p,k}$ display both physical $i_{k}, j_{k}$ and auxiliary links $\gamma$. 
	The former ones connect the operator respectively to the physical sites of $\ket{\psiqmb}$, $\bra{\psiqmb}$; the latter ones to the other local operators. 
	(c) Iterative Renormalization proceeding with respect to a given auxiliary link $\zeta$ of the TN structure. 
	For more details, see \cite{Silvi2019TensorNetworksAnthology}.}
	\label{fig_TN_variational}
\end{figure}

We then then focus on a single local optimization problem \cite{Silvi2019TensorNetworksAnthology}.
Let us consider a TTN geometry of tensors $\tensor^{[q]}$ whose nodes $q\in\qty{1\dots Q}$ are mutually joined by auxiliary links $\zeta$ of dimension $\chi_{\zeta}$. 
We suppose to use \emph{unitary gauges} and fix an arbitrary center node $\tensor_{\rm{var}}$ w.r.t. which each tensor of the environment is gauge transformed.
Correspondingly, we assume the Hamiltonian as a sum of $P$ different contributions (\eg{} the kinetic energy, the interaction potential, or an external magnetic field):
\begin{equation}
	\ham=\sum_{p=1}^{P}\ham_{p},
\label{eq_TN_hamiltonian}
\end{equation} 
where each term $\ham_{p}$ is a Tensor Product Operator (TPO) like the following (see \cref{fig_TN_variational}b):
\begin{equation}
	\qty[\ham_{p}]_{\qty{i_{s}}\qty{j_{s}}}=\sum_{\{\gamma_{\mu}\}}\prod_{s}\qty[\hop_{p,k(s)}^{[s]}]^{\qty{\gamma_{\mu}}_{k(s)}}_{i_{s}j_{s}}.
	\label{eq_TN_hamiltonian_TPO}
\end{equation}
In this expression, $\qty{i_{s}}$ and $\qty{j_{s}}$ label the sets of physical links which connect the Hamiltonian interaction $\ham_{p}$ with $\bra{\psiqmb}$ and $\ket{\psiqmb}$ respectively; $\hop_{p,k(s)}^{[s]}$ is the $k^{\text{th}}$ \emph{local operator} of the specific TPO $\ham_{p}$ acting on the physical site $s$ of the TN. 

Local operators such as $\hop_{p,k(s)}^{[s]}$ and $\hop_{p,k^{\prime}(s^{\prime})}^{[s^{\prime}]}$, concurrently acting on the physical sites $s$ and $s^{\prime}$, are connected among each other through TPO links like $\gamma_{\mu=(k,k^{\prime})}$ (see \cref{fig_TN_variational}b). 
The sum $\sum_{\{\gamma_{\mu}\}}$ runs over all the TPO links $\qty{\mu}$ and we define $\qty{\mu}_{k(s)}\subseteq\qty{\mu}$ as the set of all the links $\mu$ attached to the $k^{\text{th}}$ TPO operator $\hop_{p,k(s)}^{[s]}$.

\subsubsection{Renormalization}
At this point, once fixed a center node $c$ of the TTN structure defining $\ket{\psiqmb}$, we associate to each auxiliary link $\zeta$ of the TTN a \emph{renormalized Hamiltonian} $\hop_{p,\mathbb{K}(\zeta)}^{[\zeta]}$, which describes the effective action of the single Hamiltonian TPO in \cref{eq_TN_hamiltonian_TPO} onto the subsystem (or partition) $S_{\zeta}$, made of all the environmental sites of the TN separated by $\zeta$ from the current network center $c$.

To figure out the situation, let us consider a given node $q\neq c$ with a proper set of auxiliary links $\qty{\eta}_{q}$. 
We can suppose that, among these links, the link $\zeta$ is the one through which the node $q$ communicates with c (see \cref{fig_TN_variational}c). 
All the remaining links are then denoted with $\qty{\xi}_{q}$, in such a way that $\qty{\eta}_{q}=\zeta + \qty{\xi}_{q}$. 
Then, we define the renormalized Hamiltonian associated to the link $\zeta$ in the following recursive way:
\begin{equation}
	\qty[\hop_{p,\mathbb{K}(\zeta)}^{[\zeta]}]^{\qty{\gamma_{\mu}}_{\mathbb{K}(\zeta)}}_{\qty{\alpha_{\xi}}\qty{\beta_{\xi}}}=\sum_{\substack{\text{paired}\\ \qty{\gamma_{\mu}}}} \prod_{\xi\neq \zeta}\qty[\hop_{p,\mathbb{K}(\xi)}^{[\xi]}]^{\qty{\gamma_{\mu}}_{\mathbb{K}(\xi)}}_{\alpha_{\xi}\beta_{\xi}} \qquad \text{where} \qquad \mathbb{K}(\zeta)=\bigcup_{\xi\neq\zeta}\mathbb{K}(\xi).
	\label{renormalizedpt1}
\end{equation}
In $\mathbb{K}(\zeta)$, we collect all the physical renormalized Hamiltonians defined on all the links $\qty{\xi}_{q}$ of the TN node $q$ with $\xi\neq \zeta$ (we recall that both $\zeta$ and $\xi$ are auxiliary links of the TN structure, not of the Hamiltonian TPO). 
These terms are generally connected by some TPO links $\qty{\mu}$, over which the contractions are performed (and the sum is taken). 

The fundamental unit of a renormalized Hamiltonian $\hop_{p,\mathbb{K}(\xi)}^{[\xi]}$ is a single renormalized Hamiltonian operator $\hop_{p,\mathbb{K}}^{\xi}$, where $\mathbb{K}$ is the set of all local operators of the TPO $\hop_{p,k(s)}^{[s]}$ involved the renormalization: 
\begin{equation}
	\mathbb{K}(\xi)=\bigcup_{s\in S_{\xi}}k(s) 
\end{equation}
Properly, in order to effectively obtain the renormalized Hamiltonian over the link $\zeta$, we must additionally contract the Hamiltonian operators $\hop_{p,\mathbb{K}(\xi)}^{[\xi]}$ over the links $\xi \neq \zeta$ with the node tensor $\tensor^{[q]}$ and its complex conjugate $\tensor^{[q]*}$:
\begin{equation}
	\qty[\hop_{p,\mathbb{K}(\zeta)}^{[\zeta]}]^{\qty{\gamma_{\mu}}_{\mathbb{K}(\zeta)}}_{\alpha_{\zeta}\beta_{\zeta}}=\sum_{\qty{\alpha_{\xi}}}\sum_{\qty{\beta_{\xi}}}\tensor^{[q]}_{\alpha_{\zeta}\qty{\alpha_{\xi}}}\qty[\hop_{p,\mathbb{K}(\zeta)}^{[\zeta]}]^{\qty{\gamma_{\mu}}_{\mathbb{K}(\zeta)}}_{\qty{\alpha_{\xi}}\qty{\beta_{\xi}}}\tensor^{[q]*}_{\qty{\beta_{\xi}}\beta_{\zeta}}
	\label{renormalizedpt2}
\end{equation}
In the unitary gauge, the tensor $\tensor^{[q]}$ appears as an isometry (see \cref{eq_TN_unitary_gauge} et seq.) over the links $\xi\neq\zeta$. Therefore, it is immediately clear that if \cref{renormalizedpt1} holds 1, so \cref{renormalizedpt2} does. Obviously, such renormalized Hamiltonians are defined for each kind of interaction $p=1\dots P$.
\subsubsection{Effective Hamiltonian}
\label{effective_hamiltonian}
At this point, we exploit the variational feature of the algorithm: for a given center node $c$ (\idest{} the unitary gauge center), we define the \emph{effective Hamiltonian} $\ham_{\rm{eff}}^{[c]}=\expval{\ham}{\Psi^{[c]}_{\text{env}}}$ where $\Psi^{[c]}_{\text{env}}$ stands for the result of the TN contractions over all the tensors of the TN except $\tensor^{[c]}$, \idest{} over the environment of the center node. Then, the energy expectation value $E_{[c]}$ of the TN state with respect to the given center node $c$ reads 
\begin{equation}
E_{[c]}=\expval{\ham}{\psiqmb^{[c]}}=\expval{\expval{\ham}{\Psi^{[c]}_{\text{env}}}}{\tensor^{\qty[c]}}=\expval{\ham_{\rm{eff}}^{[c]}}{\tensor^{\qty[c]}}.
\label{effectivept1}
\end{equation}
Basically, in order to obtain $\ham_{\rm{eff}}^{[c]}$, we do not need to contract the entire environment if we have already computed the renormalized Hamiltonians over all the available links of $c$. We should have:
\begin{equation}
\ham_{\rm{eff}}^{[c]}=\sum_{p=1}^{P}\hop_{p,\rm{eff}}^{[c]}\qquad \text{where}\qquad \qty[\hop_{p,\rm{eff}}^{[c]}]_{\qty{\alpha_{\eta}}\qty{b_{\eta}}}=\sum_{\{\gamma_{\mu}\}}\prod_{\eta}\qty[\hop_{p,\mathbb{K}(\eta)}^{[\eta]}]_{\alpha_{\eta}b_{\eta}}^{\qty{\gamma_{\mu}}_{\mathbb{K}(\eta)}}.
	\label{effectivept2}
\end{equation} are the renormalized Hamiltonian terms defined over all links $\qty{\eta}_{c}$ of the node $c$ with corresponding indices $\qty{\alpha_{\eta}}$ and $\qty{b_{\eta}}$. Hence, the effective Hamiltonian involves the contraction of all renormalized Hamiltonian terms over each tensor link $\eta\in\qty{\eta}_{c}$ of the center node $c$.
Plugging together all the previous steps, we can rewrite the energy expectation value $E_{[c]}$ as following:
\begin{equation}
	\begin{split}
		&E_{[c]}\equiv\expval{\ham}{\psiqmb^{[c]}}=\expval{\sum_{p=1}^{P}\ham_{p}}{\psiqmb^{[c]}}\\
		&\underset{\eqref{effectivept1}}{=}\expval{\expval{\ham}{\Psi^{[c]}_{\text{env}}}}{\tensor^{\qty[c]}}=\expval{\ham_{\rm{eff}}^{[c]}}{\tensor^{\qty[c]}}\\
		&\;=\expval{\sum_{p=1}^{P}\hop_{p,\rm{eff}}^{[c]}}{\tensor^{\qty[c]}}
		=\sum_{\qty{\beta_{\zeta}}}\sum_{\qty{\alpha_{\zeta}}}\tensor^{[c]*}_{\qty{\beta_{\zeta}}}\qty(\sum_{p=1}^{P}\qty[\hop_{p,\rm{eff}}^{[c]}]_{\qty{\beta_{\zeta}},\qty{\alpha_{\zeta}}})\tensor^{[c]}_{\qty{\alpha_{\zeta}}}\\
		&\underset{\eqref{effectivept2}}{=}\sum_{\qty{\beta_{\zeta}}}\sum_{\qty{\alpha_{\zeta}}}\tensor^{[c]*}_{\qty{\beta_{\zeta}}}\sum_{p=1}^{P}\qty[\sum_{\{\gamma_{\mu}\}}\prod_{\zeta}\qty[\hop_{p,\mathbb{K}(\zeta)}^{[\zeta]}]^{\qty{\gamma_{\mu}}_{\mathbb{K}(\zeta)}}_{\beta_{\zeta},\alpha_{\zeta}}]\tensor^{[c]}_{\qty{\alpha_{\zeta}}}\\
		&\underset{\eqref{renormalizedpt2}}{=}\sum_{\qty{\beta_{\zeta}}}\sum_{\qty{\alpha_{\zeta}}}\tensor^{[c]*}_{\qty{\beta_{\zeta}}}\sum_{p=1}^{P}\qty[\sum_{\{\gamma_{\mu}\}}\prod_{\zeta}\qty[\sum_{\qty{\beta_{\xi}}}\sum_{\qty{\alpha_{\xi}}}\tensor^{[q]*}_{\beta_{\zeta},\qty{\beta_{\xi}}}\qty[\qty[\hop_{p,\mathbb{K}(\zeta)}^{[\zeta]}]^{\qty{\gamma_{\mu}}_{\mathbb{K}(\zeta)}}_{\qty{\beta_{\xi}},\qty{\alpha_{\xi}}}]\tensor^{[q]}_{\qty{\alpha_{\xi}},\alpha_{\zeta}}]]\tensor^{[c]}_{\qty{\alpha_{\zeta}}}\\
		&\underset{\eqref{renormalizedpt1}}{=}\sum_{\qty{\beta_{\zeta}}}\sum_{\qty{\alpha_{\zeta}}}\tensor^{[c]*}_{\qty{\beta_{\zeta}}}\sum_{p=1}^{P}\sum_{\{\gamma_{\mu}\}}\prod_{\zeta}\sum_{\qty{\beta_{\xi}}}\sum_{\qty{\alpha_{\xi}}}\tensor^{[q]*}_{\beta_{\zeta},\qty{\beta_{\xi}}}\sum_{\substack{\text{paired}\\ \qty{\gamma_{\mu}}}}\prod_{\xi\neq\zeta}\qty[\hop_{p,\mathbb{K}(\xi)}^{[\xi]}]^{\qty{\gamma_{\mu}}_{\mathbb{K}(\xi)}}_{\beta_{\xi},\alpha_{\xi}}\tensor^{[q]}_{\qty{\alpha_{\xi}},\alpha_{\zeta}}\tensor^{[c]}_{\qty{\alpha_{\zeta}}}.
	\end{split}
	\label{effectivept3}
\end{equation}
% =================================================================
\subsection{Ground state reach algorithm in details}
We can then schedule the algorithm as follows:
\subsubsection{Initialization}
\begin{enumerate}
	\item \textbf{TN geometry}
	Define a TN geometry which is compatible with both the physical system and the Hamiltonian interactions. On the same time, such a geometry should be capable of hosting (through auxiliary links) the necessary quantum correlations of the ground state (a reasonable maximal bond dimension should be assigned). 
	Moreover, dealing with loopless TNs, the TN geometry must be chosen according to these constraints:
	\begin{enumerate}
		\item The number of physical links has to be adequate for the Hamiltonian in \cref{eq_TN_hamiltonian}.
		\item The network has to be \emph{acyclic}: for each pair of nodes, namely $q$ and $q^{\prime}$, there has to exist a \emph{unique (shortest) path} connecting them.
		\item The dimension $\chi_{\eta}$ of any auxiliary link $\eta$ has not to exceed the maximal bond dimension $\chi$ of the TN.
		\item Each tensor have at least 3 links.
	\end{enumerate}
	\item \textbf{Sweep sequence} 
	Choose the sequence with which the optimization centers are used for the local optimization problem (follow the numbers in the tensors of \cref{fig_TN_variational}a). 
	Any sequence that visits every tensor of the TN is allowed. 
	However, if the TN has a proper hierarchical structure (as for TTNs), the sweep sequence is based on the distance from a given physical link $s$. 
	An \emph{optimization precedence} $p\equiv \min_{s}\qty[dist(c,s)]$ is assigned to every optimization center $c$: in this way, optimization centers with smaller $p$ take precedence over centers with larger $p$.
	\item \textbf{Initialize the TN state}
	A typical choice is to start from a \emph{random} state, as it reduces the possibilities of ending up in local energy minima leading the convergence criteria to fail.
	For other useful TN initializations, look at \cref{sec_TN_roadmap_opt_initialstates}. 
	\item \textbf{Initial optimization center}
	Select the first optimization center $c$ and transform the network with the unitary gauge w.r.t. $c$. 
	Contract the two TN environments with the Hamiltonian obtaining the effective Hamiltonian $\ham_{\rm{eff}}^{[c]}$. 
	In particular, once the optimization center has been prepared, the effective Hamiltonian is made of contributions like \cref{effectivept2}, while the expectation value of $E_{[c]}$ is given by \cref{effectivept3}.
	Repeat the local optimization problem for all the other TN nodes up to the end of the chosen sweep sequence.
\end{enumerate}
\subsubsection{Optimization loop}
\begin{enumerate}
	\item Optimize $\tensor^{[c]}$ finding the eigenvector associated to the lowest eigenvalue $E_{[c]}$ of $\ham_{\rm{eff}}^{[c]}$.
	\item Target the next optimization center $c^{\prime}$ according to the \emph{sweep sequence}; move the isometry center from $c$ to $c^{\prime}$ by gauge transforming (as in \cref{fig_TN_gauge_operations}c) and updating the renormalized Hamiltonian terms between $c$ and $c^{\prime}$. 
	When moving the effective Hamiltonian form the center $c$ to the center $c^{\prime}$, only the renormalized Hamiltonian terms defined on the links located along the path from $c$ to $c^{\prime}$ have to be updated. 
	All the other renormalized Hamiltonians are left unchanged by this movement and do not need to be computed again.
	\item Back to step 1. and repeat the optimization with respect to the new center $c^{\prime}$ up to the end of the sweep sequence.
\end{enumerate}
\subsubsection{Convergence in energy}
Perform at least 3 different sweep sequences of local optimization. Exploiting the variational definition of the algorithm, we expect that the energy of successive sweep progressively decreases. 
Then, for each successive couple among the last three sweep sequences $s$, $s^{\prime}$ and $s^{\prime}$, with the corresponding energies $E(s)$, $E(s^{\prime})$ and $E(s^{\prime})$, compare both their absolute and relative deviations to reasonable thresholds $\epsilon$ and $\delta$ respectively\footnote{In all the simulations provided in \cref{chap_SU2_groundstate}, we fixed $\epsilon=4\cdot 10^{-8}$ and $\delta=8\cdot 10^{-7}$.}. 
The simulation halts when at least one of the following constraint sets is satisfied:
\begin{subequations}
	\begin{align}
		E(s)-E(s^{\prime})&<\epsilon &\&&& E(s^{\prime})-E(s^{\prime\prime})&<\epsilon \\
		-\frac{E(s)-E(s^{\prime})}{E(s^{\prime})}&<\delta &\&&& -\frac{E(s^{\prime})-E(s^{\prime\prime})}{E(s^{\prime\prime})}&<\delta.
	\end{align}
	\label{eq_TN_variational_thresholds}
\end{subequations}

\section{Time Evolution Algorithms for Tensor Network}
\label{sec_TN_time_evolution}
Besides variational optimization for ground state searching, the previous loopless TN families can also be exploited to simulate the real-time dynamics of local Hamiltonians \cite{Jaschke2018OpenSourceMatrix}, such as the one studied in \cref{sec_scars_nonAbelianLGT} for detecting QMB scars in non-Abelian SU(2) LGTs.
Below, we briefly outline two of the most relevat methods: the Time Evolved Block Decimation (TEBD) and the Time-Dependent Variational Principle (TDVP). 
These algorithms represent important and efficient tools for simulating with TN the real-time dynamics of QMB systems.
While equilibrium states satisfy the aforementioned area law \cite{Eisert2010ColloquiumAreaLaws}, out-of-equilibrium time evolution can generate a linear growth of entanglement.
In such cases, the bond dimension of the time-evolved state grows exponentially with total time \cite{Schuch2008EntropyScalingSimulability}.
For this reason, TN methods are currently limited to studying the dynamics for low-to-moderate times, close-to-equilibrium phenomena \cite{Paeckel2019TimeevolutionMethodsMatrixproduct}, and typically on 1D systems.
In this framework, further developments are extremely important to avoid or at least mitigate this barrier, by devising new algorithms or optimizing existing strategies \cite{White2018QuantumDynamicsThermalizing,Surace2019SimulatingOutofequilibriumDynamics}.

\subsection{Time Evolved Block Decimation}
\label{sec_TN_TEBD}
One of the most widely used approaches, the Time Evolved Block Decimation (TEBD) algorithm, is based on a Suzuki-Trotter decomposition of the time evolution exponential \cite{Vidal2004EfficientSimulationOneDimensional}.
The total evolution time is discretized in small time steps.
The corresponding evolution operator is computed as products of local terms, such as two-body operators, and repeatedly applied to the TN wave function to generate the time-evolved state.
Each application can determine an increase in the bond dimension of the network, so an optimized truncation is needed to maintain an efficient and manageable description of the quantum state.
This truncation reduces the bond dimension back to $\chi$ and is performed through a singular value decomposition that minimizes the distance between the evolved and the truncated state.
In general, the method allows the simulation of the real-time dynamics for nearest-neighbor or finite-range Hamiltonians; one time-step with an MPS for a one-dimensional system with local interactions comes with a computational cost that is below a two-tensor sweep for the ground-state search algorithm.

\subsection{Time-Dependent Variational Principle}
\label{sec_TN_TDVP}
Another method for simulating the evolution of quantum states via TN is the Time-Dependent Variational Principle (TDVP), which does not rely on the Suzuki-Trotter decomposition \cite{Haegeman2011TimeDependentVariationalPrinciple, Bauernfeind2020TimeDependentVariational}.
In general, TDVP constrains the time evolution to the specific TN manifold considered, such as MPS or TTN, of a given initial bond dimension \cite{Kohn2020SuperfluidtoMottTransitionBoseHubbard}.
This is obtained by projecting the action of the Hamiltonian into the tangent space of the TN manifold and then solving the time-dependent Schr{\"o}dinger equation within this manifold.
This approach automatically preserves the energy and the norm of the quantum states during the time evolution. The TDVP algorithm and the variational ground state search rely both on a set of Krylov vectors and therefore have the same computational scaling for one time step compared to one sweep.

\section{Roadmap for advanced LGT simulations via tensor networks}
\label{sec_TN_roadmap}
As discussed in \cref{chap_LGT} and particularly in \cref{sec_dressed_site_formalism}, LGT models present some peculiar features that make TN simulations particularly challenging, especially for large system sizes and for studying the continuum limits in terms of gauge field truncation, lattice spacing, and volume.
State-of-the-art techniques, such as TTN methods, have been successfully applied for simulating ground state properties of QED in (2+1)- and (3+1)-dimensions for small-to-intermediate sizes (still well far beyond the standards of ED and quantum simulations) \cite{Felser2020TwoDimensionalQuantumLinkLattice,Magnifico2021LatticeQuantumElectrodynamics}. 
Very recently, as reported in \cref{chap_SU2_groundstate}, TTN have been also applied to the SU(2) Yang-Mills model in (2+1)-dimensions \cite{Cataldi2024Simulating2+1DSU2}, exploring a rich phase diagram at both zero and finite baryon density.
In all these simulations, small non-trivial representations of the gauge groups have been exploited, \eg{} three electric field levels for QED (see the $\spin=1$ example in \cref{sec_U1_rishondecomposition}), and the first two irreducible representations of SU(2) for the Yang-Mills model (see the \emph{harcore-gluon} approximation in \cref{sec_SU2_hardcoregluon}).

Nowadays, lattice computations with MC-based techniques are performed on large lattices, \eg{} of the order $64^{3}\times128$ sites (space and time discretization), and with no truncation of the gauge fields.
These large sizes are required to control finite-volume effects and to perform extrapolations toward the continuum limits \cite{Workman2022ReviewParticlePhysics}.
In the last decades, the impressive progress in algorithmic development, high-performance optimizations, and the availability of increasingly powerful supercomputer facilities have played a major role in the advancement of MC-based LGT computations.
Indeed, this progress has opened the doors to large-scale simulations, that currently represent the standard approach for studying non-perturbative phenomena in QFT.

However, MC techniques are generally based on computations of path integrals in which the integrand functions are overall positive.
Many physically relevant scenarios, such as finite baryon density regimes or real-time dynamics of quarks, give rise to a change in the sign of the integrands and highly oscillating behaviors .
Thus, numerical evaluations suffer from the near cancellation of the opposite-sign contributions to the integrals.
This is the essence of the infamous sign problem, a long-standing issue of LGT simulations with MC methods \cite{Troyer2005ComputationalComplexityFundamental,Li2015SolvingFermionSign, Nagata2022FinitedensityLatticeQCD,Loh1990SignProblemNumerical}.
Hence the quest lies in conceiving, developing, and optimizing alternative approaches that enable simulating these regimes, being the latter at the heart of many open problems related to our understanding of high-energy physics.

As described in the previous sections, TNs represent a promising complementary method, which found the first applications in simulating non-trivial instances of high-dimensional LGTs on small-to-intermediate lattice sizes. 
TNs are intrinsically sign-problem-free, enabling the simulation of both static properties at equilibrium, such as low-energy states, and real-time dynamics, even in the presence of finite chemical potentials or non-trivial topological terms. 
It is worth noting that, in addition to local observables and correlation functions, TNs allow the numerical computations of entanglement properties and central charges, that could potentially shield new light on LGT phenomena \cite{Rigobello2021EntanglementGenerationMathrm}.

Nevertheless, TN simulations of high-dimensional LGTs still represent a challenging problem, especially for large lattice sizes or higher representations of gauge groups needed for analyzing continuum limits. 
In this framework, further and intensive developments are required to tackle TNs' current problems related to LGTs, such as QCD's non-perturbative effects on lattices of sizes comparable with MC simulations. 
In this regard, we note that sign-problem-free TN ansätze can also be used in combination with variational MC methods to tackle high-dimensional lattice gauge theories with arbitrary gauge groups \cite{Kelman2024GaugedGaussianPEPS}.

In the following, we present a possible roadmap \cite{Magnifico2024TensorNetworksLattice} in terms of algorithmic development and optimization strategies that we foresee to be crucial for making the TN approach competitive as a complementary method to MC techniques. 
Therein, \cref{sec_TN_roadmap_opt_basistruncation,sec_TN_roadmap_opt_initialstates}, can be approached with existing TN algorithms and a good intuition on LGT problems; then, \cref{sec_TN_roadmap_opt_hpclocal} and \cref{sec_TN_roadmap_opt_hpcmpi} outline optimization for existing algorithms to leverage HPC systems; finally, we discuss new classes of ansätze to tackle finite temperature problems in \cref{sec_TN_finiteTemp}. 
In some parts, we focus on TTNs, but the vast majority of the presented concepts and techniques can be straightforwardly applied to other TN ansätze.
% ========================================================================
\subsection{Local basis truncation}        
\label{sec_TN_roadmap_opt_basistruncation}
\begin{figure}
\includegraphics[width=1\columnwidth]{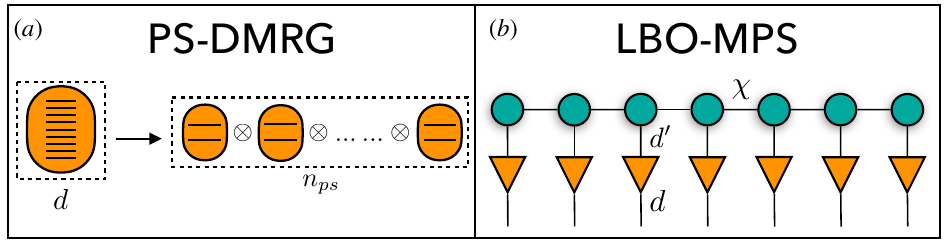}
\caption{Graphical representation of (a) the pseudo site DMRG (PS-DMRG) approach, in which a single site having large local dimension $d$ is replaced with $\Nsites_{ps}$ pseudo sites with a smaller local dimension, e.g two; (b) the local basis optimization embedded in the MPS ansatz (LBO-MPS), so that an additional layer of tensors connected to the physical legs of the MPS performs the reduction from a large local basis with dimension $d$ into an effective basis with smaller dimension $d^{\prime}$. }
\label{fig_panel_PS_LBO}
\end{figure}
One of the main issues in simulating LGTs with TN methods lies in the very large local basis dimension $d$ that one needs to handle to represent matter and gauge-field degrees of freedom properly (see \cref{sec_U1_basis_dimension}).
To some extent, this situation is very similar to some condensed matter models, \eg{} the Holstein model, involving lattice fermions coupled with phonons \cite{Holstein1959StudiesPolaronMotion}, or bosonic systems coupled to optical cavities \cite{Bloch2005UltracoldQuantumGases,Schlawin2022CavityQuantumMaterials}. 
Also in these cases, the local Hilbert space is in principle infinite-dimensional, and a truncation to a fixed cutoff is needed for performing TN simulations. For one-dimensional systems of this type, some efficient algorithms based on MPS have been developed in the last years \cite{Stolpp2021ComparativeStudyStateoftheart}, \eg{} the pseudosite DMRG (PS-DMRG) method \cite{Jeckelmann1998DensitymatrixRenormalizationgroupStudy}, and the DMRG with local basis optimization (DMRG-LBO) \cite{Guo2012CriticalStrongCouplingPhases,Brockt2015MatrixproductstateMethodDynamical,Stolpp2020ChargedensitywaveMeltingOnedimensional}.
% ========================================================================
\subsubsection{PS-DMRG}
The key idea of the PS-DMRG is to replace a single site having large local dimension $d$ with $\Nsites_{ps} \approx \logtwo(d)$ pseudo sites of local dimension $d=2$, as shown Fig. \ref{fig_panel_PS_LBO}(a). Since a large class of optimization algorithms for TNs scale at least quadratically with the local dimension but linearly with the total number of sites, as detailed in \cref{sec_TN}, one obtains a more efficient and manageable representation according to this procedure.

The price to pay lies in the range of the interactions: short-range interactions in the Hamiltonian are transformed into long-range operators due to the pseudo-site encoding of the local degrees of freedom.
As a consequence, PS-DMRG should require a larger bond dimension $\chi$ and a larger number of variational steps to converge; in these terms, the benefits offered by the pseudo-sites could progressively fade for increasing values of $\Nsites_{ps}$ at fixed $\chi$. 
PS-DMRG has been applied to one-dimensional QMB systems with $d$ up to $O(100)$ \cite{Stolpp2021ComparativeStudyStateoftheart}.
% ========================================================================
\subsubsection{DMRG-LBO}
The core of the DMRG-LBO algorithm, instead, lies in a local basis optimization protocol which enables a controlled and efficient truncation of the local Hilbert space \cite{Zhang1998DensityMatrixApproach}. 
For each site $\vecsite$ of the lattice, the optimized local basis is computed by starting from its reduced density matrix:
\begin{equation}
	\label{eq_RDM}
	\rho_{\vecsite} = \Tr_{\vecsite\ne\vecsite^{\prime}}\ketbra{\psi},
\end{equation}
where $\ket{\psi}$ is a general state of the whole system and the trace is over all the degrees of freedom which do not involve the site $\vecsite$.
By performing the diagonalization procedure, the eigenvalues $\lambda_{\alpha}$ and the eigenvectors $v_{\alpha}$ of $\rho_{\vecsite}$ can be easily determined.
The set of values $\lambda_{\alpha}$ represent the probabilities associated with the states $v_{\alpha}$.
If $\lambda_{\alpha}$ is small, \eg{} below a certain numerical threshold, the related eigenvector $v_{\alpha}$ can be discarded from the local basis of the site $\vecsite$, with a controllable loss of accuracy for the state $\ket{\psi}$.
Thus, for reducing the local basis dimension of $\vecsite$ from $d$ to a smaller value $d^{\prime}$, an optimal choice is keeping the $d^{\prime}$ eigenvectors of $\rho_{\vecsite}$ with the largest probabilities.
Then, the original state $\ket{\psi}$ can be projected on the new basis, without losing the relevant physical information. Let us notice that the site $\vecsite$ can also be a generic unit cell of the system, composed of a certain number of lattice sites.

A crucial point of the LBO procedure is the knowledge of the original state $\ket{\psi}$, generally the ground state of the system, that is not known prior.
This issue can be overcome in several ways: by performing ED of the system Hamiltonian with local dimension $d$ for small lattice sizes, to determine $\ket{\psi}$, and then truncating the basis from $d$ to $d^{\prime}$ for increasing values of the cutoff $d^{\prime}$.
From this procedure, we can obtain an optimized local basis ensuring a controlled approximation of the original ground state.
This optimized basis can then be exploited in optimization algorithms for simulating larger lattice sizes, such as DMRG.
Another strategy directly incorporates the LBO procedure in the TN ansatz, as shown \cref{fig_panel_PS_LBO}(b): by inserting an additional tensor on each physical leg of the MPS, the large local basis with dimension $d$ is transformed into an effective basis with smaller dimension $d^{\prime}$.
In this way, the MPS tensors only see the effective basis in the optimization procedures, with a significant reduction of the computational costs described in \cref{sec_TN}.
This method has been used for both static DMRG and time evolution algorithms, such as , for the study of one-dimensional quantum impurity models and correlated electron-phonon systems \cite{Guo2012CriticalStrongCouplingPhases,Brockt2015MatrixproductstateMethodDynamical,Schroder2016SimulatingOpenQuantum}.
% ========================================================================
\subsubsection{Local basis truncation for LGTs}
In the context of LGTs, the dimension of the local basis $d$ can easily go beyond values of the order of $10^6$, especially for higher-dimensional non-Abelian models and large representations of the gauge groups, as shown in \cref{sec_U1_basis_dimension} and \cref{tab_LG\tensor_localdims}.
In this scenario, numerical simulations with TNs are practically infeasible without an optimized scheme for truncating the local degrees of freedom.
Techniques like PS-DMRG might offer some benefits for small system sizes, such as unveiling the most relevant degrees of freedom in the low-energy states, but it is difficult to scale them up for large sizes due to the long-range interactions induced between the pseudo-sites.
Also, the local constraints imposed by Gauss law would become highly non-local when splitting a single site into multiple pseudo-sites representing the matter and link fields.

In principle, LBO-based procedures could instead represent a well-grounded route for addressing the problem.
Their use in condensed matter, \eg{} bosonic systems, is a rather consolidated approach, whereas their application to TN simulations of LGTs currently is an uncharted but promising territory.
The main steps that we foresee as needed and important in this direction are the following:
\begin{itemize}
\item[(i)]
    Employing ED \cite{Cataldi2024Edlgt}, testing the convergence of LBO procedures for one-dimensional systems, such as the $\varphi^4$-theory or the Schwinger model, for which analytical solutions, at least in some regimes of the phase diagram, and numerical results are widely available, also in the limit of no gauge field truncations.
    In this way, we can obtain valuable information about the scaling of the basis cutoffs concerning the final accuracy of the state representation for small system sizes.
\item[(ii)]
    Performing the same analysis on (2+1)-dimensional LGTs, such as QED or $SU(N)$ models, for one unit cell, like a single lattice-plaquette, to systematically study the effects of the magnetic interactions on the local degrees of freedom.
    Indeed, finite truncations of gauge fiels as the ones detailed in \cref{sec_U1_gaugetruncation} and \cref{sec_SU2_gaugetruncation}, generally exploits the \emph{electric field} basis, in which the electric field terms of the Hamiltonian and Gauss law are diagonal (see the dressed-site formalism in \cref{sec_dressed_site_formalism}).
    In this scheme, the magnetic interactions correspond to non-diagonal operators which can increase the number of local electric states to include for an accurate description of the system, especially for small values of the coupling constant $\coupling$ of \cref{eq_Ham_LGT}, as highlighted in the numerical analysis in \cref{sec_U1_basis_dimension} \cite{Magnifico2024TensorNetworksLattice}.
\item[(iii)]
    Exploiting the optimized local bases obtained from ED as input of TN simulations for larger sizes, testing the effects on the global ground state accuracy and in computing physically relevant quantities, such as the mass gap \cite{Clemente2022StrategiesDeterminationRunning}.
\item[(iv)]
    In the same spirit of LBO-MPS, implementing LBO protocols directly in TN ansätze that are more suitable for simulating high-dimensional LGT models, such as TTN.
    This step could be of great benefit in particular for aTTNs \cite{Felser2021EfficientTensorNetwork}, which encode the area law for the entanglement but are severely limited by large local bases (see \cref{sec_TN}).
\end{itemize}
By following and combining all these steps, we expect to reduce the effective local basis of LGT models, potentially enabling TN simulations for large representations of Abelian and non-Abelian gauge groups. 

It is worth noting that constructing optimal bases for numerical and quantum computation of LGTs is an active area of research. 
Several approaches that have been recently proposed involve performing canonical transformations of the gauge degrees of freedom before truncation \cite{Haase2021ResourceEfficientApproach,Mathur2023ExactDualityLocal,Bauer2023QuantumSimulationHighEnergy,Bauer2023QuantumSimulationFundamental}. 
By exploiting a resource-efficient protocol of this type, Ref. \cite{Haase2021ResourceEfficientApproach} has shown that the number of local states required to reach a $1\%$ accuracy level when computing the expectation value of the plaquette operators in two-dimensional (2+1)-dimensional QED can be reduced by more than $94\%$ compared to the unoptimized truncation.
Integrating these approaches into TN algorithms could greatly benefit LGT simulations.
% ========================================================================
\subsection{Tailored initial states}
\label{sec_TN_roadmap_opt_initialstates}
In TN algorithms for ground state searching, the optimization procedure generally starts from a random TN initial state $\ket*{\psi_{\mathrm{init}} }$, \idest{} the tensors in the network are filled with random coefficients at the beginning.
This strategy usually guarantees that the probability of overlapping with the true ground state $ |\braket*{\psi_{\mathrm{init}}}{\psi_{\mathrm{gs}}}| ^2 $ is not vanishing.
To reach a small error in the final energy, this procedure typically requires from 10 to 50 optimization sweeps for LGTs simulations, depending on the specific models, the Hamiltonian parameters, and the lattice size.
Since the time for completing a sweep can be very long, especially for large bond dimensions, strategies for reducing the number of needed sweeps could be beneficial for scaling up system sizes.
From this perspective, constructing appropriate states to be used as initial guesses can speed up the convergence, similar to the choice of the trial wave function for variational MC simulations. 
We consider the following options:
\begin{itemize}
\item[(i)] \emph{Physical insight:}
    Initial states can be constructed by following physical intuition, at least in those regimes in which analytical or partial numerical results are available.
    For instance, initial guesses can be constructed by starting from the TN ground states numerically obtained for a lower representation to simulate large spin representations of the gauge fields.
\item[(ii)] \emph{Machine learning:}
    Machine learning-assisted protocols can improve the construction of tailored initial states in the different regimes of the model parameters.
    For instance, feed-forward neural networks have been proposed as trial wave functions for quantum MC simulations \cite{Kessler2021ArtificialNeuralNetworks}, and machine learning techniques have been used to feed TN simulations \cite{Schroder2019TensorNetworkSimulation}. 
		Similarly, neural networks might reveal great potential in constructing initial states for LGTs to be used in large-scale TN simulations, in which reducing the number of sweeps is a key point for feasibility.
\item[(iii)] \emph{Tensor network results:}
    Following the idea of the physical insight, it is also possible to feed neighboring ground states as initial guesses into a ground state search.
    This option exists especially when scanning a phase diagram and varying parameters in a small increment such that the overlap between neighboring wave functions is sufficient; this overlap decreases for two points on the opposite sides of a quantum critical point.
    The same idea can be implemented by preparing an initial guess quenching from an easily accessible ground state to the target parameters; the quench itself does not have to be adiabatic or free of numerical errors, but must only have sufficient overlap with the ground state.
    The advantage is that one quench can generate multiple initial guesses along the quench for different parameters.
\end{itemize}
% ========================================================================
\subsection{Leverage HPC techniques for local optimization}           
\label{sec_TN_roadmap_opt_hpclocal}
We dedicate the two following sections to the numerical optimization of the TN algorithms. 
In this one, we give an overview of the topic and discuss possible strategies to improve the optimization. 
The more technical steps are discussed in the following \cref{sec_TN_roadmap_opt_hpcmpi}. 

To scale up TN simulations of LGTs, in particular regarding lattice sizes, another important factor is the number of optimization steps to be carried out. 
The number of optimization steps scales linearly with the number of sweeps as well as with the system size for MPS, PEPS, and TTN. 
The choice of the number of sweeps is set so that the algorithm reaches convergence when computing ground or low energy states. 
Let us briefly describe the general procedure for ground state searches. 
We will focus on the TTN optimization, even if the main points described here can be applied to other TN ansätze, such as MPS or PEPS. 
For a complete and technical description of the algorithms and implementation details, see Ref. \cite{Silvi2019TensorNetworksAnthology}.

Consider a generic QMB Hamiltonian $\ham$ and a generic normalized state $\ket{\psi}$, defined on the same Hilbert space. 
To numerically determine the ground state of $\ham$, the following global minimization problem has to be solved:
\begin{equation}
\label{eq_min_problem}
\min_{\ket{\psi}} \qty{E(\ket{\psi})} = \min_{\ket{\psi}} \bra{\psi}\ham\ket{\psi} \, .
\end{equation}
If $\ket{\psi}$ is written in terms of the TTN ansatz introduced in \cref{sec_TN}, the variational
parameters are the coefficients in the TTN. In this case, the global optimization problem of \cref{eq_min_problem} is broken down into a sequence of smaller optimizations, each of which involves only a minimal subset of tensors in the TTN. 
The algorithm solves the optimization via an eigenproblem searching for the eigenvector with the lowest eigenvalue.
Without any loss of generality, one single tensor at a time is optimized in the simplest case, as shown in \cref{fig_TTN_optimization}(a). 
In detail, the energy is computed by contracting the Hamiltonian between the TTN and its complex conjugate. 
Then, we start the optimization procedure from a target tensor $\tensor$, by computing its environment, \idest{} the network without the tensors $\tensor$ and $\tensor^\dagger$, which represents the effective Hamiltonian $\ham_{\rm{eff}}$ for the local problem. 
At this stage, an eigenproblem of $\ham_{\rm{eff}}$ is solved, and the tensor $\tensor$ is updated with the newly found ground state. 
The whole procedure is sequentially iterated for all the tensors in the network, performing an optimization sweep.

For each operation, efficient algorithms from linear algebra are typically used, \eg{} the Arnoldi algorithm implemented in the ARPACK library \cite{Arnoldi1951PrincipleMinimizedIterations, Lehoucq1998ARPACKUsersGuide}. 
We recall that the numerical complexity of this procedure for a single TTN-tensor is $O(d^2 \chi^2 + \chi^4)$ \footnote{As reported in \cite{Magnifico2024TensorNetworksLattice}, the scaling presented here does not consider the bond dimension of the MPO representing the hamiltonian. 
The bond dimension of the mpo depends on the dimensionality of the system. 
In a simplified argument holding for the TTN, the maximum number of interactions cut for any bipartition gives a first intuition.}.
Therefore, a single optimization can be time-consuming when $\chi$ is very large, \eg{}
$\chi \approx 1000$, as for simulating high-dimensional LGTs. 
We point out the established and promising future parallelization schemes for the single tensor optimization:
\begin{itemize}
  \item[(i)] \emph{opemMP:} 
    An efficient openMP implementation of the contraction between the effective operators with the tensor can speed up simulations. 
    Moreover, the Arnoldi algorithm of ARPACK is optimized for large-scale linear algebra operations and supports intra-node multi-core parallelization based on openMP \cite{Dagum1998OpenMPIndustryStandard}; thus, ARPACK does not become a bottleneck in the openMP implementation. 
    Nonetheless, many simulations remain expensive in computation time even with 64 or more cores available in HPC facilities; therefore, we consider more approaches beyond the well-established openMP path.
  \item[(ii)] \emph{Accelerators:} Graphics Processing Units (GPU) and Tensor
    Processing Units (TPU) offer a path to accelerate linear algebra routines,
    where both have demonstrated their usefulness: GPUs have reported speedups
    of up to a factor of 10 due to the efficient tensor
    manipulations \cite{Shi2016TensorContractionsExtended,Abdelfattah2016HighperformanceTensorContractions,Vincent2022JetFastQuantum,Pan2022SimulationQuantumCircuits};
    TPU have shown great potential in large-scale simulations
    of several quantum systems, \eg{} drastically reducing the computational time
    of DMRG calculations with very large bond dimensions from months to
    hours \cite{Jouppi2017InDatacenterPerformanceAnalysis,Hauru2021SimulationQuantumPhysics,Morningstar2022SimulationQuantumManyBody,Ganahl2023DensityMatrixRenormalization}.
    TPU are application-specific integrated circuits originally introduced
    for machine learning; we then consider the integration and tuning of TPUs as a
    step after the successful integration of GPUs. 
		While single GPUs can solve
    TTN-problems up to a bond dimension of $\chi < 1000$, multi-GPU support
    is available for libraries; HPC systems typically provide hardware with four GPUs
    per node.
  \item[(iii)] \emph{Multi-node approaches to local tensor optimizations:} Both CPU and
    GPU algorithms can be further scaled by using multiple nodes. The underlying linear
    algebra routines of the local eigenvalue problem are parallelizable via libraries
    such as ScaLAPACK or MAGMA. Both libraries provide routines for distributed memory
    machines \cite{Blackford1997ScaLAPACKUsersGuide}; MAGMA supports moreover CPU and GPUs. In this way, the workload
    of the eigenproblem procedure can be split into several computation nodes. Then, it
    is important to analyze the performances as a function of the bond dimension $\chi$,
    to test the effectiveness of this approach against the latency of the inter-node
    communications.
  \item[(iv)] \emph{Tuning of parameters and algorithms:} accelerators developed for machine learning applications have excellent support for lower and real precision. 
	Tuning parameters over the different sweeps is beneficial, \eg{} increasing the precision towards the end of the sweep. 
	This approach profits from faster single-precision implementations during the first sweeps. 
	Selecting algorithms like random SVD can also bring benefits \cite{Lu2017HighPerformanceOutofcoreBlock}.
\end{itemize}
% ========================================================================
\subsection{Sweeps and HPC parallelization}    
\label{sec_TN_roadmap_opt_hpcmpi}
So far, we have parallelized single tensor optimizations within a sweep, but the sweep itself was sequential, \idest{} serial. 
Recent works formulated parallel versions of MPS algorithms for ground state search and time evolution, \eg{} via the Message Passing Interface (MPI) \cite{Gabriel2004OpenMPIGoals,Stoudenmire2013RealspaceParallelDensity,Secular2020ParallelTimedependentVariational}.
The main difference between the serial and parallel algorithms is the effective operators used in the optimization. 
In the serial version, the effective operators contain the information of the most recent version of all other tensors. 
This update is delayed in the parallel version, \idest{} the tensor that entered the effective operator is not necessarily the one of the current sweep but can be the version of an earlier sweep.
\begin{figure}
\includegraphics[width=1\textwidth]{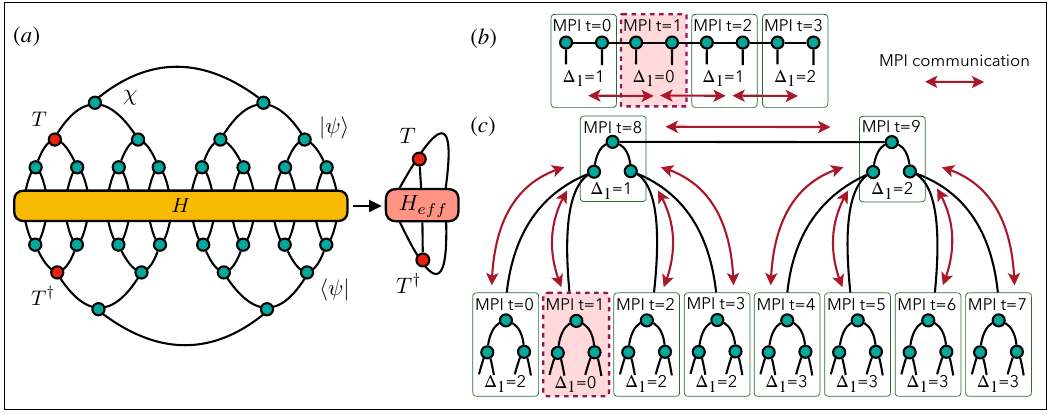}
\caption{\emph{Effective operators and parallel tensor networks.}
  (a)~Procedure for optimizing a TTN to find the ground state of a QMB system: the energy is computed by contracting the Hamiltonian $\ham$ (yellow tensor) with the TTN, representing the state $\ket{\psi}$, and its hermitian conjugate, representing $\bra{\psi}$.
  The variational optimization starts from a target tensor $\tensor$ (red tensor), by computing its effective Hamiltonian $\ham_{\rm{eff}}$ and then solving the local eigenvalue problem for the latter. 
  The tensor $\tensor$ is then updated with the newly found ground state, and the procedure iterates over all the tensors in the network (sweep).
  (b) The workload itself consists of optimizing each tensor held by the MPI thread $t$, which requires effective operators calculated by other MPI threads.
  We dub \emph{delays} $\Delta_{i}$ the number of optimization cycles needed to obtain the information of tensors in the i-th MPI thread in another MPI thread via MPI communication.
  MPS naturally split into sub-chains which communicate with one or two neighboring MPI threads to obtain updated effective operators. Delays for updates scale with the distance between two MPI threads along the chain. 
  Each MPI thread can use threading or openMP, \eg{} in a hybrid openMP-MPI approach.
  (c) Similarly, TTNs can be split into sub-trees for each MPI thread allowing for optimizing the sub-tree without communication with other MPI threads.
  Delays due to updating scale logarithmically as any distance in a tree network.}
\label{fig_TTN_optimization}
\end{figure}

If the delay becomes an obstacle to convergence, there is the option to modify parameters during the sweeps. 
Typically, ten to fifty sweeps are necessary to converge to a solution. 
As the initial state is usually random, MPI can be used especially at the beginning. 
To ensure convergence, one can consider serial steps at the end; even a gradual reduction of the MPI processes as the sweeps proceed is possible and gradually reduces the delays.

Considering the MPS scenario of a chain in \cref{fig_TTN_optimization}(b), we split the chain into equal parts of $(N / \nummpithreads)$ sites. 
Each part of the chain communicates with its two neighbors apart from the two boundaries.
The effective operators take into account the tensors of the same MPI process with zero delay as in the serial case.
The tensors of the i-th neighboring MPI process have a delay of $i$. 
The worst-case scenario of the delay scales linearly with the number of MPI processes.
The delay can be avoided by communicating the effective operators after each update through the chain, which is a quasi-serial step with no more than two MPI processes active at the same time.

The problem becomes more complicated for the TTNs suggested for LGT, but we expect a benefit for the parallelization of a TTN versus an MPS for higher dimensional systems.
\cref{fig_TTN_optimization}(c) shows an example of how each MPI process gets assigned a sub-tree within the complete TTN.
Unlike the MPS, the number of neighboring MPI processes for communication is at least three and increases with $(N / \nummpithreads)$ tensors per thread.
Assuming equally shaped subtrees for all MPI processes, the delay of the tensor update is $2 \cdot \logtwo(N)$ in the worst-case scenario.

One-dimensional systems with nearest-neighbor interactions thus exhibit a delay of $1$ in the worst-case scenario in the MPS, while the delay is up to $2 \cdot \logtwo(N)$ for the nearest neighbors in the center of the TTN. 
Rather, higher dimensional systems change this aspect, \eg{}, for an $\Nsites \times N$ two-dimensional system mapped to 1D via a zig-zag mapping. 
The MPS has a worst-case delay of the tensor update of $\Nsites$ for the slow index. 
In contrast, the TTN has the same log behavior and a maximum delay at the center of the TTN as $2 \cdot \logtwo(N^2)$. 
Thus, the worst-case delay is equal for $16 \times 16$ systems; increasing $\Nsites$ further, TTNs exhibit smaller maximum delays during parallel sweeps. 
Moreover, the TTN is unaffected by the type of mapping used; in contrast, the worst-case delay for the MPS grows to $2N$ for the snake mapping and to at least $\Nsites^2 / 2$ for the Hilbert curve \cite{Cataldi2021HilbertCurveVs}. 
Equal arguments hold for 3D systems and delays of $\Nsites^2$ (MPS, zig-zag) versus $2 \logtwo(N^3)$ (TTN, any mapping).

To get an intuition of what parallelized simulations can solve, we sketch out the specifications for a parallel simulation on the pre-exascale cluster \emph{Leonardo} hosted by \emph{Cineca}.
We choose an MPI approach together with the GPUs. Leonardo has 3456 nodes with four GPUs totaling 13824 GPUs available for the complete cluster. 
Bond dimensions on the order of $\chi = 450$ consume 54GB of memory without effective operators (assuming double complex precision, 40 Lanczos vectors) and allow to solve the eigenproblem on the GPU without temporarily storing data on the CPU.
We use the single-tensor per MPI-thread with $\chi=450$ as a baseline where we extract a rough empirical estimate with Leonardo; in detail, we use a 2D quantum Ising model with $\mathbb{Z}_{2}$ symmetry in the vicinity of the quantum critical point and the initial tensor optimizations \cite{Bacilieri2024QuantumTEAQtealeaves}.
Due to the delay of the tensor in the effective operators, the minimum number of sweeps must be beyond 24. 
Then, we consider the scaling of the TTN previously introduced and generate \cref{hpctime} with an overview of different system sizes and bond dimensions. 
These results provide a coarse-grained estimate, since plaquette terms, different symmetries, entanglement generation, and optimization time within later sweeps could further impact the computational time.

Our estimate predicts that a system of $256 \times 256$ sites takes about two months for bond dimension $\chi=450$ on 1024 GPU nodes of Cineca's \emph{leonardo}.
Future improvements are likely to bring this simulation time down, \eg{}, the next-generation GPUs in comparison to the A100 or further optimization in data movement.
In contrast, a three-dimensional system with large entanglement and many sites requires three to four orders of magnitude improvement, where cluster size and other improvements have to come together to reach this challenge.
Furthermore, \cref{hpctime} provides an estimate of the boundary for a potential quantum advantage in simulating lattice gauge theories with quantum computers or simulators.
\begin{table}[t]
\begin{center}
  \begin{tabular}{@{} lcrr @{}}
  \toprule
  System size       & $\chi$    & Factor        & Estimated walltime  \\
  \cmidrule(r){1-1} \cmidrule(r){2-2} \cmidrule(rl){3-3} \cmidrule(l){4-4}
  $64 \times 64$      & $ 450$ & $\Tbase$       & $4.16~\mathrm{days}$ \\
  $64 \times 64$      & $ 900$ & $16 \cdot \Tbase$  & $66.6~\mathrm{days}$ \\
  $256 \times 256$     & $ 450$ & $28 \cdot \Tbase$  & $116.5~\mathrm{days}$ \\
  $256 \times 256$     & $ 900$ & $448 \cdot \Tbase$  & $5.1~\mathrm{years}$ \\
  $16 \times 16 \times 16$ & $ 450$ & $4 \cdot \Tbase$   & $16.6~\mathrm{days}$ \\
  $16 \times 16 \times 16$ & $ 900$ & $64 \cdot \Tbase$  & $266~\mathrm{days}$  \\
  $64 \times 64 \times 64$ & $ 450$ & $1984 \cdot \Tbase$ & $23~\mathrm{years}$  \\
  $64 \times 64 \times 64$ & $ 900$ & $31744 \cdot \Tbase$ & $362~\mathrm{years}$ \\
  \bottomrule
  \end{tabular}
  \caption{\label{hpctime}%
  \emph{Estimated simulation time.} We derive the baseline from
  a single-tensor optimization of a $64 \times 64$ quantum Ising simulation with
  $\mathbb{Z}_{2}$ symmetry taking $7192\mathrm{s}$ on a A100 GPU. Further, we assume
  that single-tensor update, one tensor and one GPU per MPI thread, and 50 sweeps for
  the baseline. To extrapolate to larger systems, we assume a scaling with $\mathcal{O}(\chi^4 N^{D-1})$
  as well as seven (thirty-one) tensors per MPI thread for $256 \times 256$ ($64 \times 64 \times 64$) systems. The empirical scalings
  are approximately a factor of $2.3$ for doubling the system size and $13$ for doubling
  the bond dimension, which we obtain from smaller
  simulations with $\chi = 225$ and for $32 \times 32$ qubits. The times are valid for any $d < \chi$.
  }
\end{center}
\end{table}
% ========================================================================
\subsection{Finite temperature regime}
\label{sec_TN_finiteTemp}
To date, TN simulations of high-dimensional LGTs including dynamical matter are exploring zero temperature regimes, which are important to understand the low-energy properties of the models. 
To explore finite temperature phenomena, particularly relevant for open research problems such as the QCD phase diagram, technical and numerical challenges have to be tackled, \eg{} devising and testing efficient TN algorithms for targeting quantum states at equilibrium. 
As suggested in the next paragraph, we foresee two possible paths toward finite temperature TN states.

Matrix product density operators (MPDOs) and locally purified tensor networks (LPTNs) provide already today the option to tackle finite temperature regimes via an imaginary time evolution \cite{Verstraete2004MatrixProductDensity,Zwolak2004MixedStateDynamicsOneDimensional,Werner2016PositiveTensorNetwork}.
Herein, the algorithm starts at the infinite temperature state and starts \emph{cooling} the system via a specified number of time steps and specified step size to reach a given temperature. 
In its original formulation, both approaches are one-dimensional chains. 
Matrix product density operator can be formulated as TTN but faces some challenges in terms of the question of positivity \cite{Kliesch2014MatrixProductOperatorsStates} or integrating symmetries. 
In contrast, tree tensor operators (TTO) are the tree-equivalent of an LPTN; they are also a positive loopless representation of density matrices, recently introduced in \cite{Arceci2022EntanglementFormationMixed}.
However, TTOs cannot represent the infinite temperature state necessary for the imaginary time evolution approach.
Instead, a possible use of the TTO employed in LGT simulations consists of a variational algorithm to target finite-temperature states or reconstruct open-system quantum dynamics, by efficiently compressing the relevant information.
The TTO enables useful measures, \eg{} computing the entanglement of formation as already shown for representative one-dimensional models at finite temperature \cite{Arceci2022EntanglementFormationMixed}.
% ========================================================================
\section{Summary}
\label{sec_TN_summary}
In this chapter, we have provided a comprehensive exploration of Tensor Network (TN) methods applied to Lattice Gauge Theories (LGTs), highlighting their relevance as a complementary approach to traditional numerical techniques like Monte Carlo simulations. 
We began by addressing the limitations of exact diagonalization (ED) for quantum many-body (QMB) systems, emphasizing the importance of block diagonalization for computational efficiency in high-dimensional systems \cite{Cataldi2024Edlgt}.

The chapter then shifted focus to TN methods, with a detailed explanation of how they efficiently capture quantum correlations, particularly through Matrix Product States (MPS) and Tree Tensor Networks (TTN), characterized by a loopless geometry. 
We underscored the critical role of preserving locality in high-dimensional simulations, discussing the superiority of space-filling curves like the Hilbert curve in optimizing TN algorithms \cite{Cataldi2021HilbertCurveVs}. 
Additionally, we reviewed key algorithms for equilibrium and time-evolution simulations, such as the TTN ground-state variational search and TEBD/TDVP techniques.

Finally, we laid out a roadmap for improving TN simulations, particularly in large-scale LGTs beyond one dimension, focusing on addressing computational challenges such as bond dimension growth and local basis truncation \cite{Magnifico2024TensorNetworksLattice}. 
These improvements offer promising pathways for enhancing the scalability of TN methods and making them competitive with Monte Carlo techniques in simulating complex LGT models.

%% file: chapters/su2.tex
\chapter{Equilibrium properties of SU(2) LGT}
\label{chap_SU2_groundstate}
\begin{figure}
    \centering
    \includegraphics[width=1\textwidth]{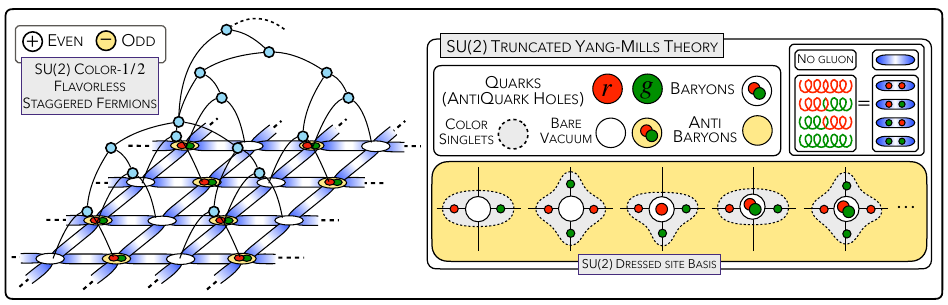}
    \caption{TTN approach to (2+1)D SU(2) Yang-Mills LGT. 
    Lattice sites host flavorless SU(2)-color-1/2 fermionic fields (red and green) in a staggered configuration (white and yellow). 
    Lattice (blue) links describe gauge degrees of freedom from a 5-dimensional truncated Hilbert space (\emph{hardcore-gluon approximation}). 
    SU(2) Gauss Law is implemented at each lattice site.}
    \label{fig_introduction_picture}
    \end{figure}
In this chapter, we collect the numerical simulations of \cite{Cataldi2024Simulating2+1DSU2} concerning the ground-state properties of the (2+1)D SU(2) Yang-Mills Hamiltonian LGT within the \emph{hardcore-gluon approximation} of \cref{sec_SU2_hardcoregluon} and obtained by using the variational ground-state search algorithm\footnote{As for the single-node TN optimization, we set the Arnoldi algorithm to discard eigenvalues smaller than $10^{-4}$. The convergence of the whole variational algorithm is established for absolute $\Delta \varepsilon_{abs}=10^{-5}$ and relative convergence thresholds $\Delta \varepsilon_{re\ell}=10^{-5}$ defined in \cref{eq_TN_variational_thresholds} for consecutive optimization sweeps $s-1$ and $s$. 
The maximal bond dimension $\chi$ adopted in the reported TN simulations is always obtained by looking at the single-site energy relative convergence of $10^{-4}$ (for more details, see \cite{Cataldi2024Simulating2+1DSU2}).} for Tree Tensor Networks (TTN) discussed in \cref{sec_TN_gs_algorithm}.
Given the inherent complexity of the model, simulations range from small to intermediate system sizes, up to 32 matter sites, well beyond the capabilities of Exact Diagonalization (ED) and state-of-the-art quantum computing. 
Indeed, due to the rich structure of the quantum degrees of freedom, the analysis of the ground state properties of the system, for lattice sizes up to $4\times 8$ as performed here, would require $>$160 qubits to describe on a quantum computer.

The results explore several regimes of the model at equilibrium, including finite baryon number density. 
We characterize the model phase diagram by evaluating multiple observables, such as energy gaps, matter/antimatter and color-charge densities, gauge field distributions, and topological invariants. 
Whenever it is possible, we highlight the role of dynamical matter out of the behavior observed in the simpler pure theory.
% ============================================================================================
\section{The model}
Let us start recalling the SU(2) Hamiltonian in \cref{eq_H_SU2_full,eq_H_SU2_pure} on a (2+1)D spatial lattice $\Lambda$ and rescale it in dimensionless energy scale units $\ham {\to} \ham^{\prime} = \frac{\lspace}{\lspeed\hbar} \ham$ so that the hopping term has constant coupling $\frac{1}{2}$. 
Correspondingly, the other Hamiltonian terms acquire the rescaled dimensionless couplings:
\begin{align}
    \mass&=\lspace\frac{\mass[0]\lspeed}{\hbar} = \frac{\lspace}{a_{\mass}}\, (\rm{mass})&
    \frac{\coupling^{2}}{2} &=\frac{\lspace}{2a_g}\, (\rm{electric})&
    \frac{1}{2\coupling^{2}}&=\frac{a_g}{2\lspace} (\rm{magnetic}),
\end{align}
where $a_{\mass}\equiv \hbar/(\mass[0]\lspeed)$, while, from the gauge coupling in \cref{eq_gauge_coupling} in two spatial dimensions, $a_{\coupling}\equiv (\lspeed \hbar \permittivity)/\charge^{2}$. 
By doing so we obtain a fixed dimensionless \emph{quark ratio} $\alpha_c = \coupling^{2}/2\mass=(a_{\mass}/2a_g)$ which does not scale with the lattice spacing and is solely determined by the color charge and the bare mass of the quark (see \cref{sec_lattice_gaugefields}).
Then, the (dimensionless) Hamiltonian reads:
\begin{equation}
    \begin{aligned}
        \ham^{\prime}=&+
        \frac{1}{2}\sum_{\alpha,\beta}\sum_{\vecsite} \qty[\text{-i} \hpsi^{\dagger}_{\vecsite, \alpha} \NApara_{\genlink_x}\hpsi_{\vecsite+\latvec[x],\beta} 
        - (-1)^{\site[x]+\site[y]}\hpsi^{\dagger}_{\vecsite,\alpha} \NApara_{\genlink_y}\hpsi_{\vecsite+\latvec[y],\beta} + \hc]\\
        &+ \mass \sum_{\vecsite}(-1)^{\site[x]+\site[y]}\sum_{\alpha} \hpsi^{\dagger}_{\vecsite,\alpha} \hpsi_{\vecsite,\alpha}+\frac{\coupling^{2}}{2}\sum_{\genlink}\eleE^{2}_{\genlink}
        -\frac{1}{2\coupling^{2}}\sum_{\square}\Tr(\plaq+\plaq*)\,,
    \end{aligned}
    \label{eq_2D_SU2_Ham_numerics}
\end{equation}
where $\alpha,\beta \in \qty{\rla,\gla}$.
Expressing in \cref{eq_2D_SU2_Ham_numerics} in terms of the dressed site operators from \cref{sec_SU2_dressedsite_operators} in the \emph{hardcore-gluon} approximation (see \cref{sec_SU2_hardcoregluon}), we obtain:
\begin{equation}
    \begin{aligned}
        \ham^{\prime}=&+
        \frac{1}{2}\sum_{\vecsite} \qty[\text{-i} \hat{Q}^{\dagger}_{\vecsite}\hat{Q}_{\vecsite+\latvec[x]} 
        - (-1)^{\site[x]+\site[y]}\hat{Q}_{\vecsite}\hat{Q}_{\vecsite+\latvec[y]} + \hc]\\
        &+ \mass \sum_{\vecsite}(-1)^{\site[x]+\site[y]}\hat{N}_{\vecsite,\rm{tot}}+\frac{\coupling^{2}}{4}\sum_{\vecsite}\hat{\Gamma}^{2}_{\vecsite}-\frac{1}{2\coupling^{2}}\sum_{\square}\Tr(\plaq+\plaq*)\,,
    \end{aligned}
\end{equation}
Assuming that the SU(2) LGT is super-renormalizable in two spatial dimensions \cite{Hamer1985SULatticeGauge} and excluding quantum corrections to the scaling (see anomalous dimension \cite{Giedt2016AnomalousDimensionsLattice}), then the continuum limit of \cref{eq_2D_SU2_Ham_numerics} is located at $\coupling^{2}=\alpha_c\mass\to0$ (more quantitatively, at $\lspace{\ll}a_g, a_{\mass}$). 
% ============================================================================================
\subsection{Local Observables}
\label{sec_SU2_local_observables}
Together with the ground-state energy density $\varepsilon=\langle\ham\rangle/\Nsites$, we evaluate the expectation values $\avg{\cdot}$ of several local observables onto the computed ground states. 
Regarding gauge fields, we track the color-electric and color-magnetic energy densities
\begin{align}
  \avg{E^{2}}&=\frac{1}{\Nsites}\sum_{\vecsite,\latvec}\avg{\hat{E}^{2}_{\vecsite,\latvec}}&
  \text{and}&&
  \avg{B^{2}}&=-\frac{1}{\Nplaqs}\sum_{\square}\avg{\Tr(\plaq+\plaq*)}+c',
\label{eq_SU2_gauge_observables}
\end{align}
where $\Nsites$ and $\Nplaqs$ correspond to the total number of sites and lattice plaquettes. 
The constant factor $c^{\prime}=\frac{1}{2}$ in \cref{eq_SU2_gauge_observables} sets the minimum of the magnetic energy density to 0.
Correspondingly, we define the electric and magnetic fluctuations as follows:
\begin{align}
    \delta E^{2}&=\sqrt{\avg{E^4} - \avg{E^{2}}^{2}}&
    \delta B^{2}&=\sqrt{\avg{B^4} - \avg{B^{2}}^{2}}\,.
  \end{align}
When considering matter fields, it is useful to separately measure the staggered fermion density for even (+) and odd (-) sites
\begin{align}
    n_{\pm}&=\frac{1}{\Nsites_{\pm}}\sum_{\vecsite\in\Lambda_{\pm}}
    \sum_{\alpha=\rla,\gla}\avg{\hpsi^{\dagger}_{\vecsite,\alpha}\hpsi_{\vecsite,\alpha}}
    \label{avg_particle_density}
\end{align}
where $\Lambda_{+}\,,(\Lambda_{-}$) is the even (odd) sub-lattice, while $\Nsites_{+}\,,(\Nsites_{-})$ is the corresponding number of sites.
Tracking these two quantities gives us immediate access to the density of quarks $(n_{+})$ and the density of anti-quarks $(2-n_{-})$ separately, according to the staggered fermion formalism.
Similarly, we can define the total \emph{particle density} (quarks plus anti-quarks)
\begin{align}
    \density &= n_{+}+(2-n_{-})&
    \text{with}&&
    0&\leq\density\leq4\,,
    \label{eq_SU2_rho_density}
\end{align}
as well as the \emph{baryon number density}, (quarks minus anti-quarks divided by two)
\begin{align}
    b &= \frac{1}{2}(n_{+}-(2-n_{-}))&
    \text{with}&&
    0&\leq b\leq1\,,
    \label{eq_SU2_b_density}
\end{align}
which is a good quantum number, as it is a conserved quantity tied to the global staggered fermion number conservation.
We stress that, unlike quantum chromodynamics, SU(2) Yang-Mills baryons $-$ colorless bound states of matter particles $-$ are made by two, not three, quarks. 
Similarly, anti-baryons are made by two anti-quarks. 
Correspondingly, mesons are made by one quark and one anti-quark as normal.

Both mesons and standalone quarks can be detected by looking at the average \emph{matter color density} $\avg{S^{2}_{\rm{matt}}}$ defined in \cref{eq_SU2_matter_casimir}: 
\begin{equation}
    \avg{S^{2}_{\rm{matt}}}=\frac{1}{\Nsites}\sum_{\vecsite}\avg{\mattercasimir}
    =\frac{1}{\Nsites}\sum_{\vecsite,a}\avg{\colormatter_{\vecsite}^{a}\colormatter_{\vecsite}^{a}}=\frac{1}{2\Nsites}\sum_{\vecsite,a}\avg{\sum_{\alpha\beta}\qty(\hpsi^{\dagger}_{\vecsite,\alpha}\hpsi_{\vecsite,\beta}\sigma^{a}_{\alpha\beta})^{2}}\,.
  \label{eq_SU2_matter_casimir_avg}
\end{equation}
Our quantitative analysis also includes the von Neumann entanglement entropy \cite{Eisert2010ColloquiumAreaLaws}
\begin{equation}
  \entropy_{A}=-\Tr \rho_{A}\log_{2}\rho_{A},
  \label{eq_entropy}
\end{equation}
where $\rho_{A}$ is the reduced density matrix of the partition $A$, which we choose exactly to be the bottom (or top) half of the system.
% ============================================================================================
\section{Magneto-electric transition in the pure theory}
\label{sec_SU2_pure_theory}
\begin{figure}
	\centering
	\includegraphics[width=1\textwidth]{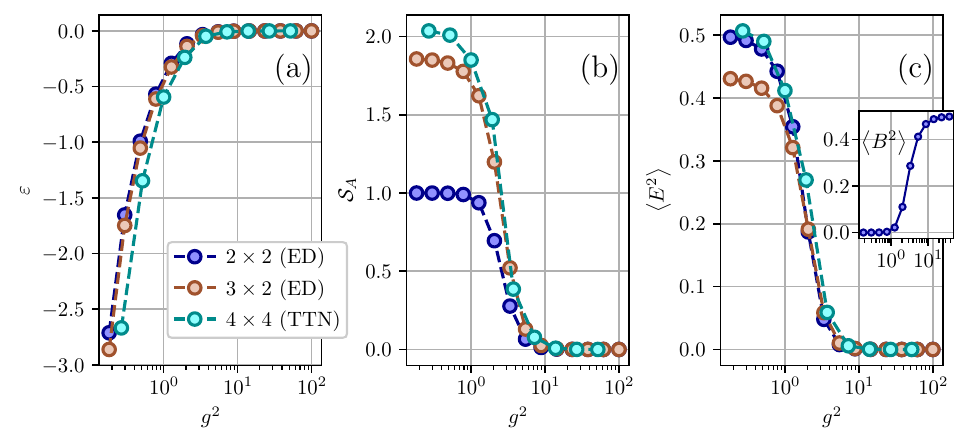}\\
    \includegraphics[width=0.5\textwidth]{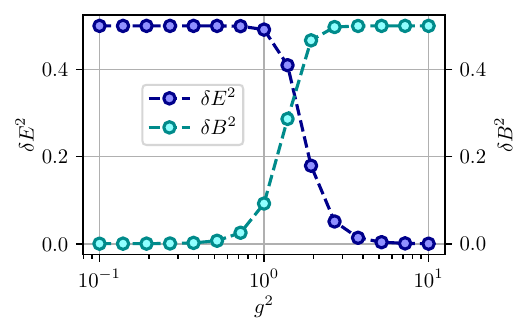}
	\caption{Numerical simulations of the pure Hamiltonian in \cref{eq_2D_SU2_Ham_numerics} with OBC for different lattice sizes. 
    The plots display respectively (a) the ground-state energy density $\varepsilon$, (b) the entanglement entropy $\entropy_{A}$ of half the system, (c) the average electric energy contribution $\avg{E^{2}}$, with the magnetic energy density $\avg{B^{2}}$ shown in the inset. 
    (bottom line) Fluctuations of the gauge observables in \cref{eq_SU2_gauge_observables} as a function of the $\coupling$-coupling for a $2\times 2 $ lattice.
    Figures from \cite{Cataldi2024Simulating2+1DSU2}.}
	\label{fig_SU2_pure_theory_simulations}
\end{figure}

We first focus on the pure theory (corresponding to the $\mass \to \infty$ of \cref{eq_2D_SU2_Ham_numerics}) under Open Boundary Conditions (OBC). 
According to the results shown in \cref{fig_SU2_pure_theory_simulations}, the pure Hamiltonian displays two phases driven by $\coupling$ and characterized by strong fluctuations of the electric and magnetic fields.

In detail, for the small-$\coupling$ (magnetic) phase, the plaquette interaction provides the largest contribution to the energy in \cref{eq_2D_SU2_Ham_numerics}. 
As such, magnetic fields are depleted, and electric fields display large quantum fluctuations $\delta \eleE^{2}$  and compensate for any electric activity. 
Conversely, in the large-$\coupling$ (electric) phase, electric fields are energetically expensive and thus depleted in the ground state, while magnetic fields show large fluctuations $\delta \magn^{2}$.

Unlike the electric phase, which displays marginal entanglement as the ground state is almost a product state, the magnetic phase reveals an entanglement that scales with the length of the bi-partition: this behavior, signaling a sharp area-law of entanglement, suggests that the magnetic phase is likely approximated by a resonant-valence bond (RVB) state of plaquettes, akin to the local structure of the ground state of the Toric Code \cite{Kitaev2006AnyonsExactlySolved}.

The entanglement entropy approximates a monotonic function along $\coupling$, without any peak in the transition between the two phases.
This observation suggests that, for large bare masses $\mass$, this quantum phase transition is either \emph{first order} or a \emph{crossover}.
Conversely, as shown in \cref{fig_ED_SU2_grid_discrete}, the small-$\mass$ scenario of the full theory peaks close to the transition, and the peak is wider and larger for smaller masses. 
We stress that the magneto-electric transition is compatible with the \emph{roughening transition} \cite{Drouffe1981RougheningTransitionLattice,Munster1981RougheningTransitionNonabelian,Kogut1983LatticeGaugeTheory} observed via MC simulations \cite{Berg1981SULatticeGauge,Ambjorn1984StochasticConfinementDimensional,Ambjorn1984StochasticConfinementDimensional-1} and Cluster Expansion Methods (CEM) \cite{Hamer1985SULatticeGauge,Arisue1984NewClusterExpansion}.
% ============================================================================================
\section{Baryonic spectrum}
\label{res_energy_gaps}
For finite $\mass$, fermionic matter displays a dynamical role in the Hamiltonian of \cref{eq_2D_SU2_Ham_numerics}. 
The baryon number density $b$ in \cref{eq_SU2_b_density} is a quantum number associated with global symmetry, and can thus be directly encoded in the TTN ansatz as well as in ED simulations. 
In this way, we directly target the ground state within a selected baryon number density sector \cite{Silvi2014LatticeGaugeTensor,Silvi2019TensorNetworkSimulation,Cataldi2024Edlgt}.

The model is symmetric under CP, that is, mirror spatial reflection ($\site[x] \to \Nsites_{x} - \site[x]$) times particle-hole exchange (${\hpsi}_{\alpha} \to i \sigma^{y}_{\alpha\beta} {\hpsi}_{\beta}^{\dagger}$) of staggered fermions. 
Then, at negative baryon densities $b{<}0$, the ground state is the CP-reflected of the ground state at positive baryon density $|b|$.

We numerically verified that the global ground state is found at null baryon density $b=0$ for any $\coupling$ and $\mass$. As we can directly tune the baryon number of each TTN simulation, we have immediate access to the inter-sector energy gap by calculating the difference
\begin{equation}
\begin{split}
    \Delta_{\abs{b}} &= \qty(\varepsilon_{b} - \varepsilon_{0})\Nsites =
    \qty(\varepsilon_{-b} - \varepsilon_{0})\Nsites
    \geq 0\\
    &=m\abs{b}\Nsites+ \Delta_{\abs{b}}^{*},
\end{split}
\label{intersector_gap}
\end{equation}
where we also defined the \emph{binding energy} $\Delta_{\abs{b}}^{*}$ by subtracting the bare mass of the corresponding excess quarks or anti-quarks ($\abs{b}\Nsites$).

A simple yet illustrative analysis is to study the energy density gap between the one-baryon sector ($b=2/\Nsites$) and the vacuum sector ($b=0$) and then approach the continuum limit $\lspace\to0$ at fixed ratio $\alpha_c = \coupling^{2}/2\mass \propto \charge^{2} /\mass[0]$ (see \cref{sec_lattice_gaugefields}).

\begin{figure}
	\centering
	\includegraphics[width=0.5\textwidth]{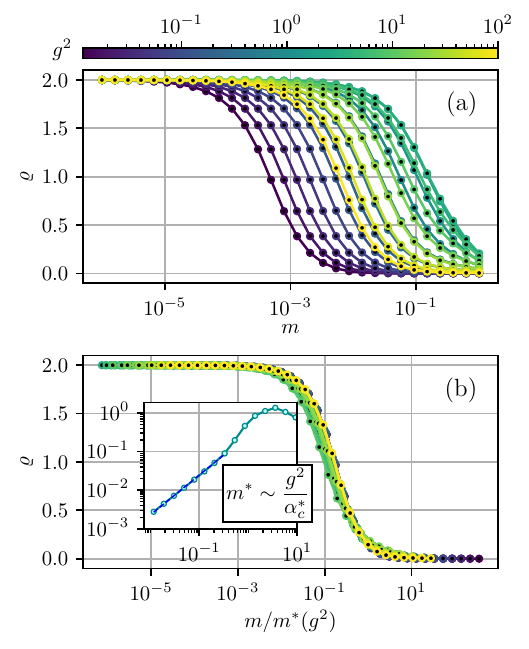}\hfill
    \includegraphics[width=0.5\textwidth]{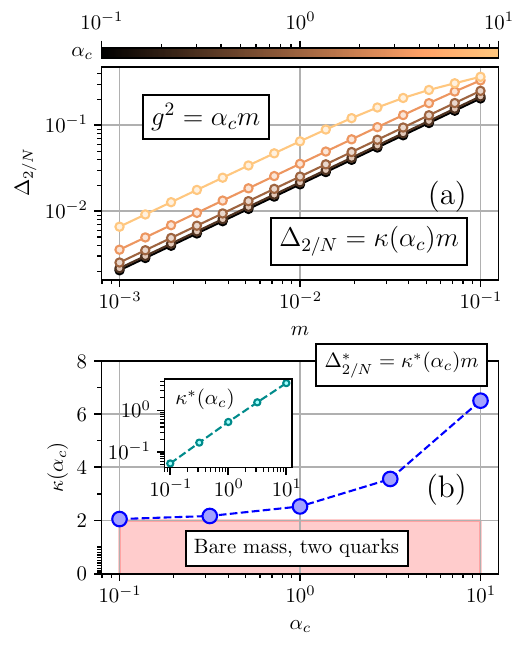}
    \caption{(a) Scaling of the particle density defined in \cref{eq_SU2_rho_density} as a function of $\mass$ for different values of the gauge coupling $\coupling$. 
    (b) All the $\density(m)$ curves of the particle density collapse on a single one simply by re-scaling the mass $\mass$ by a factor $\mass^{*}$ displaying a power-law scaling in $\coupling^{2}$ (see the inset). 
    By fitting this scaling we extract \cref{eq_powerlaw_scaling}, whose error bars have been computed exploiting error propagation onto the covariance matrix of the fit.
    Results obtained from simulations on a $2\times 2$ lattice in OBC at baryon density $b=0$.
    (a) Scaling of the inter-sector gap $\Delta_{2/\Nsites}$ in \cref{intersector_gap} as a function of $\mass$, for different choices of the $\coupling$-coupling $\coupling^{2}=\alpha_{c}m$. 
    By fitting the power-law scaling of $\Delta_{2/\Nsites}$ in the small-$\mass$ limit, we obtain the linear dependence on $\mass$ shown in \cref{baryon_mass}, whose slope $\kappa$ depends on $\alpha_{c}$ as shown in (b). The inset displays the corresponding $k^{*}$ of  the binding energy $\Delta_{2/\Nsites}^{*}$ in \cref{binding_energy}.
    Results have been obtained from simulations of a $2\times 2$ lattice in PBC, where $\Delta_{2/\Nsites}=\Delta_{\abs{b}=0.5}$.
    Figure from \cite{Cataldi2024Simulating2+1DSU2}.}
    \label{fig_local_matter_density}
\end{figure}

As shown in \cref{fig_local_matter_density}(a), the gap $\Delta_{2/\Nsites}$ displays a clear linear scaling with $\mass= \frac{\mass[0]\lspeed}{\hbar}\lspace$.
Namely:
\begin{align}
    \Delta_{2/\Nsites}=\kappa(\alpha_{c})\mass=\kappa(\alpha_{c})\frac{\mass[0]\lspeed}{\hbar}\lspace\,,
    \label{baryon_mass}
\end{align}
implying that the actual baryon mass is $\mass[b]=\kappa(\alpha_{c})\mass[0]$.
As for all hadrons, its mass is always greater than the bare mass of its quark components, thus $\kappa \geq 2$. We show this observation in \cref{fig_local_matter_density}(b), where we display $\kappa$ as a function of $\alpha_{c}$.
More interestingly, in the case of the binding energy $\Delta_{2/\Nsites}^*$ displayed in the inset of \cref{fig_local_matter_density}(b), we observe a power-law scaling of $\kappa^*$ in $\alpha_{c}$:
\begin{align}
    \kappa^*&=\frac{\Delta_{2/\Nsites}^*}{\mass}= \kappa -2 &
    \text{with}&&
    \kappa^* (\alpha_c)&{\sim} 0.13 \cdot \alpha_c^{0.96}\,,
    \label{binding_energy}
\end{align}
that is compatible with linear scaling.
Such relations confirm that baryons are actual \emph{quasi-particles} of the continuum theory and provide a connection to the bare quark properties ($\alpha_c$, $\mass[0]$). 
We carried out this analysis for a finite-size sample, but the baryon-to-quark mass ratio $\kappa$ is expected to stay finite even at the thermodynamical limit.
% ============================================================================================
\section{Baryon-liquid phase}
\label{sec_full_theory}
\begin{figure}
    \centering
    \includegraphics[width=1\textwidth]{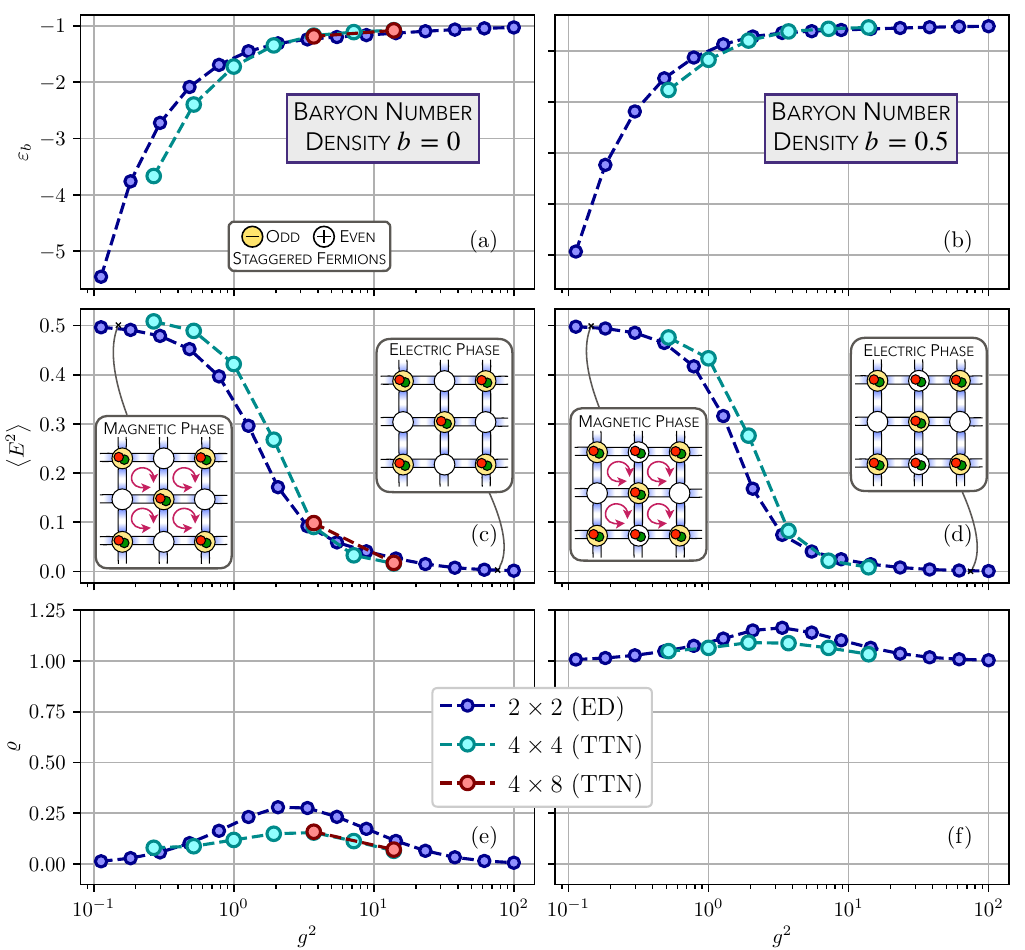}
    \caption{
    Numerical results of the full SU(2) Hamiltonian in \cref{eq_2D_SU2_Ham_numerics} with OBC and baryon number density $b=0$ (left column) and $b=0.5$ (right column). The plots display respectively: 
    (a)-(b) the ground-state energy density $\varepsilon_{b}$, (c)-(d) the average electric energy contribution $\avg{E^{2}}$ in \cref{eq_SU2_gauge_observables}, enlightening the transition between the magnetic (purple fluxes) and the electric (no fluxes) phases discussed in \cref{sec_SU2_pure_theory}, and (e)-(f) the average particle density $\density$ in \cref{eq_SU2_rho_density}, which appears peaked in the $\coupling$-transition. 
    The pictorial lattice configurations in the finite baryon density $b=0.5$ represent states with $b$ extra gapped hardcore local bosons with low dynamics compatible with the two electric/magnetic phases.
    Figure from \cite{Cataldi2024Simulating2+1DSU2}.}
    \label{fig_SU2_full_ED_TTN}
\end{figure}

\begin{figure}
	\centering
	\includegraphics[width=1\textwidth]{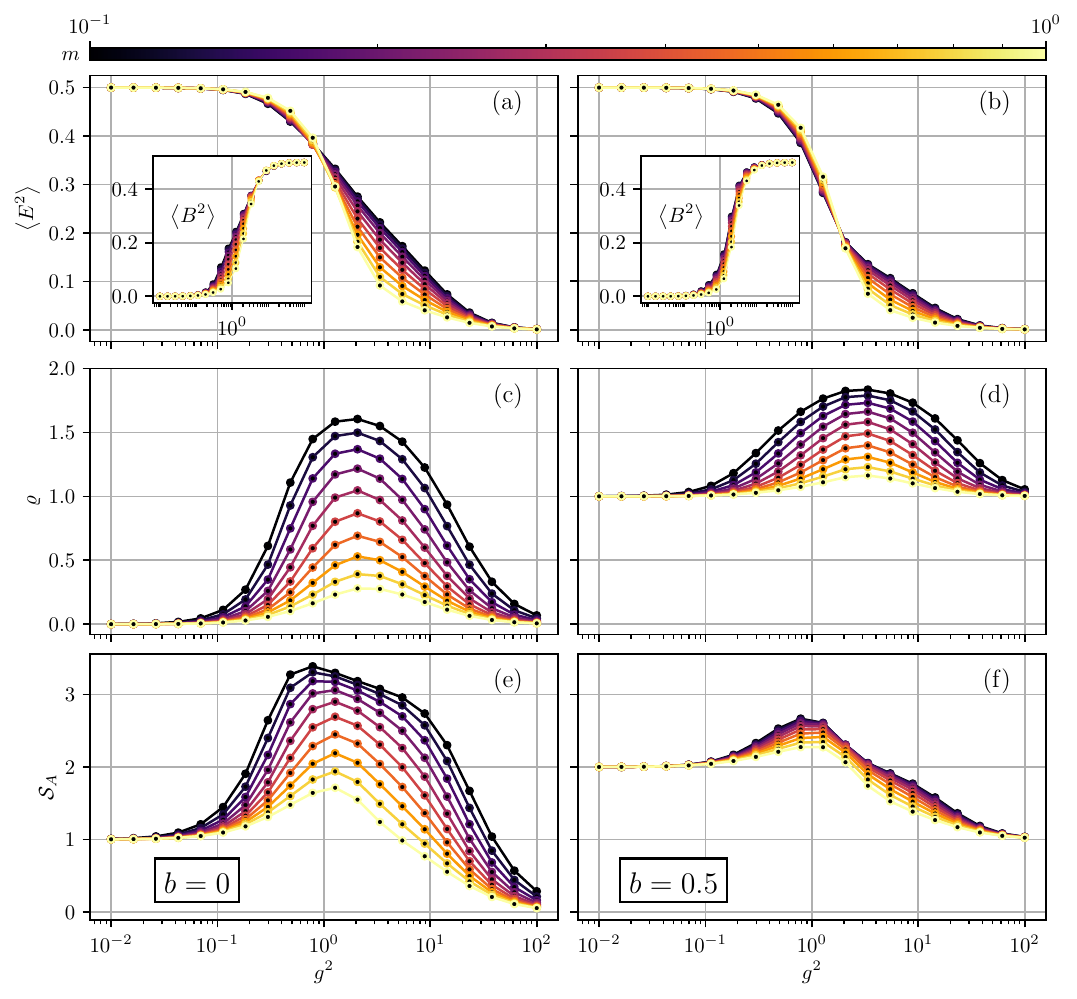}
	\caption{Simulations of the full SU(2) Hamiltonian in \cref{H_full_pt2} on a $2\times 2$ lattice with OBC in the $b=0$ (a)-(c)-(e) and $b=0.5$ (b)-(d)-(f) baryon number density sectors. 
    The plots display respectively: (a)-(b) the average electric and magnetic energy contributions $\avg{E^{2}}$ and $\avg{B^{2}}$ (inset) enlightening the transition between the magnetic and the electric phases discussed in \cref{sec_SU2_pure_theory}; (c)-(d) the average particle density $\density$ in \cref{eq_SU2_rho_density}, which appears peaked in the $\coupling$-transition; (e)-(f) the entanglement entropy $\entropy_{A}$ of half the system, with a peak in the $\coupling$-transition which is larger for smaller $\mass$ while disappearing for large ones.
    Figure from \cite{Cataldi2024Simulating2+1DSU2}.}
	\label{fig_ED_SU2_grid_discrete}
\end{figure}

Beyond energy gaps, other phase properties can be inferred when probing the observables in \cref{sec_SU2_local_observables}. 
As shown in \cref{fig_SU2_full_ED_TTN,fig_ED_SU2_grid_discrete}, by varying $m\in\qty[10^{-1},10^{0}]$, the magneto-electric transition, driven by $\coupling^{2}$, gets sharper for larger masses $\mass$, while remaining almost unalterated at finite baryon densities $b$. 

Correspondingly, for fixed $\mass$-values, the entanglement entropy of half the lattice is constant in the magnetic phase, peaked in the $\coupling$-transition, and tending to a default value ($0$ for $b=0$ and $1$ for $b=0.5$) in the electric phase. 
In the limit of large masses, this peak in the entropy progressively vanishes, and we recover the (\emph{crossover}/\emph{first-order}) transition observed in \cref{fig_SU2_pure_theory_simulations} for the pure theory. 

As for the particle density $\density$, it reveals an exciting behavior as the rescaled quark mass $\mass$ is lowered (see \cref{fig_ED_SU2_grid_discrete}). 
As long as $\mass$ is the largest energy scale of the model ($\mass \gg 1,\coupling^{2},\coupling^{-2}$) the emergent behavior is relatively trivial, as a system of gapped hardcore bosons. 

\begin{figure}
	\centering
    \includegraphics[width=1\textwidth]{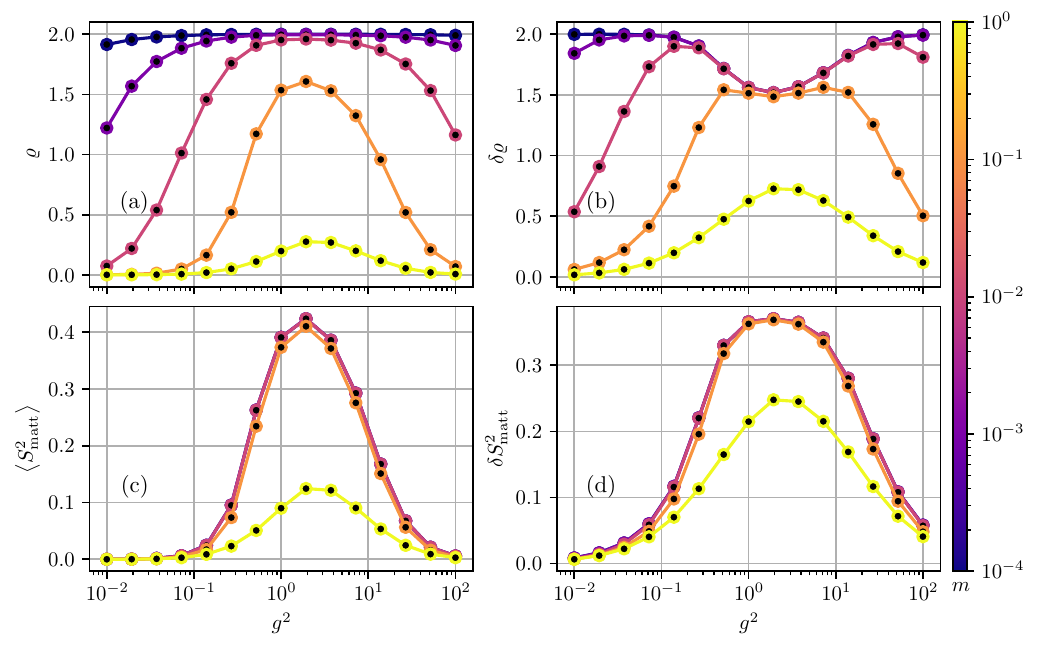}
	\caption{Simulations of a $2\times2$ lattice in PBC. The plots display respectively: (a) the average particle density $\density$ and (b) its quantum fluctuations $\delta \density$; (c) the matter color density $\avg{S^{2}_{\rm{matt}}}$ and (d) its quantum fluctuations $\delta S^{2}_{\rm{matt}}$. All the observables are studied as a function of the square coupling $\coupling^{2}$ for different mass values $m\in \qty[10^{-4},10^{1}]$.}
\label{fig_low_mass_behavior}
\end{figure}

More precisely, if $b \geq 0$ ($b \leq 0$) the antimatter (matter) sites are fully empty, while the matter (antimatter) sites host exactly $b$ quark-pair hardcore bosons, mass gapped and with almost flat-band dynamics. 
The particle density $\density$ confirms this interpretation, as it stays at its minimum possible value of $\density \simeq \density_{\text{min}}(b) = 2|b|$ and without fluctuations $\delta\density \simeq 0$ (see \cref{fig_low_mass_behavior}).

As shown in \cref{fig_local_matter_density}, this behavior drastically changes at low masses $\mass$, in relative proximity of the transition line $\coupling^{2} \sim 2(1)$. 
In fact, for $\mass$ lower than a critical value $\mass^{*}(\coupling)$, we see a sharp growth of the particle density $\density$ and its on-site fluctuations $\delta \density$, which become similar in magnitude (see \cref{fig_low_mass_behavior}). 
Even though we do not have access to long-range correlation functions at these limited system sizes, this observation is a strong hint of superfluidity of the phase, where we expect the quasi-particle excitations to be gapless (in the rescaled units).

To deeper investigate the nature of these quasi-particles we track the matter-color density $S_{\rm{matt}}$ defined in \cref{eq_SU2_matter_casimir_avg} (see \cref{fig_phase_diagram_scars,fig_low_mass_behavior}).
There is a very narrow region around the magneto-electric $\coupling$-transition where colored matter emerges (maybe a possible deconfined critical boundary) with fluctuations $\delta S_{\rm{matt}}$ of the same order of magnitude of the observable itself. 
Elsewhere, especially towards the continuum limit, the color density is vanishing $S_{\rm{matt}} = 0$. 
We must conclude that the gapless quasi-particles must be made by on-site pairs of quarks or anti-quarks. 
As such, we can regard the low-mass phase, $\mass < \mass^{*}(\coupling)$, as a \emph{gapless baryon liquid}.
Using a finite-size scaling technique shown in \cref{fig_local_matter_density}(b), we are able to characterize $\mass^{*}$ as a power-law function of $\coupling^{2}$, where a numerical regression yields
\begin{equation}
  \mass^{*}(\coupling^{2}) \simeq 0.267(4) \cdot \qty(\coupling^{2})^{1.03(2)},
  \label{eq_powerlaw_scaling}
\end{equation}
which is less than $2\sigma$ deviation from a linear scaling.
Now, if we assume that the linear scaling holds, then there must be a critical quark ratio $\alpha^{*}_c = 3.75(6)$ that determines the behavior when approaching the continuum limit (recall that $\alpha^{*}_c$ depends only on quark color-charge and bare mass, see \cref{sec_lattice_gaugefields}). 
Namely, for strong color charges $\alpha_c > \alpha^{*}_c$ the the baryon fluid at $\lspace\to0$ is gapless, while for weak charges $\alpha_c < \alpha^{*}_c$ the baryon fluid is gapped.
We recall that we are working with energy scales rescaled by $\lspace$, thus only quasiparticles that we identify as gapless at the continuum limit will survive as finite energy excitations in natural units.
% ============================================================================================
\section{Non-local/Topological properties}\label{res_topology}
Another relevant analysis that can be carried out in Yang-Mills theories is the characterization of topological properties at the critical point and the investigation of whether some form of topological order emerges within or without deconfined phases \cite{Svetitsky1982CriticalBehaviorFinitetemperature,Tagliacozzo2011EntanglementRenormalizationGauge}.
While the simplified model we considered does not support the existence of a deconfined phase in proximity to the continuum limit, it is still possible to characterize some topological properties by evaluating non-local order parameters.

In this section, we address the topological properties of the (2+1)D \emph{hardcore-gluon} SU(2) Hamiltonian in \cref{eq_2D_SU2_Ham_numerics}. 
In particular, we show that its pure theory displays a non-local $\mathbb{Z}_{2}\times\mathbb{Z}_{2}$ symmetry which exists only under periodic boundary conditions (PBC) and whose topological sectors close as approaching the $\coupling$-transition. 
Such a topological structure is completely \emph{independent} of the chosen truncation of the SU(2) gauge Hilbert space developed throughout \cref{sec_SU2_gaugetruncation}. 
The presence of dynamical matter removes such a symmetry from the full Hamiltonian in \cref{eq_2D_SU2_Ham_numerics}, but it can be recovered in the infinite mass limit.
% ============================================================================================
\subsection{Topological Invariants}
\begin{figure}
	\centering
	\includegraphics{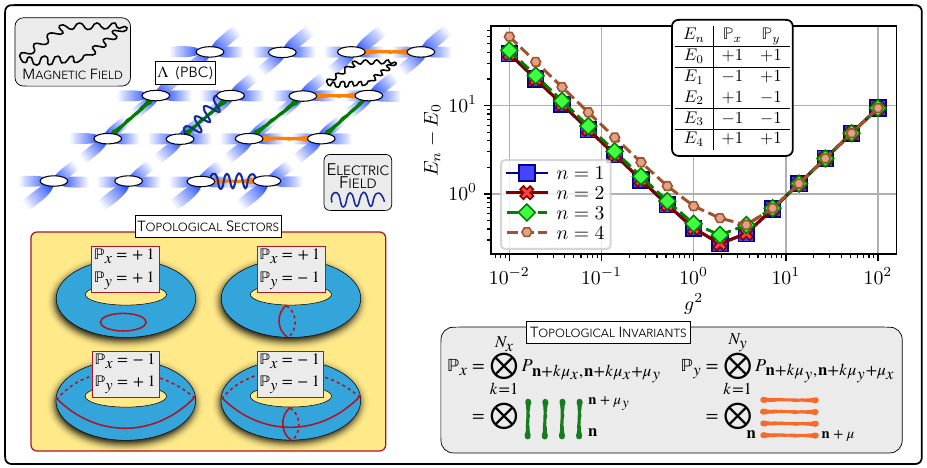}
	\caption{Pictorial representations of the topological invariants defined in \cref{eq_SU2_stringPy,eq_SU2_stringPx} on the lattice $\Lambda$ in PBC (i.e. a torus). 
    The topological sectors of \cref{topological_sectors} are sketched in the yellow panel: closed red curves on the blue torus $\Lambda$ correspond to SU(2) loop excitations.
    Energy gaps between the first excited levels and the ground state of \cref{eq_2D_SU2_Ham_numerics} in PBC, for a $2\times 2$ lattice. 
    Figure from \cite{Cataldi2024Simulating2+1DSU2}}
	\label{fig_topology_pure}
\end{figure}

In order to properly characterize a topological symmetry, we need to introduce some topological \emph{invariants}: string operators whose action commutes with pure gauge Hamiltonian terms of \cref{eq_2D_SU2_Ham_numerics}.
The right candidates involve the rishon parity operators $\parity_{\zeta}$ introduced in \cref{eq_SU2_rishon_parity}. 
Thanks to the link symmetry in \cref{SU2_linksymmetry}, we can extend such a definition to the whole link $(\genlink)$ and consider the corresponding parity operator $\parity_{\genlink}$. 
As aforementioned, it returns ($+1$) for \emph{integer} and ($-1$) for \emph{semi-integer} SU(2) representations. 
In the \emph{hardcore-gluon} approximation of \cref{sec_SU2_hardcoregluon}, such an operator reads:
\begin{equation}
    \parity_{\genlink} =\diag(+1,-1,-1,-1,-1) \quad \forall \vecsite,  \forall \latvec\,.
  \label{eq_SU2_link_Parity}
\end{equation}
Since the Casimir operator in \cref{eq_SU2_casimir_matrix} is diagonal in the link basis, it commutes with $\parity_{\genlink}$:
\begin{equation}
    \qty[\parity_{\genlink}, \hat{E}^{2}_{\genlink}]=0\quad \forall \vecsite, \forall \latvec\in \Lambda\,.
\end{equation}
Rather, $\parity_{\genlink}$ anti-commutes with the parallel transport $\hat{U}$ (and $\hat{U}^{\dagger}$), as its action on the link decreases (respectively increases) the SU(2) link-representation by $1/2$:
\begin{align}
    \qty{\parity_{\genlink}, \hat{U}^{(\dagger)}_{\genlink}}=0\quad\forall \vecsite, \forall \latvec\in \Lambda\,.
\end{align}
Then, we define the \emph{string} operator $\stringparity[y]$ as the consecutive action of the horizontal link parity operators along a vertical loop of $\Lambda$ in PBC (see orange links in \cref{fig_topology_pure}):
\begin{equation}
    \begin{split}
        \stringparity[y]&\equiv\bigotimes_{k=0}^{\abs{\Lambda_{y}}}\parity_{\vecsite+k\latvec[y],\latvec[x]}
        =\parity_{\genlink_{x}}\otimes \parity_{\vecsite+\latvec[y],\latvec[x]}\otimes \dots
    \end{split}
    \label{eq_SU2_stringPy}
\end{equation}
Correspondingly, $\stringparity[x]$ is the consecutive action of the vertical link parity operator along a horizontal loop in $\Lambda$ (see green links in \cref{fig_topology_pure}):
\begin{equation}
    \begin{split}
        \stringparity[x]&\equiv\bigotimes_{k=0}^{\abs{\Lambda_{x}}}\parity_{\vecsite+k\latvec[x],\latvec[y]}
        =\parity_{\genlink_{y}}\otimes \parity_{\vecsite+\latvec[x],\latvec[y]}\otimes \dots
    \end{split}
    \label{eq_SU2_stringPx}
\end{equation}
Within the SU(2) dressed site formalism developed in \cref{sec_SU2_model}, \cref{eq_SU2_stringPy,eq_SU2_stringPx} can be expressed just as chains of single-site operators along one of the two sides of the links. 
Indeed, as long as the SU(2) link-symmetry in \cref{SU2_linksymmetry} is satisfied, the information about every link-parity is present in both the attached neighboring sites.
% ============================================================================================
\subsection{Topological properties the pure theory}
It is clear that both the $\stringparity[x]$ and $\stringparity[y]$ operators remain unaffected by the action of the electric field along any of their steps, as their parity does not get flipped. 
Correspondingly, any plaquette term $\plaq$ of the magnetic interaction applied on the chain where $\stringparity[x]$ or $\stringparity[y]$ is evaluated flips the parity of two consecutive steps of the chain so that the overall sign is left unchanged. Namely:
\begin{align}
    \qty[\stringparity[\vecsite],\hat{E}^{2}_{\genlink}]=0
    =\qty[\stringparity[\vecsite],\plaq] 
    \qquad \forall \vecsite, \square \in \Lambda
\end{align}
Therefore, $\stringparity[x]$ and $\stringparity[y]$ are the generators of two $\mathbb{Z}_{2}$ symmetries of the pure theory in \cref{eq_2D_SU2_Ham_numerics}:
\begin{align}
    \qty[\stringparity[x],\ham_{\text{pure}}]&=0 &
    \qty[\stringparity[y],\ham_{\text{pure}}]&=0
\end{align}
and justify labeling them as \emph{topological invariants}.
The whole symmetry group is then $\mathbb{Z}_{2}\times\mathbb{Z}_{2}$, as we have $\qty[\stringparity[x],\stringparity[y]]=0$ $\forall x,y\in \Lambda$. 
Therefore, any physical state $\ket{\Psi}$ of the pure theory in \cref{eq_2D_SU2_Ham_numerics} lies in one of the sectors of $\stringparity[x]$ and $\stringparity[y]$ sketched in the yellow panel of \cref{fig_topology_pure}. 
The distinction between different symmetry sectors is in terms of the number of \emph{non-removable loop excitations} displayed by the state. 
With \emph{loop excitations}, we refer to closed magnetic strings (red circles in the blue torus of the yellow panel of \cref{fig_topology_pure}) displayed by the state on its topological geometry. 
In particular, \emph{non-removable} loops are the ones that cannot be removed through homotopies, i.e. without modifying the topology of the system.

Then, any state with an \emph{even} number of horizontal (vertical) non-removable loop excitations lies in the \emph{even} sector of the vertical $\stringparity[y]$ (horizontal $\stringparity[x]$) topological invariant. 
Correspondingly, any state with an \emph{odd} number of non-removable loop excitations lies in the \emph{odd} sector of the proper topological invariant.
Hence, $\forall k \in \qty{x,y}$:
\begin{align}
    \expval{\stringparity[k]}{\Psi}&=\lambda &\text{where}&& \lambda& \in
    \begin{array}{c|cccc}
        \stringparity[x]&+1 &+1&-1&-1 \\
        \hline
        \stringparity[y]&+1 &-1&+1&-1 \\
    \end{array}
    \label{topological_sectors}
\end{align}
By selecting each quantum number(s) for this symmetry group, we can evaluate inter-sector and intra-sector energy gaps, and verify the presence of quasi-degeneracies, signatures of a potential spontaneous breaking of the topological symmetry group, and thus of topological order.

To check numerically the previous statements, we would need to measure the topological invariants on the low energy states of the Hamiltonian in \cref{H_full_pt2}. 
As shown in \cref{fig_topology_pure}, the topological sectors of the first 4 lowest eigenstates of the pure theory in PBC belong to a different topological sector of \cref{topological_sectors}.
Moreover, the eigenstates are sorted in increasing energy according to the table in \cref{topological_sectors}. 
In particular, $E_{1}=E_{2}$ only in the case of isotropic geometries, as non-removable loop excitations along the two directions are equally expensive in energy. 
In the case of an-isotropic lattices, where $\Nsites_{x}\neq \Nsites_{y}$, non-removable loop excitations along different axes are shifted in energy.
Remarkably, when approaching the transition point from the large-$\coupling$ phase, inter-sector and intra-sector gaps reach a minimum, signaling a possible degeneracy lifted by finite-size effects. 
However, both gaps re-open while moving towards the small-$\coupling$ phase. 
This observation suggests topological order not to survive for $\coupling^{2} \ll 2$.
% ============================================================================================
\subsection{Dynamical matter removes the topological symmetry}
The addition of dynamical matter removes the topological invariants $\mathbb{P}_{x,y}$ from being symmetries of the model because of the hopping terms inverting the string parity.
Indeed, each of them includes a single parallel transport $\hat{U}$ that flips one link parity along the line where $\mathbb{P}_{x}$ or $\stringparity[y]$ are defined. 
In the large-$\mass$ limit, where the particle density vanishes, the Hamiltonian in \cref{eq_2D_SU2_Ham_numerics} reduces to the only pure theory, and the topological invariants become good quantum numbers again (at least in the ground-state).

\begin{figure}
	\centering
	\includegraphics[width=1\textwidth]{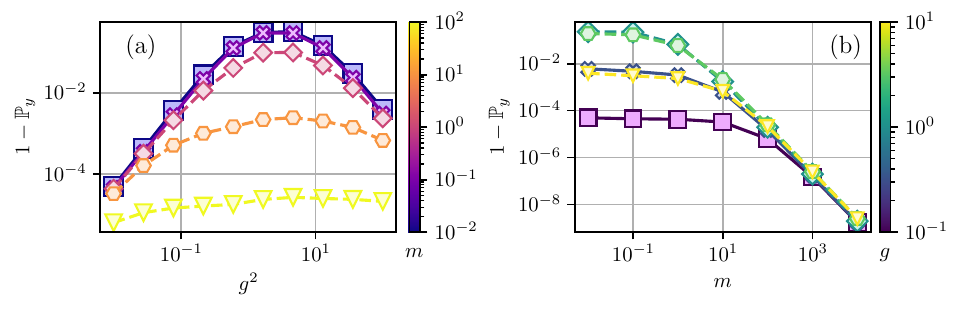}
	\caption{Distance between the ground-state $\mathbb{P}_{y}$-topological invariant in the full theory and the corresponding one of the pure theory for different $\mass$-values (a) and $\coupling$-couplings (b). Results from simulations in a $2\times 2$ lattice with PBC at $b=0$. Figure from \cite{Cataldi2024Simulating2+1DSU2}.}
	\label{fig_SU2_topology_full}
\end{figure}

Such a behavior is exactly reproduced in \cref{fig_SU2_topology_full}, where we look at the distance between the exact \emph{even} topological sector of $\stringparity[y]$ (+1) and the corresponding value measured on the ground state. 
That gap gets larger when approaching the $\coupling$-transition while vanishing far from the latter. 
Moreover, in the large-mass $\mass$ limit, we recover the full symmetry sector of the pure theory.
% ============================================================================================
\section{Large-coupling phase via perturbation theory}
\label{app_large_g_perturbation_theory}
In the large-$\coupling$ limit, the zero-density sector of the truncated SU(2) Hamiltonian can be studied via perturbation theory in $1/\coupling^{2}$. 
In this scenario, the full theory can be mapped to a good approximation into a spin-like Hamiltonian similar to an anisotropic Heisenberg model in two spatial dimensions \cite{Wang1991GroundStateTwodimensional,Wiese1994DeterminationLowEnergy}.
Let us start rewriting the Hamiltonian in \cref{H_full_pt2} as 
\begin{equation}
\begin{split}
    H&{\sim}\qty[\ham_{0}{+}\qty(\ham_{\rm{matt}}{+}\ham_{\text{x-hop}}{+}\ham_{\text{y-hop}}{+}\ham_{\text{plaq}})]
    =\sum_{\vecsite}\qty[\ham^{0}_{\vecsite}{+}\ham^{\rm{matt}}_{\vecsite}{+}\ham^{\text{x-hop}}_{\vecsite}{+}\ham^{\text{y-hop}}_{\vecsite}{+}\ham^{\text{plaq}}_{\vecsite,\square}]
\end{split}
\end{equation}
where $\ham_{0}$ is the single-site unperturbed Hamiltonian:
\begin{equation}
  \begin{aligned}
    \ham^{0}_{\vecsite}=\frac{\coupling^{2}}{4}\hat{\Gamma}_{\vecsite}\,,
  \end{aligned} 
  \label{H0_perturbative}
\end{equation}
while the perturbative terms read:
\begin{subequations}
    \begin{align}
        \ham^{\rm{matt}}_{\vecsite}&=\mass(-1)^{\site[x] + \site[y]} \hat{N}_{\vecsite,\text{tot}}\label{H_matter}\\
        \ham^{\text{x-hop}}_{\vecsite}&=\frac{1}{2}
            \qty[-i \hat{Q}^{\dagger}_{\vecsite,+\latvec[x]}\hat{Q}_{\vecsite+\latvec[x],-\latvec[x]}+ \hc]\label{xhop}\\
        \ham_{\vecsite}^{\text{y-hop}}&=\frac{1}{2}\qty[- (-1)^{\site[x]+\site[y]}\hat{Q}^{\dagger}_{\vecsite,+\latvec[y]} \hat{Q}_{\vecsite+\latvec[y],-\latvec[y]}+ \hc]\label{yhop}\\
        \ham^{\text{plaq}}_{\vecsite,\square}&=
            - \frac{1}{2\coupling^{2}}\qty(\begin{matrix}
            \hat{C}_{\ulcorner} &\hat{C}_{\urcorner}\\
            \hat{C}_{\llcorner} &\hat{C}_{\lrcorner}
            \end{matrix}
            )\,.
            \label{Hplaq}
        \end{align}    
\end{subequations}
where for simplicity, we replaced the two species of \emph{arrival operators} in \cref{eq_arrival_operators} and \emph{corner operator} in \cref{eq_corner_operators} with a unique version, $\hat{Q}$ and $\hat{C}$ respectively.

In the large-$\coupling$ limit, we expect the $0^{th}$ order ground-state $\ket{E_{0}}$ not to display gauge activity, as the electric interaction is energetically penalized. 
Then, the effective Hilbert state of the dressed sites reduces just to states with singlets in the matter fields. 
Namely, in terms of the sectors of the local charge density operator, we have only
\begin{align}
   \ket{0}&\equiv \five{0}{0}{0}{0}{0} &
   \text{and}&&
   \ket{2}&\equiv \five{\rla\gla}{0}{0}{0}{0}\,.
   \label{eq_SU2_singlesite_states}
\end{align}
Therefore, at the $0^{th}$-order, the single-site ground-state $\ket{E_{0}}$ can be expressed as a linear combination of \cref {eq_SU2_singlesite_states} with energy $E_{0}=0$:
\begin{equation}
\ket{E_{0}}=\alpha\ket{0}+\beta\ket{2} \qquad \text{with}\qquad \sqrt{\alpha^{2}+\beta^{2}}=1
\label{ground_state_order0}
\end{equation}
\begin{figure}
	\centering
	\includegraphics{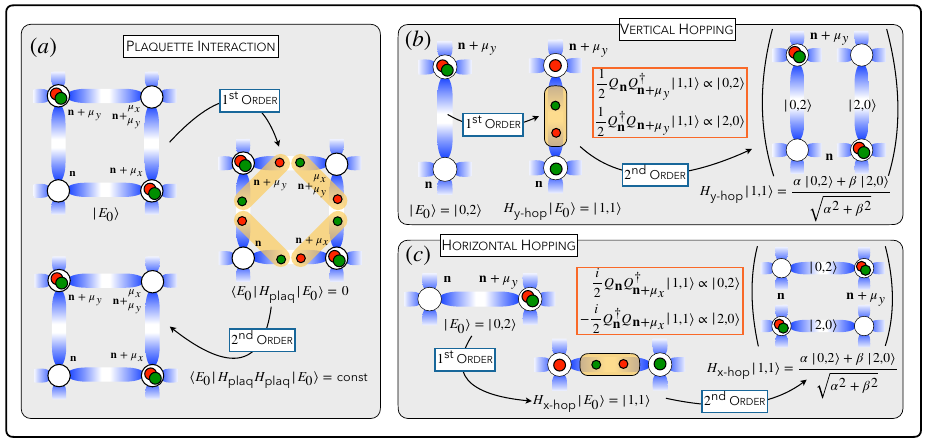}
	\caption{Graphical representation of the $1^{st}$ and $2^{nd}$ order perturbative effects of the magnetic (a) and the hopping terms (b)-(c) to the ground state of \cref{H0_perturbative}.}
	\label{fig_SU2_perturbation_theory}
\end{figure}
At the $1^{st}$ perturbative order, we have to separately consider the action of every single term in \cref{H_matter}-\cref{Hplaq}. 
As for the plaquette term in \cref{Hplaq}, we expect it to yield a vanishing contribution. Indeed, if we refer to $\ket{E_{0}}$ as a single-plaquette ground state, then we have:
\begin{equation}
\begin{split}
    \bra{E_{0}}\ham^{\text{plaq}}\ket{E_{0}}
    &=\bra{E_{0}}\qty(-\frac{1}{2\coupling^{2}})\qty(\frac{1}{\sqrt{2}})^{4}\ket{E_{1}}=-\frac{1}{8\coupling^{2}}\braket{E_{0}}{E_{1}}=0
\end{split}
\end{equation}
since the plaquette-state $\ket{E_{1}}$ is orthogonal to the ground state $\ket{E_{0}}$, as all its links are electrically active (see \cref{fig_SU2_perturbation_theory}).
The factor $1/\sqrt{2}$ is due to each single corner operator $\hat{C}$ acting on the corresponding empty corner of the plaquette state $\ket{E_{0}}$.  

As for the hopping terms, we focus on the effective Hilbert space of the neighboring sites $\vecsite$ and $\siteplus$:
\begin{equation}
\mathcal{H}^{\text{eff}}_{\genlink}=
\qty{\ket{0,0},\ket{0,2},\ket{2,0},\ket{2,2}}\qquad \forall \vecsite,\forall \latvec
\end{equation}
where we labeled the states $\ket{\vecsite, \vecsite+\latvec}$ in terms of the only two possible single-site states in \cref{eq_SU2_singlesite_states}. First of all, we notice that $\ket{0,0}$ and $\ket{2,2}$ are completely decoupled from the other two states, since $\forall \latvec$:
\begin{equation}
\hat{Q}^{\dagger}_{\vecsite,+\latvec}
\hat{Q}_{\siteplus,-\latvec}\ket{0,0}=
\hat{Q}^{\dagger}_{\vecsite,+\latvec}
\hat{Q}_{\siteplus,-\latvec}\ket{2,2}=0
\label{hop_decoupled_states}
\end{equation}
Then, the only relevant matrix elements of the effective (perturbed) hopping-Hamiltonian are:
\begin{equation}
    \begin{aligned}
\hat{Q}^{\dagger}_{\vecsite,+\latvec}
\hat{Q}_{\siteplus,-\latvec}\ket{0,2}=&(-1)^{2}\ket{1,1}\\
\hat{Q}_{\vecsite,+\latvec}
\hat{Q}^{\dagger}_{\siteplus,-\latvec}\ket{2,0}=&(-1)^{2}\ket{1,1}
    \end{aligned}
    \qquad \forall \vecsite,\forall \latvec
    \label{hop_coupled_states}
\end{equation}
where $\ket{1,1}$ is figured in \cref{fig_SU2_perturbation_theory}, while the $(-1)$ factor is due to the action of a single arrival operator $\hat{Q}_{\vecsite,\latvec}^{(\dagger)}$ defined in \cref{eq_arrival_operators} on the states in \cref{eq_SU2_singlesite_states}. 
Since $\braket{0,2}{1,1}={0}=\braket{2,0}{1,1}$, none of the hopping Hamiltonians \cref{xhop}-\eqref{yhop} do provide any $1^{st}$-order correction to $\ham_{0}$ in \cref{H0_perturbative}.

The only relevant $1^{st}$ order term is the one related to $\ham_{\rm{matt}}$, as it acts just on the matter fields without yielding any gauge activity. 
Moreover, it removes the ground-state degeneracy of \cref{ground_state_order0} by favoring a staggered configuration to the lattice, namely:
\begin{equation}
\ket{E_{1}(\site)}=\delta_{1,(-1)^{\site[x]+\site[y]}}\ket{0}+\delta_{-1,(-1)^{\site[x]+\site[y]}}\ket{2}
\label{ground_state_order1}
\end{equation}
where $\delta_{ij}$ is the Kronecker delta function.
However, for sufficiently small values of the mass $\mass$, the staggering effect is irrelevant, and the degeneracy of \cref{ground_state_order0} is restored. 
Therefore, in the small-$\mass$ limit, the first relevant perturbative order is the $2^{nd}$ one. 

As for the plaquette interaction, the $2^{nd}$ order does not remove the ground-state degeneracy, as it completely restores $\ket{E_{0}}$ providing just an energy shift.
Namely, the $2^{nd}$ order perturbative corrections to the single-site ground-state energy reads:
\begin{equation}
\begin{split}
E_{2}^{\text{plaq}}&=\frac{1}{4}\expval{\ham_{\text{plaq}}
[E_{0}-H_{0}]^{-1}\ham_{\text{plaq}}}{E_{0}}\\
&=\frac{1}{4}\qty(-\frac{1}{2\coupling^{2}})\bra{E_{0}} \ham_{\text{plaq}}[E_{0}-H_{0}]^{-1}\ket{E_{1}}\\
&=-\frac{1}{8\coupling^{2}}\bra{E_{0}} \ham_{\text{plaq}}\qty[-\frac{\coupling^{2}}{4}\sum_{\vecsite\in\square}\hat{\Gamma}_{\vecsite}]^{-1}\ket{E_{1}}\\
&=-\frac{1}{8\coupling^{2}}\qty(-\frac{\coupling^{2}}{4}\cdot\frac{3}{4})^{-1}\bra{E_{0}} 
\ham_{\text{plaq}}\ket{E_{1}}=-\frac{2}{3g^{4}}\qty(-\frac{1}{2\coupling^{2}})\braket{E_{0}}=\boxed{\frac{1}{3g^{6}}}
\end{split}
\end{equation}
where $[\obs]^{-1}$ is the Moore-Penrose inverse and the initial $1/4$ factor is put to get the single-site energy out of the one of a plaquette.

As for the hopping terms, because of \cref{hop_decoupled_states}-\eqref{hop_coupled_states}, $\forall k\in\qty{x,y}$, the only relevant terms are the diagonal ones
\begin{equation}
\begin{split}
    \frac{1}{2}&\bra{0,2}\ham^{\text{hop}}[E_{0}-\ham_{0}]^{-1}\ham^{\text{hop}}\ket{0,2}
    =\frac{1}{2}\bra{2,0}\ham^{\text{hop}}[E_{0}-\ham_{0}]^{-1}\ham^{\text{hop}}\ket{2,0}
\end{split}
\end{equation}
and the off-diagonal ones:
\begin{equation}
\begin{split}
    \frac{1}{2}&\bra{0,2}\ham^{\text{hop}}[E_{0}-\ham_{0}]^{-1}\ham^{\text{hop}}\ket{2,0}\
    =\frac{1}{2}\bra{2,0}\ham^{\text{hop}}[E_{0}-\ham_{0}]^{-1}\ham^{\text{hop}}\ket{0,2}.
\end{split}
\end{equation}
The factor $1/2$ is put to take into account just the single-site energy out of the corresponding two-site energy. As for the hopping along the $x$-axis, we have:
\begin{equation}
    \begin{split}
    &{\frac{1}{2}}\bra{0,2}\ham^{\text{x-hop}}
    [E_{0}-\ham_{0}]^{-1}\ham^{\text{x-hop}}\ket{0,2}\\
    &=\frac{1}{2}\bra{0,2}h^{\text{x-hop}}[E_{0}-\ham_{0}]^{-1}\qty(-\frac{i}{2})\ket{1,1}\\
    &=\qty(\frac{-i}{4})\bra{0,2}\ham^{\text{x-hop}}\qty[-\frac{1\coupling^{2}}{4}\qty(\hat{\Gamma}_{\vecsite}{+}\hat{\Gamma}_{\vecsite+\latvec[x]})]^{-1}\ket{1,1}\\
    &=\qty(\frac{-i}{4})\bra{0,2}\ham^{\text{x-hop}}\qty(-\frac{8}{3\coupling^{2}})\ket{1,1}\\
    &=\frac{2i}{3\coupling^{2}}\bra{0,2}\qty(\frac{i}{2})\ket{0,2}=\boxed{-\frac{1}{3\coupling^{2}}}.
    \end{split}
\end{equation}
Analogously proceeding, we have:
\begin{equation}
    \begin{split}
    &\frac{1}{2}\bra{2,0}\ham^{\text{x-hop}}[E_{0}-\ham_{0}]^{-1}\ham^{\text{x-hop}}\ket{0,2}=\boxed{\frac{1}{3\coupling^{2}}}.
    \end{split}
\end{equation}
Then, the $2^{nd}$-order perturbative $x$-hopping term reads:
\begin{equation}
\begin{split}
    \ham_{\text{x-hop}}^{\text{eff}}&=-\frac{1}{3\coupling^{2}}
    \begin{pmatrix}
    0&0&0&0\\
    0&+1&-1&0\\
    0&-1&+1&0\\
    0&0&0&0\\
    \end{pmatrix}
    =-\frac{1}{6\coupling^{2}}\qty[\Sx_{\vecsite}\Sx_{\vecsite+\latvec[y]}+\Sy_{\vecsite}\Sy_{\vecsite+\latvec[y]}-\Sz_{\vecsite}\Sz_{\vecsite+\latvec[y]}].
\end{split}
\end{equation}
As for the $y$-hopping Hamiltonian, one can prove that:
\begin{equation}
\begin{split}
    \ham_{\text{y-hop}}^{\text{eff}}&=\frac{1}{3\coupling^{2}}
    \begin{pmatrix}
    0&0&0&0\\
    0&+1&+1&0\\
    0&+1&+1&0\\
    0&0&0&0\\
    \end{pmatrix}
    =-\frac{1}{6\coupling^{2}}\qty[\Sx_{\vecsite}\Sx_{\vecsite+\latvec[y]}+\Sy_{\vecsite}\Sy_{\vecsite+\latvec[y]}+\Sz_{\vecsite}\Sz_{\vecsite+\latvec[y]}].
\end{split}
\end{equation}
Summarizing, in the large-$\coupling$ limit, the Hamiltonian in \cref{H_full_pt2} can be approximated as:
\begin{equation}
  \begin{aligned}
    \ham^{\text{eff}}\sim -\frac{1}{6\coupling^{2}}\sum_{\vecsite}\qty[
    \qty[\Sx_{\vecsite}\Sx_{\vecsite+\latvec[x]}{+}\Sy_{\vecsite}\Sy_{\vecsite+\latvec[x]} -\Sz_{\vecsite}\Sz_{\vecsite+\latvec[x]}]
    + \qty[\Sx_{\vecsite}\Sx_{\vecsite+\latvec[y]}{+}\Sy_{\vecsite}\Sy_{\vecsite+\latvec[y]}{+}\Sz_{\vecsite}\Sz_{\vecsite+\latvec[y]}]]
    \end{aligned}
\end{equation}
which looks similar to a 2D quantum Heisenberg Hamiltonian apart from the staggering factor in the kinetic term $\Sz\Sz$ \cite{Wang1991GroundStateTwodimensional,Wiese1994DeterminationLowEnergy}.
% ============================================================================================
\section{Phase Diagram}
\label{sec_SU2_phase_diagram}
\begin{figure}
	\centering
	\includegraphics[width=1\textwidth]{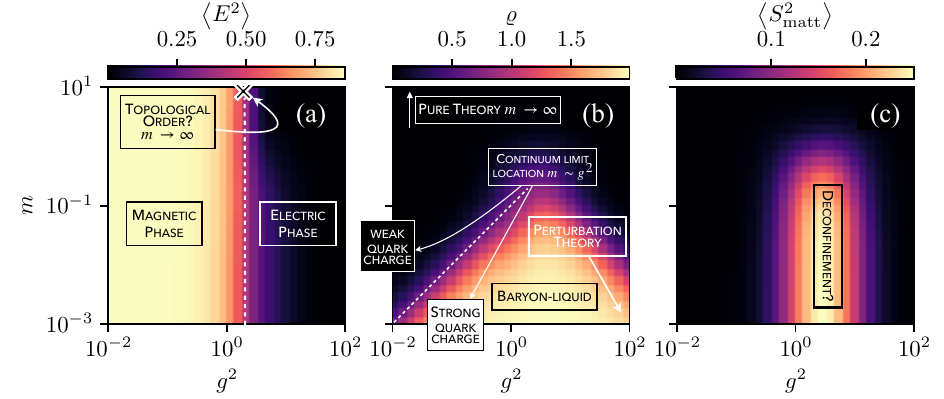}
 \caption{Phase diagram $(\coupling^{2},\mass)$ of the full SU(2) Hamiltonian in \cref{eq_2D_SU2_Ham_numerics} in the sector with zero baryon number density from (a) the average electric energy density in \cref{eq_SU2_gauge_observables}, (b) the average particle density in \cref{eq_SU2_rho_density}, and (c) the matter color density defined in \cref{eq_SU2_matter_casimir_avg}. 
 Phases are marked according to all the discussions throughout the entire chapter.}
\label{fig_phase_diagram_scars}
\end{figure}

By collecting and summarizing all the previous observations, we can outline in \cref{fig_phase_diagram_scars} the full phase diagram of the (2+1)D SU(2) Yang-Mills Hamiltonian in \cref{eq_2D_SU2_Ham_numerics} around zero baryon density $b=0$ (where the baryon mass gap opens).

We observed that the presence of fermionic degrees of freedom affects only marginally the behavior of the gauge degrees of freedom of \cref{eq_SU2_gauge_observables}, albeit the magneto-electric transition becomes smoother at lower $\mass$ values.

For sufficiently large masses, $\mass{>}\mass^{*}(\coupling)$, matter fields play a minor role (trivial phase). 
In the infinite mass limit, the Hamiltonian recovers the topological properties of the pure theory of \cref{res_topology}, but no spontaneous topological order survives outside the magneto-electric transition $\coupling^{2}{\sim} 2(1)$.

Conversely, for small masses, $\mass{<}\mass^{*}(\coupling)$ \cite{Engelhardt2000DeconfinementSUYangMills}, we observe an emergent color-density of the matter fields, only in the proximity of the magneto-electric transition. 
Such observation is compatible with the existence of a deconfined critical phase (i.e. where color particles may appear) in the region where electric and magnetic fields are maximally frustrated.
Elsewhere, the system shows the free emergence of a liquid of colorless baryons and anti-baryons.
The collective behavior towards the continuum limit is particularly intriguing, as it can exhibit both trivial or baryon superfluid phase depending on the quark ratio $\alpha_c$.

% ============================================================================================
\section{Summary}
\label{sec_conclusions}
In this chapter, we reviewed the numerical results obtained from TN simulations of a non-Abelian SU(2) Yang-Mills LGT in two spatial dimensions, with dynamical matter and hardcore gluons \cite{Cataldi2024Simulating2+1DSU2}.
Our focus on this physical setting is motivated by the wide use of the latter as a paradigmatic model to address fundamental properties that could be relevant for high-dimensional QCD.
For instance, standard MC simulations of such an LGT have highlighted intriguing effects, such as the \emph{dimensional reduction} \cite{Ambjorn1984StochasticConfinementDimensional,Ambjorn1984StochasticConfinementDimensional-1}, the compatibility with \emph{string theory} \cite{Ambjorn1984ObservationStringThreedimensional,Ambjorn1984ThreedimensionalLatticeGauge}, and the possibility of accessing features of the \emph{continuum theory} already at small correlation lengths \cite{Berg1981SULatticeGauge}.

In summary, we have investigated in detail both the zero and finite baryon number density regimes, where MC methods are severely limited due to the sign problem.
Our results confirm TN methods as a reliable approach to addressing the non-perturbative phenomena of LGTs, capable of accessing strong coupling regimes as well as finite baryon number densities.

Despite the truncation of the gauge field, by exploiting numerical estimations of various observables, we inferred quite a few qualitative and quantitative observations concerning the zero-temperature phase diagram of the model.
First, when approaching the continuum limit ($\lspace\to0$ at fixed $\mass[0]$, $\alpha_c$) SU(2) baryons and anti-baryons become the actual quasiparticles of the theory. 
Interestingly, if their color charge is strong enough, \idest{} for a sufficiently large quark ratio $\alpha_c{\geq}\alpha_c^{*}(\mass[0])$, baryons seem to be able to condense into a superfluid phase.

In the parameter regime at $\coupling^{2}{\sim}2(1)$, where the electric term and the magnetic term are maximally frustrated, and electric and magnetic field fluctuations are proportional, we witnessed more exotic physics: at low quark masses, the system manifests colorful matter sites, possibly indicating a quark-deconfined regime, such as a quark-gluon plasma. 
At high quark masses, the system encounters a degeneracy between topological sectors (string symmetries in periodic boundary conditions), possibly signaling the emergence of a topological order reminiscent of the Toric code.

From a theoretical perspective, the simulated Hamiltonian describes the interaction between flavorless 2-color fermionic matter and gauge fields in the \emph{hardcore-gluon} approximantion of \cref{sec_SU2_hardcoregluon}.
Considering larger representations in the gauge Hilbert space (following the prescription detailed in \cref{sec_SU2_model}) would be a natural extension of these simulations and an improved approximation of the continuous gauge field theory.
A larger truncation becomes substantial in the small coupling limit, where the Hamiltonian is dominated by the magnetic interaction, which is non-local and non-diagonal in the representation basis developed in \cref{sec_SU2_gaugetruncation}. 
This makes the model significantly entangled and challenging to be numerically attacked via TNs.

As an outlook of this study, we plan to develop an analogous formalism in the magnetic basis, where plaquette terms are diagonal \cite{Kaplan2020GaussLawDuality,Haase2021ResourceEfficientApproach,Paulson2021Simulating2DEffects}.
This change of basis should ease TN simulations, which in our case are limited to finite system sizes, but anyway larger than the state-of-the-art of quantum-inspired or quantum simulations of non-Abelian LGTs \cite{Atas2021SUHadronsQuantum,Ciavarella2021TrailheadQuantumSimulation,Ciavarella2023QuantumSimulationLattice}. 
Accessing larger system sizes would be a substantial advantage, as it would enable the characterization of correlation functions not distorted by finite-size effects. 
Correspondingly, larger sizes would allow for studying magnetic effects at small coupling, as in MC simulations \cite{Kiskis1983NumericalStudyFlux,Kiskis1984IllustratedStudyFlux,Hietanen2006PlaquetteExpectationValue}.

To overcome these limitations (finite gauge representation and finite system sizes), further developments of the numerical simulation architecture are also required: on the hardware side, the possibility of running the computation on a (pre)exascale HPC environment, while on the software side the development of new and improved TN-based algorithms. 
The latter will be achieved by exploiting the augmented TTN ansatz, which drastically enhances the capability of representing area law-states in high dimensions \cite{Felser2021EfficientTensorNetwork}. 
These steps will be fundamental for the long-term goal of applying TN methods to large-scale (3+1) lattice QCD and ultimately address open, secular research problems, such as confinement and asymptotic freedom.

From an experimental viewpoint, similarly to the one-dimensional case \cite{Calajo2024DigitalQuantumSimulation}, the dressed-site formalism developed for the (2+1)D SU(2) Hamiltonian could be encoded on quantum hardware. 
In this perspective, the results presented in this chapter and in \cite{Cataldi2024Simulating2+1DSU2} represent essential benchmarks for validating current and future experimental implementations \cite{Meurice2022TensorNetworksHigh,DiMeglio2023QuantumComputingHighEnergy,Zhang2023ManybodyHilbertSpace,Su2023ObservationManybodyScarring}. 
% ============================================================================================

%% file: chapters/scars.tex
\chapter{Non-Ergodic dynamics of LGTs}
\label{chap_scars}
The nature of equilibration in a isolated interacting quantum many-body (QMB) systems is a fundamental question in physics \cite{Reimann2008FoundationStatisticalMechanics}. 
Whereas open quantum systems are expected to thermalize due to exchanging energy with a bath \cite{Reichental2018ThermalizationOpenQuantum}, the conditions under which an isolated QMB system thermalizes are less intuitive.
While the classical counterpart can be easily figured out by using the Liouville theorem and applying the statistical mechanics ensemble on the phase space \cite{Deutsch2018EigenstateThermalizationHypothesis}, thermalization in quantum systems is stipulated by the Eigenstate Thermalization Hypothesis (ETH) \cite{Deutsch1991QuantumStatisticalMechanics,Srednicki1994ChaosQuantumThermalization,DAlessio2016QuantumChaosEigenstate,Deutsch2018EigenstateThermalizationHypothesis}.
According to the ETH, for an \textit{ergodic} Hamiltonian, the expectation values of observables in individual energy eigenstates match thermal ensemble averages, and through the process of \emph{dephasing}, the system’s long-time behavior averages out to these thermal values, leading to thermalization.

So far, ETH has been widely confirmed and numerically verified in several contexts, from strongly correlated models of condensed matter and ultracold atom gases of (softcore and hardcore) bosons \cite{Rigol2008ThermalizationItsMechanism,Vidmar2015DynamicalQuasicondensationHardCore,Biroli2010EffectRareFluctuations,Beugeling2014FinitesizeScalingEigenstate,Beugeling2015OffdiagonalMatrixElements}, spins \cite{Rigol2009BreakdownThermalizationFinite,Santos2010LocalizationEffectsSymmetries,Khatami2013FluctuationDissipationTheoremIsolated,Steinigeweg2014PushingLimitsEigenstate}, and  fermions \cite{Rigol2009QuantumQuenchesThermalization,Khatami2012QuantumQuenchesDisordered,Genway2012ThermalizationLocalObservables,Neuenhahn2012ThermalizationInteractingFermions}.
However, in recent years, a new paradigm of weak ergodicity breaking has emerged violating the ETH. 
In certain ergodic Hamiltonians, there is a polynomial (in system size) number of special non-thermal \textit{scar} eigenstates that are roughly equally spaced in energy over the whole spectrum \cite{Turner2018WeakErgodicityBreaking,Moudgalya2018ExactExcitedStates,Schecter2019WeakErgodicityBreaking} and exhibit anomalously low bipartite entanglement entropy \cite{Lin2019ExactQuantumManyBody}. 
Upon initializing the system in an initial state with a high overlap with these scar eigenstates, the subsequent quench dynamics gives rise to \textit{quantum many-body scarring} (QMBS), which manifests as prominent oscillations in local observables and persistent revivals in the (local and global) Loschmidt fidelity that last longer than all relevant timescales, thereby circumventing expected thermalization \cite{Bernien2017ProbingManybodyDynamics,Moudgalya2018ExactExcitedStates,Zhao2020QuantumManyBodyScars,Jepsen2022LonglivedPhantomHelix,Serbyn2021QuantumManybodyScars,Moudgalya2022QuantumManybodyScars,Chandran2023QuantumManyBodyScars}. 
QMBS has received a lot of attention since its initial discovery in \cite{Bernien2017ProbingManybodyDynamics} and has been the subject of several quantum simulation experiments \cite{Bluvstein2021ControllingQuantumManybody,Bluvstein2022QuantumProcessorBased,Su2023ObservationManybodyScarring,Zhang2023ObservationMicroscopicConfinement,Dong2023DisordertunableEntanglementInfinite}.

After the discovery of QMBS, it has been shown that the effective model quantum-simulated in \cite{Bernien2017ProbingManybodyDynamics} is a spin-$1/2$ $\mathrm{U(1)}$ lattice gauge theory (LGT), where the electric field is represented by a spin-$1/2$ $z$-operator \cite{Surace2020LatticeGaugeTheories}. 
In detail, the robustness of QMBS has been shown to depend on the stability of an underlying gauge symmetry of this model \cite{Halimeh2023ColdatomQuantumSimulators}. 
Later, QMBS have been further observed in spin-$S$ $\mathrm{U}(1)$ LGTs \cite{Desaules2023ProminentQuantumManybody,Desaules2023WeakErgodicityBreaking} and other Abelian LGTs \cite{Iadecola2020QuantumManybodyScar,Aramthottil2022ScarStatesDeconfined,Desaules2024MassAssistedLocalDeconfinement}, even in two spatial dimensions \cite{Banerjee2021QuantumScarsZero,Biswas2022ScarsProtectedZero,Sau2024SublatticeScarsTwodimensional,Osborne2024QuantumManyBodyScarring,Budde2024QuantumManyBodyScars}. 
Nonetheless, none of these studies tried to explore QMBS in the non-Abelian scenario. 
Recently, Ref. \cite{Ebner2024EntanglementEntropyBoldsymbol} has shown the emergence of scarring dynamics on a pure (without dynamical matter) (2+1)D SU(2) LGT. 
In detail, looking at the bipartite entanglement entropy, \cite{Ebner2024EntanglementEntropyBoldsymbol} identified a few scar-eigenstate candidates in a certain parameter regime. 
However, no experimentally feasible scarred initial states have been proposed, and the type of exhibited scarring vanishes beyond the crudest truncation of the electric field basis.

In this chapter, we bridge this gap discussing the emergence of QMBS in non-Abelian LGTs with dynamical matter and connect the presence of the latter at the origin of such an exotic dynamics.
We start from \cref{sec_classical_thermalization}, with a pedagogical introduction to classical thermalization. 
Then, in \cref{sec_quantum_thermalization}, we switch to the corresponding quantum version and expose in \cref{sec_ETH} the Eigenstate Thermalization Hypothesis (ETH) for isolated QMB systems in terms of their spectral and dynamical properties.
After this introduction, we devote \cref{sec_QMB_scars} to the first experimental and numerical observations of \emph{scarring} dynamics in Rydberg atom chains.
In \cref{sec_scars_AbelianLGT}, we highlight the strong connection QMB scars and fundamental mechanisms appearing in Abelian LGTs.
The original part is in \cref{sec_scars_nonAbelianLGT}, where we extend such a relation to non-Abelian LGTs by simulating a one-dimensional truncated SU(2) Yang-Mills LGT and observing all the spectral and dynamical properties of a scarring dynamics \cite{Cataldi*2025QuantumManybodyScarring}.
% =========================================================================================
\section{Thermalization in Classical Systems}
\label{sec_classical_thermalization}
Before introducing the concept of thermalization in quantum physics, it is useful to first consider its classical counterpart and gain a physical intuition of the mechanism. 
Thermalization in isolated classical systems refers to the process by which a system evolves over time to reach a state where its macroscopic properties become consistent with those predicted by statistical mechanics. 
This concept is closely tied to the behavior of the system in its phase space, particularly when the system’s total energy is conserved.

Let us consider a classical system of $\Nsites$ particles without any exchange of energy or matter with the environment. 
The corresponding phase space $\Sigma$ of the system is defined by a set of $2\Nsites$ variables $\Gamma=\qty{\vecpos,\vecmom}=\qty{\pos[1],\dots,\pos[N],\mom[1],\dots, \mom[N]}$ identifying position and momentum of each particle on a point of the phase space $\Gamma$.
Since the system is isolated, $\Gamma(t)$ moves on a hypersurface of the whole phase space $\Sigma_{E}$ characterized by constant energy $E$.

Then, in principle, the dynamics of the system requires to follow the evolution of $\Gamma(t)$ at each time step along that hypersurface. 
Namely, for any observable $\obs$, we need to evaluate $\obs(\Gamma(t))$ $\forall t$.
Correspondingly, the long-time average of such an observable is given by:
\begin{equation}
    \bar{\obs}=\lim_{T\to\infty}\frac{1}{T}\int^{T}_{0}\obs(\Gamma(t))dt\,.
    \label{eq_classical_longtimeavg}
\end{equation}
If the system is \emph{ergodic}, we expect that, given enough time, the trajectory $\Gamma(t)$ will pass arbitrarily close to every point of the hypersurfase $\Sigma_{E}$ at energy $E$. 
More in detail, thanks to the Liouville theorem, we then assume that the time spent in any small region of $\Sigma_{E}$ is proportional to the phase space volume of that region.
This implies that the time average of a function over a long period is statistically equivalent to the predictions made by statistical mechanics, where all accessible microstates at a given energy are equally probable.

In statistical mechanics, the microcanonical ensemble (ME) describes a system with fixed total energy $E$ and number of particles $\Nsites$. 
Theen, the ensemble average of an observable $\obs$ corresponds to the uniform average over all possible microstates with energy $E$, because all accessible microstates are equally probable:
\begin{equation}
    \avg{\obs}_{\rm{ME}}=\frac{\int_{\Sigma_{E}}\obs(\Gamma)d\Gamma}{\int_{\Sigma_{E}}d\Gamma}\,.
    \label{eq_classical_MEavg}
\end{equation}
Let us then assume the system to be in a non-equilibrium initial state where its macroscopic properties do not match the predictions of statistical mechanics.
According to classical ergodicity, as the system evolves, the interactions between particles lead to a redistribution of energy and momentum, causing the system to explore the whole hypersurface $\Sigma_{E}$ of the phase space.
Over time, the system “forgets” its initial conditions (except for conserved quantities like energy) and its macroscopic observables settle down to values that match the microcanonical ensemble predictions \cite{Deutsch2018EigenstateThermalizationHypothesis}.
Once thermalization is complete, the long-time average in \cref{eq_classical_longtimeavg} matches the microcanonical ensemble average in \cref{eq_classical_MEavg} and $\bar{\obs}=\avg{\obs}_{\rm{ME}}$. 
The system is then said to be in \emph{thermal equilibrium}.

While a robust proof of ergodicity is typically challenging, a few examples have been demonstrated to display thermalization, such as gases in hard spheres in some volumes, \aka{} Sinai billiards \cite{Sinai1970DynamicalSystemsElastic}, and the Bunimovich Stadium \cite{Bunimovich1979ErgodicPropertiesNowhere}, describing a free particle inside a stadium with hard walls that are circular on the sides and straight in the middle.

Of course, for an \emph{integrable} system, \idest{} with an extensive number of conserved quantities $\sim \Nsites$, its motion is constrained in a very small subportion of the hypersurface $\Sigma_{E}$ which strongly depends on the initial conditions, and the whole dynamics reveals (quasi-)periodic. 
In such cases, ergodicity does not hold, as the statistical mechanics average is not applicable.
% =========================================================================================
\section{Thermalization in Quantum Systems}
\label{sec_quantum_thermalization}
While for classical systems the connection between ergodicity of isolated systems and statistical mechanics is quite straightforward, a more accurate analysis is needed to understand how isolated quantum systems reaches thermal equilibrium, as in these case position $\pos$ and momentum $\mom$ do not commute and we cannot rely on the phase space.

Let us consider a finite non-integrable\footnote{If the system has a set of symmetries, we assume that all of them are resolved.} $\Nsites$-body quantum system with $\Nsites$ sufficiently large and a non-degenerate Hamiltonian spectrum:
\begin{equation}
    \ham=\sum_{\alpha=1}^{\Nsites}E_{\alpha}\ketbra{\Phi_{\alpha}}\,,
\end{equation}
where $\qty{E_{\alpha}}_{\alpha=1\dots N}$ and $\qty{\ket{\Phi_{\alpha}}}_{\alpha=1\dots \Nsites}$ are the corresponding eigenvalues and eigenvectors.
Let us then assume to quench the system with an initial pure\footnote{Generalizations of ETH to mixed states are discussed in \cite{Deutsch2018EigenstateThermalizationHypothesis}.} state $\ket{\Psi(0)}=\sum_{\alpha}C_{\alpha}\ket{\Phi_{\alpha}}$, whose energy is well defined, extensive in $N$\footnote{This corresponds to assume that each \dof{} has a finite amount of energy, and the system energy is far above the ground state energy.} and robut under (subextensive) fluctuations: 
\begin{align}
    E&=\avg{H}=\expval{H}{\Psi(0)}\sim N&
    \Delta E &=\sqrt{\avg{H^{2}}-\avg{H}^{2}} \sim N^{\nu} \qquad \rm{with}\; \nu<0.
    \label{eq_QMB_energy}
\end{align}
In the Schrödinger picture, the system can be described in terms of the state
\begin{align}
    \ket{\Psi(t)}&=\sum_{\alpha}C_{\alpha}e^{-itE_{\alpha}/\hbar}\ket{\Phi_{\alpha}}&
    \text{where}&&
    \sum_{\alpha}\abs{C_{\alpha}}^{2}&=1\,.
\end{align}
Correspondingly, any generic (local) observable $\obs$ can be instantaneously measured as follows
\begin{equation}
    \begin{split}
        \obs(t)&= \expval{\obs}{\Psi(t)}=\sum_{\alpha,\beta}C_{\alpha}^{*}C_{\beta}e^{it(E_{\alpha}-E_{\beta})}\obs_{\alpha\beta}\\
        &=\sum_{\alpha}\abs{C_{\alpha}}^{2}\obs_{\alpha\alpha} + \sum_{\alpha, \alpha\neq \beta}C_{\alpha}^{*}C_{\beta}e^{it(E_{\alpha}-E_{\beta})}\obs_{\alpha\beta}
    \end{split}
\end{equation}
where $\obs_{\alpha\beta}=\expval{\Phi_{\alpha}|\obs|\Phi_{\beta}}$.
In analogy to the classical counterpart, we can formally define the \emph{long time average} resulting from the unitary dynamics of the observable, which is well described by the average value predicted by the \emph{diagonal} ensemble \cite{Rigol2008ThermalizationItsMechanism}:
\begin{equation}
    \begin{split}
        \bar{\obs}
        &\equiv\lim_{T\to\infty} \frac{1}{T}\int_0^T dt \expval{\obs}{\Psi(t)}\\
        &=\lim_{T\to\infty} \frac{1}{T}\int_0^T dt
        \sum_{\alpha,\beta}C^{*}_{\alpha}C_{\beta}\expval{\Phi_{\alpha}|e^{it\ham}\obs e^{-it\ham}|\Phi_{\beta}}\\
        &=\sum_{\alpha}\abs{C_{\alpha}}^2\obs_{\alpha\alpha}+ \sum_{\alpha,\alpha\neq\beta}C^{*}_{\alpha}C_{\beta}\obs_{\alpha\beta}\cancel{\lim_{T\to\infty}\frac{1}{T}\int_0^T e^{i(E_{\alpha}-E_{\beta})t} dt}\\
        &=\sum_{\alpha}\abs{C_{\alpha}}^2\obs_{\alpha\alpha}\equiv\avg{\obs}_{\rm DE},
    \end{split}
    \label{eq_diagonal_avg}
\end{equation}
where the canceled term is exactly due to the dephasing, while $\avg{\obs}_{\rm DE}$ is the average from the diagonal ensemble.
Therefore, after enough time, we expect the observable $\obs(t)$ to approach long-time average $\avg{\obs}_{\rm DE}$ which however depends on the initial quench of the system through the probabilities $\abs{C_{\alpha}}^{2}$, as the latters are conserved in time.

This clearly differs from the expected \emph{thermalization}, according to which the system relaxes to states whose macroscopic quantities are stationary, universal with respect to different initial conditions, and predictable using statistical mechanics \cite{Rigol2008ThermalizationItsMechanism}.
Namely, thermalization in quantum system requires the long-time average in \cref{eq_diagonal_avg} to match the prediction of the \emph{microcanonical ensemble} (ME), which relies on a superposition within the energy shell $[E-\Delta E , E+\Delta E ]$ around the system energy $E$ in \cref{eq_QMB_energy} and containing $N_{E,\Delta E}$ eigenstates:
\begin{equation}
    \avg{\obs}_{\rm ME}=\frac{1}{N_{E,\Delta E}}\sum_{\substack{\alpha;\\|E_{\alpha}-E|<\Delta E}}O_{\alpha\alpha}\,.
    \label{eq_microcanonical_avg}
\end{equation}
% =========================================================================================
\subsection{The Eigenstate Thermalization Hypothesis (ETH)}
\label{sec_ETH}
In order to prove the equivalence between the two ensemble averages (diagonal and microcanonical) and show ergodicity as an emerging properity out of the microscopical properties of realistic QMB systems, we rely on the Eigenstate Thermalization Hypothesis (ETH), originally proposed by Deutsch \cite{Deutsch1991QuantumStatisticalMechanics} and Srednicki \cite{Srednicki1994ChaosQuantumThermalization} and later widely confirmed \cite{Rigol2008ThermalizationItsMechanism}.
For a detailed review, we recommend the reader to see \cite{DAlessio2016QuantumChaosEigenstate,Deutsch2018EigenstateThermalizationHypothesis}.

According to the ETH, the matrix elements of the observables displaying thermalization can be expressed in the Hamiltonian eigenbasis by the following ansatz:
\begin{equation}
    \obs_{\alpha\beta}=\obs(E_{\alpha\beta})\delta_{\alpha\beta}+e^{-\entropy(E_{\alpha\beta})/2}f_{\obs}(E_{\alpha\beta},\omega_{\alpha\beta})R_{\alpha\beta}
    \label{eq_ETH}
\end{equation}
where $E_{\alpha\beta}=(E_{\alpha}+E_{\beta})/2$, $\omega_{\alpha\beta}=(E_{\alpha}-E_{\beta})$, and $\entropy(E)\sim N$ is the thermodynamic entropy at energy $E$.
Correspondingly, $\obs(E_{\alpha\beta})$ and $f_{\obs}(E_{\alpha\beta},\omega_{\alpha\beta})=f_{\obs}(E_{\alpha\beta},-\omega_{\alpha\beta})$ are smooth functions of their arguments, and $\obs(E_{\alpha\beta})$ equals the expectation value of the microcanonical ensemble at energy $E_{\alpha\beta}$.
Ultimately, $R_{\alpha\beta}=R_{\beta\alpha}^{(*)}$ is a real (complex) variable with zero mean and unit variance ($\avg{\abs{R_{\alpha\beta}}^{2}}=1$).
Right candidates for observables that are proven to satisfy \cref{eq_ETH} are $n$-body observables, with $n\ll N$.

No matter of $C_{\alpha}$, if the fluctuations of the system energy (and correspondingly the energy window of the microcanonical ensemble) are sufficiently small, \idest{} $\Delta E\ll 1$, then, applying \cref{eq_ETH}, we can assume $\obs(E_{\alpha\beta})$ to be almost constant $\sim \obs(E)$ and approximate the long-time average in \cref{eq_diagonal_avg} as:
\begin{equation}
    \avg{\obs}_{\rm DE}=\sum_{\alpha}^{N}\abs{C_{\alpha}}^{2}O_{\alpha\alpha}
    \sim \obs(E)\sum_{\alpha}^{N}\abs{C_{\alpha}}^{2}=\obs(E)\,.
\end{equation}
Similarly, the thermal average given by the microcanonical ensamble reads:
\begin{equation}
    \avg{\obs}_{\rm{ME}}=\sum_{\alpha}^{N_{E,\Delta E}}\frac{O_{\alpha\alpha}}{N_{E,\Delta E}}
    \sim \obs(E)\sum_{\alpha}^{N_{E,\Delta E}}\frac{1}{N_{E,\Delta E}}
    =\obs(E)\,,
\end{equation}
so that, in the end, the long-time average matches the microcanonical ensemble precition.
Furthermore, using ETH, we can compute the long-time average of the temporal fluctuations of the observable expectation value:
\begin{equation}
    \begin{split}
        \sigma_{\obs}^{2}&=\lim_{T\to\infty} \frac{1}{T}\int_0^T dt \qty[\obs(t)^{2}]- \avg{\obs}_{\rm DE}^{2}\\
        &=\lim_{T\to\infty}\frac{1}{T}\int_0^T dt \sum_{\alpha,\beta,\gamma,\delta}\obs_{\alpha\beta}\obs_{\gamma\delta}C^{*}_{\alpha}C_{\beta}C^{*}_{\gamma}C_{\delta}e^{it(E_{\alpha}-E_{\beta}+E_{\gamma}-E_{\delta})}-\avg{\obs}_{\rm DE}^{2}\\
        &=\sum_{\alpha,\alpha\neq\beta}\obs_{\alpha\beta}C^{*}_{\alpha}C_{\beta}\sum_{\gamma,\gamma\neq\delta}\obs_{\gamma\delta}C^{*}_{\gamma}C_{\delta}\lim_{T\to\infty} \frac{1}{T}\int_0^T dte^{it(E_{\alpha}-E_{\beta}+E_{\gamma}-E_{\delta})}\\
        &=\sum_{\alpha,\alpha\neq\beta}\abs{\obs_{\alpha\beta}}^{2}\abs{C_{\alpha}}^{2}\abs{C_{\beta}}^{2}\leq \max \abs{\obs_{\alpha\beta}}^{2}\sum_{\alpha,\beta}\abs{C_{\alpha}}^{2} = \max \abs{\obs_{\alpha\beta}}^{2} \propto e^{-\entropy(E)}\,.
    \end{split}
\end{equation}
Therefore, time fluctuations of the observable expectation values are exponentially small in the system size. 
Namely, $\obs(t)$ is almost equal at any time to the diagonal ensamble average and no time average is needed. 
In other words, ETH ansatz implies the ergodicity in the strong sense.

The equivalence between the long-time average and the statistical mechanics prediction is  similar to the classical counterpart, where ergodicity can be explained in terms of particle collisions and energy redistribution between different \dof.
However, in isolated quantum systems, chaos, ergodicity, and thermalization are hidden in the nature of the Hamiltonian eigenstates.
In other words, the information of the eventual thermal state is already encoded in the system from the initial quench, and relaxation of observables to their equilibrium values results from the \emph{dephasing} originated in time evolution \cite{Rigol2008ThermalizationItsMechanism,DAlessio2016QuantumChaosEigenstate,Deutsch2018EigenstateThermalizationHypothesis}.

Numerical evidence of ETH has been found in several strongly correlated nonintegrable lattice models of condensed matter and ultracold atom gases, such as hard-core bosons \cite{Rigol2008ThermalizationItsMechanism,Vidmar2015DynamicalQuasicondensationHardCore}, soft-core bosons \cite{Biroli2010EffectRareFluctuations,Beugeling2014FinitesizeScalingEigenstate,Beugeling2015OffdiagonalMatrixElements}, interacting spin chains \cite{Rigol2009BreakdownThermalizationFinite,Santos2010LocalizationEffectsSymmetries,Khatami2013FluctuationDissipationTheoremIsolated,Steinigeweg2014PushingLimitsEigenstate}, and spinless and spinful fermions \cite{Rigol2009QuantumQuenchesThermalization,Khatami2012QuantumQuenchesDisordered,Genway2012ThermalizationLocalObservables,Neuenhahn2012ThermalizationInteractingFermions}.
Nowadays, ETH is also applied in contexts like quantum gravity and warmholes \cite{Maldacena2013CoolHorizonsEntangled,Marolf2013GaugeGravityDualityBlack}.

Nevertheless, despite a wide variaty of QMB systems were observed displaying ergodicity and thermalization via ETH, some experiments have recently pointed out strong and weak violations of ETH.
Apart from integrable systems, strong violations of ETH have been displayed by more interesting phenomena such as \emph{many-body localization} (MBL) \cite{Nandkishore2015ManyBodyLocalizationThermalization}, where thermalization is strongly avoided by means of localized quench disorder.

\section{Quantum Many-Body Scars and Lattice Gauge Theories}
\label{sec_QMB_scars}
Here, we are interested in cases of \emph{weak} ergodicity breaking, \idest{} refferred to a small portion of the spectrum violating ETH, while the overall system behaving ergodically. 
Such peculiar eigenstates are called \emph{scars} (like defects of the spectrum) and have been shown to be at the origin of slow dynamics for cold atoms experiments for some intial states \cite{Bluvstein2021ControllingQuantumManybody,Bernien2017ProbingManybodyDynamics,Serbyn2021QuantumManybodyScars}.
In the following, we review the main features of such an exotic dynamics, starting from the observations in Ryderg atoms chains in \cref{sec_rydberg_chains} to the intimate connection with Abelian LGTs in \cref{sec_scars_AbelianLGT}.
% =========================================================================================
\subsection{Experimental observations of scars in Rydberg atoms}
\label{sec_rydberg_chains}
Experimental observation of slow dynamics, have been remarkably encountered in 1D Rydberg atom arrays of $N$ optical traps, each of them hosting a atom \cite{Bernien2017ProbingManybodyDynamics}.
Every atom is trapped in its electronic ground state (denoted with $\ket{\downarrow}_{\vecsite}$), which is quasirenonantly coupled to a single Rydberg state, \idest{} a highly excited electronic level (denoted with $\ket{\uparrow}_{\vecsite}$). 
Overall, the Rydberg atom array behaves like a chain of qubits $\qty{\ket{\uparrow,\downarrow}}_{\vecsite =1\dots N}$, whose dynamics is ruled by the following Hamiltonian:
\begin{equation}
    \ham_{\rm{Ryd}}=\sum_{\vecsite}\qty(\Omega \Sx_{\vecsite} + \delta \Sz_{\vecsite})+\sum_{\vecsite, \vecsite^{\prime}\neq \vecsite}V_{\vecsite, \vecsite^{\prime}}\nop_{\vecsite}\nop_{\vecsite^{\prime}}\,,
    \label{eq_Ham_Rydberg}
\end{equation}
where $\hat{\sigma}^{\alpha}_{\vecsite}$ are Pauli matrices at site $\vecsite$, while the operator $\nop_{\vecsite}=(\Sz_{j}+1)/2$ measures the presence of a Rydberg excitation ($\ket{\uparrow}_{\vecsite}$) at site $\vecsite$.
The coefficients $\Omega$ and $\delta$ are the Rabi frequency and the detuning of the laser excitation scheme respectively; the former one describes the coupling between the ground and Rydberg states due to an external driving field, while the second one represents the difference between the driving field frequency and the atomic transition frequency. 
Correspondingly, $V_{\vecsite, \vecsite^{\prime}}$ couples the interaction between atoms in their Rydberg states at sites $\vecsite$ and $\vecsite^{\prime}$.

In typical experimental realizations of Rydberg arrays made by optical lattices or optical tweezers, the interaction is strong at short range and decays as $1/\abs{\vecsite-\vecsite^{\prime}}^{6}$ as in the Van der Waals scenario \cite{Bernien2017ProbingManybodyDynamics,Zeiher2017CoherentManyBodySpin,Barredo2018SyntheticThreedimensionalAtomic}. 
Such an interaction is typically approximated with the Rydberg-blockade effect, which assumes $V_{\vecsite, \vecsite^{\prime}}$ as the largest energy scale of the system, avoiding atoms on neighboring sites to be simoultaneously excited to Rydberg states $\nop_{\vecsite}\nop_{\vecsite+\latvec}=0$ \cite{Bernien2017ProbingManybodyDynamics}.
Within this approximation, the Hamiltonian in \cref{eq_Ham_Rydberg} can be approximated as done by Fendley, Sengupta, and Sachdev (FSS) in \cite{Fendley2004CompetingDensitywaveOrders}:
\begin{equation}
    \ham_{\rm{FSS}}=\sum_{\vecsite}\qty(\Omega\Sx_{\vecsite}+2\delta \nop_{\vecsite})
    \label{eq_Ham_FSS}
\end{equation}
plus the blockade constraint on neighboring sites. 
In the limit of zero detuning, \cref{eq_Ham_FSS} is equivalent to the PXP model \cite{Hudomal2022DrivingQuantumManybody}:
\begin{equation}
    \ham_{\rm{PXP}}=\sum_{\vecsite}\hat{P}_{\vecsite-\latvec}\Sx_{\vecsite}\hat{P}_{\vecsite+\latvec}\,,
    \label{eq_ham_pxp}
\end{equation}
where $\hat{P}_{\vecsite}$ is a projector enforcing the state in $\vecsite$ to be in the groundstate. 
In both these realizations of the Rydberg Hamiltonian, the experimental and numerical observations of the corresponding dynamics pointed out two different behaviors according to the intial quench of the time evolution \cite{Bernien2017ProbingManybodyDynamics,Bluvstein2021ControllingQuantumManybody,Serbyn2021QuantumManybodyScars,Surace2020LatticeGaugeTheories}.
Namely, starting from a configuration with all the Rydberg atoms in their groundstate, \idest{} $\ket{\downarrow,\downarrow,\dots,\downarrow,\downarrow}$ (empty), the system thermalize, as expected from the ETH discussed in \cref{sec_ETH}, and the average value of observables such as the local density $\nop$ matches the corresponding statistical mechanics predictions.
Correspondingly, the return fidelity with the initial state $\abs{\braket{\Psi(0)}{\Psi(t)}}^{2}$ quickly vanishes.
Conversely, quenching the dynamics from staggered configurations like the following charge-density wave (CDW) states \cite{Surace2020LatticeGaugeTheories}
\begin{align}
    \ket{\downarrow,\uparrow,\downarrow,\uparrow,\downarrow,\uparrow,\dots} \rm{(CDW1)}&&
    \ket{\uparrow,\downarrow,\uparrow,\downarrow,\uparrow,\downarrow,\dots} \rm{(CDW2)}&\,,
    \label{eq_CDW_states}
\end{align}
the system undergoes an oscillatory behavior of the local density operator $\nop(t)$ (the return fidelity) between the two states $\qty{\downarrow,\uparrow}$ (between CDW1 and CDW2) and escapes quantum thermalization (see Fig6b of \cite{Bernien2017ProbingManybodyDynamics} and Fig1 of \cite{Surace2020LatticeGaugeTheories}).
Such a pattern has also been detected by measuring the half-chain entanglement entropy, whose growth in time is slow and oscillating with the same period of density oscillations (see Fig10 of \cite{Bernien2017ProbingManybodyDynamics}).
Moreover, the energy spectrum has revealed the presence of some scar states characterized by a high overlap with the initial quench state \cite{Serbyn2021QuantumManybodyScars}. 
As we will see, the oscillating behavior between CDW1 and CDW2 passing from the full ground state, can be mapped into a process of particle-antiparticle pairs creation and annihilation, which is a foundational mechanism occurring in high-energy \cite{Surace2020LatticeGaugeTheories}.
% =========================================================================================
\subsection{Scars in Abelian Lattice Gauge Theories}
\label{sec_scars_AbelianLGT}
Indeed, it has been recently demonstrated that the experimental model of Ryderg atoms can be mapped into a spin-$1/2$ Quantum Link Model (QLM) version of a U(1) LGT \cite{Surace2020LatticeGaugeTheories} and correspondingly, the observation of scarring dynamics can be associated to high-energy physics phenomena such as the Schwinger mechanism \cite{Calzetta2009NonequilibriumQuantumField}. 
In particular, we can then identify the computational basis of configurations allowed by the Rydberg blockade with the classical configurations of the electric field allowed by Gauss law.

In detail, a Rydberg atom chain in the Rydberg-blockade approximation can be mapped into a (1+1)D realization of the QED Hamiltonian in \cref{eq_U1_Hamiltonian}:
\begin{equation}
    \ham = -w\sum_{\vecsite}\qty[\hpsi_{\vecsite}^{\dagger}\Apara_{\genlink}\psi_{\vecsite+\latvec}+\hc]+\mass\sum_{\vecsite}(-1)^{\vecsite}\hpsi^{\dagger}_{\vecsite}\hpsi_{\vecsite}+ J\sum_{\vecsite}\casimir_{\genlink}\,,
    \label{eq_U1_ham_numerics}
\end{equation}
where $w=1/2$, $\mass=\mass[0]\lspeed/\hbar$, and $J=\coupling^{2}/2$ are adimensional parameters\footnote{Similarly to \cref{eq_2D_SU2_Ham_numerics}, for numerical simulations, we made the Hamiltonian in \cref{eq_U1_Hamiltonian} adimensional by multiplying it by $\lspace/(\lspeed\hbar)$.}.
In the finite truncation of U(1) discussed in \cref{sec_U1_gaugetruncation}, the gauge fields can be represented by spin-$\spin$ variables \cite{Chandrasekharan1997QuantumLinkModels,Horn1981FiniteMatrixModels}:
\begin{align}
    \eleE_{\genlink}&=\spinmat[z]_{\genlink}(\spin)&
    \Apara_{\genlink}&=\spinmat[+]_{\genlink}(\spin)&
    \comm{\eleE_{\genlink}}{\Apara_{\genlink}}&=\spinmat[+]_{\genlink}(\spin)\,.
\end{align}
At any finite truncation, U(1) Gauss law always requires that $\hat{G}_{\vecsite}\ket{\Psi_{\rm{phys}}}=0$ $\forall \vecsite$, where $\hat{G}_{\vecsite}$ is the local generator of the U(1) gauge symmetry defined in \cref{eq_U1_gausslaw} and constrains the electric fields incoming and outcoming from a local site to match the corresponding amount of charge:
\begin{equation}
    \hat{G}_{\vecsite}=\eleE_{\genlink}-\eleE_{\genlinkm}-\hpsi_{\vecsite}^{\dagger}\hpsi_{\vecsite}+\frac{1-(-1)^{\vecsite}}{2}\,.
\end{equation}
In particular, due to the use of staggered fermions, an empty \emph{odd} site corresponds to the presence of an anti-particle of charge $\bar{\charge}$, while occupied \emph{odd} sites implies to the presence of a particle of charge $\charge$ (assuming that U(1) $\charge<0$, as it refers to the electron). 
Namely, indicating with $\ket{\Omega}_{\vecsite}$ an empty matter state in site $\vecsite$, and with $\ket{\varnothing}_{\vecsite}$ the corresponding absence of charge, we have:
\begin{equation}
    \begin{aligned}
        \ket{\Omega}_{\vecsite,\emph{even}} & = \ket{\varnothing}_{\vecsite,\emph{even}}\qquad\qquad &
        \ket{\Omega}_{\vecsite,\emph{odd}} & = \ket{\bar{\charge}}_{\vecsite,\emph{odd}}\\
        \hpsi^{\dagger}\ket{\Omega}_{\vecsite,\emph{even}}&= \ket{\charge}_{\vecsite,\emph{even}}\qquad\qquad&
        \hpsi^{\dagger}\ket{\Omega}_{\vecsite,\emph{odd}}&= \ket{\varnothing}_{\vecsite,\emph{odd}}\,.
    \end{aligned} 
\end{equation}
\subsubsection{Mapping the Rydberg blockade into the U(1) Gauss law}
To understand the connection between the two models, let us reconsider from the Rydberg blockade.
When the latter holds, such as in \cref{eq_Ham_FSS,eq_ham_pxp}, the total Hilbert space is restricted to those states without double occupancies on neighboring sites. 
Then, the effective Hilber space of each pair of neighboring Rydberg atoms along the chain within the Rydberg-blockade approximation is
\begin{align}
    \ket{1}_{\rm{Ryd}}&=\ket{\uparrow,\downarrow}&
    \ket{2}_{\rm{Ryd}}&=\ket{\downarrow,\downarrow}&
    \ket{3}_{\rm{Ryd}}&=\ket{\downarrow,\uparrow}\,.
    \label{eq_rydberg_states}
\end{align}
Correspondingly, in the $\spin=1/2$ finite truncation of U(1) gauge fields, the effective Hilbert space of a matter site coupled with its attached (left and right) gauge links is made of configurations like $\ket{\eleE_{\vecsite,-\latvec},\charge_{\vecsite}, \eleE_{\vecsite,+\latvec}}$, Denoting $\eleE=+1/2$ ($\eleE=-1/2$) with $\rightarrow$ ($\leftarrow$), in the staggered fermion solution we have
\begin{subequations}
    \begin{align}
        \ket{1}^{\rm{even}}_{\rm{U(1)}}&=\ket{\rightarrow, \varnothing, \rightarrow}&
        \ket{2}^{\rm{even}}_{\rm{U(1)}}&=\ket{\leftarrow, q, \rightarrow}&
        \ket{3}^{\rm{even}}_{\rm{U(1)}}&=\ket{\leftarrow, \varnothing, \leftarrow}\\
        \ket{1}^{\rm{odd}}_{\rm{U(1)}}&=\ket{\rightarrow, \varnothing, \rightarrow}&
        \ket{2}^{\rm{odd}}_{\rm{U(1)}}&=\ket{\rightarrow, \bar{\charge}, \leftarrow}&
        \ket{3}^{\rm{odd}}_{\rm{U(1)}}&=\ket{\leftarrow, \varnothing, \leftarrow}\,.
    \end{align}
    \label{eq_U1_gaugestates}
\end{subequations}
One can then associate the effective Hilbert space of neighboring Rydberg atoms in \cref{eq_rydberg_states} with the U(1) gauge invariant configurations of \cref{,eq_U1_gaugestates} (for even and odd sites respectively).
Correspondingly, states violating Gauss law are exactly mapped into the nearest-neighbor occupied sites which are strongly suppressed by the Rydberg blockade.
A pictorial representation of such a correspondence is given in Fig1 of \cite{Surace2020LatticeGaugeTheories}. 

We can further identify the Hamiltonian parameters $\Omega =-w$, $\delta=-m$ and the operators
\begin{align}
    \Sz_{\vecsite}& \leftrightarrow (-1)^{\vecsite}2\spinmat[z]_{\genlinkm}&
    \Sx_{\vecsite}& \leftrightarrow \qty[\hpsi_{\vecsite-\latvec}^{\dagger}\spinmat[+]_{\genlinkm}\hpsi_{\vecsite} +\hc]&
    \Sy_{\vecsite}& \leftrightarrow -\init(-1)^{\vecsite}\qty[\hpsi_{\vecsite-\latvec}^{\dagger}\spinmat[+]_{\genlinkm}\hpsi_{\vecsite} +\hc]\,,
    \label{eq_rydberg_U1_mapping}
\end{align}
so that the two Hamiltonians in \cref{eq_U1_ham_numerics,eq_Ham_FSS} are equivalent.
% =========================================================================================
\subsection{Gauge theory interpretation of slow dynamics}
Remarkably, beyond providing a direct link between Gauss law and the Rydberg blockade mechanism, the connection between \cref{eq_Ham_FSS} and \cref{eq_U1_ham_numerics} with the mapping in \cref{eq_rydberg_U1_mapping} provides an immediate connection between Rydberg-blockade chains experiments and the U(1) LGT.
We can then interpret the slow dynamics observed in Rydberg atoms \cite{Bernien2017ProbingManybodyDynamics} in terms of particle-antiparticle pairs production in high-energy physics.
In particular, the two charge density wave states CDW1 and CDW2 of \cite{Surace2020LatticeGaugeTheories} can be mapped into the two uniform electric field states $\eleE_{\genlink}=\pm 1/2$.
Namely, we can map the empty Rydberg configuration and the ones in \cref{eq_CDW_states} into the following states:
\begin{equation}
    \begin{aligned}
        \rm{(CDW1)}&& \longleftrightarrow&& &\ket{\rightarrow, \varnothing, \rightarrow, \varnothing, \rightarrow, \varnothing, \rightarrow,\varnothing, \rightarrow}& &\rm{(String)}\\
        \rm{(empty)}&& \longleftrightarrow&& &\ket{\leftarrow, \charge\,, \rightarrow, \bar{\charge}\,, \leftarrow, \charge\,, \rightarrow,\bar{\charge}\,, \leftarrow}& &\rm{(Pairs)}\\
        \rm{(CDW2)}&& \longleftrightarrow&& &\ket{\leftarrow, \varnothing, \leftarrow, \varnothing, \leftarrow, \varnothing, \leftarrow,\varnothing, \leftarrow}& &\rm{(anti-String)}\,,
    \end{aligned}
\end{equation}
where the \emph{string} (\emph{anti-string}) refers to a uniform rightward (leftward) electric flux, while \emph{pairs} refers to the state filled by adjacent particle-antiparticle pairs.

Then, the long-lived oscillations between the CDW1 and CDW2 states observed in the Rydberg-blockade chain are related to the rightward-leftward switch of the electric flux passing throught the intermediate production of particle-antiparticle pairs.
Such a pattern is called \emph{string inversion} and is directly tied to string breaking \cite{Calzetta2009NonequilibriumQuantumField,Bali2005ObservationStringBreaking,Hebenstreit2013RealTimeDynamicsString} prototypical of gauge theories including dynamical matter. 

Remarkably, this mechanism is disturbed by quantum fluctuations, which are controlled by the dimensionless ratio $\mass/w$.
In detail, for $\mass< 0.655w=\mass[c]$, particle-antiparticle pairs are created, which accelerates in an electric field and gradually screens it. 
Eventually, coherent pair annihilation takes place, leading to a state with an opposite electric flux.
Such an inversion happens several times, causing the slow dynamics in the system. 
This leads to a significant slowdown of thermalization and limits quantum information propagation.
Conversely, when the particle mass exceeds a critical value ($\mass>\mass[c]$), pair production becomes a virtual process and string inversion does not occur. 
This marks a different phase of the system’s evolution, preventing the formation of real particle pairs.
This transition is related to a quantum phase transition observed in the FSS model in \cref{eq_Ham_FSS} at a critical detuning $\delta_c$ \cite{Fendley2004CompetingDensitywaveOrders} and corresponds to the spontaneous breaking of chiral symmetry in the U(1) LGT \cite{Surace2020LatticeGaugeTheories,Rico2014TensorNetworksLattice}.
% =========================================================================================
\section{Scars in Non-Abelian Lattice Gauge Theories}
\label{sec_scars_nonAbelianLGT}
Given the intimate connection between QMBS and LGTs, and the current large effort to quantum simulate the latter \cite{Dalmonte2016LatticeGaugeTheory,Banuls2020SimulatingLatticeGauge,Zohar2015QuantumSimulationsLattice,Alexeev2021QuantumComputerSystems,Aidelsburger2021ColdAtomsMeet,Zohar2021QuantumSimulationLattice,Klco2022StandardModelPhysics,Bauer2023QuantumSimulationHighEnergy,Bauer2023QuantumSimulationFundamental,DiMeglio2023QuantumComputingHighEnergy,Halimeh2023ColdatomQuantumSimulators,Cheng2024EmergentGaugeTheory}, it is important to investigate the origin of QMBS in connection to gauge symmetry by exploring its possible occurrence in non-Abelian LGTs with dynamical matter. 
In particular, investigating a high-energy context of QMBS requires the inclusion of dynamical matter, as demonstrated in the Abelian scenario.

In this section, we report and discuss the exact diagonalization (ED) and matrix product state (MPS) results of \cite{Cataldi*2025QuantumManybodyScarring} that showcase robust QMBS dynamics in a non-Abelian SU(2) LGT with dynamical matter starting in simple initial product states (\cref{sec_scars_model}). 
We show how the QMBS dynamics involve state transfer through meson and baryon-anti-baryon bare states, highlighting the non-Abelian nature of this scarred behavior.
In addition to the exotic dynamics observed in \cref{sec_scars_dynamics} and the spectral properties of the model discussed in \cref{sec_scars_spectralproperties}, we enrobust our results with further analyses. 
In \cref{sec_scars_ergodic_initialstates}, we observe that log-time average of local observables displayed in \cref{fig_scars_dynamics} are not compatible with the thermal relaxation expected for other initial states at the same energy.
In \cref{sec_scars_ergodic_paramsregimes}, we demonstrate how, for other parameter regimes, the spectral and dynamical properites of the two initial product states are compatible with the ergodic behavior predicted by statistical mechanics (microcanonical ensemble). 
Conversely, we further illustrate how the scars-induced revivals of \cref{fig_scars_dynamics} persist for larges gauge-field trunctations (\cref{sec_scars_spin1}), at larger system sizes (\cref{sec_scars_systemsize}), and over longer timescales (\cref{sec_scars_longtime}) for the candidate initial states.
These results confirm the overall ergodic nature of the model and regard the observed scarring behavior as a weak violation of ETH and the expected thermalization.
% =========================================================================================
\subsection{The model}
\label{sec_scars_model}
As a prototypical model to probe non-ergodic dynamics in non-Abelian gauge theories, we consider a (1+1)D matter-coupled \emph{hardcore-gluon} SU(2) LGT \cite{Calajo2024DigitalQuantumSimulation, Cataldi*2025QuantumManybodyScarring, Cataldi2024Simulating2+1DSU2, Rigobello2023HadronsHamiltonianHardcore, Silvi2019TensorNetworksAnthology,Zohar2019RemovingStaggeredFermionic} such as the one derived in \cref{sec_SU2_hardcoregluon}.  
The model describes a matter quark field of mass $\mass$, living on lattice sites $\vecsite$,
coupled to a truncated SU(2) gauge field, defined on lattice links | see \cref{fig_scars_dynamics}.
The quark field is a staggered fermion doublet $\hpsi_{\vecsite,\alpha}$, obeying the anticommutation relations in \cref{eq_SU2_matter_commutation_rules}:
$\acomm*{\hpsi_{\vecsite,\alpha}}{\hpsi^{\dagger}_{\vecsite^{\prime},\beta}}{=}\hbar \delta_{\vecsite,\vecsite^{\prime}}\delta_{\alpha,\beta}$ \cite{Susskind1977LatticeFermions}, where $\alpha,\beta\in\qty{\rla,\gla}$ span SU(2) colors.
A basis for matter sites is
\begin{equation}
    \qty{
    {\ket{0}},\,
    {\ket{\rla} = \hpsi^{\dagger}_{\rla}\ket{0}},\,
    {\ket{\gla} = \hpsi^{\dagger}_{\gla}\ket{0}},\,
    {\ket{2}    = \hpsi^{\dagger}_{\rla} \hpsi^{\dagger}_{\gla}\ket{0}}
    }.
\end{equation}
Gauge link states can be expanded in the basis $\ket{j, \mL, \mR}$,
where $\mL,\mR\in\{-j,\ldots,+j\}$ index spin-$j$ states,
and the link energy density operator is diagonal and coincides with the quadratic Casimir, $\hat{E}^2\ket{j, \mL, \mR}=j(j\mathop+1)\ket{j, \mL, \mR}$ \cite{Zohar2015FormulationLatticeGauge}.
Adopting a hardcore-gluon truncation, we restrict $j$ to $\{0,1/2\}$, namely keep only basis states reachable from the singlet $\ket{00}$ via at most a single application of the parallel transporter $\NApara$: $\qty{\ket{00},\ket{\rla\rla},\ket{\gla\gla},\ket{\gla\rla},\ket{\rla\gla}}$, where $\rla(\gla)=+(-)1/2$.

Correspondingly, non-Abelian Gauss law mandates that physical states are local gauge singlets, $\hat{G}^{\nu}_{\vecsite} \ket{\Psi_{\rm{phys}}} \equiv 0, \; \forall \vecsite,\,\nu\in\{x,y,z\}$, where $\hat{G}^\nu_{\vecsite}$ are the generators of local rotations at site $\vecsite$ and defined in \cref{eq_SU2_gausslaw}.
In numerical simulations, we enforce this constraint using a dressed site approach, which yields a defermionized qudit model with a $6$-dimensional local basis discussed of \cref{eq_SU2_hardcoregluon_basis1D} \cite{Calajo2024DigitalQuantumSimulation}. 

The dynamics is governed by the Kogut-Susskind (adimensional) Hamiltonian \cite{Kogut1975HamiltonianFormulationWilson} employing staggered fermions and defined in \cref{eq_SU2_1D_Hamiltonian} on an $\Nsites$-site lattice with spacing $\lspace$
\begin{equation}
    \ham = \frac{1}{2}\sum_{\vecsite}\sum_{\alpha,\beta}
    \qty[-i\hpsi^{\dagger}_{\vecsite,\alpha}\NApara_{\genlink}\psi_{\vecsite+\latvec,\beta}+\hc]
    +\mass\sum_{\vecsite,\alpha}
    (-1)^{\vecsite}\hpsi^{\dagger}_{\vecsite,\alpha}\psi_{\vecsite,\alpha}
    {+}\frac{g^2}{2}\sum_{\vecsite}\hat{E}^2_{\genlink}.
    \label{eq_SU2_scars_Hamiltonian}
\end{equation}
where, similarly to \cref{eq_2D_SU2_Ham_numerics}, for the sake of numerical simulations, we rescaled the Hamiltonian with dimensionaless couplings such as $\mass = \mass[0]\lspeed/\hbar$ and $\coupling^{2} =\charge^{2}\lspace^{2}\cdot(\lspeed\hbar \permittivity)^{-1}$. 
Values reported in the Figures refer to these rescaled quantities.
% =========================================================================================
\subsection{Observing scarred dynamics}
\label{sec_scars_dynamics}
\begin{figure}
    \includegraphics[width=0.5\textwidth]{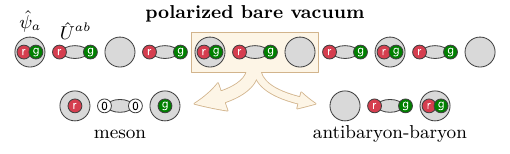}\hfill
    \includegraphics[width=0.5\textwidth]{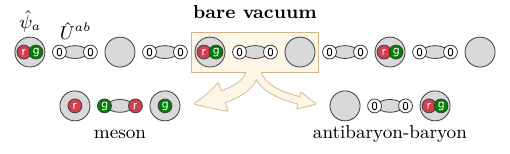}
    \includegraphics[width=0.5\textwidth]{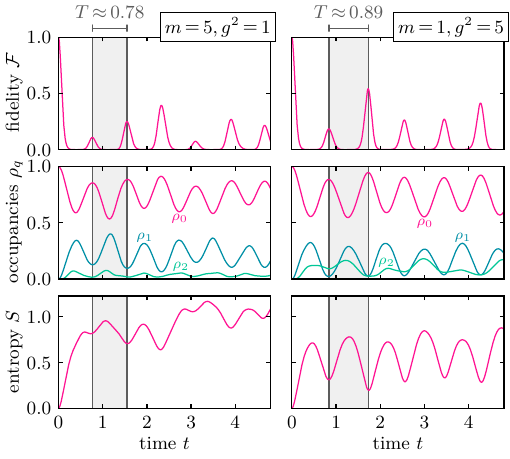}\hfill
    \includegraphics[width=0.5\textwidth]{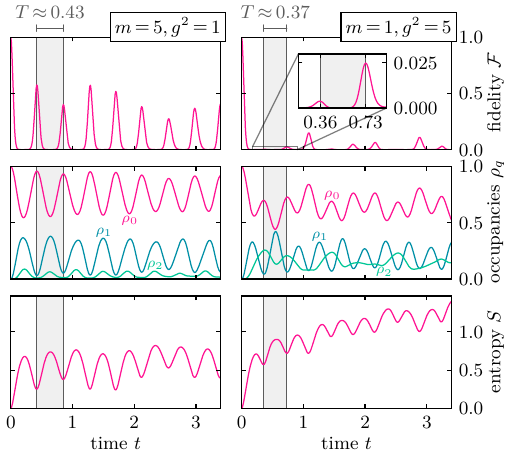}
    \caption{Many-body scarring dynamics of the polarized bare vacuum (left) and the bare vacuum (right) initial states.
    \emph{Top:} cartoons depicting the classical configurations that give the leading contribution to the dynamics; with circles and ellipses denoting matter sites and gauge links respectively.
    \emph{Bottom:} return fidelity, average quark occupancy, and bipartite entanglement entropy as a function of time,
    for different mass and coupling regimes, reported in Figure.
    MPS simulation with $\Nsites=30$ sites in open boundary conditions (OBC), maximum bond dimension $\chi_{\max}=350$, and truncation tolerance $\rm{tol} = 10^{-7}$.
    }
    \label{fig_scars_dynamics}
\end{figure}
To investigate the presence of non-ergodic behavior in this model, we consider the Schrödinger time evolution of different initial states $\ket{\Psi(t\mathop=0)}$, across various parameter regimes.
The time evolution is performed on a chain of $\Nsites=30$ sites via MPS methods\footnote{In detailed, simulations we obtained from the iTensor Library \cite{Fishman2022ITensorSoftwareLibrary,Fishman2022CodebaseReleaseITensor}.}.
As initial configuration, we first consider the \emph{polarized bare vacuum} (PV), which consists of the matter sites in the bare vacuum configuration, while the gauge sites in the excited electric field state,
\begin{equation}
    \ket{\Psi_{\rm{PV}}}{=}
    \ket{\dots2020\dots}_{\mathrm{m}}
    \ket*{\cdots 
        \frac{\rla\gla-\gla\rla}{\sqrt{2}}
        \cdots
        \frac{\rla\gla-\gla\rla}{\sqrt{2}}               
        \cdots
    }_{\mathrm{g}},
    \label{eq_pol_vacuum_state}
\end{equation}
where $\ket{\:\cdot\:}_\mathrm{m}$ and $\ket{\:\cdot\:}_\mathrm{g}$ refer to the matter and gauge subsystems, respectively.
Here $(\mR)_{\vecsite-\latvec,\vecsite}$ and $(\mL)_{\genlink}$ around each bulk site $\vecsite$ are valence-bonded in a singlet.
In the qudit formulation of the model (see \cref{sec_SU2_hardcoregluon} and \cite{Calajo2024DigitalQuantumSimulation}), the state in \cref{eq_pol_vacuum_state} becomes a product state:
\begin{equation}
\ket{\Psi_{\rm{PV}}}=\ket{6}\ket{2}\dots \ket{6}\ket{2},  
\end{equation}
We identify two regimes avoiding thermalization:
the large-mass one, $\mass\gg (g^2,1)$, and the large-coupling one, $g^2\gg (\mass,1)$.
For each regime, we compute the return fidelity between the initial state and the evolved state, $\mathcal{F}(t)=\abs{\braket{\Psi(t)}{\Psi(0)}}^2$, shown in \cref{fig_scars_dynamics}.
After an initial complete system relaxation, we observe persistent revivals, signaling a clear deviation from the expected ergodic behavior.
The revivals occur approximately periodically, with the period indicated in the figure for both the considered regimes.

The observed dynamics can be understood as a process of persistent particle pair creation out of an initial false vacuum \cite{Banerjee2013AtomicQuantumSimulation,Kuhn2014QuantumSimulationSchwinger,Magnifico2020RealTimeDynamics}.
To see this, we define the (staggered) site occupancy operator,
\begin{equation}
    \hat{\mathcal{N}}_{\vecsite} = (-1)^{\vecsite}\sum_{\alpha}\qty[\hpsi^{\dagger}_{\vecsite,\alpha}\hpsi_{\vecsite,\alpha}-\frac{1-(-1)^{\vecsite}}{2}]\,,
\end{equation}
which counts the number of quarks (on even sites) or antiquarks (on odd sites).
Then, we compute the average occupancy $\density_{\charge}(t)$ of each $\mathcal{N}_{\vecsite}$ eigenvalue (single site quark occupancy, $q\in\{0,1,2\}$),
projecting on the corresponding eigensubspace of $\mathcal{N}_{\vecsite}$ and averaging over all sites $n$:
\begin{equation}
    \density_{\charge}(t)=\frac{1}{\Nsites}\sum_{\vecsite}\ev{\delta_{\mathcal{N}_{\vecsite}, q}}{\Psi(t)}\,.
\end{equation}
As lately discussed in \cref{sec_scars_ergodic_initialstates}, they oscillate around average values that do not coincide with the ones predicted by thermal ensembles, confirming deviations from ergodic behavior.
Large values of either mass or coupling confine the system dynamics close to the initial state, creating few particle pair excitations in the system.
In the large-mass regime, most of the excitations are meson-like (quark and antiquark adjacent pairs with an excited shared gauge link).
In the large-coupling regime, the formation of pairs of baryon- (quark pair) and antibaryon- (antiquark pair) like excitations becomes likely as well, as shown by the enhanced double occupancy with respect to the large-mass regime.

Finally, the last row of \cref{fig_scars_dynamics} shows the evolution of the  bipartite entanglement entropy defined as $\entropy=-\Tr[\density_A\log\density_A]$, where $A$ and $B$ indicate the two halves of the chain and $\density_A=\Tr_B[\density_{AB}]$ is the reduced density operator of subsystem $A$.
In both cases, we observe a slow growth of the entanglement entropy after an initial fast increase up to the first fidelity revival peak. The oscillations on top of the growth are driven by the successive fidelity revivals, which indeed have the same period.

In addition to the polarized bare vacuum initial state, we also observe similar scarred-like dynamics for the \textit{bare vacuum} (V) state: the ground state of the Hamiltonian in \cref{eq_SU2_scars_Hamiltonian} without the hopping term, made of alternated doubly occupied and empty sites, and trivial links,
\begin{equation}
    \ket{\Psi_{\rm{V}}(t=0)}=\ket{\dots2020\dots}_{\mathrm{m}} \ket{\dots0000\dots}_{\mathrm{g}}\,.
    \label{eq_vacuum_state}
\end{equation}
which in the qudit formulation (see \cref{sec_SU2_hardcoregluon} and \cite{Calajo2024DigitalQuantumSimulation}) reads:
\begin{equation}
    \ket{\Psi_{\rm{V}}}=\ket{5}\ket{1}\dots \ket{5}\ket{1}.
\end{equation}
The resulting non-ergodic dynamics is displayed on the right-side of \cref{fig_scars_dynamics}.
Compared to the polarized vacuum case, the oscillation's period in the fidelity revivals and occupancy is smaller.
As discussed in the following section, it can be explained by the many-body spectrum.
The large coupling regime is characterized by a faster entanglement growth, consistently with a corresponding lower return fidelity.
Although in the regimes considered this product state is energetically close to the true ground state of the model, the observed dynamics is driven by nontrivial physics. This is evident from the significant oscillations observed in the average occupancy, which lead the system significantly away from its initial configuration.
% =========================================================================================
\subsection{Tower of scar states}
\label{sec_scars_spectralproperties}
\begin{figure*}
    \includegraphics[width=0.5\textwidth]{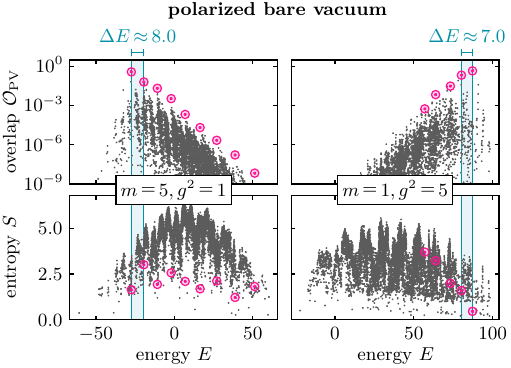}\hfill
    \includegraphics[width=0.5\textwidth]{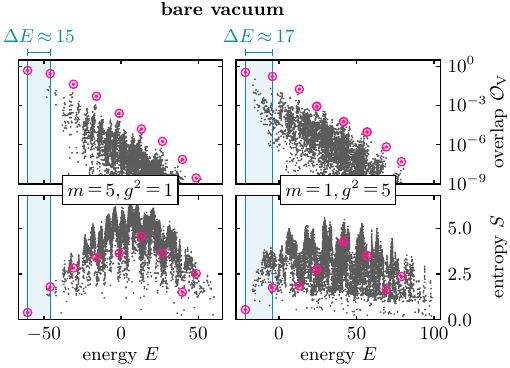}
    \caption{Spectrum analysis for a lattice chain of $N\mathop=10$ sites in OBC, in the two parameter regimes of \cref{fig_scars_dynamics}.
    Red circles highlight the scar states.
    \emph{First row:} the overlap of the many-body spectrum with the polarized bare vacuum (left) and the bare vacuum (right);
    only states with $\mathcal{O}_{\mathrm{(P)V}}>10^{-9}$ are included.
    \emph{Second row:} bipartite entanglement entropy of each eigenstate.
    }
    \label{fig_tower}
\end{figure*}
To determine whether the observed non-ergodic dynamics originates from QMB scars, we perform ED up to $\Nsites=10$ sites and, for each eigenstate in the many-body spectrum $\{\ket{\Phi_s}\}$,
we compute the overlap
\begin{equation}
    \mathcal{O}_{\mathrm{PV/V}}
    = \abs*{\braket{\Psi_{\mathrm{PV/V}}}{\Phi_s}}^2
\end{equation}
and the bipartite entanglement entropy $S$ 
with the two considered initial states (polarized, PV, and unpolarized bare vacua, V).
These quantities are plotted in \cref{fig_tower}, for both regimes where we observe revivals in the return fidelity.
For each candidate QMBS regime, we find a `tower of scar states' (highlighted by red circles) characterized by a high overlap with the initial product states and entanglement low enough (compared to the rest of the many-body spectrum) to give rise to scarring behavior.
In each case, the energy gap between the scar states in the tower, $\Delta E$ (reported in \cref{fig_tower}), approximately matches the frequency of the revivals observed in the return fidelity from \cref{fig_scars_dynamics}, $\Delta E \approx 2\pi/T$.
The correspondence between spectral and dynamical features allows us to conclude that QMBS is indeed the underlying mechanism behind the non-ergodic dynamics shown in \cref{fig_scars_dynamics}.
% =========================================================================================
\subsection{Ergodic dynamics for other initial states}
\label{sec_scars_ergodic_initialstates}
In order to prove the scarring behavior as a specific feature of the configurations in \cref{eq_pol_vacuum_state,eq_vacuum_state} and not of all the initial states, we consider the \textit{microcanonical} ensemble (ME) state constructed as a uniform coherent superposition of all eigenstates lying within a small energy window around the quench energy of the polarized vacuum state $E_{\rm PV}{=} \ev*{\ham}{\Psi_{\rm PV}}$.
Namely, we define the thermal state
\begin{equation}
    \label{eq_thermal_state}
    \ket{\Psi^{\rm ME}_{\rm PV}} = \frac{1}{\sqrt{N_{E_{\rm PV},\delta E}}} \sum\nolimits_{s,\: |E_s-E_{\rm PV}|<\delta E} \ket{\Phi_s},
\end{equation}
where $\delta E{=}\sqrt{\ev*{\ham^2}{\Psi_{\rm PV}} - E_{\rm PV}^2}$ defines the energy shell $[E_{\rm PV}{-}\delta E, E_{\rm PV}{+}\delta E]$,
and $N_{E_{\rm PV},\delta E}$ is the number of eigenstates in this energy shell \cite{Rigol2008ThermalizationItsMechanism}.
\begin{figure}
    \centering
    \includegraphics{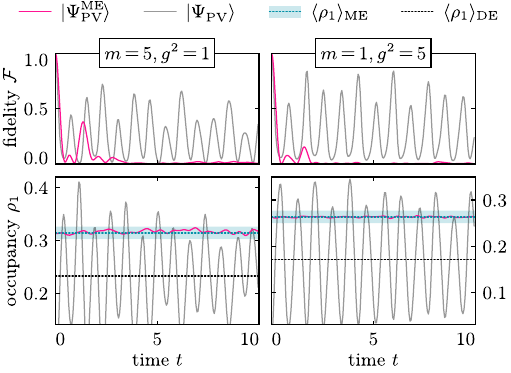}
    \caption{Dynamics of the \textit{microcanonical} state in \cref{eq_thermal_state} for the same parameter regimes of \cref{fig_scars_dynamics}.
    Gray curves reproduce the polarized bare vacuum dynamics, for comparison.
    The blue horizontal lines reproduce the thermal $\density_{1}$ occupancy, obtained via microcanonical ensemble (ME).
    The black ones show the diagonal ensemble (DE) results for the long-time average of $\density_{1}$ on $\ket{\Psi_{\rm PV}(t)}$, highlighting its lack of thermalization.
    ED results for $\Nsites=10$ sites in PBC.}
    \label{fig_thermal_dyn}
\end{figure}

In \cref{fig_thermal_dyn}, we compares the evolution of this state to that of $\ket{\Psi_{\rm PV}}$, in the identified scarred regimes.
As the plot illustrates, the \textit{microcanonical} state exhibits ergodic behavior:
return fidelities relax toward zero (up to minor oscillations attributable to the finite lattice size) and the single occupancy $\density_{1}$ remains close to its microcanonical ensemble average $\ev*{\density_{1}}_{\rm ME}$ \cite{Rigol2008ThermalizationItsMechanism}.
In contrast, for the polarized vacuum, $\density_{1}$ widely oscillates around its long-time average (diagonal ensemble) $\ev*{\density_{1}}_{\rm DE}$ largely deviates from the thermal result.
We thus conclude that the observed non-ergodic dynamics strictly depends on the initial conditions and escapes an otherwise thermal behavior through the occurrence of scar states in the many-body spectrum.
% =========================================================================================
\subsection{Ergodic dynamics for other parameter regimes}
\label{sec_scars_ergodic_paramsregimes}
\begin{figure}
    \centering
    \includegraphics[width=0.49\textwidth]{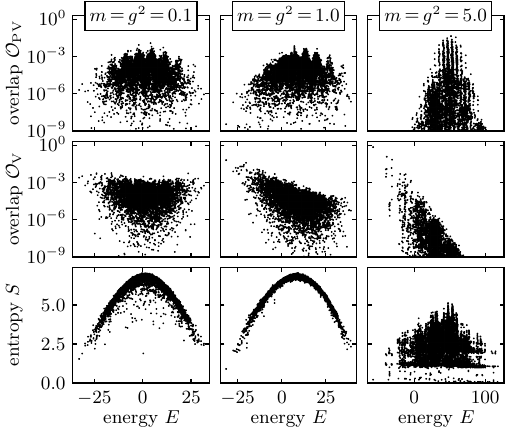}\hfill
    \includegraphics[width=0.49\textwidth]{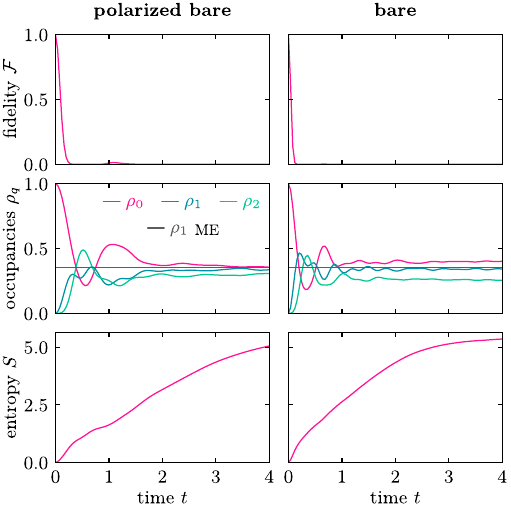}
    \caption{[Left]: \emph{Many-body spectrum.} Spectrum analysis for a lattice chain of $N = 10$ sites with open boundary conditions in a few paradigmatic regimes (corresponding to different columns) where we observe ergodic dynamics.
    The first and second rows show the overlap of the many-body spectrum with the bare vacuum and polarized bare vacuum state, respectively.
    Only states with $\mathcal{O}_{\mathrm{PV}}>10^{-9}$ are included in the plot.
    In the third row, we display the bipartite entanglement entropy of all eigenstates in the energy spectrum as a function of their respective eigenvalues. 
    [Right]: \emph{Ergodic dynamics.} First to third row: return fidelity, average quark occupancy, and bipartite entanglement entropy as a function of time for the two considered initial states, polarized bare vacuum (PV), and bare vacuum (V) at $\mass=g^2=1$.
    The horizontal lines in the occupancy panels indicate the thermal expectation value of the single particle occupancy density obtained through ED of a lattice of $\Nsites=10$ with PBC for the microcanonical ensemble.
    The simulation was performed with MPS for $\Nsites=20$ sites by setting open boundary conditions, maximum bond dimension $\chi_{\max}=350$, and truncation tolerance $\rm{tol}=10^{-7}$.}
    \label{fig_various_spectra}
\end{figure}

We have shown that the two considered initial product states polarized bare vacuum (PV) and bare vacuum (V), escape thermalization for the two regimes $(m=5, g^2=1)$ and $(m=1, g^2=5)$.
To confirm the scarring nature of the observed non-ergodic dynamics, we investigate the absence of scars states in the many-body spectrum for other parameter regimes.
In \cref{fig_various_spectra}[Left], we look at the overlap of the full many-body spectrum with the two considered initial states PV and V and the corresponding bipartite entanglement entropy for different values of $g^2=\mass$.
Rather than in \cref{fig_tower}, we do not observe a clear emergence of a tower of states, and thus, we expect ergodic dynamics for the two considered states.
This is also confirmed by the time evolution of the two initial states for $g^2=m=1$ displayed in \cref{fig_various_spectra}[Right].
In this case, the fidelity does not exhibit persistent revivals, and the entanglement entropy shows rapid growth compared to the previously observed scarred dynamics.
Moreover, as expected for thermalization, the occupancy relaxes to a steady value compatible with the microcanonical ensemble averages.

We leave any further investigations on scarring dynamics for other candidate initial states to future work.
To this extent, it could be interesting to investigate the $g^2=m=0.1$ case, whose many-body spectrum contains extremely low entanglement entropy states.
% =========================================================================================
\subsection{Scars at higher gauge link truncation}
\label{sec_scars_spin1}
\begin{figure*}
    \includegraphics[width=1\textwidth]{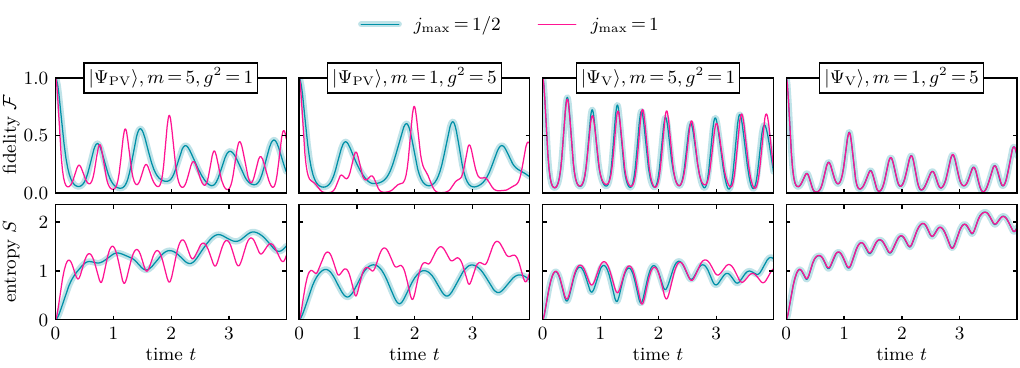}
    \caption{\textit{Scarring at higher gauge truncation.} Many body scarring dynamics for the polarized bare vacuum (PV) and the bare vacuum (V) for the truncated SU(2) YM LGT at $\jmax=1$ (pink line) in comparison with the corresponding one at $\jmax=1/2$ (cyan line).
    Each column reports the fidelity and bipartite entanglement entropy as a function of time for the same two cases $(\mass,g)$ considered in \cref{fig_scars_dynamics}.
    The simulations are obtained via ED for $\Nsites=10$ sites.}
    \label{fig_scars_spin1}
\end{figure*}

It is interesting to highlight that the scarring dynamics observed for the minimal truncation in \cref{fig_scars_dynamics}, is not a feature of the truncation itself.
Indeed, a similar behavior is displayed in the next gauge truncation $\jmax=1$, which leads to a dressed-site basis of 10 gauge invariant states.
As shown in \cref{fig_scars_spin1}, where we return the fidelity and the entanglement entropy for the same two cases discussed in the main text.
Analogously with what has been observed in Abelian LGTs, these results remarkably suggest that the scars dynamics of non-Abelian theories also persist at higher gauge truncations.
% =========================================================================================
\subsection{Scaling with the system size}
\label{sec_scars_systemsize}
\begin{figure}
    \centering
    \includegraphics{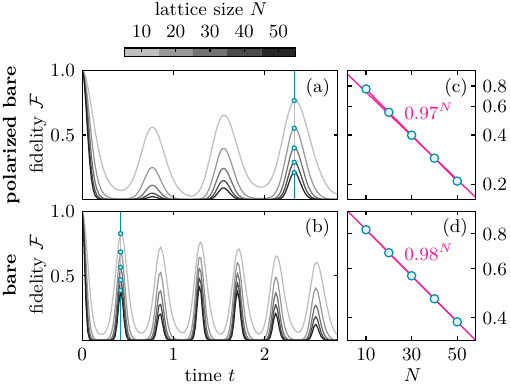}%
    \caption{\textit{Finite-size scaling.} Return fidelity during the evolution of the polarized bare vacuum (a) and bare vacuum (b) initial states, for $\mass=5$, $g^2=1$,  open boundary conditions, and various systems sizes $\Nsites$.
        The side plots (c) and (d) show the finite-size scaling of the highest fidelity peak, highlighted by a blue line in (a) and (b).
        Notice the log $y$-scale.
        The pink line is an exponential fit $\mathcal{F}_{\Nsites}=f^{\Nsites}$, and the extrapolated peak fidelity per site $f$ is reported in the Figure.
        Simulations performed with maximum MPS bond dimension $\chi_{\max}=350$, and truncation tolerance $\rm{tol} = 10^{-7}$.}
    \label{fig_FvsN}
\end{figure}

Many-body scars represent a polynomial (in number) subset of special non-thermal eigenstates within the full many-body spectrum.
This implies that the overlap with the initially considered product states decreases as the system size increases.
To investigate this behavior, in \cref{fig_FvsN}(a), we plot the return fidelity between the evolved state and two initial scarring states: the polarized bare vacuum (PV) and the bare vacuum (V).
Exploiting the slow entanglement growth in the many-body scars regime, we considered system sizes ranging from $\Nsites=10$ to $\Nsites=50$.
Although the fidelity peaks decrease with increasing $\Nsites$ as expected, pronounced revivals are still visible even for large chains with $\Nsites=50$ sites.
The peaks become sharper, with the fidelity minima approaching zero, accordingly with the expected revivals induced by scars.
The maximum amplitude of these peaks follows an exponential scaling in the system size, $\mathcal{F}_\Nsites=f^{N}$, with the extrapolated peak fidelity per site being $f=0.97$ and $f=0.98$ for the PV and V states, respectively.
This estimate further demonstrates the robustness of the observed scar dynamics even at large system sizes.
% =========================================================================================
\subsection{Long-time dynamics}
\label{sec_scars_longtime}
\begin{figure}
    \includegraphics[width=1\textwidth]{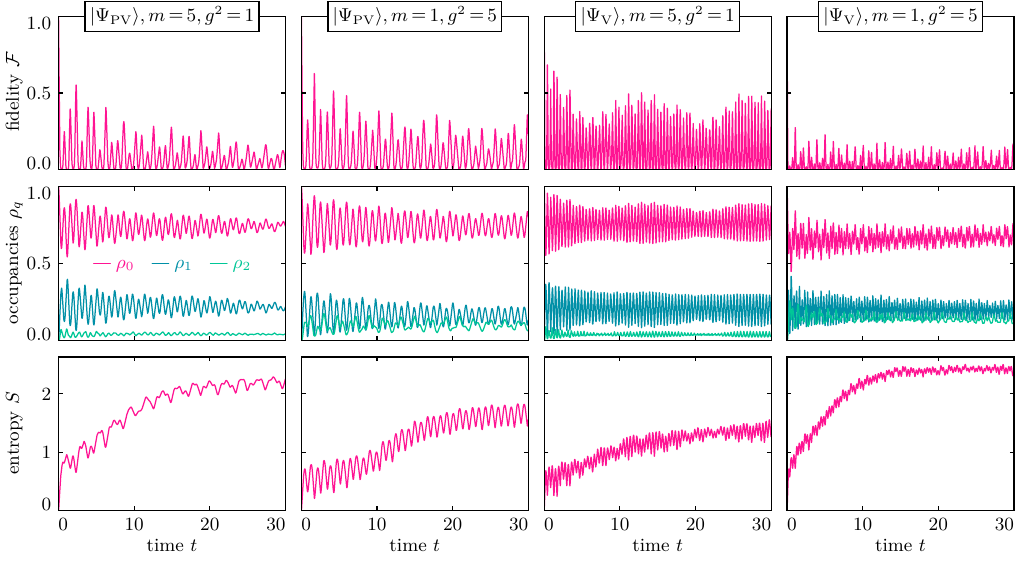}%
    \caption{\textit{Long time dynamics.} First to third row: return fidelity, average quark occupancy, and bipartite entanglement entropy as a function of time for the two initial states: polarized bare vacuum, and bare vacuum.
    We considered different values of mass and coupling as indicated in the Figure. The simulation was performed with MPS for $\Nsites=30$ sites by setting open boundary conditions, maximum bond dimension $\chi_{\max}=350$, and truncation tolerance  $\rm{tol} = 10^{-7}$.}
    \label{fig_Flongt}
\end{figure}

To illustrate how the observed non-ergodic dynamics persist over longer times, we conducted simulations for an extended duration.
In \cref{fig_Flongt}, we present the same dynamics presented in \cref{fig_scars_dynamics}, now up to $t=30$, nearly ten times longer than previously shown.
At this extended duration, we continue to observe persistent revivals and oscillations in both the return fidelity and occupancy.
Moreover, in addition to the rapid oscillations with a period $T<1$ previously discussed, we observe a beating with frequency $T\sim 1$ attributed to the non-uniform distribution of scars in energy spacing.
Finally, we still observe a slow growth in entanglement at such extended times, consistent with dynamics constrained to the many-body scars subset of the spectrum.
% =========================================================================================
\section{Summary}
In this chapter, we reviewed the foundational concepts of thermalization in isolated quantum many-body (QMB) systems. 
We discussed how quantum thermalization is governed by the Eigenstate Thermalization Hypothesis (ETH) \cite{Deutsch2018EigenstateThermalizationHypothesis}, which connects the equilibration over time to the dephasing mechanism of the spectrum eigenstates, as seen in a wide variety of models including bosonic, fermionic, and spin systems.

After this introduction, we highlighted the emergence of weak ergodicity breaking in recent studies \cite{Bernien2017ProbingManybodyDynamics,Bluvstein2021ControllingQuantumManybody,Serbyn2021QuantumManybodyScars}, where a novel class of special non-thermal eigenstates, known as quantum many-body scars (QMBS), was identified. These scarred eigenstates exhibit anomalously low entanglement entropy and result in the system displaying long-lived oscillations and persistent revivals in quench dynamics, effectively violating the predictions of ETH.

This behavior was first observed experimentally in systems of Rydberg atoms, where scarred dynamics manifest in the form of coherent oscillations in local observables and revivals in state fidelity. 
Furthermore, we explored how this phenomenon extends to LGTs, particularly in U(1) models, where QMBS arise in both spin-1/2 \cite{Surace2020LatticeGaugeTheories} and spin-S formulations \cite{Desaules2023WeakErgodicityBreaking,Desaules2023ProminentQuantumManybody,Halimeh2023RobustQuantumManybody,Hudomal2022DrivingQuantumManybody,Osborne2024QuantumManyBodyScarring}. 
These Abelian gauge theories show robust scarring dynamics, tied to the stability of the underlying gauge symmetry. 
Scars in these systems are deeply connected to the gauge structure and reveal an interplay between the gauge constraints and non-thermal eigenstates.

The original part was the extension of these results to the non-Abelian scenario.
In detail, we presented strong numerical evidence for the occurrence of QMB scars in a non-Abelian SU(2) lattice gauge theory with dynamical matter \cite{Cataldi*2025QuantumManybodyScarring}.
In particular, starting from simple experimentally friendly vacuum product states, scarred dynamics manifests as persistent oscillations in local observables and periodic revivals in the state fidelity. 
This dynamics arises from a state transfer through meson and baryon-antibaryon bare states, which allow the system to escape otherwise expected thermalization.

Even though the results were obtained for a non-Abelian lattice gauge theory model with the gauge-link truncated at the lowest nontrivial irreducible representations, we provided evidence that the observed scar dynamics persists even at higher gauge-link truncations, similar to what has recently been observed for Abelian models \cite{Desaules2023WeakErgodicityBreaking}, hinting at the possible survival of QMBS in the Kogut-Susskind limit of the SU(2) model under consideration.
These findings demonstrate that the strong connection between QMBS and LGTs persists even for non-Abelian gauge groups, raising the intriguing question of whether such behavior represents an intrinsic feature of more fundamental theories such as quantum chromodynamics.

All the results we presented open the door to exciting future directions.
Given the numerical evidence of scar eigenstates with very low bipartite entanglement entropy throughout the entire spectrum, even in regimes where both $\mass$ and $\coupling$ are quite small, it would be interesting to find simple product states showing large overlap with these eigenstates. 
Additionally, it would be interesting to map the scarring phase diagram for this model as has been done for the PXP model \cite{Daniel2023BridgingQuantumCriticality}, and see how driving protocols can enhance non-Abelian scarring dynamics \cite{Hudomal2022DrivingQuantumManybody}. 
Another venue to pursue is exploring the robustness of the non-Abelian scarring uncovered here in two spatial dimensions, as it has been recently done for Abelian gauge theories \cite{Osborne2024QuantumManyBodyScarring,Budde2024QuantumManyBodyScars}. 
This can shed further light on the nature of this non-Abelian scarring and its possible occurrence in the QFT limit. 
Furthermore, scarring is known to give rise to complex periodicities in the corresponding dynamical quantum phase transitions of Abelian LGTs \cite{VanDamme2022DynamicalQuantumPhase,VanDamme2023AnatomyDynamicalQuantum}, and it would be interesting to see if this carries over to the non-Abelian case.

Finally, we stress that, as addressed in \cite{Calajo2024DigitalQuantumSimulation}, the qudit formulation of the proposed model is well-suited for conducting a digital quantum simulation on a recently demonstrated trapped-ion qudit quantum processor \cite{Ringbauer2022UniversalQuditQuantum}. 
This opens up the intriguing possibility to experimentally observe this exotic non-ergodic dynamics in the near future.

%% file: chapters/hubbard.tex
\chapter{Simulating Lattice Fermion Theories}
\label{chap_defermionization_lattice_fermion}
The quest for quantum simulation of interacting fermionic models \cite{Fradkin2013FieldTheoriesCondensed,Auerbach1994InteractingElectronsQuantum,Giamarchi2003QuantumPhysicsOne} is necessary to reach a novel understanding of collective phenomena both at low and high energies \cite{Lee2006DopingMottInsulator,Hartke2023DirectObservationNonlocal,Halperin2020FractionalQuantumHall,Hemery2023MeasuringLoschmidtAmplitude,Hofrichter2016DirectProbingMott}, but it is hindered by fundamental challenges \cite{Gattringer2016ApproachesSignProblem,Troyer2005ComputationalComplexityFundamental,Altman2021QuantumSimulatorsArchitectures}. 
While analog quantum simulation, e.g.~with optical lattices, has advanced greatly in recent decades \cite{Esslinger2010FermiHubbardPhysicsAtoms,Bloch2012QuantumSimulationsUltracold,Tarruell2018QuantumSimulationHubbard, Bohrdt2021ExplorationDopedQuantum,Cheuk2015QuantumGasMicroscopeFermionic,Duarte2015CompressibilityFermionicMott,Edge2015ImagingAddressingIndividual,Greif2013ShortRangeQuantumMagnetism,Haller2015SingleatomImagingFermions,Hart2015ObservationAntiferromagneticCorrelations,Hofstetter2002HighTemperatureSuperfluidityFermionic,Jordens2008MottInsulatorFermionic,Messer2015ExploringCompetingDensity,Murmann2015AntiferromagneticHeisenbergSpin,Omran2015MicroscopicObservationPauli,Parsons2015SiteResolvedImagingFermionic,Schneider2008MetallicInsulatingPhases,Taie2012SUMottInsulator,Uehlinger2013ArtificialGrapheneTunable}, it presents limitations in tailoring exotic interactions. 
And while fermionic digital quantum processors are still at an early stage of development \cite{Zache2023QuantumClassicalSpin,Gonzalez-Cuadra2023FermionicQuantumProcessing}, the well-established \emph{conventional} digital quantum simulation platforms (e.g.~superconducting qubits, trapped ions, Rydberg arrays, quantum dots) \cite{Arute2020ObservationSeparatedDynamics,Barends2015DigitalQuantumSimulation,Salathe2015DigitalQuantumSimulation,OMalley2016ScalableQuantumSimulation,Stanisic2022ObservingGroundstateProperties} are built on distinguishable, spatially localized qubits (qudits).
In this framework, a \emph{Fermion Encoding} is the analytical process to exactly convert a fermionic algebra (mutually-anticommuting operations) into a genuinely local algebra (mutually-commuting operations) of qudits.

Such encoding can not be carried out free of cost. 
Traditional strategies focused on encoding $\Nsites$ Dirac orbitals into $\Nsites$ qubits. 
By construction, these strategies can not preserve locality and, as a result, end up encoding the ubiquitous two-body interactions into cumbersome $W$-body interactions (with $W$ often addressed as \emph{Pauli weight}).
Alongside recent efforts, which were able to reduce $W$ from linear to logarithmic in $\Nsites$ \cite{Jordan1928UeberPaulischeAequivalenzverbot,Bravyi2002FermionicQuantumComputation}, a separate sector of strategies arose: encodings attempting to preserve locality \cite{Fradkin1980FermionRepresentationLattice,Srednicki1980HiddenFermionsTheories,Verstraete2005MappingLocalHamiltonians,Chen2018ExactBosonizationTwo}. 
They exhibit a flat Pauli weight ($W$ does not scale with $\Nsites$) at the price of requiring a number of qubits larger (but still linear scaling) than the number $\Nsites$ of fermion orbitals.
For non-gauge lattice fermion theories, Ref. \cite{Kitaev2006AnyonsExactlySolved} showed that it is sufficient to add one qubit for each lattice bond to achieve local encoding (a generalization to lattice gauge fermionic theories was also recently introduced \cite{Zohar2018EliminatingFermionicMatter,Zohar2019RemovingStaggeredFermionic}). 
The extra qubits play the role of an effective, discrete lattice gauge field with pure gauge dynamics akin to the Toric Code Hamiltonian \cite{Kitaev2003FaulttolerantQuantumComputation}. 
The formal mapping is well understood \cite{Chen2023EquivalenceFermiontoQubitMappings}, but evidence of practical feasibility of the local encoding for numerical simulation or digital quantum simulation is still the subject of active research \cite{Pardo2023ResourceefficientQuantumSimulation,Irmejs2023QuantumSimulationMathbb}.

\begin{figure}
    \centering
    \includegraphics[width=\textwidth]{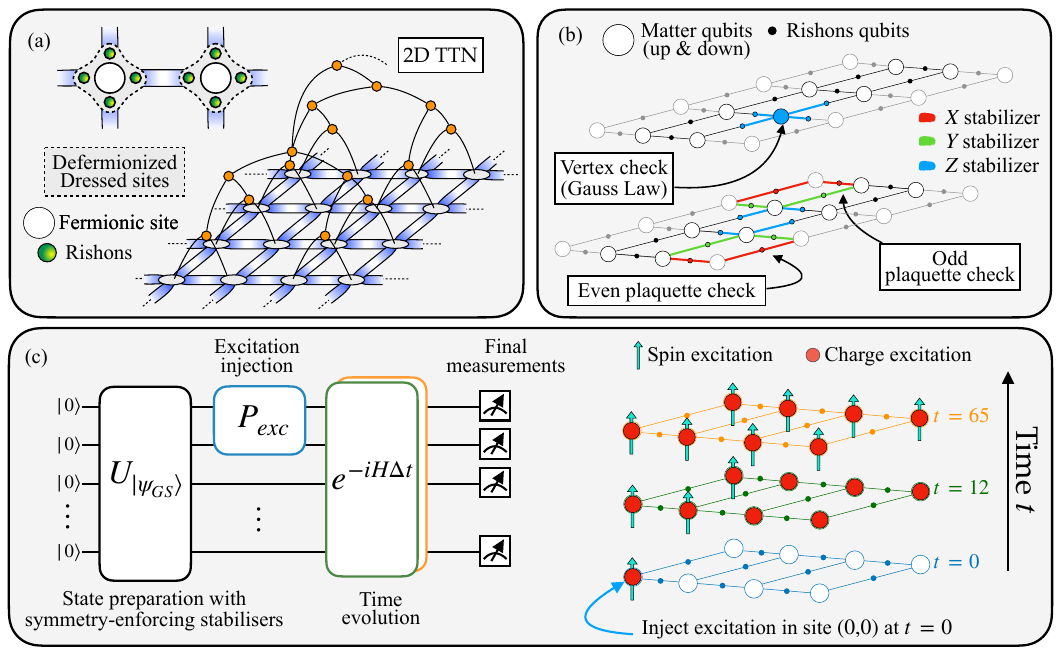}
    \caption{Simulation of the Hubbard model in 2D with tensor networks (equilibrium) and quantum circuits (out-of-equilibrium). (a) Schematic of the tree-tensor network (TTN) installed on the 2D lattice: the fermionic degrees of freedom are removed with the gauge defermionization and encoded in a dressed site (dashed closed line) comprising fermionic matter and rishons. With TTN, the ground state of the defermionized model is computed for up to $4\times4$ lattice. (b) To run the digital quantum simulation, dressed sites are decomposed into qubits and Hamiltonian terms in Pauli strings. The symmetries of the system are here enforced through the stabilizer formalism. (c) (left) Quantum circuit for simulating out-of-equilibrium dynamics: adiabatic preparation of the initial state (half-filling, repulsive $U$), injection of charge (spin) excitation, and time evolution with Hubbard Hamiltonian. (right) Spin-charge dynamics schematic: the charge excitation propagates faster than the spin one.}
    \label{fig_graph_abstract}
\end{figure}

This chapter is based on the twofold study of Ref. \cite{Cataldi*2024DigitalQuantumSimulation} on the feasibility of the local fermion encoding for classical and quantum simulation, as summarized in \cref{fig_graph_abstract}: on one hand side, we investigate its performance for ground-state numerical simulations based on tensor network (TN) ansatz states. 
On the other side, we numerically test its applicability and scalability for digital quantum simulation on a generic platform (considering standard 1-qubit and 2-qubit programmable dynamical resources), by numerically emulating the noiseless quantum simulation processing in real-time.
Equipped with these methods, we investigate the equilibrium and out-of-equilibrium properties of the spin-$\frac{1}{2}$ Hubbard model on a two-dimensional square lattice, with TN simulations. 
At zero temperature, we can access system sizes that allow us to identify the transition between liquid and insulating (crystalline spin-lattice) phases. 
In real-time dynamics, for a ladder in the $t-J$ model limit, we observe spin and charge time-evolution displaying distinct time-scales, an effect that in 1D Hubbard chains is a precursor of the spin-charge separation phenomenon, see for example the recent experiments \cite{Hilker2017RevealingHiddenAntiferromagnetic,Arute2020ObservationSeparatedDynamics,Vijayan2020TimeresolvedObservationSpincharge}, whereas in 2D it only governs short time scales before the polaron picture and strongly-correlated effects set in \cite{Ji2021CouplingMobileHole,Koepsell2021MicroscopicEvolutionDoped,Bohrdt2021ExplorationDopedQuantum}. 

The chapter is structured as follows: in \cref{sec_defermion}, we revisit the gauge-based encoding to locally remove the fermionic matter, or gauge defermionization. 
Such an approach can be seen as a generalization of the dressed-site formalism developed in \cref{sec_dressed_site_formalism}, as in this case there are no gauge fields.
We devote \cref{sec_hubbard_model} to the implementation of the encoding on a tree-tensor network (TTN) ansatz and show ground state simulation results of the 2D Hubbard model. 
Finally, in \cref{sec_def_digitalqs}, we detail our prescription for digital quantum simulation of the 2D Hubbard model, based on gauge defermionization, and numerically emulate the digital quantum simulation to observe spin-charge separation effects in two-dimensional lattice samples.
% ==========================================================================================
\section{Gauge Defermionization of Lattice Fermion Theories}
\label{sec_defermion}
In this section, we revisit and extend a known technique, based on lattice gauge theories, to eliminate fermionic matter from 2D lattice models \cite{Zohar2018EliminatingFermionicMatter,Zohar2019RemovingStaggeredFermionic}.
The strategy consists of developing an analytical mapping from an input fermion lattice Hamiltonian (here we consider it to be a pure non-gauge fermion theory) to a lattice gauge Hamiltonian, with a $\mathbb{Z_2}$ gauge symmetry, equipped with appropriate gauge constraints. 
Then, by applying the dressed-site formalism developed in \cref{sec_dressed_site_formalism}, the resulting theory can be manipulated so that the fermionic parity at each (dressed) site is protected, thus resulting in a theory where operator algebras at different sites always commute, as in a local spin theory.

Our formulation of the defermionization technique can be applied to any 2D fermion lattice Hamiltonian, regardless of the lattice system (it works on Honeycomb and Kagome lattices as well as irregular lattices), as long as few general constraints are satisfied:
\emph{(i)} the global parity of fermions is protected;
\emph{(ii)} Hamiltonian terms that flip the local fermion parity must be nearest-neighbor;
\emph{(iii)} the full system must be under open boundary conditions (OBC), \idest{} every closed path on the lattice must be topologically shrinkable to a point.
These conditions are often satisfied in solid state Hamiltonians and cold atoms systems \cite{Lieb1968AbsenceMottTransition}

Therefore, without any loss of generality, a general fermion Hamiltonian on a two-dimensional lattice $\Lambda$ with $\Nsites=\Nsites_{x}\times \Nsites_{y}$ sites $\vecsite$ reads:
\begin{equation}
\label{eq_def_originalhamiltonian}
\begin{split}
    \ham_{0}=& \sum_{\vecsite, k}\sum_{\alpha,\beta}\qty[
    \qty[h_{\alpha,\beta}\hpsi^{\dagger}_{\vecsite,\alpha} \hpsi_{\vecsite+\latvec[k],\beta}+\hc]+ 
    \qty[\Delta_{\alpha,\beta} \, \hpsi_{\vecsite,\alpha} \hpsi_{\vecsite+\latvec[k],\beta}+\hc]]
    + V\qty( \{ \nop_{\vecsite,\alpha} \}),
\end{split}
\end{equation}
where the fermion fields satisfy the usual Dirac anti-commutation relations
\begin{align}
    \qty{ \hpsi^{\dagger}_{\vecsite,\alpha}, \hpsi_{\vecsite^{\prime},\beta}} &= i\hbar\delta_{\vecsite,\vecsite^{\prime}} \delta_{\alpha,\beta}&
    \qty{ \hpsi_{\vecsite,\alpha}, \hpsi_{\vecsite^{\prime}, \beta}} &= 0,
\end{align}
and can exhibit an internal degree of freedom, here labeled by the flavor indices $\alpha,\beta$. 

The interaction potential $V$ term, which is a function of the fermion densities $\nop_{\vecsite,\alpha} = \hpsi^{\dagger}_{\vecsite,\alpha} \hpsi_{\vecsite,\alpha}$, can in principle have any range and shape, and even include chemical potentials.
For instance, it can also account local terms that do not conserve the spin, like $b_{\alpha,\beta}\hpsi^\dagger_{\vecsite,\alpha}\hpsi_{\vecsite,\beta}$, or local s-wave superconducting terms as $\Delta\hpsi^\dagger_{\vecsite,\alpha}\hpsi^\dagger_{\vecsite,\beta}$.

Conversely, the hopping $h_{\alpha,\beta}$ and the double creation/annihilation $\Delta_{\alpha,\beta}$ processes (the latter often seen in superconductors) are the terms that break local fermion parity: their single-site algebras do not commute between distant sites (\idest{} they are not \emph{genuinely} local), and this is the source of all the numerical difficulty when simulating lattice fermions.

While in one spatial dimension, the Jordan-Wigner transformation \cite{Jordan1928UeberPaulischeAequivalenzverbot} provides an easy solution to this problem, tackling the mapping in higher dimensions requires sophisticated, often cumbersome techniques \cite{Verstraete2005MappingLocalHamiltonians,Corboz2010SimulationStronglyCorrelated}. 
In this perspective, converting a fermion algebra into a genuinely local (spin-like) algebra using introducing a lattice gauge field is an elegant strategy, which is also practical from a numerical simulation standpoint, as we discuss later on.
% ==========================================================================================
\subsection{Mapping into a \texorpdfstring{$\mathbb{Z}_2$}{Z2} gauge theory}
We now perform a set of exact algebraic manipulations to the Hamiltonian of \cref{eq_def_originalhamiltonian} until we reach a defermionized form. 
The very first step is to promote the total fermion parity operator at site $\vecsite$ to a gauge transformation (for the matter sites $M$) that behaves like a parity operator, or an element of the $\mathbb{Z}_2$ group: 
\begin{equation}
    G_{\vecsite}^{[M]} 
    = \rm{exp}\qty(i \pi {\textstyle{\sum_{\alpha}}}\hpsi^{\dagger}_{\vecsite,\alpha} \hpsi_{\vecsite,\alpha})
    = G_{\vecsite}^{[M] \dagger} = (G_{\vecsite}^{[M]})^{-1}.
\end{equation}
Under $G_{\vecsite}^{[M]}$, operators that preserve total fermion parity, such as densities $\nop_{\vecsite, \alpha}$, are left invariant, while operators that flip the total fermion parity, such as $\hpsi_{\vecsite,\alpha}^{(\dagger)}$, acquire a sign. 
Namely:
\begin{align}
    G_{\vecsite}^{[M]} \nop_{\vecsite,\alpha} G_{\vecsite}^{[M]} &= \nop_{\vecsite,\alpha}&
    G_{\vecsite}^{[M]} \hpsi_{\vecsite,\alpha} G_{\vecsite}^{[M]}&= - \hpsi_{\vecsite,\alpha}
\end{align}
At this stage, we promote the gauge transformation into a gauge symmetry. 
This task is performed by adding an auxiliary quantum lattice field, the gauge field, on the \emph{bonds} $(\vecsite,\latvec)$ of the lattice. 
The local Hilbert space for a gauge site should correspond to the regular representation space for the $\mathbb{Z}_{2}$ group \cite{Zohar2015FormulationLatticeGauge}, so it should be a two-level system, or a qubit, equipped with the (genuinely local) algebra of Pauli matrices $\qty{\Sx,\Sy,\Sz}$. 

Being $\mathbb{Z}_{2}$ an Abelian gauge group, the left- $G_{\vecsite,\latvec}^{[L]}$ and right- $G_{\vecsite,\latvec}^{[R]}$ groups of transformations of the gauge field must commute and square to the identity.
Without loss of generality, we can set them both to the same operator: $G_{\vecsite,\latvec}^{[L]} = G_{\vecsite,\latvec}^{[R]} = \Sz_{\vecsite,\latvec}$.
Thanks to this auxiliary field, we equip the gauge-violating terms of the Hamiltonian with a parallel transporter operator. 
We replace
\begin{equation}
    \hpsi^{\dagger}_{\vecsite,\alpha} \hpsi_{\vecsite+\latvec,\beta}\quad \longrightarrow \quad
    \hpsi^{\dagger}_{\vecsite,\alpha} \Apara_{\vecsite,\latvec} \hpsi_{\vecsite+\latvec,\beta},
\end{equation}
that transforms covariantly under the gauge field groups: 
\begin{equation}
    G_{\vecsite,\latvec}^{[L]}  \Apara_{\vecsite,\latvec} G_{\vecsite,\latvec}^{[L]}{=}
    G_{\vecsite,\latvec}^{[R]}  \Apara_{\vecsite,\latvec} G_{\vecsite,\latvec}^{[R]}{=}- \Apara_{\vecsite,\latvec}
\end{equation}
Again, without loss of generality, we set $\Apara_{\vecsite,\latvec} = \Sx_{\vecsite,\latvec}$.
Overall, this procedure translates into modifying the fermion parity-flipping terms from \cref{eq_def_originalhamiltonian} according to
\begin{equation} 
\label{eq_def_gauge_modification}
  \begin{aligned}
   h_{\alpha,\beta} \, \hpsi^{\dagger}_{\vecsite,\alpha} \hpsi_{\vecsite+\latvec,\beta}
   \quad &\longrightarrow \quad
   h_{\alpha,\beta} \, \hpsi^{\dagger}_{\vecsite,\alpha} \Sx_{\vecsite,\latvec} \hpsi_{\vecsite+\latvec,\beta}
\\
   \Delta_{\alpha,\beta} \, \hpsi_{\vecsite,\alpha} \hpsi_{\vecsite+\latvec,\beta}
   \quad &\longrightarrow \quad
   \Delta_{\alpha,\beta} \, \hpsi_{\vecsite,\alpha} \Sx_{\vecsite,\latvec} \hpsi_{\vecsite+\latvec,\beta}\, .
  \end{aligned}
\end{equation}
With this modification, the Hamiltonian satisfies a gauge symmetry around every site,
precisely $G_{\vecsite}^{[\rm{total}]} = G_{\vecsite}^{[M]} \prod_{\latvec} G_{\vecsite,\latvec}^{[L/R]}$. Consequently, we have to choose a symmetry sector (for each of these symmetries) to play the role of physical space. 
As it is common for lattice gauge theories, we consider the quantum states \emph{invariant} under all the $G_{\vecsite}^{[\rm{total}]}$, \idest{} which satisfy
\begin{equation} 
    \label{eq_def_physpace}
 \qty[\exp \left(i \pi \sum_{\alpha} \hpsi^{\dagger}_{\vecsite,\alpha} \hpsi_{\vecsite,\alpha}\right)
 \prod_{\latvec} \Sz_{\vecsite,\latvec}]\ket{\Psi_{\rm{phys}}}{=}
 \ket{\Psi_{\rm{phys}}},
\end{equation}
in order to be the physical states $\ket{\Psi_{\rm{phys}}}$. 
This equation plays the role of effective Gauss' Law of the resulting gauge theory, and the condition can be satisfied at every site \emph{only} if the total number of fermions in the system is even. To simulate an odd number of fermions, one or more virtual bonds, going out of the system, must be added to keep track of the appropriate parity gauge flux.
% ==========================================================================================
\subsection{Restoring equivalence with Plaquette Operators}
It is then possible to check that, under the Gauss' Law of \cref{eq_def_physpace}, the gauge-theory modification in \cref{eq_def_gauge_modification} does not change the actual dynamics given by the original Hamiltonian as in \cref{eq_def_originalhamiltonian}, \emph{but only} when the lattice is a non-cyclic graph (such as a 1D chain, or a dendrimer lattice). For each (product state) fermion configuration, the extra degrees of freedom introduced by the gauge fields are completely locked by Gauss' Law; thus all residual degeneracy is removed, and all the matrix elements of $\ham_{0}$ are unaltered.

Conversely, when lattice cycles are present, the modification changes the dynamics.
In fact, for each elementary closed cycle of sites, or \emph{plaquette} $\square\in \Lambda$, one Gauss' Law operator becomes linearly dependent on others, thus contributing to degeneracy with a 2-fold space. 
Moreover, when a fermion winds around a closed cycle, all of the gauge fields in its path undergo a $\Sx$ flip, thus the final state may differ from the original one.
If we want to restore the dynamics of the original model, we have to add more physical contents to the plaquette, either in terms of an additional \emph{pure gauge field} Hamiltonian or in terms of \emph{extra symmetries}.

Fortunately, a simple recipe to do either is inspired by Kitaev's Toric Code \cite{Kitaev2003FaulttolerantQuantumComputation}.
The idea is to add a plaquette Hamiltonian term with a $\Sx$ for each gauge link of the plaquette, that is 
\begin{equation}
\label{eq_def_plaquette_term}
\begin{split}
    \ham_{p} &= - \sum_{\square \in\Lambda}\prod_{\langle\vecsite,\vecsite^{\prime} \rangle \in \square}\Sx_{\vecsite,\vecsite^{\prime}}=- \sum_{\square \in\Lambda}    \qty(\begin{array}{ccc}
    \ulcorner& \Sx & \urcorner\\
    \Sx & & \Sx \\
    \llcorner&  \Sx   & \lrcorner\\
    \end{array}).
    \end{split}
\end{equation} 
Notably, each of these plaquette operators commutes with the lattice gauge Hamiltonian and can be equivalently cast as a (gauge) symmetry. 
Moreover, plaquettes also commute with Gauss' laws, because a vertex and a plaquette always share either zero or two bonds, regardless of the lattice system, and $[\Sx_1 \Sx_2, \Sz_1 \Sz_2] = 0$. 

We can then map the original lattice fermion Hamiltonian, \cref{eq_def_originalhamiltonian}, to a dynamically-equivalent, $\mathbb{Z}_2$-invariant lattice gauge model, which reads
\begin{equation} 
    \label{eq_def_torichamiltonian}
  \begin{split}
   \ham_{1}= \sum_{\vecsite,k}\sum_{\alpha,\beta}
   \qty[h_{\alpha,\beta}\hpsi^{\dagger}_{\vecsite,\alpha}\Sx_{\vecsite,\latvec} \hpsi_{\vecsite+\latvec,\beta}
   +\Delta_{\alpha,\beta} \hpsi_{\vecsite,\alpha} \Sx_{\vecsite,\latvec}  \hpsi_{\vecsite+\latvec,\beta} \hc]
   + V\qty(\{ \nop_{\vecsite,\alpha} \}),
  \end{split}
\end{equation}
with the constraints
\begin{align}
    \label{eq_def_allgauges}
    \qty[e^{i \pi \sum_{\alpha} \nop_{\vecsite,\alpha}}
    \prod_{\latvec}^{\rm{vertex}} \Sz_{\vecsite,\latvec}]\ket{\Psi_{\rm{phys}}} &= \ket{\Psi_{\rm{phys}}},& 
    \qty[\prod_{\avg{\vecsite,\vecsite^{\prime}}}^{\rm{plaquette}} \Sx_{\vecsite, \vecsite^{\prime}}]\ket{\Psi_{\rm{phys}}}&=\ket{\Psi_{\rm{phys}}}.
\end{align}
Indeed, the combined constraints in \cref{eq_def_allgauges} completely resolve the 2-fold degeneracy introduced by the plaquette, and each closed path of moving fermions returns to its original state with the correct amplitude and phase. Therefore, the whole mapping holds for the subspace of the Hilbert space satisfying \cref{eq_def_allgauges}. 
Any excitations from the ground state of \cref{eq_def_plaquette_term} break the encoding validity.
A rigorous proof of the equivalence between \cref{eq_def_originalhamiltonian} and \cref{eq_def_torichamiltonian,eq_def_allgauges} was provided for a number-conserving theory on the square lattice in \cite{Zohar2018EliminatingFermionicMatter}; the generalization to other cases is fairly straightforward.

Notice that the degeneracy is not fully removed in periodic boundary conditions (PBC), where each winding dimension introduces an additional closed cycle. This case requires the additional constraint of a full $\Sx$ string along each winding dimension, but we will not treat this case here, because the theory becomes non-local.
% ==========================================================================================
\subsection{Fermionic Rishons}
So far, it seems that we have increased the complexity of the fermion lattice theory. In this last step, we will locally manipulate the gauge fields and achieve defermionization explicitly, ending with an algebra of genuinely local operators only. 

Similarly to the rishon decomposition discussed in \cref{sec_SU2_rishondecomposition,sec_U1_rishondecomposition}, we split each gauge field, living on the $(\vecsite,\latvec)$ lattice link, in a pair of spinless fermion modes (rishons) $\cop_{\vecsite,\latvec}^{(\dagger)}$ and $\cop_{\vecsite+\latvec,-\latvec}^{(\dagger)}$, equipped with another symmetry. 
These new Dirac fermion operators belong to dressed sites $\vecsite$ and $\vecsite+\latvec$ respectively and satisfy the usual anti-commutation relations between themselves and with the physical matter fermions
\begin{align}
    \qty{c^{\dagger}_{\vecsite, \latvec}, \cop_{\vecsite^{\prime}, \latvec^{\prime}}} &= \delta_{\vecsite \vecsite^{\prime}} \delta_{\latvec \latvec^{\prime}}&
    \qty{\cop_{\vecsite, \latvec}, \cop_{\vecsite^{\prime}, \latvec^{\prime}}} &= 0&
    \qty{\hpsi_{\vecsite},\cop_{\vecsite^{\prime},\latvec}}&=0\,.
\end{align}
The combined 4-dimensional space of the two modes is then reduced back to the 2-dimensional space of a qubit by imposing that the total number of rishon fermions on a bond must be an even number. 
This constraint can again be cast as a \emph{link symmetry}, requiring that 
\begin{equation}
    \exp[i \pi( \cop_{\vecsite,\latvec}^{\dagger} \cop_{\vecsite,\latvec}{+}\cop_{\vecsite+\latvec,-\latvec}^{\dagger} \cop_{\vecsite+\latvec,-\latvec})]{\ket{\Psi_{\rm{phys}}}}{=}\ket{\Psi_{\rm{phys}}}.
\end{equation}
We then convert the gauge fields Pauli algebra into an operator algebra acting on the rishon pair: 
\begin{align}
    \label{eq_def_sigmasplit}
    \Sz_{\vecsite,\latvec} &\to 1 - 2 \cop_{\vecsite,\latvec}^{\dagger} \cop_{\vecsite,\latvec} = 1 - 2 \cop_{\vecsite+\latvec,-\latvec}^{\dagger} \cop_{\vecsite+\latvec,-\latvec}&
    \Sx_{\vecsite,\latvec} &\to \id \majo_{\vecsite,\latvec} \majo_{\vecsite+\latvec,-\latvec},
\end{align}
where now $\majo_{\vecsite,\latvec}{=}\cop_{\vecsite,\latvec}{+}c^{\dagger}_{\vecsite,\latvec}{=}\majo_{\vecsite,\latvec}^{\dagger}$ is a Majorana operator on the rishon fermion, which squares to the identity $\majo_{\vecsite,\latvec}^2{=}1$ but still anti-commutes with other fermions $\{\majo_{\vecsite,\latvec},\majo_{\vecsite^{\prime},\latvec^{\prime}}\}{=}2 \delta_{\vecsite, \vecsite^{\prime}} \delta_{\latvec,\latvec^{\prime}}$. 
These operators defined at \cref{eq_def_sigmasplit} respect the fermion parity symmetry on the bond, and on the even parity sector, they act exactly as Pauli matrices.

We can then plug this exact manipulation into the gauge theory Hamiltonian, resulting in
\begin{equation} 
\label{eq_def_Hamiltonian}
  \begin{split}
   \ham_{2}&{=} \sum_{\vecsite,k}\sum_{\alpha,\beta}
    \qty[i h_{\alpha,\beta}\hpsi^{\dagger}_{\vecsite,\alpha} \majo_{\vecsite,\latvec} \majo_{\vecsite+\latvec,-\latvec} \hpsi_{\vecsite+\latvec,\beta} 
    {+} i\Delta_{\alpha,\beta}\hpsi_{\vecsite,\alpha} \majo_{\vecsite,\latvec} \majo_{\vecsite+\latvec,-\latvec} \hpsi_{\vecsite+\latvec,\beta} +\hc]+ V\,,
  \end{split}
\end{equation}
where we have to add the new link parity symmetry to the constraints, thus
\begin{subequations} 
    \label{eq_trigauges}
\begin{align}
    &\qty[ e^{i \pi \sum_{\alpha} \nop_{\vecsite,\alpha}}
 \prod_{\latvec}^{\rm{vertex}} e^{i \pi c^{\dagger}_{\vecsite, \latvec} \cop_{\vecsite, \latvec}}]{\ket{\Psi_{\rm{phys}}}}=\ket{\Psi_{\rm{phys}}}\label{eq_def_vertex}\\
  &\qty[\prod_{\vecsite,\latvec}^{\rm{plaquette}} \id \majo_{\vecsite,\latvec} \majo_{\vecsite+\latvec,-\latvec}]\ket{\Psi_{\rm{phys}}} = \ket{\Psi_{\rm{phys}}} \label{eq_def_plaq}\\
 & e^{i \pi \cop_{\vecsite,\latvec}^{\dagger} \cop_{\vecsite,\latvec}} e^{i \pi \cop_{\vecsite+\latvec,-\latvec}^{\dagger} \cop_{\vecsite+\latvec,-\latvec} } \ket{\Psi_{\rm{phys}}} =  \ket{\Psi_{\rm{phys}}}\label{eq_def_link}\,,
  \end{align}
\end{subequations}
which must hold for every vertex, every plaquette, and every bond respectively.

We have finally reached the final form of our model: now, every term of the Hamiltonian and every vertex, link, or plaquette symmetry \emph{protects total fermion parity} at each dressed site, counting together matter and rishon fermions (for plaquettes, remember that each dressed site contributes with two rishon modes in a closed lattice path).
Therefore, it is possible to think of each dressed site as a large spin, and both Hamiltonian and extra symmetries can be written in terms of \emph{genuinely local} operator algebras, which commute on different dressed sites. The theory has been effectively defermionized.
% ==========================================================================================
\subsection{The price of defermionization}
Defermionization of the lattice model using LGT, \idest{} going from
\cref{eq_def_originalhamiltonian} to \cref{eq_def_Hamiltonian,eq_trigauges} is an exact mapping. And while it provides the clear benefits of eliminating fermionic operator algebras, it does not come free of costs.

First of all, the transformation increases the local space dimension, only to shrink the local dimension again once the gauge symmetries are installed. 
If the model has $\Nsites$ sites, $f$ flavors and coordination number $v$,
we increase the dimension from $2^{fN}$ to $2^{(f+v)N}$.

Secondly, we impose a new interaction term in the form of the plaquettes. 
While for typical 2D lattices plaquettes are rather small, it is still an interaction involving three to six sites depending on the lattice geometry. 
If the original interaction $V$ was on-site, or nearest-neighbor, the defermionization effectively increased the interaction supports.
Moreover, there are no more components in the Hamiltonian that are quadratic in the Fermi operators (except chemical potentials within $V$). This means that Green's function perturbative approaches, which start from the free Fermi gas propagator of the quadratic theory, are no longer viable.

Finally, introducing an auxiliary field whose pure gauge dynamics is analogous to a Toric Code Hamiltonian has the drawback of increasing the entanglement. 
Even for product states of fermions, where fermions are locked in a specific integer filling configuration, and the original entanglement is zero (e.g.~for a Jordan-Wigner encoding), adding a Toric Code field on top of that raises the entanglement entropy to an exact area-law \cite{Kitaev2009TopologicalPhasesQuantum}, carrying one e-bit of entanglement per plaquette that is cut by the bi-partition. 
This observation has substantial implications for TN simulations of said models, as we will discuss in detail later on.
% ==========================================================================================
\section{Hubbard Model, Defermionized for tensor networks}
\label{sec_hubbard_model}
Despite its Hamiltonian's apparent simplicity, the Hubbard model has eluded physicists for decades \cite{Hubbard1963ElectronCorrelationsNarrow,Hubbard1964ElectronCorrelationsNarrow,Hubbard1964ElectronCorrelationsNarrow-1,Arovas2022HubbardModel}. 
While exact solutions are available for both one-dimensional \cite{Lieb1968AbsenceMottTransition} and infinite-dimensional cases \cite{Metzner1989CorrelatedLatticeFermions,Muller-Hartmann1989CorrelatedFermionsLattice}, addressing finite-size systems in higher dimensions at arbitrary temperature has required the development of various computational techniques. 
Although Monte Carlo methods can handle substantially large systems, they suffer from the well-known sign problem when computing physically significant quantities within specific regimes \cite{Loh1990SignProblemNumerical,Li2015SolvingFermionSign,Gattringer2016ApproachesSignProblem}. 
Comprehensive reviews detailing and comparing the accomplished computational outcomes for the 2D case can be found in the following references \cite{Qin2022HubbardModelComputational,LeBlanc2015SolutionsTwoDimensionalHubbard}.

In this section, we join the effort by performing TN simulations of the Hubbard model at equilibrium at zero temperature. 
We start reviewing the model and manipulating its defermionized formulation to be ready for a numerical approach.
Subsequently, we focus on a 2D \emph{square lattice} geometry, fermionic matter with 2 flavors (e.g. $\uparrow$ and $\downarrow$), and open boundary conditions.
On a rectangle lattice $\Lambda$ of $\Nsites=\Nsites_{x}\times \Nsites_{y}$ sites, the Hubbard Hamiltonian reads:
\begin{equation} 
\label{eq_Hubbard}
\begin{split}
   \ham_{\rm{Hub}} =& 
   -t\sum_{\vecsite,k}\sum_{\alpha=\uparrow,\downarrow}
    \qty[\hpsi^{\dagger}_{\vecsite,\alpha} \hpsi_{\vecsite+\latvec[k],\alpha} +\hc]
    + U \sum_{\vecsite} \qty(\nop_{\vecsite,\uparrow} -\frac{1}{2})
    \qty( \nop_{\vecsite,\downarrow} -\frac{1}{2}),
  \end{split}
\end{equation}
where, apart from an energy rescaling, its ground state properties only depend on the dimensionless parameter $U/t$. 
The Hubbard model is regarded as the simplest theory of strongly correlated electrons, where band electrons interact via a two-body repulsive Coulomb interaction \cite{Fradkin2013FieldTheoriesCondensed}. 
This model enables the description of a wide range of phenomena including metal-insulator transitions, superconductivity, and magnetism \cite{Giamarchi2003QuantumPhysicsOne}.  
The Hubbard Hamiltonian in \cref{eq_Hubbard}  belongs to the Hamiltonian class of \cref{eq_def_originalhamiltonian}, with flavor-transparent hopping terms ($h_{\alpha,\beta} = -t \,\delta_{\alpha,\beta}$) and no pair creation processes ($\Delta_{\alpha,\beta} = 0$), so it can be readily defermionized.

The form of \cref{eq_Hubbard} includes the $- \frac{1}{2}$ terms in the interaction component, which are equivalent to setting a specific uniform chemical potential, which explicitly reveals the rich symmetry content of the Hubbard model. 
In fact, besides the obvious lattice symmetries (translation by 1 site in $x$ and $y$, $ \frac{\pi}{2}$ rotations, vertical and horizontal reflection, and compositions thereof) Hubbard dynamics exhibits two useful \emph{glocal} symmetries, where the transformation is global but comprised of separate single-site non-Abelian operations.

The first symmetry, an SU(2) group, represents rotational invariance in the flavors. 
Since the hopping is flavor-transparent and the interaction is based on double occupancy, the model has to be flavor-invariant. 
This symmetry is a Lie group, generated by the operator algebra
\begin{align}
    \hat{S}_{\rm{tot}}^{a} &= \sum_{\vecsite} \hat{S}^{a}_{\vecsite},& 
    \rm{where}&&
    \hat{S}^{a}_{\vecsite} &= \frac{1}{2} \sum_{\alpha,\beta}\hat{\sigma}^{a}_{\alpha,\beta} \hpsi^{\dagger}_{\vecsite,\alpha} \hpsi_{\vecsite,\beta}
\end{align}
are the single-site flavor (spin) operators previously introduced in \cref{eq_SU2_gausslaw}, acting non-trivially only on the singly-occupied sites.

The second symmetry, assembling total fermion number conservation and particle-hole inversion, is an O(2) group, the symmetry group of the circle (rotations of a scalar angle plus one reflection). 
Despite being a continuous group, it is not Lie, but any element can be written as a rotation $\hat{R}_{\rm{tot}}(\theta)$ eventually followed by a reflection $\hat{F}_{\rm{tot}}$:
\begin{align}
    \hat{R}_{\rm{tot}}(\theta)&= \prod_{\vecsite} \hat{R}_{\vecsite}(\theta)=\prod_{\vecsite}e^{i \theta (\nop_{\vecsite,\uparrow} + \nop_{\vecsite,\downarrow})} &
    \hat{F}_{\rm{tot}} &= \prod_{\vecsite} \hat{F}_{\vecsite}\,.
\end{align} 
It is indeed possible to write the particle-hole transformation $\hat{F}$ in a way that commutes with the flavor-rotations $\vec{S}$, thus the two symmetries are independent. 
By doing so, we get
\begin{align} 
\label{eq_trigaugesagain}
    \hat{R}_{\vecsite}(\theta)\hpsi_{\vecsite,\alpha}\hat{R}_{\vecsite}^{\dagger}(\theta) &= e^{-i \theta} \hpsi_{\vecsite,\alpha}&
    \hat{F}_{\vecsite}\hpsi_{\vecsite,\alpha}\hat{F}_{\vecsite}^{\dagger} &=  (-1)^{\vecsite} \sum_{\beta} \Sy_{\alpha,\beta} \hpsi_{\vecsite,\beta}^{\dagger}\,.
\end{align}
Other symmetries, such as the pseudospin conservation \cite{Lieb1989TwoTheoremsHubbard,Arovas2022HubbardModel}, are indeed present, but not practically useful for our numerical simulation purposes.

When defermionizing the Hubbard model for TN simulation, we find it convenient to enforce the \emph{vertex gauge constraint} as an exact, actual symmetry. 
By the prescription discussed in Ref.~\cite{Silvi2014LatticeGaugeTensor}, this constraint will allow us to select a reduced canonical basis for the dressed site, made by only and all the vertex-gauge invariant states.
Conversely, the \emph{link constraint} and especially the \emph{plaquette constraint} are cumbersome to treat as exact symmetries (their local algebras do not commute) can be equivalently included in the Hamiltonian as penalty terms, increasing the energy of gauge symmetry-violating sectors. 
In conclusion, we have
\begin{equation} 
\label{eq_HubbarDef_pt1}
  \begin{split}
   \ham^{\prime}_{\rm{Hub}} &= 
   -t\sum_{\vecsite,k}\sum_{\alpha}\qty[i \hpsi^{\dagger}_{\vecsite,\alpha} \majo_{\vecsite,\latvec[k]} \majo_{\vecsite+\latvec[k],-\latvec[k]} \hpsi_{\vecsite+\latvec[k], \alpha}]
    + U \sum_{\vecsite}\qty(\nop_{\vecsite, \uparrow} -\frac{1}{2})
    \qty(\nop_{\vecsite, \downarrow} -\frac{1}{2}),
  \end{split}
\end{equation}
plus a penalty
\begin{equation}
\label{eq_HubbarDef_pt2}
    \begin{split}
    \ham^{\prime}_{\rm{pen}}&= 
    -\alpha_{b}\sum_{\vecsite,k} 
    \qty[(-1)^{c^{\dagger}_{\vecsite,\latvec[k]}\cop_{\vecsite,\latvec[k]} + c^{\dagger}_{\vecsite+\latvec[k],-\latvec[k]} \cop_{\vecsite+\latvec[k],-\latvec[k]}} -1]
    -\alpha_p
    \sum_{\square\in \Lambda} 
    \qty[\begin{array}{@{\hskip 0.05cm}c@{\hskip 0.05cm}c@{\hskip 0.05cm}c@{\hskip 0.05cm}}
    \ulcorner& \majo_{+\latvec[x]} \majo_{-\latvec[x]} & \urcorner\\
    \majo_{-\latvec[y]} & & \majo_{-\latvec[y]}  \\
    \majo_{+\latvec[y]} & & \majo_{+\latvec[y]} \\
    \llcorner&  \majo_{+\latvec[x]} \majo_{-\latvec[x]}  & \lrcorner\\
    \end{array}-1],
    \end{split}
\end{equation}
with separate penalty couplings $\alpha_p > 0$ and $\alpha_b > 0$ for the plaquette and bond violations respectively. The added $-1$ constants ensure that the correct gauge symmetry sector provides no energy contribution.
Notice that, for the mapping to be perfectly equivalent to the original model, the penalty terms of \cref{eq_HubbarDef_pt2} must represent the largest energy-scale contribution of the Hamiltonian. 
Therefore, the penalties $\alpha_p$ and $\alpha_b$ must be sufficiently larger than $t$ and $U$.
% ==========================================================================================
\subsection{Dressed-site Hamiltonian}
Once the vertex gauge symmetry is successfully installed, we are left with a dressed-site vertex gauge-invariant canonical basis of dimension 32 (4 matter states and four 2-dimensional rishon modes, divided in half by the vertex constraint). Within this canonical basis, the algebra of 1-site operators is genuinely local and can be expressed in terms of the following quadratic operators
\begin{align}
    \label{eq_def_operators}
    \hat{Q}_{\vecsite,\latvec,\alpha} &= \majo_{\vecsite,\latvec} \hpsi_{\vecsite,\alpha} &
    \cop_{\vecsite,\latvec[1],\latvec[2]} &= \majo_{\vecsite,\latvec[1]} \majo_{\vecsite,\latvec[2]}&
    \hat{W}_{\vecsite,\latvec} &= 1 - 2 \cop^{\dagger}_{\vecsite,\latvec}\cop_{\vecsite, \latvec},
\end{align}
each one preserving local fermion parity by design.
After this re-formatting, we have a final expression for the defermionized Hubbard model, and it reads
\begin{equation} 
\label{eq_HubbardFinal}
\begin{split}
   \ham^{\prime\prime}_{\rm{Hub}} =& 
   -t\sum_{\vecsite,k}\sum_{\alpha}
    \qty[i \hat{Q}^{\dagger}_{\vecsite,\latvec[k],\alpha} \hat{Q}_{\vecsite+\latvec[k],-\latvec[k],\alpha} +\hc]
    + U \sum_{\vecsite} \qty( \nop_{\vecsite,\uparrow} -\frac{1}{2})
    \qty( \nop_{\vecsite,\downarrow} -\frac{1}{2}),
  \end{split}
\end{equation}
plus a penalty
\begin{equation} 
\label{eq_HubbarDef}
  \begin{split}
    \ham^{\prime\prime}_{\rm{pen}} =
     &-\alpha_{b}\sum_{\vecsite,k} 
     \qty(\hat{W}_{\vecsite,\latvec[k]}\hat{W}_{\vecsite+\latvec[k],-\latvec[k]} -1)
     -\alpha_p
    \sum_{\square \in \Lambda}\qty(\begin{matrix}
      \corner_{\ulcorner} & \corner_{\urcorner}\\
      \corner_{\llcorner} & \corner_{\lrcorner}
    \end{matrix}-1),
    \end{split}
\end{equation}
where $\corner$ are the \emph{corner} operators defined in \cref{eq_def_operators} and form the plaquette interaction previously defined in \cref{eq_def_plaquette_term} in terms of links.
Using tensor networks, we simulate the model as it exactly appears in these expressions.
% ==========================================================================================
\subsection{Numerical Results}
\label{sec_def_ttnresults}
\begin{figure}
    \centering
    \includegraphics[width=\textwidth]{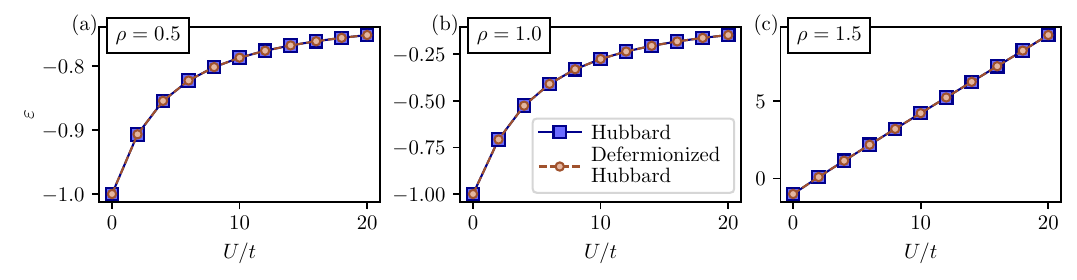}
    \caption{Exact Diagonalization comparison on a $2\times 2$ lattice between the ground state energy density $\varepsilon$ of the original 2D Hubbard model and its defermionized version as a function of $U/t$, and for three values of the particle density $\rho$: (a) below half-filling with $\rho=0.5$, (b) at half-filling with $\rho=1.0$, (c) above half-filling with $\rho=1.5$.}
    \label{fig_ed_comparison_2x2}
\end{figure}
\begin{figure}
    \centering
    \includegraphics[width=\textwidth]{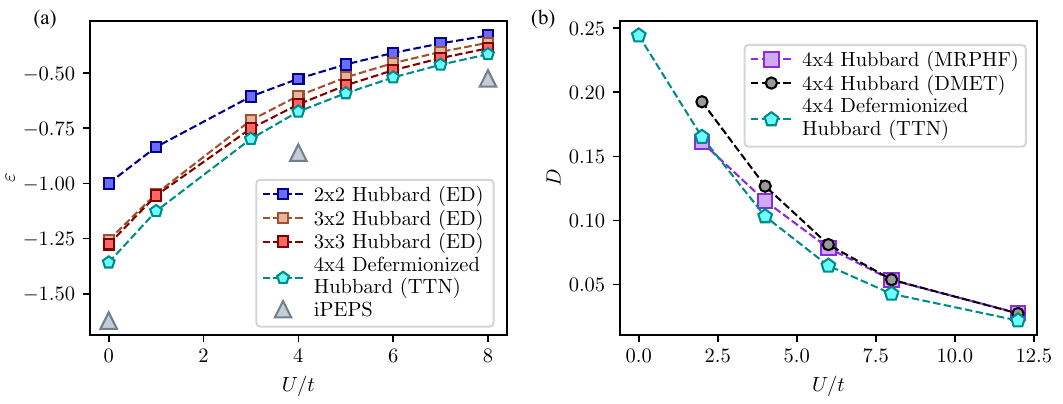}
    \caption{(a) Ground state energy density of the 2D Hubbard Hamiltonian at $\rho=1$ for different lattice sizes. 
    Results concerning $2\times 2$, $3\times 2$, and $3\times 3$ lattices are obtained via exact diagonalization (ED) of the original Hubbard Hamiltonian in \cref{eq_Hubbard}. 
    Energies of the $4\times 4$ lattice result from TTN simulations of the defermionized Hamiltonian in Eqs.~\eqref{eq_HubbardFinal}-\eqref{eq_HubbarDef} with bond dimension $\chi=350$ and energy penalties $\alpha_p = \alpha_b = 15$.
    (b) Double occupancy $D$ as a function of $U/t$ is computed with different numerical methods. 
    Data for MRPHF and DMET are taken from \cite{LeBlanc2015SolutionsTwoDimensionalHubbard}.}   
    \label{fig_comparison_4x4}
\end{figure}
In this section, we present the numerical results obtained using exact diagonalization (ED) and tree tensor networks (TTN) algorithms \cite{Cataldi2021HilbertCurveVs,Ferrari2022AdaptiveweightedTreeTensor}. 
Our numerical computations concern only open boundary conditions (OBC).

To numerically check the validity of our mapping, we compare the original 2D Hubbard model of \cref{eq_Hubbard} and its defermionized version described in Eqs. \eqref{eq_HubbardFinal}-\eqref{eq_HubbarDef}. For both Hamiltonians, we perform ED on a $2 \times 2$ lattice. 

Since the two Hamiltonians do not fix a specific number of particles on the lattice, we add to them an extra term of the form 
$\ham_{\Nsites} = \Tilde{\latvec} \qty(\sum_{\vecsite}\qty( \nop_{\vecsite, \uparrow} + \nop_{\vecsite, \downarrow}) - \Nsites_{0})^2$,
where $\Tilde{\latvec}$ plays the role of a large penalty coefficient that increases the energy of all the states with a number of particles differing from $\Nsites_{0}$. In this way, by tuning $\Nsites_{0}$ and setting $\alpha_p =\alpha_b = 20$ and $\tilde{\latvec} = 20$, we perform ED at fixed values of the particle density $\rho=\Nsites_{0}/\Nsites$, comparing the ground-state energy densities of the two models as a function of the ratio $U/t$. 
The chosen values of the penalty coefficients are enough to satisfy both link and plaquette constraints with a precision larger than $10^{-7}$.

As displayed in \cref{fig_ed_comparison_2x2} for three values of the particle density, \idest{} $\rho\in\qty{0.5, 1.0, 1.5}$, the relative distance between the energy densities of the two models ($\varepsilon_{H}$ for the original Hubbard model, $\varepsilon_{def. H}$ for its defermionized version) $\Delta \varepsilon=\abs{\varepsilon_{H}-\varepsilon_{def. H}}/\abs{\varepsilon_{H}}<10^{-12}$ confirms the exactness of our mapping. 
As expected, the above half-filling particles are forced to form at least a single-site pair which determines a linear dependence of the energy with $U/t$. 

Aware of the equivalence between the two Hamiltonians, we focus on the defermionized model for larger lattice sizes by using TTN simulations. Our TTN algorithm for variational ground state search exploits the global $U(1)$ symmetry of the Hubbard model and the Krylov subspace expansion \cite{Silvi2019TensorNetworksAnthology}. 
Thus, the additional term $H_{N}$ is no longer needed, as the algorithm directly conserves the total number of particles by encoding the $U(1)$ symmetry sector in the TTN. 

\begin{figure}
    \centering
    \includegraphics[width=1\textwidth]{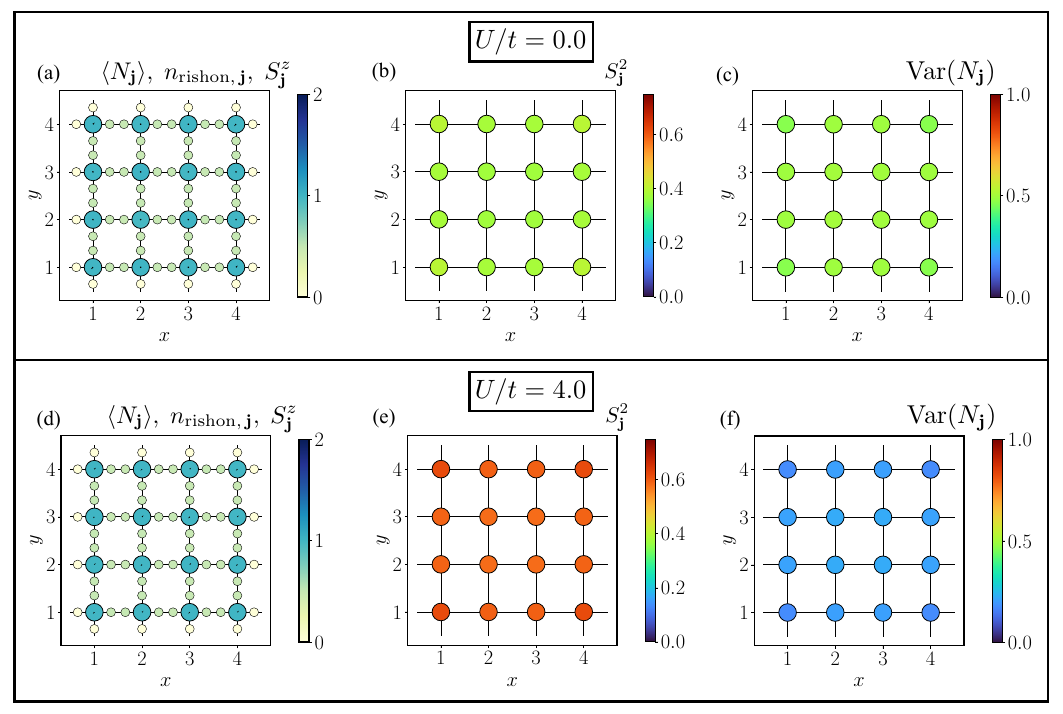}
    \caption{Ground state characterization for the defermionized Hamiltonian with $U/t=0.0, 4.0$ numerically simulated via TTNs. Computed observables: (a),(d) local configurations of fermionic matter for each lattice site $\avg{\nop_{\vecsite}}$, local spin along the $z$-axis $\spinmat[z]_{\vecsite}$ represented by arrows in the center of the lattice sites (values are close to zero), rishon modes occupation on the links $ \nop_{\rm{rishon}, \vecsite}$; (b),(e) spin modulus squared $\spinmat[2]_{\vecsite}$; (c),(f) variance of the matter occupation number $\mathrm{Var}(\nop_{\vecsite})$.}
    \label{fig_local_configurations}
\end{figure}
Given the defermionized Hubbard Hamiltonian of \cref{eq_HubbardFinal,eq_HubbarDef}, the ground state at a specified bond dimension $\chi$ is determined by iteratively optimizing each of the tensors in TTN, gradually reducing the energy expectation value. 
This procedure is iterated several times to reach the desired convergence. For a detailed and exhaustive description of the algorithms, please see \cref{sec_TN_gs_algorithm} and the technical reviews and textbooks \cite{Silvi2019TensorNetworksAnthology,Montangero2018IntroductionTensorNetwork}. 
In our TTN simulations, we use a maximum bond dimension $\chi=350$, which is sufficient to reach a relative error on the energy density of the order $10^{-6}$, ensuring the stability of our findings.

In \cref{fig_comparison_4x4}(a), we show the TTN results concerning the defermionized Hamiltonian on a $4 \times 4$ lattice at half-filling, \idest{} $\rho=1$. 
We also report the corresponding energy densities obtained via ED for the original Hubbard Hamiltonian at smaller lattice sizes, and the results of the extrapolation of the infinite-size limit obtained by using iPEPS methods \cite{Corboz2016ImprovedEnergyExtrapolation}. 
The energies obtained with the TTN simulations for the defermionized Hamiltonian are in agreement with the overall scaling shown by the exact energies as a function of $U/t$ and the lattice sizes.

To further test the equivalence between the original Hubbard model and its defermionized version, we compute the ground-state pair density $D\equiv \sum_{\vecsite}\avg{\nop_{\vecsite,\uparrow}\nop_{\vecsite,\downarrow}}/\Nsites$. 
In \cref{fig_comparison_4x4}(b), we show the results obtained from the TTN simulations of the defermionized model on the $4\times 4$ lattice. 
For the sake of comparison, we report the data of the original Hubbard Hamiltonian, obtained in Ref. \cite{LeBlanc2015SolutionsTwoDimensionalHubbard} by using two independent numerical methods, \idest{} the multireference projected Hartree-Fock method (MRPHF) and the Density matrix embedding theory (DMET). 
A good agreement and consistency of our results with the reference data is visible for all the simulated values of $U/t$. 

TTN simulations allow for an extra characterization of the ground state in terms of local observables, such as the occupation of fermionic matter and its variance 
\begin{align}
    \avg{\nop_{\vecsite}} &= \avg{\nop_{\vecsite,\uparrow} + \nop_{\vecsite, \downarrow}}&  
    \mathrm{Var}(\nop_{\vecsite}) &= \langle\qty(\nop_{\vecsite})^{2}\rangle - \avg{\nop_{\vecsite}}^2,
\end{align}
the local spin along the $z$-axis and its square modulus,
\begin{align}
    \spinmat[z]_{\vecsite} &= \frac{1}{2}\avg{\nop_{\vecsite,\uparrow} - \nop_{\vecsite, \downarrow}} &
    \spinmat[2]_{\vecsite} = \frac{3}{4}\avg{(\nop_{\vecsite,\uparrow} - \nop_{\vecsite, \downarrow})^2},
\end{align}
but also the rishon mode occupation on lattice links $ \nop_{\rm{rishon}, \vecsite} = \langle c^{\dagger}_{\vecsite, \latvec}\cop_{\vecsite, \latvec}\rangle$.

In \cref{fig_local_configurations}, we show the results for $U/t=0$ and $U/t=4$. 
In both cases (see \cref{fig_local_configurations}(a) and \cref{fig_local_configurations}(d)), the fermionic occupation number $\nop_{\vecsite} \sim 1$, while the local spin along $z$, represented by arrows in the center of the lattice sites, is close to zero; correspondingly, all the rishon modes occupations on the links are $ \nop_{\rm{rishon}, \vecsite} \sim 0.5 $,  as a consequence of the satisfied link and plaquette penalty constraints that encode the fermion parity. 
By looking at Figs. \ref{fig_local_configurations}(b) and \ref{fig_local_configurations}(e), we observe that the squared spin modulus $\spinmat[2]_{\vecsite}$ increases for all lattice sites by varying $U/t$ from $0.0$ to $4.0$, whereas the variance of the fermionic occupation number $\mathrm{Var}(\nop_{\vecsite})$ decreases, as shown in Figs. \ref{fig_local_configurations}(c) and \ref{fig_local_configurations}(f). 

These configurations agree with the expected ground state of the original Hubbard model, which in the half-filling case, maps to the Heisenberg model \cite{Auerbach1994InteractingElectronsQuantum}, and its ground state mimics an antiferromagnetically long-range ordered state.
% ==========================================================================================
\section{Digital quantum simulation}
\label{sec_def_digitalqs}
\begin{table}
    \centering
    \begin{tabular}{l|c|c|c|c}
        & Qubit-fermion ratio & Fermion parity weight & Hopping weight & Stabilizers weight \\
       Our & 3 & 1 & 6 & 6 \\
       \cite{Nielsen2006QuantumComputationGeometry} & 1 & 1 & $O(L)$ & -\\
       \cite{Bravyi2002FermionicQuantumComputation} & 1 & $O(\log_2 L^2)$ & $O(\log_2 L^2)$ & - \\
       \cite{Jiang2019MajoranaLoopStabilizer}       & 1 & $O(\log_3{(L +1)})$ & $O(\log_3{(L +1)})$ & - \\
       \cite{Zohar2018EliminatingFermionicMatter} * & 3 & 1 & 5-7 & 5-6 \\
       \cite{Chen2018ExactBosonizationTwo}          & 2 & 4 & 2-6 & 6\\
       \cite{Chen2023EquivalenceFermiontoQubitMappings} & 1.25 & 1-2 & 2-6 & 12\\
       \end{tabular}
    \caption{Comparison of different encoding mapping a single species of fermions to qubits for a 2D square lattice of size $L\times L$: Jordan-Wigner~\cite{Nielsen2006QuantumComputationGeometry} Bravyi-Kitaev~\cite{Bravyi2002FermionicQuantumComputation} Optimal fermion-to-qubit mapping~\cite{Jiang2019MajoranaLoopStabilizer}  Zohar-Cirac~\cite{Zohar2018EliminatingFermionicMatter} Exact bosonization~\cite{Chen2018ExactBosonizationTwo} Supercompact fermion-to-qubit mapping~\cite{Chen2023EquivalenceFermiontoQubitMappings}. 
    The mapping with * is specifically developed for LGTs, similarly to \cite{Pardo2023ResourceefficientQuantumSimulation}.}
    \label{tab_def_encodings}
\end{table}
The gauge \emph{defermionization} offers a natural encoding of fermionic degrees of freedom to qubits, thus allowing for the digital quantum simulation of fermionic models on quantum computers.
Here, we break down the procedure to express the spin-$\frac{1}{2}$ Hubbard defermionized model on a square lattice in terms of qubits and Pauli operators on a generic platform, see \cref{sec_def_digital_fqm}, \cref{sec_def_digital_lc}, and \cref{sec_def_digital_spc}.  
This gauge-field-based encoding is genuinely local, $\id$est{} the lattice support of each Hamiltonian term is not increased in the encoding, and thus each Pauli weight $W$ is conserved and does not depend on the system size $\Nsites$.
However, locality comes at the cost of including auxiliary qubits representing the defermionized modes, \idest{} fermionic modes and  $\mathbb{Z}_{2}$ gauge fields. 
There currently exists a few local mappings \cite{Bravyi2002FermionicQuantumComputation,Verstraete2005MappingLocalHamiltonians,Whitfield2016LocalSpinOperators,Chen2018ExactBosonizationTwo,Steudtner2019QuantumCodesQuantum,Setia2019SuperfastEncodingsFermionic,Jiang2019MajoranaLoopStabilizer,Derby2021CompactFermionQubit} and the relation between them is analyzed in \cite{Chen2023EquivalenceFermiontoQubitMappings}. 
Fermion-to-qubit mappings need resources that are usually quantified in terms of the number of qubits to simulate one fermion on average (qubit-fermion ratio), the number of qubits to express the parity of a fermionic state (parity weight), the length of the hopping operators expressed as Pauli strings, and, ultimately, the maximum length of Pauli strings for the stabilizers.
The resources required by the gauge-field-based encoding are comparable with state-of-the-art methods: values for qubit-fermion ratio, fermion parity weight, hopping weight, and stabilizer weight are $(3, 1, 6, 6)$ respectively. 
Building on Table 1 of \cite{Chen2023EquivalenceFermiontoQubitMappings}, we present 
a comparison of different fermionic encodings in \cref{tab_def_encodings}. 

Within these prescriptions, we test our construction in out-of-equilibrium scenarios that will be relevant to investigate nontrivial dynamical effects.
In particular, the Hubbard model offers the opportunity to explore the dynamics of spin-like and charge-like excitations, which in one dimension manifest as distinct degrees of freedom with independent propagation velocities \cite{Giamarchi2003QuantumPhysicsOne,Kollath2005SpinChargeSeparationCold}, and have been recently observed in cold gases experiments \cite{Vijayan2020TimeresolvedObservationSpincharge,Bohrdt2021ExplorationDopedQuantum} and digital quantum simulations \cite{Arute2020ObservationSeparatedDynamics}. 
In two dimensions, the dynamics of spin and charge degrees of freedom is highly nontrivial \cite{Grusdt2018PartonTheoryMagnetic} as strongly-correlated effects become relevant, still a subject of ongoing theoretical and experimental analysis with quantum simulation platforms \cite{Ji2021CouplingMobileHole,Koepsell2021MicroscopicEvolutionDoped}.

For these reasons, we simulate the digital dynamics of spin and charge excitations over a half-filled ($\rho=1$) 2 $\times$ 4 system in the antiferromagnetic phase with $U/t=10$.
In \cref{sec_def_digital_methods}, we lay out the protocols to adiabatically prepare the antiferromagnetic ground state, and then inject the excitations. 
Finally, in \cref{sec_def_digital_scs}, we observe the corresponding out-of-equilibrium dynamics for spin and charge excitations. 
While protocols are designed for a general digital quantum computer, the presented results are obtained from tensor networks  CITE to emulate the evolution of the quantum circuit.
% ==========================================================================================
\subsection{Fermion to qubits mapping of the defermionized Hubbard Hamiltonian} \label{sec_def_digital_fqm}
We introduce a qubit for each flavor (up $u$ and down $d$) and one for each rishon (north $\Nsites$, west $w$, east $e$, south $s$). To minimize the Pauli weight of the Hamiltonian terms, a specific ordering for the qubits composing each dressed site is defined:
\begin{align} \label{eq_even_odd_order}
    \begin{cases}
    \{u, d, w, s, e, n\} & \text{if $(-1)^{\site[x]+\site[y]}=+1$ (even site)} \\
    \{d, u, s, w, n, e\} & \text{if $(-1)^{\site[x]+\site[y]}=-1$ (odd site)}. 
    \end{cases}
\end{align}
All the operators in \cref{eq_HubbardFinal} and penalties in \cref{eq_HubbarDef} are first mapped to Majorana fermions and then to spin-$\frac{1}{2}$ algebra, while preserving all the commutation relations.
We recall that a hopping operator from an odd ($O$) and an even ($E$) site for the up species can be written as:
\begin{align}
     \hpsi_{\uparrow E}^\dagger&\majo_{eE}\majo_{wO}\hpsi_{\uparrow O} - \hpsi_{\uparrow E}\majo_{eE}\majo_{wO}\hpsi_{\uparrow O}^\dagger
     =\majo_{eE}\majo_{wO}\left( \hpsi_{\uparrow E}^\dagger\hpsi_{\uparrow O} - \hpsi_{\uparrow E}\hpsi_{\uparrow O}^\dagger \right).
\end{align}
Then, by defining the Majorana degrees of freedom $d_{x_{\uparrow E, \uparrow O}}, d_{y_{\uparrow E, \uparrow O}}$ as:
\begin{align}
    \dop_{x_{\uparrow E, \uparrow O}}&=\hpsi_{\uparrow E, \uparrow O} + \hpsi_{\uparrow E, \uparrow O}^\dagger &
    \dop_{y_{\uparrow E, \uparrow O}}&=i\qty(\hpsi_{\uparrow E, \uparrow O}-\hpsi^\dagger_{\uparrow E,\uparrow O}),
\end{align}
we can trivially prove that
\begin{align}
    \dop_{x_{\uparrow E}}\dop_{y_{\uparrow O}} - \dop_{y_{\uparrow E}}\dop_{x_{ \uparrow O}} 
    = 2i\left( \hpsi_{\uparrow E}^\dagger\hpsi_{\uparrow O} - \hpsi_{\uparrow E}\hpsi_{\uparrow O}^\dagger \right).
\end{align}
We can thus express the hopping in terms of the Majorana fermions.
Following the order defined in \cref{eq_even_odd_order}, the operators acting on an even site $\vecsite$ can be written as:
\begin{align}
    \label{eq_qub_map}
    \dop_{x_{\vecsite,\uparrow }} &= \X\Z\Z\Z\Z\Z &
    \dop_{x_{\vecsite,\downarrow }}& = \id\X\Z\Z\Z\Z &
    \dop_{y_{\vecsite,\uparrow }} &= \Y\Z\Z\Z\Z\Z &
    \dop_{y_{\vecsite,\downarrow }} &= \id\Y\Z\Z\Z\Z
\end{align}
\begin{equation}
    \label{eq_qub_map_2}
    \majo_{\vecsite,r} = \id_{u}\id_{d}\bigotimes_{k=w}^{r-1}\id_{k} \otimes \X_r \bigotimes_{k=r+1}^{n}\Z_k,
\end{equation}
where $\X,\Y,\Z$ are Pauli matrices, while $\id$ is the identity operator. 
In the two equations above we dropped the index of species and rishons for clarity.
The string of $\Z$ operators applies to all the qubits forming a site. 
The index $r$ runs over the different rishon species of a site, as defined in \cref{eq_even_odd_order}. 
The mapping for an odd sites can be obtained analogously using \cref{eq_even_odd_order}, and is equivalent to swapping the first two operators and using the odd-ordered sites.

Now, we compute the explicit form of the Hamiltonian terms, hopping and on-site interaction, in terms of Pauli operators. 
Following the notation defined in \cref{eq_even_odd_order}, the horizontal hopping operator from an odd to an even site for flavor $up$ reads:
\begin{equation}
   \begin{tabular}{c|cccccc|cccccc}
   & u & d & w & s & e & n & d & u & s & w & n & e \\ 
   \hline
    $\dop_{x_{\uparrow E}}$& $\X$ & $\Z$ & $\Z$ & $\Z$ & $\Z$ & $\Z$ & $\id$ & $\id$ & $\id$ & $\id$ & $\id$ & $\id$ \\
    $\majo_{eE}$& $\id$ & $\id$ & $\id$ & $\id$ & $\X$ & $\Z$ & $\id$ & $\id$ & $\id$ & $\id$ & $\id$ & $\id$ \\
    $\majo_{wO}$& $\id$ & $\id$ & $\id$ & $\id$ & $\id$ & $\id$ & $\id$ & $\id$ & $\id$ & $\X$ & $\Z$ & $\Z$\\
    $\dop_{y_{\uparrow O}}$& $\id$ & $\id$ & $\id$ & $\id$ & $\id$ & $\id$ & $\id$ &$\Y$& $\Z$ & $\Z$ & $\Z$ & $\Z$ \\
    \hline
    hopping & $\X$ & $\Z$ & $\Z$ & $\Z$ & i$\Y$ &$\id$ & $\id$ &$\Y$& $\Z$ & -i$\Y$ & $\id$ & $\id$ 
\end{tabular}
\end{equation}
Matrix multiplication goes from bottom to top. By computing also the hermitian conjugate, the hopping term results:
\begin{align}\label{eq_hopping_qubs}
    \dop_{x_{\uparrow E}}
    &\majo_{eE}\majo_{wO}\dop_{y_{\uparrow O}} - \dop_{y_{\uparrow E}}\majo_{eE}\majo_{wO}\dop_{x_{\uparrow O}}= \qty(\X\Z\Z\Z\Y\id\id\Y\Z\Y\id\id-\Y\Z\Z\Z\Y\id\id\X\Z\Y\id\id).
\end{align}
Terms for vertical hopping or hoppings involving different species can be derived analogously.

Similarly, we derive the on-site interaction term. 
The number operator acting on site $\vecsite$ can be written as $\nop_{\vecsite,\alpha}=\frac{1}{2}(1-\Z_{\vecsite,\alpha})$, then the on-site interaction reads:
\begin{align}
    \label{eq_onsite_qubs}
    \qty(\nop_{\vecsite,\uparrow}-\frac{1}{2})\qty(\nop_{\vecsite,\downarrow}-\frac{1}{2})=\frac{1}{4}\Z_{\vecsite,\uparrow}\Z_{\vecsite, \downarrow}.
\end{align}
Once we obtain the Hamiltonian terms described as Pauli strings, we decompose the relative time propagator in single and two-qubit gates following Ref.~\cite{Whitfield2016LocalSpinOperators}. 

\subsubsection{Mapping the time propagator to a quantum circuit}
\begin{figure}
    \centering
    \includegraphics[width=0.5\textwidth]{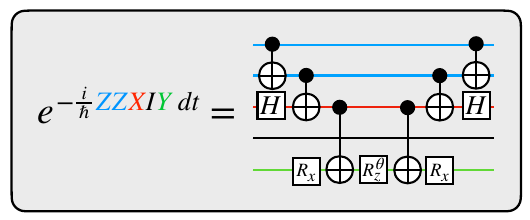}
    \caption{Mapping of the time propagator of a generic Pauli string to a quantum circuit using only single and two-qubit gates.}
    \label{fig_def_ham_to_gate}
\end{figure}

To simulate the dynamics of our system on a quantum computer it is important to map the evolution operator to a quantum circuit using single and two-qubit gates since those are the available operations. 
First, we Trotterize the evolution treating each Hamiltonian term separately. 
Then, it has been shown that the propagator $e^{-\frac{i}{\hbar} J \ham_{i} dt}$ of a $\ham_{i} =\Z\Z\dots \Z$ interaction with strength $J$ can be compiled by combining a cascade of CNOTs gates and a rotation along the $z$ axis~\cite{Whitfield2016LocalSpinOperators}:
\begin{align}
    \hat{R}_{z}^{\theta} = \begin{pmatrix}
    1 & 0 \\
    0 & e^{i\theta}
    \end{pmatrix}, \quad \theta = Jdt.
\end{align}
In general, we can treat an arbitrary Pauli string by moving to the $Z$ basis along the CNOTs cascade. 
Practically, we apply the following single-qubit basis change gate before and after the CNOT, respectively for the $\X$ and $\Y$ Pauli matrices:
\begin{align}
    \ham = \frac{1}{\sqrt{2}}\begin{pmatrix}
    1 & 1 \\
    1 & -1
    \end{pmatrix},
    \quad
    \hat{R}_{x}\qty(-\frac{\pi}{2}) = \frac{1}{\sqrt{2}}\begin{pmatrix}
    1 & \id \\
    \id & 1
    \end{pmatrix}.
\end{align}
In \cref{fig_def_ham_to_gate}, we report the form of the CNOTs cascade with the basis change, showing an example for a general Pauli string.
% ==========================================================================================
\subsection{Vertex and plaquette constraints}
\label{sec_def_digital_spc}
Following the prescription of the gauge defermionization, we have to constrain the state to a specific subspace of the Hilbert space, a condition set by \cref{eq_def_vertex,eq_def_plaq}.
Each symmetry can be cast into a language that directly translates into a quantum circuit scenario and is equivalent to a stabilizer of a quantum error correcting code; to respect the gauge symmetry, the state is restrained to the $+1$ eigenstate of the stabilizers~\cite{Gottesman1997StabilizerCodesQuantum}.
The Pauli representation of the stabilizers can be obtained following the mapping presented in \cref{eq_qub_map}. As expected, all the stabilizers commute with the Hamiltonian. 

The vertex stabilizer term reads from \cref{eq_def_vertex}. 
As the exponentials are counting the parity of a given species/rishon, these operators correspond to $\Z$ operators in the qubit language. 
Thus, the stabilizer acting on each lattice vertex $\vecsite$ results in
\begin{align}
    \spinmat_{\vecsite} = \bigotimes_{\alpha=u, d}\Z_{\alpha} \bigotimes_{k}\Z_{\vecsite+\latvec[k]},
\end{align}
where $\latvec[k]$ spans on all rishons connected to the vertex $\vecsite$. 

The plaquette stabilizer reads from \cref{eq_def_plaq}. 
As the product always runs on four terms, the imaginary unit $i$ drops. 
Following the representation of the $\majo$ operators reported in \cref{eq_qub_map_2}, the final form of each plaquette term respects the qubits ordering of the dressed site. 
Thus, the form of an even plaquette stabilizer is different from an odd one, where a plaquette is said even (odd) if the lower-left corner of the plaquette lies on an even (odd) site. We report the explicit computation for an even plaquette in \cref{tab_def_plaquette_stabilizer}.

Finally, we report the size of the Hilbert space to perform a digital quantum simulation of the defermionized Hubbard model over a $x\times y$ grid, and compare it to the size of the accessible subspace:
\begin{subequations}
    \begin{align}
        \rm{dim}(\mathcal{H}) &= 2^{ 2xy+x(y-1)+y(x-1) }=2^{4xy-x-y} \\
        \rm{dim}(\mathcal{H}_{eff}) &= 2^{ 2xy+x(y-1)+y(x-1)-(x-1)(y-1)-xy }=2^{2xy-1}.
    \end{align}
\end{subequations}
We notice that constraining the system to the physical subspace quadratically decreases the dimension of the Hilbert space, showing that we are working with a highly redundant space.
\begin{table}
    \centering
    \setlength{\tabcolsep}{3pt} 
    \begin{tabular}{c|cccccc|cccccc|cccccc|cccccc}
    & u & d & w & s & e & n & d & u & s & w & n & e & d & u & s & w & n & e & u & d & w & s & e & n \\ \hline
  $\majo_{00e}$ & $\id$ & $\id$ & $\id$ & $\id$ & $\X$ & $\Z$ & $\id$ & $\id$ & $\id$ & $\id$ & $\id$ & $\id$ & $\id$ & $\id$ & $\id$ & $\id$ & $\id$ & $\id$ & $\id$ & $\id$ & $\id$ & $\id$ & $\id$ & $\id$ \\
  $\majo_{00n}$ & $\id$ & $\id$ & $\id$ & $\id$ & $\id$ & $\X$ & $\id$ & $\id$ & $\id$ & $\id$ & $\id$ & $\id$ & $\id$ & $\id$ & $\id$ & $\id$ & $\id$ & $\id$ & $\id$ & $\id$ & $\id$ & $\id$ & $\id$ & $\id$ \\
  $\majo_{01s}$ & $\id$ & $\id$ & $\id$ & $\id$ & $\id$ & $\id$ & $\id$ & $\id$ & $\X$ & $\Z$ & $\Z$ & $\Z$ & $\id$ & $\id$ & $\id$ & $\id$ & $\id$ & $\id$ & $\id$ & $\id$ & $\id$ & $\id$ & $\id$ & $\id$ \\
  $\majo_{01e}$ & $\id$ & $\id$ & $\id$ & $\id$ & $\id$ & $\id$ & $\id$ & $\id$ & $\id$ & $\id$ & $\id$ & $\X$ & $\id$ & $\id$ & $\id$ & $\id$ & $\id$ & $\id$ & $\id$ & $\id$ & $\id$ & $\id$ & $\id$ & I\\
  $\majo_{10w}$ & $\id$ & $\id$ & $\id$ & $\id$ & $\id$ & $\id$ & $\id$ & $\id$ & $\id$ & $\id$ & $\id$ & $\id$ & $\id$ & $\id$ & $\id$ & $\X$ & $\Z$ & $\Z$  & $\id$ & $\id$ & $\id$ & $\id$ & $\id$ & $\id$\\
  $\majo_{10n}$ & $\id$ & $\id$ & $\id$ & $\id$ & $\id$ & $\id$ & $\id$ & $\id$ & $\id$ & $\id$ & $\id$ & $\id$ & $\id$ & $\id$ & $\id$ & $\id$ & $\X$ & $\Z$ & $\id$ & $\id$ & $\id$ & $\id$ & $\id$ & $\id$\\
  $\majo_{11w}$ & $\id$ & $\id$ & $\id$ & $\id$ & $\id$ & $\id$ & $\id$ & $\id$ & $\id$ & $\id$ & $\id$ & $\id$ & $\id$ & $\id$ & $\id$ & $\id$ & $\id$ & $\id$ & $\id$ & $\id$ & $\X$ & $\Z$ & $\Z$ & $\Z$\\
  $\majo_{11s}$ & $\id$ & $\id$ & $\id$ & $\id$ & $\id$ & $\id$ & $\id$ & $\id$ & $\id$ & $\id$ & $\id$ & $\id$ & $\id$ & $\id$ & $\id$ & $\id$ & $\id$ & $\id$ & $\id$ & $\id$ & $\id$ & $\X$ & $\Z$ & $\Z$\\ \hline
    stabilizer & $\id$ & $\id$ & $\id$ & $\id$ & $\X$ & i$\Y$ & $\id$ & $\id$ & $\X$ & $\Z$ & $\Z$ & i$\Y$ & $\id$ & $\id$ & $\id$ & $\X$ & i$\Y$ & $\id$ & $\id$ & $\id$ & $\X$ & i$\Y$ & $\id$ &$ \id$ \\
\end{tabular}
    \caption{Explicit definition of the stabilizer for an even plaquette. This calculation does not take into account the link symmetry: thus it is necessary to do that further computation to obtain the final expression of the qubit mapping.}
    \label{tab_def_plaquette_stabilizer}
\end{table}
% ==========================================================================================
\subsection{Link constraint}\label{sec_def_digital_lc}
\begin{figure}[ht]
    \centering
    \includegraphics[width=0.5\textwidth]{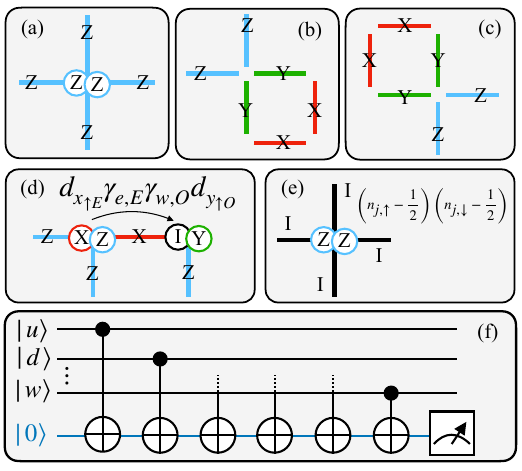}
    \caption{Graphical representation of the Hamiltonian terms in the defermionized Hubbard model and of the stabilizers for installing the gauge symmetries. These terms result from applying \cref{eq_from2_to1_rishon}, where the states of the two rishons satisfying the link constraint are mapped to a single qubit for each link. In (a), we depict the vertex stabilizer, while (b) corresponds to the even plaquette stabilizer, and (c) represents the odd plaquette stabilizer. Figures (d) and (e) illustrate the hopping term and the onsite interaction respectively. 
    As dictated by \cref{eq_even_odd_order}, in (d), the $u$ qubit is the first on the even site and the second on the odd site. 
    In (f) we report the quantum circuit used to measure the stabilizer in figure (a) using CNOT gates and a projective measurement on an ancilla qubit (depicted in blue). The qubits $\ket{n}, \ket{e}, \ket{s}$ are not reported in the figure, even if their CNOT gate is present with a dashed line.}
    \label{fig_def_stabilizers}
\end{figure}

The state of the two rishons sharing the same link is also subject to a constraint set by the third line of \cref{eq_def_link}. 
The only qubits states fulfilling this constraint are $\ket{00}$ and $\ket{11}$. The effective Hilbert space for a pair of rishons is $2$-dimensional; for this reason, both rishons can be described with a single qubit, by mapping $\ket{00}\rightarrow \ket{0}, \ket{11}\rightarrow\ket{1}$. After this procedure, we need to project all the operators acting on the two rishons on the new subspace. Given an operator $A_{r_1r_2}$ we can write the new operator living on the two-dimensional space as:
\begin{align}\label{eq_from2_to1_rishon}
    A_{\overline{r}} = \rm{Tr}_{\ket{01},\ket{10}} A_{r_1r_2},
\end{align}
where we trace out the states violating the constraint. 
We report the new form of some operators of interest as an example: $\Z\Z\rightarrow \id$, $\X\X\rightarrow \X$. 
This way, each dressed site shares a rishon with its nearest neighbors, and thus there is no longer a clear separation between the dressed sites.
In \cref{fig_def_stabilizers}, we report the final form of the stabilizers and the Hamiltonian terms, once the link constraint is applied.
% ==========================================================================================
\subsection{Preparation of spin and charge excitations}
\label{sec_def_digital_methods}
Here, we lay out the protocols to adiabatically prepare the antiferromagnetic ground state, and to inject spin- and charge-excitations.
% ==========================================================================================
\subsubsection{Adiabatic ground state preparation} 
The adiabatic state preparation targets the ground state of the defermionized Hubbard model over a $4\times 2$ lattice, at half-filling ($\rho=1$) with $t=0.1$ and $U=1$. 
The adiabatic evolution starts from the ground state of \cref{eq_HubbardFinal} with $t=0$, here renamed $\ham_{0}=\ham^{\prime\prime}_{\rm{Hub}}(t=0)$; the hopping terms are slowly switched on to reach $\ham_{1}=\ham^{\prime\prime}_{\rm{Hub}}$. Then, the time-dependent Hamiltonian reads:
\begin{align}
    \ham &= (1-\beta)\ham_{0}+\beta \ham_{1},
\end{align}
where $\beta \in [0,1]$ is the adiabatic parameter that increases linearly for $100$ steps; for each $\beta$, the system evolves for ten time-steps $d\tau$ to ensure a smooth convergence, for a total of $1000$ evolution steps. 
The real-time evolution is simulated by decomposing the evolution operator via a first-order Trotterization with time step $dt=d\tau=0.01$.

Since the initial state of $\ham_{0}$ is hugely degenerate, we select as initial state for the adiabatic process is the ground state of $\ham_{0}$ at half-filling ($\rho=1$) and respects the spin-flip symmetry. 
While neglecting the rishons' state, we consider only the state of the matter qubits ($u$ and $d$) for each dressed site over the $4\times 2$ lattice, where $\ket{10}=\ket{\uparrow}$ and $\ket{01}=\ket{\downarrow}$. 
It results in
\begin{align}\label{eq_adiabatic_initial_state}
    \frac{1}{\sqrt{2}}\qty(\ket{
    \begin{matrix}
    \uparrow & \downarrow & \uparrow & \downarrow \\
    \downarrow & \uparrow & \downarrow & \uparrow
    \end{matrix}} +
    \ket{
    \begin{matrix}
    \downarrow & \uparrow & \downarrow & \uparrow \\
    \uparrow & \downarrow & \uparrow & \downarrow
    \end{matrix}}).
\end{align}
Notice that the state in \cref{eq_adiabatic_initial_state} is a GHZ state, thus it can be easily prepared using only Hadamard, NOT, and controlled-NOT gates.
Finally, the rishons' state is uniquely determined once the stabilizers are measured, after which we apply a conditional operation to ensure the state lies in the correct symmetry sector. For a depiction of the stabilizers see \cref{fig_def_stabilizers}, while, for an example of the measurement of the stabilizer depicted in \cref{fig_def_stabilizers}(a), see \cref{fig_def_stabilizers}(f). 
This measurement is performed through a projective measurement of an ancilla qubit.

At the end of the adiabatic evolution, we checked that local spin projection $\spinmat[z]_{\vecsite}$ and local charge $ \nop_{\vecsite}$ are stationary up to oscillations compatible with the cut of the singular values in the TN emulator ($10^{-8}$). 
For a discussion about convergence see \cite{Cataldi*2024DigitalQuantumSimulation}.

While an adiabatic state preparation is performed in this case, we do not think this approach to be scalable on real quantum devices. However, there are multiple possibilities for preparing the initial state with shallower circuits. 
If a TN state can efficiently represent the target state, we can variationally map the state to a quantum circuit~\cite{Rudolph2023DecompositionMatrixProduct}. 
Another possibility is to implement a Quantum Approximate Optimization Algorithm (QAOA), which can encode the Hamiltonian interactions in the quantum circuit structure \cite{Wauters2020PolynomialScalingQuantum}. 
Both approaches have the advantage of being able to tune the fidelity of the algorithm with the depth of the quantum circuit.
\begin{figure}
    \centering
    \includegraphics[width=1\textwidth]{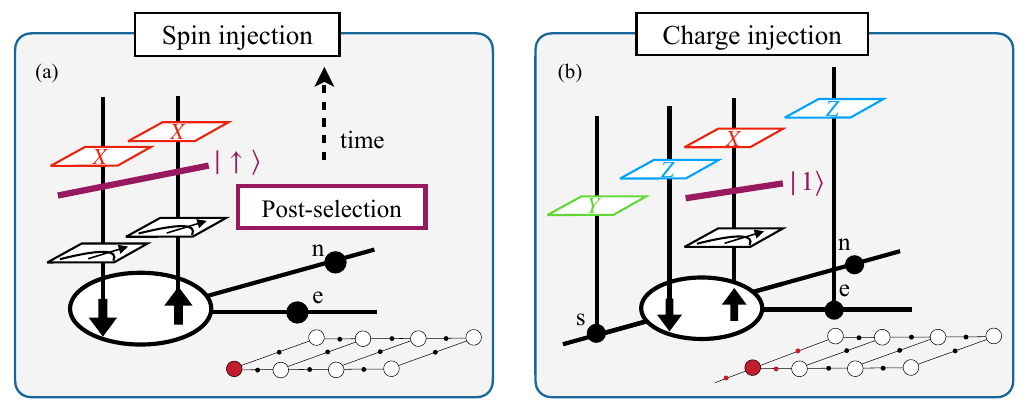}
    \caption{Schematic of injecting spin- and charge-excitations on the $(0,0)$ corner. 
    Both figures depict the qubit lattice, highlighting up and down qubits within the matter site. 
    Rishons are represented as black dots and labeled with respect to the corner. 
    In the bottom right of each subfigure, we report the entire lattice, highlighting in red the site and rishons involved in the operation. 
    All applied operations commute with the stabilizers. (a) \emph{Spin excitation:} To inject the spin excitation, we initially measure the qubit states for up and down, post-selecting only the $\ket{01}=\ket{\uparrow}$ state. Then, the $\X\X$ operator flips the spin state. 
    (b) \emph{Charge excitation:} Injecting the charge excitation involves measuring the up qubit and post-selecting the state $\ket{1}$. We then apply the Pauli operators as depicted in the figure, equivalent to causing the matter qubit $u$ to jump outside the lattice. To achieve this, an additional rishon (the $s$ rishon in this case) must be added.}
    \label{fig_excitation_sketch}
\end{figure}
% ==========================================================================================
\subsubsection{Excitations} 
Spin- and charge-excitations are injected in the ground state of $H_1$ using local operations and classical communication (LOCC)s: this procedure is, in principle, completely reproducible on state-of-the-art quantum computers. 
Both protocols preserve the symmetries of the system, namely every constraint previously defined is satisfied. In \cref{fig_excitation_sketch}, we provide a graphical representation of the two methods.
% ==========================================================================================
\paragraph{Spin excitation} 
Without losing any generality, the spin excitation is created on the site $(0, 0)$. 
Recalling that qubits $u,d$  represent up and down flavors, we choose a general superposition of the two states; after a projective measurement on both qubits, only the flavor up is post-selected, \idest{} the qubits state $\ket{10}=\ket{\uparrow}$.
Then, a spin excitation is introduced by flipping the spin state:
\begin{align}
    \ket{\uparrow} &\longrightarrow \ket{\downarrow},&
    \ket{1_u0_d} &\xrightarrow{\X\X}\ket{0_u1_d}.
\end{align}
This operation corresponds to applying the operator $\X\X$ on the qubits $u, d$. 
Notice that $\X\X$ commutes with all the stabilizers, since it shares support only with the vertex stabilizer in $(0,0)$, and they commute.
% ==========================================================================================
\paragraph{Charge excitation}
In principle, the charge excitation can be created anywhere on the lattice. However, for improved efficiency, we restrict the protocol to the sites along the border and in particular to the site $(0, 0)$. A projective measurement is performed on the $u$ qubit, then post-selecting the state $\ket{1}$.
Being left with a single charge on the site, the excitation is obtained by removing that very charge. However, this operation does not commute with the vertex stabilizer in $(0,0)$, thus breaking that constraint. To preserve Gauss' law we introduce an additional qubit, labeled as an extra rishon of site $(0,0)$, in this case, $s$ (or $w$). The excitation is implemented by flipping the state of the qubits $u, s$: 
\begin{align}
    \ket{1_u0_s} &\stackrel{\X\X}{\longrightarrow}\ket{0_u1_s}.
\end{align}
This procedure, however, does not commute with the stabilizers, and it would bring the state outside the subspace of the physical states. To avoid this issue, we implement
this operation by applying the hopping operator $i\majo_{\vecsite,s}\hpsi_{\vecsite,\uparrow}$ with $\vecsite=(0,0)$, which makes the particle in $u$ hop through $s$ outside the lattice. The operator reads $g_{charge}=\X_u\Z_d\Z_w\Y_s$, and it is equivalent to the bit-flip just discussed.
We stress that qubit $d$ is not projected during this protocol.
% ==========================================================================================
\subsection{Propagation of spin and charge excitations}
\label{sec_def_digital_scs}
By employing the local encoding, we simulate the dynamics of spin and charge excitations over a half-filled ($\rho=1$) $2\times4$ system in the antiferromagnetic phase $U/t=10$. 
To this aim, we use quantum matcha TEA \cite{Ballarin2024QuantumTEAQmatchatea}, an emulator of quantum circuits based on MPS. 
All the simulations converge with bond dimension $\chi=1024$ - below the maximum achievable for $27$ qubits, where the maximum bond dimension is bound by $\chi_{\mathrm{max}}=8192$ - and Trotter step $dt=0.01$. 
In this regime with $U\gg t$, the Hubbard model dynamics is effectively described by the $t-J$ model \cite{Fradkin2013FieldTheoriesCondensed}, where high-energy doublon states are perturbatively removed and lead to an antiferromagnetic spin-exchange coupling $J=2t^2/U$ of the Heisenberg type. 
In this regime, we therefore expect to observe slower spin dynamics governed by the $J$ coupling and faster hole dynamics governed by $t$.
\begin{figure}[t]
    \centering
    \includegraphics[width=\textwidth]{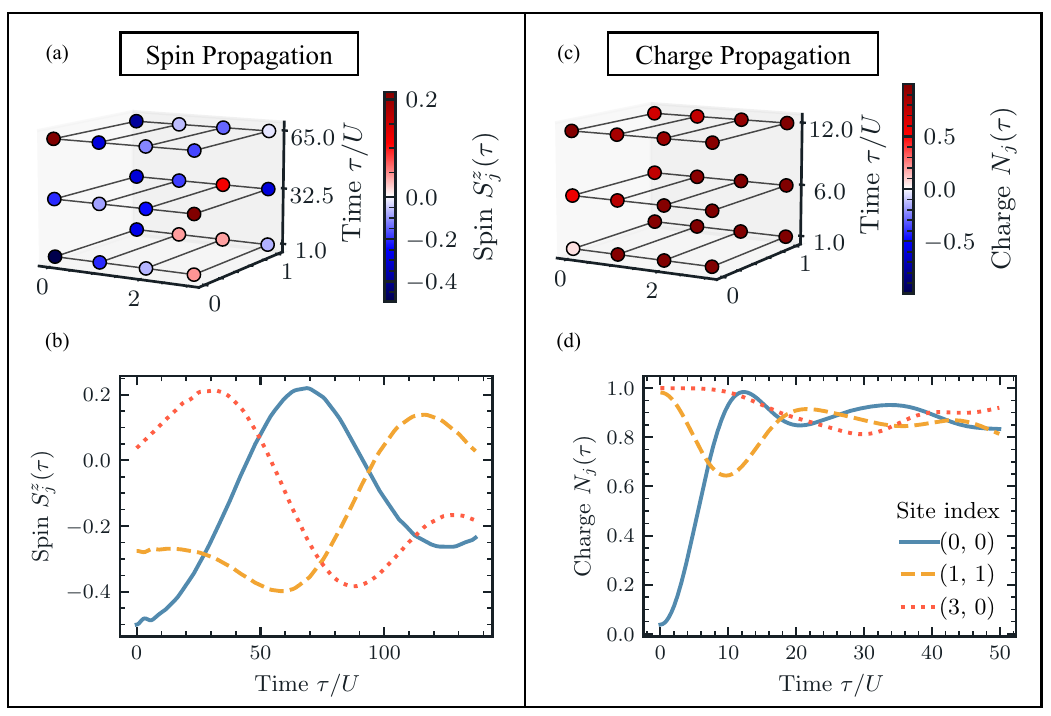}
    \caption{Digital quantum simulation of spin-charge dynamics in the $t-J$ model limit: 
    (a) Propagation of the spin $S^z_j$ profile on each site at different times in the evolution $\tau/U=1, 32.5, 65$. 
    (b) Spin profile for selected sites on the lattice.
    (c) Propagation of the charge $N_j$ profile on each site at different times in the evolution $\tau/U=1, 6, 12$. 
    (d) Charge profile for selected sites on the lattice.
    Notice that different timescales for spin and charge dynamics occur as $t\gg J$.
    }
    \label{fig_excitations}
\end{figure}

The first column of \cref{fig_excitations} shows the evolution of the spin excitation.
By measuring the local spin along the $z$ direction $\spinmat[z]_{\vecsite}(\tau)$, one can monitor in time the deviations from the initial condition, \idest{} the excitation injection:
\begin{align}
    \spinmat[z]_{\vecsite}(\tau)=\frac{1}{2}\qty(\avg{\nop_{\vecsite, \uparrow}}(\tau) - \avg{\nop_{\vecsite, \downarrow}}(\tau)).
\end{align}
This assessment requires only the expectation values of local observables, since
\begin{align}\label{eq_n_to_qubits}
    \avg{\nop_{\vecsite, \uparrow(\downarrow)}} &= \frac{1-\avg{\Z_{\vecsite, u(d)}}}{2}, &
    \rm{where} &&\avg{\Z_{\vecsite, u(d)}}
\end{align}
is the expectation value of the $u(d)$ qubits over the $\Z$ basis.
The second column of \cref{fig_excitations} shows the evolution of the injected charge excitation.
Similarly to the previous case, we measure the local charge $\nop_{\vecsite}(\tau)$: 
\begin{align}
    \nop_{\vecsite}(\tau) = \avg{ \nop_{\vecsite, \uparrow}}(\tau) + \avg{\nop_{\vecsite, \downarrow}}(\tau),
\end{align}
where $\tau$ is the time dependence.
The numerical findings are in agreement with the $t-J$ model description, namely show much faster hole dynamics ($\tau_h \sim 1/t$) and slower spin dynamics ($\tau_s \sim 1/J$). 
This is also quantitatively confirmed for the parameters chosen in these simulations by the corresponding monitored observables. 
Considering the injection site $\vecsite = (0,0)$, the first peak in the charge sector, $N_j(\tau)$, occurs at $\tau_h\approx 12/U = 1.2 /t$, whereas the first peak in the spin sector, $S^z_j(\tau)$, occurs at $\tau_s \approx 65/U = 6.5/t$. 
The corresponding ratio $\tau_s/\tau_h = 5.4$ is close to the ratio $t/J=U/2t = 5$, as expected.
Notice that this analysis only provides evidence of the short timescale propagation of the excitations and is additionally affected by the small system size.
While in 1D spin and charge would manifest as independent degrees of freedom also at longer times due to the well-known spin-charge separation phenomenon, this will not be the case in two-dimensional lattices.
Spin and charge would indeed display strongly-correlated dynamics at longer times, as a consequence of spinon-holon coupling (see, for example, Refs.~\cite{Grusdt2018PartonTheoryMagnetic,Bohrdt2021ExplorationDopedQuantum} and references therein).
 % ==========================================================================================
\section{Summary}
\label{sec_def_conclusion}
We have generalized a technique for local fermion encoding to any 2D lattice configurations eliminating the fermionic degrees of freedom by absorbing them into an auxiliary $\mathbb{Z}_2$ gauge ﬁeld (gauge defermionization). 
We have successfully tested this method against the 2D spin-$\frac{1}{2}$ Hubbard model. 
The ground state properties have been computed for varying particle densities utilizing both tree-tensor network ansatz and exact diagonalization methods. 
Here, we have observed the expected transition from the liquid to the anti-ferromagnetically ordered phase, accessing lattice sizes up to $4\times4$. 
Furthermore, gauge defermionization introduces a fermion-to-qubit mapping with comparable resource requirements to state-of-the-art encodings. 
We have shown that this mapping offers a scalable pathway for the digital quantum simulation of fermionic theories when the available fault-tolerant quantum computers~\cite{Bluvstein2024LogicalQuantumProcessor,daSilva2024DemonstrationLogicalQubits} will be scaled. 
For the defermionized 2D Hubbard Hamiltonian, hopping terms and gauge constraints, here included as stabilizers, result in a maximum Pauli weight of 6. 
We have then demonstrated the feasibility of this approach by simulating the digital out-of-equilibrium dynamics of the defermionized 2D Hubbard Hamiltonian over a $4\times2$ lattice in the antiferromagnetic phase. 
Our protocol entails the adiabatic preparation of the ground state (half-filling, $U/t=10$), the injection of a charge (spin)-excitation in the system, and the time evolution of the perturbed state. 
Finally, we have observed a faster propagation for the charge excitation compared to the spin one as expected from the low-energy description based on the $t-J$ model.

Our collected results show numerical evidence that the local encoding is a scalable and feasible pathway toward the digital quantum simulation of fermionic lattice theories. We have highlighted that is indeed feasible to provide physically relevant results with the current technology of digital quantum processing platforms.

Tensor networks also showed positive results, although specifically the tree tensor network ansatz state manifests some limitations to accommodate the area-law of entanglement introduced by the auxiliary, resonant gauge fields. 
We expect tensor network geometries capable of capturing a wider entanglement distribution, such as ATTN \cite{Felser2021EfficientTensorNetwork} or iPEPS \cite{Jordan2008ClassicalSimulationInfiniteSize}, to yield even better results.
% ==========================================================================================
\paragraph{Code and Data availability}
The code to map the fermionic Hamiltonian to a defermionized one and run the digital quantum simulation is available at \cite{Ballarin2024SimulationScriptsDigital}. 
The engine of the simulations are distributed through the quantum tea leaves \cite{Bacilieri2024QuantumTEAQtealeaves} and quantum matcha tea \cite{Ballarin2024QuantumTEAQmatchatea} python packages of \emph{Quantum TEA}.
% ==========================================================================================

%% file: chapters/conclusions.tex
\chapter{Conclusions and Outlook}
This thesis presents advances in the numerical and theoretical study of Lattice Gauge Theories (LGTs) by exploiting Tensor Network (TN) simulations. 
The use of TN methods aims to overcome longstanding challenges in non-perturbative quantum field theory (QFT), particularly those those arising in traditional Monte Carlo (MC) simulations, like the sign problem \cite{Loh1990SignProblemNumerical}. 
By leveraging TNs and gauge-invariant formulations, we have made substantial progress in addressing LGTs, especially non-Abelian theories, and outlined pathways for exploring them in higher-dimensional systems and real-time dynamics, which remain some of the most challenging tasks in lattice high-energy physics. 

\paragraph{Theoretical advancements}
The main theoretical contribution of this thesis is the dressed-site formalism, introduced in \cref{sec_dressed_site_formalism}. 
This formalism enables the encoding of gauge and matter fields into local, bosonic, and gauge-invariant degrees of freedom, thus simplifying the complexities typically encountered when simulating LGTs on both classical and quantum hardware. 
It has been reviewed for Abelian U(1) LGTs (see \cref{sec_U1_model} and \cite{Felser2020TwoDimensionalQuantumLinkLattice,Magnifico2021LatticeQuantumElectrodynamics,Magnifico2024TensorNetworksLattice}) and successfully extended to the non-Abelian SU(2) Yang-Mills case with dynamical matter (see \cite{Cataldi2024Simulating2+1DSU2} and \cref{sec_SU2_model}), allowing us to explore challenging regimes of these models in two spatial dimensions.

The dressed-site formalism facilitated the development of a more general fermion-to-qubit encoding based on gauge defermionization \cite{Cataldi*2024DigitalQuantumSimulation}, which addresses the issue of fermion statistics in LGTs and other fermionic lattice models, such as those found in condensed matter physics. 
Among other proposals \cite{Nielsen2006QuantumComputationGeometry,Bravyi2002FermionicQuantumComputation,Jiang2019MajoranaLoopStabilizer,Zohar2018EliminatingFermionicMatter,Chen2018ExactBosonizationTwo,Chen2023EquivalenceFermiontoQubitMappings}, this approach efficiently maps fermionic degrees of freedom onto bosonic ones, making them easier to handle in both classical TN methods and quantum simulations. 
We applied this encoding to the two-dimensional Fermi-Hubbard model (\cref{chap_defermionization_lattice_fermion}), demonstrating its effectiveness for both equilibrium and non-equilibrium simulations.

The numerical implementation of the dressed-site scheme and its generalizations is available in the Python library ED-LGT \cite{Cataldi2024Edlgt}, designed for simulating LGT Hamiltonians in arbitrary spatial dimensions. 
It offers scalable precision for gauge field truncation, making it a versatile tool for pre-processing complex theories in future simulations on various platforms, including classical and quantum hardware.

\paragraph{Numerical results}
Within this theoretical scheme, we have provided the first TN simulations of the SU(2) Yang-Mills LGT in two spatial dimensions \cite{Cataldi2024Simulating2+1DSU2}, successfully exploring zero and finite baryon number densities. 
Our results revealed new insights into the ground-state properties of this systems with a rich phase-diagram, including the emergence of a baryon liquid phase in the proximity of the continuum limit location.
We extended the study to non-equilibrium dynamics, where we uncovered Quantum Many-Body Scars (QMBS) in non-Abelian LGTs \cite{Cataldi*2025QuantumManybodyScarring}. 
These exotic states, which weakly violate the Eigenstate Thermalization Hypothesis (ETH) \cite{Deutsch2018EigenstateThermalizationHypothesis}, offer a novel way to study weak violations of ergodic dynamics occurring in LGTs. 
Our findings indicate that the connection between QMBS and gauge invariance persists even in the more complex non-Abelian cases, hinting at possible deeper links between gauge theory and non-thermal states of matter.

\paragraph{Methodological Innovations} 
Aside from the theoretical and numerical achievements, a key aspect of this thesis lies in the methodological improvements made to TN algorithms for high-dimensional LGTs. The first advancement was the development of an optimal encoding for quantum correlations and distances within the TN structure by leveraging space-filling curves, such as the Hilbert ordering (see \cref{sec_hilbert_curve}). This mapping significantly enhanced the precision and efficiency of TN simulations for large-scale high-dimensional quantum many-body (QMB) systems, such as the 2D Ising model near its critical point \cite{Cataldi2021HilbertCurveVs}.

In parallel, we outlined a broader roadmap aimed at enabling large-scale TN simulations of LGT, incorporating innovative strategies for high-performance and parallelized computing \cite{Magnifico2024TensorNetworksLattice} (see \cref{sec_TN_roadmap}). 
These improvements are designed to manage the computational demands of simulating complex gauge groups with high gauge-field truncations and large bond dimensions.

\subsection*{Outlook}
While the accomplishments reported here demonstrate the effectiveness of TN methods for LGTs, they also highlight the remaining challenges in tackling cutting-edge research problems, such as reaching the continuum limit or high-dimensional QCD simulations on large-scale lattices. 
In focusing on algorithmic improvements and numerical strategies, we have presented a roadmap to make TN algorithms competitive in contemporary research problems in high-energy physics and strongly correlated fermionic systems. 

Despite the progress made, several challenges and open questions remain, particularly regarding the finite truncation of gauge fields and the scalability of simulations to larger lattice sizes. 
Although we have addressed these issues with innovative truncation techniques, further improvements are necessary to fully capture the gauge dynamics, especially in high-dimensional systems. 
Additionally, while TN methods have proven successful in simulating intermediate system sizes, expanding these methods to larger scales comparable to MC state-of-the-art will require advancements in both algorithms and computational infrastructure.
To these extents, the results and methods developed in this thesis open up several exciting avenues for future research, some of them representing ongoing projects.

\paragraph{Improved truncation schemes for gauge fields}
One of the key challenges in simulating LGTs is the finite representation of continuous gauge fields, which motivated several truncation strategies, such as Quantum Link Models (QLM) \cite{Horn1981FiniteMatrixModels,Orland1990LatticeGaugeMagnets,Chandrasekharan1997QuantumLinkModels,Brower1999QCDQuantumLink,Tagliacozzo2014TensorNetworksLattice}, finite subgroups \cite{Ercolessi2018PhaseTransitionsGauge,Magnifico2020RealTimeDynamics,Haase2021ResourceEfficientApproach},
digitization of gauge fields \cite{Hackett2019DigitizingGaugeFields}, and fusion-algebra deformation \cite{Zache2023QuantumClassicalSpin}. 
As shown in \cref{sec_U1_basis_dimension}, extending the dressed-site formalism to include larger gauge-representations in the gauge-invariant Hilbert space would bring the model closer to the continuum limit, particularly in the weak coupling regime \cite{Magnifico2024TensorNetworksLattice}. 
However, approaching this region of the couplings would requires a large computational effort in terms of resources (due to the increase of the local Hilbert space) and entanglement scaling (see \cref{fig_QED_convergence}), as the continuum limit corresponds to a quantum critical point of the underlying lattice theory \cite{HernAndez2011LatticeFieldTheory,Eisert2013EntanglementTensorNetwork}. 
Developing new truncation schemes that preserve gauge invariance and simultaneously reduce computational costs is crucial for accurately simulating systems at finer lattice spacings.
Along this direction, there are ongoing efforts on a new algorithm for an optimal and scalable reduction of large dressed-site local Hilbert spaces that could significantly reduce these foundational bottlenecks.

\paragraph{Scaling to large lattice sizes}
While our TN simulations have provided valuable insights into SU(2) Yang-Mills LGTs, accessing larger lattice sizes is necessary to investigate long-range correlations and magnetic effects. 
Advanced TN geometries, such as the augmented Tree Tensor Network (aTTN) \cite{Felser2021EfficientTensorNetwork} or Infinite Projected Entangled Pair States (iPEPS) \cite{Jordan2008ClassicalSimulationInfiniteSize,Orus2019TensorNetworksComplex}, could significantly enhance the capability of TN methods to handle the large entanglement inherent in high-dimensional systems. 
Furthermore, running these simulations on pre-exascale high-performance computing (HPC) platforms \cite{Abdelfattah2016HighperformanceTensorContractions,Gerster2014UnconstrainedTreeTensor,Jouppi2017InDatacenterPerformanceAnalysis,Lu2017HighPerformanceOutofcoreBlock,Shi2016TensorContractionsExtended} would enable larger-scale studies of LGTs. 
Along this direction, we could profit from the numerical optimization schemes developed in \cite{Magnifico2024TensorNetworksLattice} and detailed in \cref{sec_TN_roadmap}.

\paragraph{Applications to quantum chromodynamics (QCD)}
A natural extension of the work presented here is the application of TN methods to full lattice QCD, which involves the SU(3) gauge group.
Progress in simulating SU(2) systems provides a foundation for tackling the more complex SU(3) theory, with the long-term goal of addressing open questions such as confinement \cite{Greensite2010QCDVacuumWave,Kogut1985FurtherEvidenceFirstorder}, and the QCD phase diagram at finite density \cite{Nagata2022FinitedensityLatticeQCD}. 
Despite the inherent complexity of the model, recent strategies have been proposed to attack QCD via quantum hardware \cite{Ciavarella2021TrailheadQuantumSimulation,Brower1999QCDQuantumLink} and TN methods, the latter being already successful in one spatial dimension \cite{Rigobello2023HadronsHamiltonianHardcore,Silvi2019TensorNetworkSimulation}. 

\paragraph{Real-Time Dynamics and Quantum Simulation}
The ability of TN methods to access real-time dynamics in LGTs is particularly promising for simulating out-of-equilibrium phenomena such as scattering processes \cite{Rigobello2021EntanglementGenerationMathrm}. 
These methods, combined with the dressed-site formalism, provide a framework for studying the non-equilibrium behavior of gauge fields and dynamical matter. 
Furthermore, our results serve as benchmarks for experimental quantum simulations \cite{Calajo2024DigitalQuantumSimulation}, which are rapidly evolving with platforms such as ultra-cold atoms and trapped-ion systems. 
Extending our formalism to quantum hardware will enable experimental observation of LGT dynamics.

\paragraph{Quantum Many-Body Scarring and Exotic Dynamics in Gauge Theories}
The discovery of QMBS in non-Abelian LGTs \cite{Cataldi*2025QuantumManybodyScarring} opens a new field of exploration in both fundamental physics and quantum computing. 
Future research could explore the role of gauge symmetry in protecting scarred states and investigate whether such scarring persists in more complex models like lattice QCD. 
Additionally, driving protocols could be developed to enhance scarring behavior in non-Abelian systems, offering new insights into the relationship between gauge theories and non-ergodic quantum phases. 
Our dressed-site scheme reveals further suitable for exploring different gauge invariant sectors and eventual background charges, allowing for numerical studies of other exotic dynamical behaviors such as Disorder Free Localization \cite{Brenes2018ManyBodyLocalizationDynamics} in non-Abelian LGTs, which is currently under study.

\paragraph{Defermionization and Quantum Simulation of Fermionic Theories}
The fermion-to-qubit mapping developed in this thesis \cite{Cataldi*2024DigitalQuantumSimulation} offers a scalable approach to simulating fermionic theories on quantum hardware. 
As fault-tolerant quantum processors become available, this mapping could be applied to more complex fermionic models, such as the Hubbard model in higher dimensions or lattice gauge theories with dynamical fermions. These quantum simulations would provide valuable insights into strongly correlated fermionic systems, which are currently difficult to study using classical methods.

In conclusion, the advancements made in this thesis not only contribute to the theoretical and numerical understanding of LGTs, but also pave the way for future developments in quantum simulations and numerical techniques. 
By extending TN methods and gauge invariant encodings, we are well-positioned to address some of the most pressing challenges in high-energy and condensed matter physics, providing a foundation for future breakthroughs in understanding the fundamental interactions that govern the behavior of quantum systems.